\documentclass[12pt]{article}
\usepackage{amsmath,amsthm,amssymb, mathtools}
\usepackage{booktabs}
\usepackage[table]{xcolor}
\usepackage{color}
\usepackage{enumerate}
\usepackage{tabularx}
\usepackage{footnote}
\usepackage{authblk}
\usepackage{xr-hyper}
\usepackage[nodisplayskipstretch]{setspace}
\usepackage[numbers]{natbib}

\usepackage{graphicx}
\graphicspath{ {./figs/} }

\usepackage{natbib}

\usepackage{url} 

\definecolor{lgray}{gray}{0.65}

\newtheorem{ass}{Assumption}
\newtheorem{prop}{Proposition}

\newcommand{\diag}{\text{diag}}

\newcommand{\convP}{\stackrel{p}{\longrightarrow}}

\newcommand{\eps}{\varepsilon}

\renewcommand{\epsilon}{\varepsilon}

\DeclareMathOperator*{\argmin}{arg\,min}

\newcommand{\blind}{1}

\allowdisplaybreaks

\theoremstyle{plain}
\newtheorem{assumption}{\protect\assumptionname}

\theoremstyle{plain}
\newtheorem{thm}{\protect\theoremname}

\theoremstyle{plain}
\newtheorem{cor}{\protect\corollaryname}

\theoremstyle{plain}
\newtheorem{lem}{\protect\lemmaname}

\theoremstyle{definition}
\newtheorem*{rmk}{\protect\remarkname}

\providecommand{\assumptionname}{Assumption}
\providecommand{\lemmaname}{Lemma}
\providecommand{\theoremname}{Theorem}
\providecommand{\corollaryname}{Corollary}
\providecommand{\remarkname}{Remark}

\newcolumntype{L}{>{\raggedright\arraybackslash}X}%
\newcolumntype{R}{>{\raggedleft\arraybackslash}X}%
\newcolumntype{C}{>{\centering\arraybackslash}X}%
\newcommand{\tabfnsymbol}{\renewcommand*{\thefootnote}{\fnsymbol{footnote}}}
\newcommand{\tabfootnote}[4]{
\multicolumn{#3}{>{\raggedright\arraybackslash}p{#4\textwidth}}{\spacingset{1}\footnotesize \footnotemark[#1]\ #2}
}
\newcommand{\tabfootnotex}[4]{
\multicolumn{#3}{>{\hsize = #4\textwidth \raggedright\arraybackslash}X}{\spacingset{1}\footnotesize \footnotemark[#1]\ #2}
}
\newcommand{\captionx}[2][]{\caption[#1]{\spacingset{1.2}#2}}

\global\long\def\Acal{\mathcal{A}}%
\global\long\def\Gcal{\mathcal{G}}%
\global\long\def\Qcal{\mathcal{Q}}%

\global\long\def\RR{\mathbb{R}}%

\global\long\def\hL{\hat{L}}%
\global\long\def\hF{\hat{F}}%
\global\long\def\hV{\hat{V}}%

\global\long\def\hr{\hat{r}}%

\global\long\def\hbeta{\hat{\beta}}%
\global\long\def\heps{\hat{\varepsilon}}%
\global\long\def\hpi{\hat{\pi}}%
\global\long\def\hDel{\hat{\Delta}}%
\global\long\def\hGamma{\hat{\Gamma}}%
\global\long\def\hsigma{\hat{\sigma}}%
\global\long\def\hOme{\hat{\Omega}}%
\global\long\def\home{\hat{\omega}}%

\global\long\def\tL{\tilde{L}}%
\global\long\def\tF{\tilde{F}}%
\global\long\def\tY{\tilde{Y}}%
\global\long\def\tbeta{\tilde{\beta}}%
\global\long\def\tGamma{\tilde{\Gamma}}%
\global\long\def\ttheta{\tilde{\theta}}%
\global\long\def\tW{\tilde{W}}%
\global\long\def\tDel{\tilde{\Delta}}%
\global\long\def\tdel{\tilde{\delta}}%
\global\long\def\tu{\tilde{u}}%

\global\long\def\bH{\bar{H}}%
\global\long\def\br{\bar{r}}%
\global\long\def\rank{{\rm rank}\,}%
\global\long\def\diag{{\rm diag}\,}%
\global\long\def\trace{{\rm trace}}%
\global\long\def\IC{{\rm IC}}%
\global\long\def\eIC{{\rm eIC}}%
\global\long\def\mse{{\rm mse}}%
\global\long\def\MSE{{\rm MSE}}%
\global\long\def\rmse{{\rm RMSE}}%
\global\long\def\cov{{\rm cov}}%

\global\long\def\dL{\dot{L}}%
\global\long\def\dF{\dot{F}}%
\global\long\def\dW{\dot{W}}%
\global\long\def\dU{\dot{U}}%
\global\long\def\dD{\dot{D}}%
\global\long\def\dV{\dot{V}}%

\global\long\def\jj{\cdot , j}%
\global\long\def\ii{i , \cdot}%

\global\long\def\texti{\text{(i)}}%

\global\long\def\textii{\text{(ii)}}%

\global\long\def\lessp{\underset{P}{\lesssim}}%

\global\long\def\sumi{\sum_{i=1}^{n}}%
\global\long\def\sumj{\sum_{j=1}^{m}}%

\let\olditemize=\itemize \let\endolditemize=\enditemize \renewenvironment{itemize}{\olditemize \itemsep0pt}{\endolditemize}

\begin{document}

\def\spacingset#1{\renewcommand{\baselinestretch}%
{#1}\small\normalsize} \spacingset{1}


\if1\blind
{
  \title{\bf Statistical Inference For Noisy Matrix Completion Incorporating Auxiliary Information\thanks{The authors are alphabetically ordered. Shujie Ma is Professor,
Department of Statistics, University of California-Riverside. E-mail: shujie.ma@ucr.edu. Po-Yao Niu is a PhD student,
Department of Statistics, University of California-Riverside. E-mail: pniu002@ucr.edu. Yichong Zhang is Associate Proefessor, School of Economics, Singapore Management University. E-mail: yczhang@smu.edu.sg. Yinchu Zhu is Assistant Professor, Department of Economics, Brandeis University. E-mail: yinchuzhu@brandeis.edu. Niu and Ma's research was supported by NSF grant DMS-2014221, DMS-2310288 and UCR Academic Senate CoR grant.  Yichong Zhang acknowledges the financial support from the NSFC under grant No. 72133002.}}

\author[$^\dagger$]{Shujie Ma}
\author[$^\dagger$]{Po-Yao Niu }
\author[$^\star$]{Yichong Zhang}
\author[$\ddagger$] {Yinchu Zhu }

\affil[$\dagger$]{Department of Statistics,
	University of California at Riverside}

\affil[$\star$]{School of Economics, Singapore Management University}

\affil[$\ddagger$]{Department of Economics, Brandeis University}

  \maketitle
} \fi

\if0\blind
{
  \bigskip
  \bigskip
  \bigskip
  \begin{center}
    {\LARGE\bf Statistical Inference For Noisy Matrix Completion Incorporating Auxiliary Information}
\end{center}
  \medskip
} \fi

\bigskip
\begin{abstract}
This paper investigates statistical inference for noisy matrix completion in a semi-supervised model when auxiliary covariates are available.  The model consists of two parts. One part is a low-rank matrix induced by unobserved latent factors; the other part models the effects of the observed covariates through a coefficient matrix which is composed of high-dimensional column vectors. We model the observational pattern of the responses through a logistic regression of the covariates, and allow its probability to go to zero as the sample size increases.
We apply an iterative least squares (LS) estimation approach in our considered context. The iterative LS methods in general enjoy a low computational cost, but deriving the statistical properties of the resulting estimators is a  challenging task.  We show that our method only needs a few iterations, and the resulting entry-wise estimators of the low-rank matrix and the coefficient matrix are guaranteed to have asymptotic normal distributions. As a result, individual inference can be conducted for each entry of the unknown matrices. We also propose a simultaneous testing procedure with multiplier bootstrap for the high-dimensional coefficient matrix. This simultaneous inferential tool can help us further investigate the effects of covariates for the prediction of missing entries.
\end{abstract}

\noindent%
\vfill

\newpage
\spacingset{1.9} 
\section{Introduction}
\label{sec:intro}
Advances in modern technology have facilitated us to collect large-scale data that are naturally presented in the form of a matrix with both dimensions increasing vastly. 
 Recovering an intact matrix from partial observations, known as the matrix completion problem, has received considerable attention in different fields. 
Most existing methods for estimating missing entries of a matrix only use information from its partial observations. In many real applications, auxiliary information is often available in addition to the observed entries. For example, in a recommender system that aims to predict ratings of users based on the observed ratings from others, the data often contain additional information such as user demographical profiles, apart from the observed ratings by users. 
Indeed, such auxiliary information can be exploited to enrich the basic model and improve prediction accuracy, especially when only a few entries are observed. Because of the increased availability of auxiliary covariates in real-world datasets, there is a pressing need to develop matrix completion techniques that can make good use of the auxiliary information. As a result, a few computational algorithms have been recently proposed to tackle this problem, see, e.g., \cite{XuJinZhou2013}, \cite{Chiang2015}, \cite{ZhuShenYe2016}, \cite{AlayaKlopp2019} and \cite{JinMaJiang2022}, and see \cite{Ibriga2023} for tensor completion with covariate information.

In this paper, we consider a semi-supervised model for the matrix completion problem with row-feature information, in which a target matrix $\Theta \in \RR^{n\times m}$ can be written as
\begin{equation}
\setlength{\belowdisplayskip}{0pt}
\setlength{\abovedisplayskip}{0pt}
\Theta =X\beta^{\prime }+\Gamma ,  \label{eq: model theta}
\end{equation}
where $X\in \RR^{n\times d}$ is an observable row-feature matrix, $\beta\in \RR^{m\times d}$ is an unknown coefficient matrix for $X$, and $\Gamma \in \RR^{n\times m}$ is an unknown low-rank matrix driven by unobserved latent factors, so that it can be decomposed as $\Gamma =LF^{\prime }$ with $L\in \RR^{n\times r}$ and $F\in \RR^{m\times r}$. As a result, we have $\Theta=X\beta^{\prime }+\Gamma =[X,L][\beta,F]^{\prime }$. Therefore, the target matrix $\Theta $ can be learned from both observed covariates in $X$ and unobserved latent variables in $L$ of the subjects, while $\beta$ and $F$ can be considered unknown coefficients for $X$ and $L$, respectively. For example, in a recommendation system, we use the observed baseline characteristics in $X$ and the unobserved variables in $L$ of users to predict their ratings.
This model was also mentioned in \cite{fithian2018} and \cite{Mao2019}, and they proposed different penalized methods for estimating the parameters. Moreover, \cite{Mao2019} has investigated the convergence rates of their proposed penalized estimators.

Unlike \cite{fithian2018} and \cite{Mao2019} that focus on the estimation of the target matrix, we aim to perform statistical inferences for the unknown matrix $\Theta $ and the high-dimensional coefficient matrix $\beta$ in model \eqref{eq: model theta}. Our goal is to provide an interval estimator associated with a given confidence level rather than a point estimator for each entry in $\Theta $ and to test the significance of the covariates for the prediction of the missing entries. For the matrix completion problem with an incomplete and noise-corrupted data matrix, estimation error bounds in terms of entry-wise, Euclidean and spectral norm losses have been established for the estimators of the unknown low-rank matrix, obtained from various convex and nonconvex optimization algorithms \citep[e.g.][]{Candes2010,koltchinskii2011nuclear,negahbanWainwright2012,ChenChi2019,athey2017matrix}. However, confidence intervals derived directly from such bounds are expected to be too conservative, which is mainly caused by the presence of a nonnegligible bias.

To quantify the uncertainty associated with a parameter estimator, one needs to characterize the (asymptotic) distribution of the estimator. In general, this is a challenging task to accomplish when fitting a high-dimensional statistical model, as it involves non-linear and non-explicit parametric estimation procedures \citep{Javanmard2014}. It can be even more difficult to derive the asymptotic distribution when the data matrix has a large number of missing entries. Thus, the literature on inference for matrix completion is still scarce.
There is some recent development in statistical inference for matrix completion without the observed auxiliary covariates based on either a de-biased strategy or singular value decomposition (SVD) estimation. \cite{Carpentier2018}, \cite{Chen2019} and \cite{XiaYuan2021} proposed de-biased estimators to
construct confidence intervals of the unknown underlying matrix with a low-rank structure. The de-biased estimators are built upon initial estimates that can be obtained from nuclear norm penalization. A sample splitting step is needed in the approach considered in \cite{Carpentier2018} and \cite{XiaYuan2021}. \cite{su2019factor} proposed an iterative SVD method with the missing entries replaced by the SVD estimates from the previous step. Their estimator requires that the number of iterations diverges with the sample size to have asymptotic normality. Under a block structure assumption for the observed entries, \cite{bai2021matrix} and \cite{cahan2022factor} proposed to impute the missing values using the estimated factors and loadings obtained from applying SVD on fully observed sub-matrices. Moreover, \cite{xiong2022large} applied SVD to an adjusted covariance matrix computed from observed data.


Unlike the aforementioned works, we consider an iterative least squares (LS) estimation procedure and provide an inferential analysis for the parameters of model \eqref{eq: model theta} with auxiliary information. The iterative LS method has become a popular approach for matrix completion due to its computational advantages \citep{zhouwilkinson2008, Hastie2015, sunluo2016}. However, the literature on the asymptotic distributions of iterative LS estimators is still scarce.
Our algorithm starts from the initial estimates of $\beta$ and $\Gamma $, which are obtained from ordinary LS regression and SVD of the residual matrix, respectively. Based on these initial estimates, we show that we only need to iterate the LS estimation a finite number of times, and the resulting entry-wise estimators of $\beta$, $\Theta $ and $\Gamma $ are guaranteed to have asymptotic normality. As a result, a pointwise confidence interval and individual inference can be conducted for each entry of the unknown matrices. The iterative LS method enjoys low computational cost compared to the iterative SVD approach \citep{su2019factor}, but the development of its statistical properties is quite challenging.  We show that without including the covariate matrix in the model, our iterative LS estimator of the unknown low-rank matrix $\Gamma $ has the same asymptotic distribution as the iterative SVD estimator proposed in \cite{su2019factor}. Because our method only requires finite iterations of LS estimation,  it is computationally more efficient and much faster than their method which needs to iterate the SVD procedure a diverging number of times. This computational advantage becomes more significant as the data matrices are larger. Moreover, we allow that the observational pattern of the responses depends on the baseline covariates and its probability goes to zero as the sample size increases, whereas the existing works on inference for matrix completion require the observational probability of the responses to be independent of the baseline covariates and/or be bounded below by a constant.

   It is worth noting that each column of the coefficient matrix $\beta$ is a high-dimensional vector when $m$ is large. It is of practical interest to conduct simultaneous inference for these high-dimensional column-vectors in $\beta$, which correspond to the effects of the covariates for the prediction of all missing entries jointly. To achieve this goal, we develop a Gaussian multiplier bootstrap inferential procedure, and provide theoretical justification for our bootstrap-based simultaneous inference in this high-dimensional setting. Gaussian multiplier bootstrap that involves empirical processes is considered a powerful tool for conducting tests in classical statistical problems, and has recently been successfully applied to high-dimensional regression settings \citep{chernozhukov2013gaussian,ChernozhukovChetverikov2017}. Our work is the first to apply this technique to the matrix completion problem with a thorough theoretical investigation. The proposed multiplier bootstrap inferential method can help us identify the important auxiliary covariates for the prediction of all missing entries.

   In model \eqref{eq: model theta}, the rank of matrix $\Gamma $, which is $r$, is unknown \textit{a priori}. We propose a new information criterion ($\eIC$) method for estimating $r$ based on our iterative LS method, and show that the proposed $\eIC$ approach can consistently estimate $r$ with a high probability. This method has better finite sample performance than the commonly used singular-value-based approaches for rank selection in matrix completion, and its advantage becomes more significant when the data have more missing entries.

The rest of this paper is organized as follows. The proposed estimators and the theoretical results are given in Section~\ref{sec: model and estimators} and Section~\ref{sec: theory}. Section~\ref{sec: rank est} provides the information criterion method for rank estimation. The simultaneous inference for $\beta$ is given in Section \ref{sec: bootstrap inference of beta}. Sections \ref{sec: simulation} and \ref{sec: application} provide simulation studies and analysis of the MovieLens 1M dataset using the proposed method, respectively. A conclusion is given in Section \ref{sec: conclusion}. All technical proofs and additional numerical results are provided in the Supplementary Materials.

\noindent {\bf Notations.} Throughout the paper, $\|\cdot\|$ denotes the spectral norm, $\|\cdot\|_{*}$ the nuclear norm, $\|\cdot\|_{F}$ the Frobenius norm, and $\|\cdot\|_\infty$ the maximum absolute value of the entries of a matrix. Let $A\circ B$ be the Hadamard product of two matrices $A,B$ of the same dimensions. Let $n\wedge m$ ($n\vee m$) denote the minimum (maximum) of $n$ and $m$.
For two sequences of positive numbers $a_n$ and $b_n$, $a_n\ll b_n$ means $a_n=o(b_n)$ and $a_n\lesssim b_n$ means that $a_n=O(b_n)$.

\vspace{-0.7cm}
\section{Model and Estimation}\label{sec: model and estimators}
\vspace{-0.5cm}
\subsection{The Model}\label{sec: model}
\vspace{-0.3cm}
We consider the following model:
\begin{equation}
\setlength{\belowdisplayskip}{0pt}
\setlength{\abovedisplayskip}{-0pt}
    Y = \Theta + \varepsilon = X\beta' + \Gamma + \varepsilon, \label{eq: model Y}
\end{equation}
where $Y,\Gamma,\varepsilon\in\RR^{n\times m}$, $X=(X_{1},...,X_{n})'\in \RR^{n\times d}$ in which $X_{i}=(1,\tilde{X}'_i)'$, and $\tilde{X_i}\in\RR^{(d-1)\times 1}$  is the vector of baseline covariates for the $i^{th}$ subject. Moreover, $\beta=(\beta_{1},...,\beta_{m})'\in\RR^{m\times d}$ with $\beta_{j}\in\RR^{d}$, so model (\ref{eq: model Y}) allows the unknown coefficients of the covariates to be different across $j$. We assume that $\Gamma=\{\Gamma _{ij}\}=LF'$ with $L=(L_{1},...,L_{n})'\in\RR^{n\times r}$ and $F=(F_{1},...,F_{m})'\in\RR^{m\times r}$. We let $r$ and $d$ be fixed. To identify $\beta$, we assume that $E(L_i|\tilde{X_i})=0$, and $F_j$ are independent of $\tilde{X_i}$ and $L_i$. We do not observe all entries in $Y=(Y_{i,j})$, so let $\Xi=(\xi_{i,j}) \in\RR^{n\times m}$ with each entry $\xi_{i,j}\in\{0,1\}$ denoting the status of $Y_{i,j}$: $\xi_{i,j}=1$ if and only if $Y_{i,j}$ is observed.

We assume that $P(\xi_{i,j}=1|\tilde{X_i})=\eta (  \gamma_{0,n} +\tilde{X_i}'\gamma_1)=\pi _{i}$, where $\gamma_{0,n} = \log(\alpha_n) + \gamma_0$ and $\eta(\cdot)$ is the logit link function for logistic regression, so the probability of the observed rate depends on the baseline characteristics of each
subject.  We allow $\alpha _{n}\rightarrow 0$ as $n\rightarrow \infty $, so $ \pi _{i}\rightarrow 0$ as $n\rightarrow \infty $. The probability of the observed responses $ \pi _{i}$ can be written as $\pi _{i}=\alpha _{n}e^{(\gamma _{0}+\tilde{X_{i}}'\gamma
_1)}/\{1+e^{(\gamma _{0,n}+\tilde{X_{i}}'\gamma _1)}\}$, so the rate of $\alpha _{n}$ determines how fast $ \pi _{i}$ can go to zero, which will be discussed in Section \ref{sec: theory}.

\vspace{-0.5cm}
\subsection{The Estimation Procedure}
\subsubsection{Initial Estimators} \label{sec: init est}
To obtain an initial estimator of $\beta$, we compute the ordinary LS estimator $\hbeta = (\hbeta_1, \cdots, \hbeta_m)'$ without considering the latent matrix $\Gamma$, so each $\hbeta_j$ is obtained by
\begin{equation}
\setlength{\belowdisplayskip}{0.5pt}
\setlength{\abovedisplayskip}{0.5pt}
\hbeta_{j}=\left(\sum\nolimits_{i=1}^n X_{i}X_{i}'\xi_{i,j}\right)^{-1} \left(\sum\nolimits_{i=1}^n X_{i}Y_{i,j}\xi_{i,j}\right).\label{eq: initial est beta}
\end{equation}
Next, we obtain an SVD estimate of $\Gamma$ as follows. Define
\begin{equation}
\setlength{\belowdisplayskip}{0pt}
\setlength{\abovedisplayskip}{0pt}
 W_{i,j}=\hpi_i^{-1}\xi_{i,j}(Y_{i,j}-X_{i}'\hbeta_{j}).\label{eq: est W}
\end{equation}
where $\hat{\pi}_i = \eta (\hat{\gamma}_{0,n}+\tilde{X_i}' \hat \gamma_1)$ is the estimated observation rate for the $i$-th subject, in which
\begin{equation*}
\setlength{\belowdisplayskip}{0pt}
\setlength{\abovedisplayskip}{0pt}
\hat{\gamma} = (\hat{\gamma}_{0,n},\hat{\gamma}_1) = \arg\min_{r_0,r_1} \sum\nolimits_{i=1}^n \sum\nolimits_{j=1}^{m} Y_{ij}(r_0 + \tilde{X}'_ir_1) - \log(1+\exp(r_0 + \tilde{X}_i'r_1)).
\end{equation*}
We perform SVD on $W$ such that  $W=UDV'=\sum_{s=1}^{m\wedge n}  d_s u_s v_s'$ where $d_s$'s are the singular values in $D$ in decreasing order and $u_s$'s, $v_s$'s are the corresponding left and right singular vectors in $U$ and $V$. Then for a given rank $r$, the SVD estimator of $\Gamma$ is $\hGamma = (\hGamma_{i,j})=\hL \hF = \sum_{s=1}^{r}  d_s u_s v_s'$ where
$\hL= \sqrt{n}\left(u_1, \cdots, u_r\right) \label{eq: initial est L}$ and $\hF = 1/\sqrt{n}\left(d_1\cdot v_1, \cdots, d_r\cdot v_r\right)$.

\vspace{-0.3cm}
\subsubsection{The Iterative LS Estimators}\label{sec: est beta gamma}
The initial estimator $\hGamma=\hL\hF'$ is actually the minimizer of the following function:
\begin{equation}
\setlength{\belowdisplayskip}{0.7pt}
\setlength{\abovedisplayskip}{0.7pt}
    f(\hbeta, L,F)=\left\|LF'-\diag(\hpi)^{-1}(Y-X\hbeta')\circ \Xi\right\|_F^2,
    \end{equation}
where
$\hat \pi = (\hat \pi_1,\cdots,\hat \pi_n)'$, and $\diag(\hpi)$ is an $n \times n$ diagonal matrix with the diagonals being $\hat \pi_1,\cdots,\hat \pi_n$ and the off-diagonal entries equal to zeros.

In the above function, the missing values are treated as zeros and contribute to the residuals while fitting, so the resulting estimates may not be optimal as they ignore the information about the missing positions. To solve this problem, we consider another objective function in which the missing entries do not contribute to the residuals, and propose an updating procedure (algorithm) that iteratively updates the estimates using the estimates given in Section \ref{sec: init est} as the initial values.

We define the following objective function where the missing values do not contribute to the residuals:
\vspace{-0.5cm}
\begin{equation}
\setlength{\belowdisplayskip}{0pt}
\setlength{\abovedisplayskip}{0pt}
    f^*(\beta, L,F)=\left\|\Xi \circ[\,LF'-(Y-X\beta')\,]\, \right\|_F^2 \label{eq: obj function}
    \vspace{-0.3cm}
\end{equation}
Although it is hard to find out the joint minimizers of \eqref{eq: obj function} explicitly, we can easily obtain its minimizer of each $\beta$, $L$ and $F$ if the other two are fixed at their current values by solving an LS problem. Therefore, we can consider the following updating procedure:
\vspace{-0.3cm}
\begin{align}
    \tbeta^{(g)} &= \arg\min\nolimits_\beta f^*\left(\beta, \tL^{(g-1)}, \tF^{(g-1)}\right); \notag\\
    \tF^{(g)} &= \arg\min\nolimits_F f^*\left(\tbeta^{(g)}, \tL^{(g-1)}, F\right); 
    \tL^{(g)} = \arg\min\nolimits_L f^*\left(\tbeta^{(g)}, L,\tF^{(g)}\right) \label{eq: est beta L F at step g}
    \vspace{-0.5cm}
\end{align}
for any given $g \geq 1$, where $g$ is the step index in the iterative algorithm. This algorithm requires initial values for $\tL$ and $\tF$ to start with. An obvious option is to use $\tL^{(0)}=\hL$, and $\tF^{(0)}=\hF$ given in Section \ref{sec: init est}. Then the resulting estimator of $\Gamma$ at the $g^{th}$ step is
\begin{equation}
\setlength{\belowdisplayskip}{0pt}
\setlength{\abovedisplayskip}{0pt}
    \tGamma^{(g)} = \tL^{(g)}\tF^{(g)'}.\label{eq: est gamma at step g}
\end{equation}
We call the corresponding estimators $\tbeta^{(g)}$ and $\tGamma^{(g)}$ the {\bf iterative LS estimators}.

\vspace{-0.7cm}
\section{Asymptotic Theory} \label{sec: theory}
\vspace{-0.3cm}
\subsection{Assumptions}
\vspace{-0.2cm}
We make the following assumptions to investigate the asymptotic theories about the proposed iterative LS estimator.
\vspace{-0.2cm}
\begin{assumption}\label{assu: moment cond}\
\vspace{-0.3cm}
\begin{itemize}
    \item[i)] $Y_{i,j}\perp \xi _{i,j}|(\tilde{X}_i,L_{i},F_{j})$.
    \item[ii)]  Assume $\xi_{i,j} = 1\{\gamma_{0,n} + \tilde{X}_i'\gamma_1 \geq v_{i,j}\}$, where $\{v_{i,j}\}_{1\leq i \leq n,1 \leq j \leq m}$ is a sequence of i.i.d. logistic random variables independent of $(\epsilon, X, L, F)$. Denote $P(\xi_{ij} = 1|\tilde{X}_i) = \eta (  \gamma_{0,n} + \tilde{X}_i'\gamma_1) = \pi_i$, where $\gamma_{0,n} = \log(\alpha_n) + \gamma_0$, $\eta (\cdot)$ is the standard logistic cdf, and $\alpha_n \leq 1$ is some deterministic sequence.
  \item[iii)]  $n^{1/2}\log(m+n)\lesssim m\lesssim n^{2}$ and $(n \vee m)^{\varrho}\ll (n \wedge m) \alpha_n^2$ for an arbitrarily small constant $0<\varrho<1$.
   \item[iv)] There exists a constant $C>0$ such that
    $\sup\limits_{||u|| \leq C}		\left\Vert \frac{1}{n\alpha_n}\sum\limits_{i=1}^n \Lambda_{i}(u)(1-\Lambda_{i}(u))X_iX_i' - H_0\right\Vert =o_P(1),$
    where $\Lambda_{i}(u) = \Lambda_{i}(u_0,u_1) = \eta((\gamma_{0,n} + u_0(nm \alpha_n)^{-1/2})+ \tilde{X}_i' ( \gamma_1+u_1(nm \alpha_n)^{-1/2}) )$, and $H_0 = E \{\exp(\gamma_{0} + \tilde X_i'\gamma_1) X_iX_i'\}$ is positive definite. 		
    \item[v)] Entries of $\tilde{X}_i$ have sub-Gaussian norms bounded by a constant, $\Sigma_{X}=E\left(X_{i}X_{i}'\right)$ and $\alpha_n^{-1} E\left( \pi_i X_{i}X_{i}'\right)$ have eigenvalues bounded away from zero and infinity for $i=1, \cdots, n$.
    \item[vi)] Entries of $L_{i}$, $F_{j}$ and $\varepsilon_{i,j}$ have sub-Gaussian norms bounded by a constant.
    \item[vii)]
    For some constant $c>0$,
    \vspace{-0.5cm}
    \begin{align*}
    &        P\left(\sigma_r(L'L/n)>c\right)\rightarrow 1, \quad
             P\left(\sigma_r(F'F/m)>c\right)\rightarrow 1, \\
    &        P\left(\sigma_r\left(\sumi L_iL_i' \pi_i / (\alpha_n n)\right)>c\right)\rightarrow 1, \quad         P\left(\sigma_r\left(\sumi X_iX_i' \pi_i / (\alpha_n n)\right)>c\right)\rightarrow 1,
    \end{align*}
    where $\sigma_r(A)$ is the $r$-th largest singular value of $A$.
    \item[viii)]  Conditional on $(X,L,F)$, $\varepsilon_{i,j}$ is independent across $(i,j)$ with $E(\varepsilon_{i,j}\mid L,F,X)=0$.  $\{\tilde{X}_i\}_{i=1}^{n}$, $\{L_{i}\}_{i=1}^{n}$ and $\{F_{j}\}_{j=1}^{m}$ are sequences of $
i.i.d.$  random variables, respectively.
\end{itemize}
\end{assumption}
With the above assumptions and the model identification assumption that $E(\Gamma _{ij}|X_{i})=0$, we can first show that $\hbeta$ and $\hGamma$ are consistent estimators of $\beta$ and $\Gamma$ (see Supplement B).
\begin{rmk}[Comments about Assumption \ref{assu: moment cond}]
Assumption (i) assumes that the response $Y_{i,j}$ and the variable for missingness $\xi_{i,j}$ are independent conditional on the observed covariates and the latent variables. Assumption (ii) assumes that the missingness of each response depends on the observed baseline covariates of each individual, and the probability of the missing pattern is modeled through a logistic regression model, which is called the propensity score function \citep{rosenbaumrubin1983}. This assumption is more relaxed and practical than the ``missing uniformly at random" condition imposed in \cite{Carpentier2018}, \cite{Chen2019} and \cite{su2019factor} for statistical inference.  For example, in the MovieLens data in Section \ref{sec: application}, whether users rate a movie or not often depends on their baseline characteristics, including gender, age, etc. It is worth noting that when the baseline characteristics for movies, denoted by $Z_j$ for the $j^{th}$ movie, are observed, it is possible to include both $\tilde{X}_{i}$ and $Z_j$ in the logistic model such that $\xi_{i,j} = 1\{\gamma_{0,n} + \tilde{X}_i'\gamma_1+ Z_j'\gamma_2 \geq v_{i,j}\}$. Our proposed estimation procedure and its statistical properties can be extended to this model. We can also consider a logistic model for missing probabilities by including the entries of a latent low-rank matrix, denoted by $A_{i,j}$, so $\xi_{i,j} = 1\{\gamma_{0,n} + \tilde{X}_i'\gamma_1+ A_{i,j} \geq v_{i,j}\}$. The estimation of the unobserved $A_{i,j}$ in this nonlinear model and the development of the associated statistical properties are nontrivial. We leave the study of extending our method to these two models for future work.

Assumption (iii) provides the order requirement for $n$ and $m$. It is typically assumed to ensure the asymptotic properties of the estimators of $L$ and $F$ in factor models; see \cite{bai2002determining} and \cite{su2019factor}. {\color{black} Moreover, we allow $\alpha_n$ to decay to zero in polynomial order of $n \vee m$. Given the sub-Gaussianity of $\tilde X_i$ in Assumption (v), one has $\pi:=E (\pi_i) =O(\alpha_n)$, so $\alpha_n$ is the observation rate that controls how fast the probability of the observed responses goes to zero. If $n$ and $m$ are of the same order, then the main restriction in Assumption (iii) is that $\alpha_n$ cannot decay to zero faster than $n^{-1/2}$. If only concerning the estimation in matrix completion models, $\alpha_n$ can decay to zero faster than the order given in Assumption (iii). For example, \cite{klopp2014} 
provided a Frobenius norm-based estimation error bound for the nuclear-norm penalized estimators under the condition that the observation rate is $\text{polylog}(n)/n$. To establish the distributional theory and uniform convergence rate of our iterative LS estimators in matrix completion, we require a higher observation rate, as the higher-order terms in our estimator involve the term $O(\alpha_n^{-2})$. To make the higher-order terms negligible so the resulting estimator can have an asymptotic linear expansion, $\alpha_n$ needs to satisfy the condition given in Assumption (iii).}

Assumptions (iv)-(vii) are the moment and distribution conditions on the covariates, latent variables, and error terms. These are typical conditions for convergence rates and asymptotic analysis; see similar assumptions in \cite{bai2002determining}, \cite{Chen2019} and \cite{su2019factor}. {\color{black} Specifically, Assumption (iv) can be directly verified by the uniform law of large numbers. One sufficient condition for $H_0$ being a positive definite matrix is that $X_i$ has compact support and $E X_iX_i'$ is of full rank. This condition is common for sparse logistic regressions; see, for example, \citet[Assumption 3]{graham2020}.} Under the first condition in Assumption (viii), model (\ref{eq: model theta}) is correctly specified for the conditional mean of the responses. The second condition can be relaxed to that $\{\tilde{X}_i\}_{i=1}^{n}$, $\{L_{i}\}_{i=1}^{n}$ and $\{F_{j}\}_{j=1}^{m}$ are sequences of independent random variables. Our theoretical results still hold under this relaxed condition.
\end{rmk}
\vspace{-0.2cm}
The following two theorems provide the asymptotic representations of $\tbeta^{(g)}$, $\tGamma^{(g)}$ and their proofs are left in Supplement B-D.
\vspace{-0.3cm}
\begin{thm}\label{thm: beta est}
    Let Model \eqref{eq: model Y} and Assumption \ref{assu: moment cond} hold and $\kappa_n = (1/n+1/m)\alpha_n^{-3/2}n^{1/q}$ for a constant $q>0$ that can be arbitrarily large. The estimator $\tbeta^{(g)}$ obtained from the updating procedure \eqref{eq: est beta L F at step g} has the following asymptotic representation for any finite $g\geq 1$,
    \begin{equation*}
\setlength{\belowdisplayskip}{0.3pt}
\setlength{\abovedisplayskip}{0.3pt}
\Big\|\tbeta^{(g)} -\beta - n^{-1}\left(\Xi\circ \varepsilon + \diag(\pi)LF^{\prime }\right)^{\prime }X(E\pi _{i}X_{i}X_{i}^{\prime })^{-1}\Big\|_{2,\infty}
    =O_P\left(\kappa_n\right),
    \end{equation*}
   where $\pi = (\pi_1,\cdots,\pi_n)'$, $\diag(\pi)$ is a diagonal matrix with $\pi$ as the diagonal, and $\|\cdot\|_{2,\infty}$ is the maximum row 2-norm of a matrix.
\end{thm}


\begin{thm}\label{thm: Gamma est}
    Let Model \eqref{eq: model Y} and Assumption \ref{assu: moment cond} hold. The estimator $\tGamma^{(g)}$ defined in \eqref{eq: est gamma at step g} has the following asymptotic representation, for any finite $g \geq 1$,
    $ \left\|\tGamma^{(g)}-\Gamma- \Delta
        \right\|_\infty   =  O_{P}\left(\kappa_n \right),$
 where $\Delta$ is a $n \times m$ matrix with its $(i,j)$th entry being
 \vspace{-0.5cm}
\begin{align*}
 \Delta_{i,j} & =  \frac{1}{n} L'_i (EL_iL_i' \pi_i)^{-1} \sum\nolimits_{k=1}^n L_k \xi_{k,j}\varepsilon_{k,j} + \frac{1}{m \pi_i}\sum\nolimits_{t=1}^m F'_t\xi_{i,t}\varepsilon_{i,t} \Sigma^{-1}_F F_j \\
    & -  \frac{1}{n}X'_i E(X_iX_i'\pi_i)^{-1} \sum\nolimits_{k=1}^n \pi_k X_k L_k' F_j\text{,
where }\Sigma _{F}=E\left( F_{j}F_{j}^{\prime }\right).
 \vspace{-0.5cm}
\end{align*}
\end{thm}

\begin{rmk}
 \vspace{-0.3cm}
Point-wise confidence intervals or inference for each component in $\beta_j$ and $\Gamma$ can be constructed based on the asymptotic representations given in Theorems \ref{thm: beta est} and \ref{thm: Gamma est}. In addition, we propose a multiplier bootstrap statistic in Section \ref{sec: bootstrap inference of beta} for conducting simultaneous inference on the high-dimensional matrix $\beta$.

{\color{black}
\begin{rmk}
 \vspace{-0.3cm}
The iterative estimation algorithm starts from the initial ordinary LS
estimator $\hbeta$ given in (\ref{eq: initial est beta}). Under Assumption 1
(i), according to the derivation given in Lemma 1 of the supplement, we can obtain the asymptotic variance of $\hbeta_{j}$,
denoted by $\widetilde{var}\left( \hbeta_{j}\right) $, as
$\widetilde{var}\left( {\normalsize \hbeta_{j}}\right) =n^{-1}(E\pi
_{i}X_{i}X_{i}^{\prime })^{-1}\{var(\xi _{i,j}X_{i}\varepsilon
_{i,j})+var(\xi _{i,j}X_{i}L_{i}^{\prime }F_{j})\}(E\pi
_{i}X_{i}X_{i}^{\prime })^{-1}.$
Moreover, from Theorem \ref{thm: beta est}, we obtain the asymptotic
variance of the iterative estimator $\tbeta_{j}^{(g)}$, for $g\geq 1$, denoted by $\widetilde{var}\left( \tbeta_{j}^{(g)}\right) $, as
$\widetilde{var}\left( {\normalsize \tbeta_{j}^{(g)}}\right) =n^{-1}(E\pi
_{i}X_{i}X_{i}^{\prime })^{-1}\{var(\xi _{i,j}X_{i}\varepsilon
_{i,j})+var(\pi _{i}X_{i}L_{i}^{\prime }F_{j})\}(E\pi _{i}X_{i}X_{i}^{\prime
})^{-1}.$
Given that $E(\xi _{i,j}X_{i}L_{i}^{\prime
}F_{j}|X_{i},L_{i},F_{j})=\pi _{i}X_{i}L_{i}^{\prime }F_{j}$, one has
$var(\xi _{i,j}X_{i}L_{i}^{\prime }F_{j})=E\{\pi _{i}(1-\pi _{i})\Gamma
_{ij}^{2}X_{i}X_{i}^{\prime }\}+var(\pi _{i}X_{i}L_{i}^{\prime }F_{j})$.
Thus,
 \vspace{-0.3cm}
\begin{equation*}
\widetilde{var}\left( {\normalsize \hbeta_{j}}\right) -\widetilde{var}\left(
{\normalsize \tbeta_{j}^{(g)}}\right) =n^{-1}(E\pi _{i}X_{i}X_{i}^{\prime
})^{-1}E\{\pi _{i}(1-\pi _{i})\Gamma _{ij}^{2}X_{i}X_{i}^{\prime }\}(E\pi
_{i}X_{i}X_{i}^{\prime })^{-1}\geq 0.
 \vspace{-0.3cm}
\end{equation*}%
This means that $\tbeta_{j}^{(g)}$ always has a smaller asymptotic variance
than $\hat{\beta}_{j}$. In fact, when the observation rate $\alpha _{n}=o(1)$%
, $var(\pi _{i}X_{i}L_{i}^{\prime }F_{j})$ in  $\widetilde{var}\left( \tbeta_{j}^{(g)}\right) $ is asymptotically negligible
compared to $var(\xi _{i,j}X_{i}L_{i}^{\prime }F_{j})$ in  $\widetilde{var}\left( \hbeta_{j}\right) $, and thus the difference between the asymptotic variances of
the initial and iterative estimators is larger when more observations are missing.
\end{rmk}
}
\begin{rmk}{\color{black}
 \vspace{-0.5cm}
Without the existence of the covariate matrix $X$, model \eqref{eq: model Y} becomes $Y=\Gamma + \varepsilon$. When $\pi_i=\pi$ such that missingness does not depend on covariates as considered in \cite{su2019factor}, our iterative LS estimator $\tGamma^{(g)} - \Gamma$ at any finite $g \geq 1$ has asymptotic representation: $\pi^{-1}\left(n^{-1}L\Sigma_L^{-1} L'(\varepsilon\circ\Xi) +
                  m^{-1}(\varepsilon\circ\Xi) F\Sigma_F^{-1}F'\right)$. Therefore, it achieves the same efficiency as the iterative PCA estimator given in \cite{su2019factor}, and it has been shown in \cite{su2019factor} that the iterative estimator has smaller asymptotic variance than the initial estimator $\hat \Gamma$ obtained from one-step PCA  when $\pi<1$. We also note that to achieve such efficiency improvement, \cite{su2019factor} need the number of iterations to go to infinity, while our iterative LS estimator only requires a few iterations. }
\end{rmk}

\end{rmk}

{\color{black}
\begin{rmk}
 \vspace{-0.5cm}
 Based on the asymptotic linear expansions given in Theorems \ref{thm: beta est} and \ref{thm: Gamma est}, one can immediately obtain the error bounds of our iterative LS estimators in Frobenius norm:
  \vspace{-0.3cm}
  \begin{align}\label{LSrate}
  \resizebox{.9\hsize}{!}{
    $||\tbeta^{g} - \beta||_F^2/m = O_P(((n\wedge m)\alpha_n)^{-1} \log n) \text{; }||\tGamma^{g} - \Gamma||_F^2/(nm) = O_P(((n\wedge m)\alpha_n)^{-1}\log n).$}
\end{align}
Since our iterative LS estimators are asymptotically unbiased, the rate in (\ref{LSrate}) comes from the asymptotic variance.
Under a similar model as ours, \cite{Mao2019} proposed a regularized estimation method penalizing the nuclear and Frobenius norms, and derived the convergence rate of the estimators for  $\beta$ and $\Gamma$. Without incorporating covariates, regularization methods based on different norms have been studied in the matrix completion problems; see, for example, \cite{klopp2014} and \cite{cai2016matrix}.  In general, the regularized estimators have an inherent bias term from the penalties that can go into the convergence rate in addition to the rate from the asymptotic variance. To conduct inference, a debiasing procedure is often needed for the regularized estimation, which is nontrivial in matrix completion problems. Our iterative estimators are asymptotically unbiased and have an asymptotic linear representation based on which we can conduct inference. Moreover, the iterative LS estimation enjoys computational convenience, which is important for modern large-scale data analysis.

\end{rmk}

}

\vspace{-1cm}
\section{\label{sec: rank est}Rank Estimation}
\vspace{-0.5cm}
In practice, $r=\rank(\Gamma)$ is often unknown and needs to be estimated. In this section, we introduce a mean-square-error (MSE)-based approach to estimating the rank. This method fully takes advantage of the proposed iterative LS estimates, and it is described as follows. We compute $\hbeta$ following \eqref{eq: initial est beta}. Recall $W$ defined in \eqref{eq: est W} and its SVD $\sum_{s=1}^{m\wedge n}  d_s u_s v_s'$. We then define $\hL^k\hF^{k'} = \sum_{s=1}^{k}  d_s u_s v_s'$ as the analogues of $\hL, \hF$ in Section \ref{sec: init est} with a superscript $k$ denoting the rank used. Note that the true rank is unknown, and thus $k$ could vary and is not necessarily equal to $r$. We then consider an estimation procedure similar to \eqref{eq: est beta L F at step g} but without updating $\beta$:
\begin{equation}
\setlength{\belowdisplayskip}{0.3pt}
\setlength{\abovedisplayskip}{0.3pt}
    \tF^{k,(g+1)} = \arg\min_F f^*\left(\hbeta, \tL^{k,(g)}, F\right)\text{; }
    \tL^{k,(g+1)} = \arg\min_L f^*\left(\hbeta, L,\tF^{k,(g)}\right), 
\label{eq: est beta L F at step g'}
\end{equation}
where $f^*(\cdot)$ is defined in \eqref{eq: obj function}. The initial value $\tL^{k,(0)}$ is set as $\hL^k$.

Given a fixed positive integer $g$ and for any $k\ll n\wedge m$, we define the following function

   \centerline{ $\mse\left(k, g\right) = \frac{1}{nm}\left\|\Xi\circ\left(Y - X\hbeta - \tGamma^{k,(g)}\right)\right\|_F^2,$}

\noindent where $\tGamma^{k,(g)}= \tL^{k,(g)}\tF^{k,(g)'}$ is the rank $k$ iterative LS estimator of $\Gamma$ at step $g\geq 1$.

We define the MSE-based rank estimating criterion and the resulting estimator of the rank given as follows.
\begin{equation}
\setlength{\belowdisplayskip}{0.5pt}
\setlength{\abovedisplayskip}{0.5pt}
     \eIC(k\,|\,g) = \log \mse\left(k, g\right) + k\cdot h(n,m) \text{, }
    \hr^{\eIC(g)}=\arg\min\nolimits_{1\leq k\leq\br}\ \eIC(k\,|\,g),\label{eq: est r eIC}
\end{equation}
 for $k\geq1$ and a predetermined upper bound $\bar r$,  where $k\cdot h(n,m)$ is a penalty function that depends on $n,m$.
The theorem for the statistical guarantee of $\hr^{\eIC(g)}$ is stated below, and its proof is in Supplement E.
\begin{thm}
\label{thm: r est eIC}
Let Model \eqref{eq: model Y} and Assumption \ref{assu: moment cond} hold. Assume that $\br$ is fixed and satisfies $\br\geq r$. The rank estimator $\hr^{\eIC(g)}$ defined in \eqref{eq: est r eIC} satisfies $P\left(\hr^{\eIC(g)}=r\right)\rightarrow1$ if $h(n,m) = o(1)$ and $\sqrt{\frac{mn \alpha_n}{(m+n) }}h(n,m)\rightarrow\infty$ in a polynomial rate in $n \vee m$.
\end{thm}
{\color{black}
\begin{rmk}
 \vspace{-0.5cm}
Theorem \ref{thm: r est eIC} shows that the MSE-based rank estimator $\hr^{\eIC(g)}$  can consistently estimate the true rank $r$ when  $h(n,m)$ satisfies certain conditions.
Section \ref{sec: sim rank est} provides a formula for calculating $h(n,m)$ in our numerical analysis.  

\end{rmk}
}
\begin{rmk}
 \vspace{-0.5cm}
We have an interesting finding that the MSE-based method for rank selection cannot be constructed based on the initial estimates $\hbeta$, $\hL^k$ and $\hF^k$, where $\hL^k$ and $\hF^k$ are rank $k$ SVD estimates of $L$ and $F$, because the MSE value may not be decreasing as $k$ increases when the observation rate is small. A heuristic argument and the numerical illustration are given in Section G of the Supplementary Materials.
\end{rmk}
\vspace{-1cm}
\section{\label{sec: bootstrap inference of beta}Bootstrap Inference of $\beta$}
\vspace{-0.3cm}
In this section, we provide a testing procedure for the null hypothesis:
\begin{equation}
\setlength{\belowdisplayskip}{0pt}
\setlength{\abovedisplayskip}{0pt}
    H_{0}:\ A_j\beta_j=a_j^0\qquad\forall j\in\Gcal \label{eq: simu test null}
\end{equation}
where each $A_j\neq 0$ is a given matrix with dimension $q \times k$, and $q\leq k$, each $a_j^0$ is a $q$-dimensional vector, and $\Gcal$ is a subset of $\{1,...,m\}$. By Theorem \ref{thm: beta est},

\centerline{$\tbeta^{(g)}_{j}-\beta_{j}=n^{-1}\sumi \omega_{i,j}+\text{smaller terms},$}

\noindent {\color{black} where $\omega_{i,j}=E(\pi_iX_iX_i')^{-1}X_{i}\left( \xi_{i,j}\varepsilon_{i,j}+\pi_i\Gamma_{i,j}\right)$.} Therefore, a simple test statistic is
\begin{equation}
\setlength{\belowdisplayskip}{0.7pt}
\setlength{\abovedisplayskip}{0.7pt}
\label{eq: test stat beta}
    T=\max_{j\in\Gcal}\|A_j\tbeta^{(g)}_j-a_j^0\|_{\infty}.
\end{equation}
We can use a simple multiplier bootstrap procedure to compute the $p$-value. {\color{black} Define

\centerline{$\home_{i,j}=\left(n^{-1}\sumi \hat{\pi_i}X_{i}X_{i}'\right)^{-1}X_{i} \left(\xi_{i,j}\heps_{i,j}+\hat{\pi_i}\tGamma_{i,j}\right),$}

\noindent where $ \heps_{i,j}=Y_{i,j}-X_{i}'\tbeta_{j}-\tGamma_{i,j}$,} in which $\tbeta_{j}$ and $\tGamma_{i,j}$ are the iterative LS estimates of $\beta_{j}$ and $\Gamma_{i,j}$ at the last step.

Let $\{\iota_{i}\}_{i=1}^{n}$ be random variables generated from $N(0,1)$ that are independent of the data. The bootstrapped test statistic is
 \vspace{-0.2cm}
\begin{equation}
\setlength{\belowdisplayskip}{0pt}
\setlength{\abovedisplayskip}{0pt}
\label{eq: bootstrap test stat beta}\textstyle
    T^{*}=\max_{j\in\Gcal}\left\| n^{-1}\sumi \iota_{i}A_j\home_{i,j}\right\|_{\infty}.
\end{equation}
Conditional on the data, the randomness of $T^{*}$ comes from the generated variables $\{\iota_i\}_{i=1}^{n}$. By generating many realizations of $T^{*}$, we can compute the $(1-\alpha)$ quantile of $T^{*}$ conditional on the data, i.e., $\Qcal(T^{*},1-\alpha)$ satisfies $P(T^{*}\leq\Qcal(T^{*},1-\alpha)\mid{\rm data})=1-\alpha$.
\begin{assumption}\label{assu: inference}
Suppose that the following conditions hold:
 \vspace{-0.3cm}
\begin{itemize}
\item[i)] There exists a constant $M_{1}>0$ such that $\min_{i,j}E(\varepsilon_{i,j}^{2}\mid X,L,F)\geq M_{1}$ almost surely.
\item[ii)] $(\log m)^{5/2}\ll\min\{nm^{-1/2},mn^{-1/2}\}$.
\end{itemize}
\end{assumption}

\begin{thm}
\label{thm: beta hypothesis}
Let Model \eqref{eq: model Y} and Assumptions \ref{assu: moment cond} and \ref{assu: inference} hold. Under the null hypothesis \eqref{eq: simu test null}, if $|\Gcal|\leq m$, then
$P\left(T>\Qcal(T^{*},1-\alpha)\right)=\alpha+o(1),$
where $T$ and $T^*$ are defined in \eqref{eq: test stat beta}, \eqref{eq: bootstrap test stat beta}, and $\Qcal$ is the quantile function.
\end{thm}

\vspace{-0.8cm}
\section{Simulation Studies}\label{sec: simulation}
\vspace{-0.2cm}
In this section, we conduct simulation studies to illustrate the finite sample performance of our proposed iterative LS method. We generate the responses by model \eqref{eq: model Y}: $Y=X\beta^{\prime}+\Gamma+\epsilon$,  where $\Gamma = LF'$, in which $L\in\RR^{n\times r}$ and $F\in \RR^{m\times r}$. We then generate the covariates, the coefficients, and the latent matrices as follows. For $i=1,2,\cdots,n$ and $j=1,2,\cdots,m$, we independently generate the covariates by $X_i\sim N(0, \Sigma_X)$, the hidden matrix by $L_i\sim N(0, \Sigma_L)$, $F_j\sim N(0, 4\Sigma_F)$, and the noise by $\varepsilon_{i,j}\sim N(0,1)$. The covariance matrices are $(\Sigma_X)_{k,k'}=\cov(X_{i,k}, X_{i,k'})=0.5^{|k-k'|}$, $(\Sigma_L)_{k,k'}=\cov(L_{i,k}, L_{i,k'})=0.5^{|k-k'|}$ and $(\Sigma_F)_{k,k'}=\cov(F_{j,k}, F_{j,k'})=0.2^{|k-k'|}$.  We generate the coefficients by $\beta_j\sim N(0, 4 I_d)$. We regenerate $(X,L,F,\varepsilon)$ for each simulation replicate while $\beta$ remains fixed. The dimension of $X_i$ and $\beta_j$ is $d=3$  while the rank, $r=3$, is considered for the latent factor matrix.

Next, we generate the observed entries of the responses according to the two data-generating processes (DGPs) with constant and covariate-dependent observation rates, respectively. For each type of DGP, we consider $n=m=200,\,500,\,1000$ to see how the estimators and their asymptotic properties behave in different sample sizes.

\textbf{DGP 1 (Constant observation rate $\pi$).} In this design, we consider constant observed rates and run simulations for $\pi=1,\,0.8,\,0.5,\,0.2$ to see how the observed rate would affect the performance (the data is fully observed when $\pi=1$).

\textbf{DGP 2 (Covaraiate-dependent observation rate $\pi_i$).} In this design, we let the observational rates of the response variables depend on the observed covariates of each individual, so we generate $\pi_i$ from the logistic model: $P(\xi_{i,j}=1|X_{i})=\pi _{i}=\eta(\gamma_{0,n} +  X_i'\gamma_1)$, where  $\gamma_{0,n}=\log (\alpha_n)$ with $\alpha_n=Cn^{-1/2}\log n$ and $\gamma_1=(0.2,...,0.2)'$. We see that $\alpha_n$ controls the sparseness of the observed values of each response, and we allow that  $\alpha _{n}\rightarrow 0$ as $n\rightarrow \infty $, so that $ \pi _{i}\rightarrow 0$ as $n\rightarrow \infty $.  We let $C=1.0,1.5,2.0$. When the $C$ value is larger, it corresponds to larger observation rates of the responses.

When data are generated from DGP 1, we compare the performance of our proposed iterative LS method with that of the iterative PCA method given in \cite{su2019factor}. In \cite{su2019factor}, they assume that the observed rate is a constant $\pi$ which is bounded below by a constant. As a result, their setting only satisfies the condition on the observation rate in DGP1, not the one in DGP2.

Without the presence of the covariates $X$,~\cite{su2019factor} proposed to estimate $\Gamma $ using an iterative PCA method with the missing values of $Y$ replaced by the PCA estimate of $\Gamma$ from the previous step. To make the iterative PCA method in \cite{su2019factor} be accommodated to our model \eqref{eq: model Y}, once we obtain the estimate of $\Gamma$ by PCA, we use the same LS method to obtain the estimate for $\beta$.
To distinguish the estimators from our proposed method and the one from \cite{su2019factor}, we denote our $g^{\text{th}}$ step iterative LS estimator by $\tGamma^{(g)}_{ls}$, and their iterative PCA estimator by $\tGamma^{(g)}_{pca}$. The estimator $\tGamma^{(g)}_{ls}$ is obtained as described in Section \ref{sec: est beta gamma}. To adapt the iterative PCA method given in~\cite{su2019factor} for our model, at the $g^{\text{th}}$ step, $g\geq 1$, we replace the missing values in $W$ by the corresponding values of the estimates obtained from the previous step, and then $\tGamma^{(g)}_{pca}$ is the rank $r$ SVD of the updated $W$. Once the estimate of $\Gamma$ is obtained, the estimate of $\beta$ is obtained by the same LS method. The same initial estimator $\tGamma^{(0)}_{pca}=\hGamma$ is used. In each simulation, we obtain $\tGamma_{pca}^{(g)}$ and $\tGamma_{ls}^{(g)}$ at the steps $g=1,2,3$ and $g\rightarrow \infty $. The estimate at convergence denoted by $\tGamma^{(c)}$ is obtained by iterating the algorithm until convergence, i.e., the maximum difference between the estimates from two consecutive steps, $\|X\tbeta^{(g)} + \tGamma^{(g)} - X\tbeta^{(g-1)} - \tGamma^{(g-1)} \|^2_\infty$, is smaller than the small threshold $10^{-6}$.

The iterative PCA method in~\cite{su2019factor} requires that the number of iterations go to infinity to have the desired convergence rate and the asymptotic distribution of the estimator for $\Gamma $. We will show that our iterative LS estimator for $\Gamma $ only needs a finite number of iterations to achieve the same asymptotic distribution, so our method enjoys great computational advantage, especially in the large dimensional setting. Moreover, we will illustrate the performance of our proposed multiplier bootstrap inferential method for testing the high-dimensional coefficient matrix and the rank estimation methods.

In the following subsections, we show partial simulation results due to the space limit. For the complete numerical results, we refer to Section H of the Supplementary Materials\footnotemark[1].

\vspace{-0.5cm}
    \subsection{Performance of The Estimators}\label{sec: sim performance}
    To evaluate the performance, we repeat the simulation under each setting $500$ times and, for any estimator $\ttheta$ for a parameter $\theta_0$, we calculate the average mean-square-error:
        $\MSE(\ttheta) = \frac{1}{500|\theta_0|}\sum_{s=1}^{500}\left\|\ttheta_s -\theta_{0,s}\right\|^2_F,$
    where $\ttheta_s$, $\theta_{0,s}$ are the estimator and the true parameter in $s^{\rm th}$ repetition, and $|\theta_0|$ is the number of elements in $\theta_0$.

    We first compare the performance of our iterative LS estimator with that of the iterative PCA estimator using DGP 1. Table~\ref{tab: mse dgp1} shows the $\MSE$ of different estimators obtained with the true rank based on the 500 simulation replicates in each setting of DGP 1 for $g=3$ and $g=c$ (at convergence), and $r=3$. Results for other cases are similar, and are provided in the Supplementary Materials\footnotemark[1].

  For larger sample sizes $n,m=500,\,1000$, we see that $\tGamma_{ls}^{(3)}$ has much smaller $\MSE$ than the initial estimator $\hGamma$, and it has the same MSE as $\tGamma_{ls}^{(c)}$ at all values of $\pi$. It indicates that our LS estimate of $\Gamma$ at a finite step performs better than the initiate estimate, and it has a similar performance as the LS estimate at convergence. Moreover,  $\tGamma_{ls}^{(3)}$ and $\tGamma^{(c)}_{pca}$ have similar MSE values, both of which are significantly smaller than the MSE obtained from $\tGamma^{(3)}_{pca}$. The difference between the $\MSE$ values of $\tGamma_{ls}^{(3)}$ and $\tGamma^{(3)}_{pca}$  becomes more dramatic as the observation rate $\pi$ is smaller. This result corroborates our theoretical finding that the proposed iterative LS estimator at a finite step $g\geq 1$ achieves the same convergence rate and asymptotic property as the iterative PCA estimator at $g\rightarrow \infty $. For small sample size $n,m=200$, we can observe the same pattern for $\tGamma^{(3)}_{ls}$ at $\pi=0.5,\,0.8$. The $\MSE$ of $\tGamma^{(3)}_{ls}$ is almost the same as that of $\tGamma^{(c)}_{ls}$ and $\tGamma^{(c)}_{pca}$ at $\pi=0.5,\,0.8$, but it is slightly worse at $\pi=0.2$. However, the $\MSE$ of $\tGamma^{(3)}_{pca}$ is much larger than that of the other three estimates.  This result further shows that the iterative PCA method needs a diverging number of iterations to achieve the desired convergence rate as proven in \cite{su2019factor}. The performance of the estimators of $\beta$ is similar for both methods. Only in the case $n=m=200$ and $\pi=0.2$, $\tbeta^{(3)}_{pca}$ is slightly worse than $\tbeta^{(3)}_{ls}$.

    Next, we compare the computing time of the iterative LS and the iterative PCA methods.  When missing values exist, our proposed iterative LS\ method has a great computational advantage over the iterative PCA method in two aspects. First, for one complete iteration, the computational complexity of PCA on the updated matrix $W$ is $O(mn^{2}+m^{3})$, and it is only $O(r^{2}\pi mn)$ for solving the two LS systems defined in \eqref{eq: est beta L F at step g} for $L$ and $F$. Since we have the low-rank assumption, $r$ is fixed and $r\ll\min (m,n)$, we see that our LS method is much more computationally efficient than the PCA method for one complete update. Second, our estimator only needs a finite number of iterations, while the iterative PCA estimator requires a diverging number of iterations to have the same asymptotic properties. This result was already demonstrated by the performance comparison in Table~\ref{tab: mse dgp1}.

    To test the actual computing time of the two estimators, we run simulations using the data generated from DGP 1 when the true rank $r=3$, and the sample sizes, $n=m= 200,\, 500,\, 1000$, with observation rate $\pi =0.8,\, 0.5,\, 0.2$, respectively. Based on $100$ simulation replications of each setting, Table \ref{tab: cputime dgp1} reports the average computing time and the number of iterations needed to obtain the converged estimate for each method, and the average computing time of one iteration (one complete update). For a fair comparison, all simulations are run on a regular laptop with specs: Intel(R) Core(TM) i7-8750H CPU, 2667MHz 16 GB RAM without the help of GPU or CPU parallel computing. 

    The last two columns in Table \ref{tab: cputime dgp1} show the average time for one update by both methods at different sample sizes $n,m$. We see that the iterative PCA method has a more dramatic increase (from 0.04 for $n,m=200$ to 3 seconds for $n,m=1000$) than our proposed iterative LS method (from 0.04 to 0.3 instead) when the sample size increases. We can also see that the number of iterations needed to converge increases as the observation rate $\pi$ decreases. From the ``Number of iterations" columns, we observe that the iterative PCA method in general needs more iterations to converge, and the difference between the PCA and the LS methods becomes more prominent as the $\pi$ value becomes smaller even in the settings with large sample sizes. For instance, in the case with $n,m=1000$, the iterative LS method needs around $3$ iterations at $\pi=0.8$ and $5$ iterations at $\pi=0.2$, whereas the number of iterations for the iterative PCA method grows from $7$ to $49$.

    Next, we show in Table \ref{tab: mse dgp2} the $\MSE$ of our iterative LS estimators based on the 500 simulation replicates in each setting of DGP 2 for $g=3$ and $r=3$. We can observe similar patterns as shown in DGP 1; the estimators at $g=3$ have almost the same $\MSE$ as the converged estimators in every case when $n,m=500$ or $1000$. Even for $C=1$, $n,m=200$, the estimator at $g=3$ performs quite well. When the $C$ value is larger, the response matrix is more densely observed, so the estimators are expected to have better performance. The last column shows the average number of iterations to obtain the converged estimator, and we can see that the numbers are all small. With the low computational complexity, it is possible to use the converged solution in practice, or use the estimate at $g=3$ if the algorithm is implemented on large datasets and we need a faster computational speed.


    In the last of this section, we construct pointwise confidence intervals for $\Gamma_{i,j}$ and $\mu_{i,j}=E(Y_{i,j}\,|\,L_i, X_i, F_j)$ for some given $i,j$ based on the asymptotic representations in Theorem \ref{thm: beta est} and Theorem \ref{thm: Gamma est}. For $g\geq 1$, let $\tY_{i,j}=X'_i\tbeta^{(g)}_j + \tGamma^{(g)}_{i,j}$ and $\sigma^2=E(\varepsilon^2_{i,j})$, then $(\tY_{i,j} - \mu_{i,j})/\sigma_{n,m}(\tY_{i,j})$ and $(\tGamma^{(g)}_{i,j} - \Gamma_{i,j})/\sigma_{n,m}(\tGamma^{g)}_{i,j})$ asymptotically follow $N(0,1)$, where{\color{black}
     \vspace{-0.3cm}
    \begin{align}
        \sigma^2_{n,m}(\tGamma^{(g)}_{i,j})&=\sigma^2\left[\,
        n^{-1} L'_iE(\pi_iL_iL_i')^{-1}L_i +  (m\pi_i)^{-1} F_j' \Sigma_F^{-1}F_j
        \,\right] + n^{-1}\zeta_{i,j}^2 \notag\\
        \sigma^2_{n,m}(\tY_{i,j})&=\sigma^2 \left[\,
        n^{-1} \left(L'_iE(\pi_iL_iL_i')^{-1}L_i + X'_iE(\pi_iX_iX_i')^{-1}X_i\right) +  (m\pi_i)^{-1} F_j' \Sigma_F^{-1}F_j\right] \label{asympvar},
             \vspace{-0.5cm}
    \end{align}}
    and $\zeta_{i,j}^2= X'_iE(\pi_iX_iX_i')^{-1}E(\pi_k^2X_k L'_k F_j F_j' L_k X'_k\,|\, F_j)E(\pi_iX_iX_i')^{-1}X_i$. The unknown terms in the above representations can be replaced by empirical estimators so that we can get the estimated standard error $\hsigma_{n,m}(\cdot)$.

    In the literature, the latent factor model is often considered for matrix completion without considering the covariates. In this model, the matrix $\Theta$ is directly decomposed as $\Theta=L^*F^{*\prime}$, where $L^*$ and $F^*$ are latent factors and their loadings, both of which are unknown and need to be estimated.     Compared to model (\ref{eq: model theta}), we can write $L^*=[X,L]$ and $F^*=[\beta,F]$, but both $X$ and $\beta$ are treated as unknown variables in the latent factor model. When  $E(X_iL_i')=0$ and $E(\beta_jF_j')=0$, it can be shown that the asymptotic variance of the estimator $\tY_{i,j}:=\tilde{\Theta}_{i,j}$ based on the latent factor model with $\Theta$ having rank $r+d$ is
$\sigma^2_{n,m}(\tY_{i,j})=\sigma^2 \left[\,
        n^{-1} \left(L'_iE(\pi_iL_iL_i')^{-1}L_i + X'_iE(\pi_iX_iX_i')^{-1}X_i\right) +  (m\pi_i)^{-1} (F_j' \Sigma_F^{-1}F_j +\beta_j' \Sigma_\beta^{-1}\beta_j)\right],$ where \newline $\Sigma_\beta=E\left(\beta_{j}\beta_{j}'\right)$, and $\tilde{\Theta}$ is the iterative LS estimator of $\Theta$  and has the rank of $d+r$.  We can see that this asymptotic variance has one additional term  $(m\pi_i)^{-1}\beta_j' \Sigma_\beta^{-1}\beta_j$ compared to the one given in (\ref{asympvar}), so it is larger than the asymptotic variance of the estimator for model  (\ref{eq: model theta}) that incorporates the observed covariates. When the estimator $\tilde{\Theta}$ has a rank smaller than $d+r$, it has an asymptotically nonnegligible bias.

\begin{figure}[ht]
        \tabfnsymbol \centering
        \includegraphics[width=.9\linewidth]{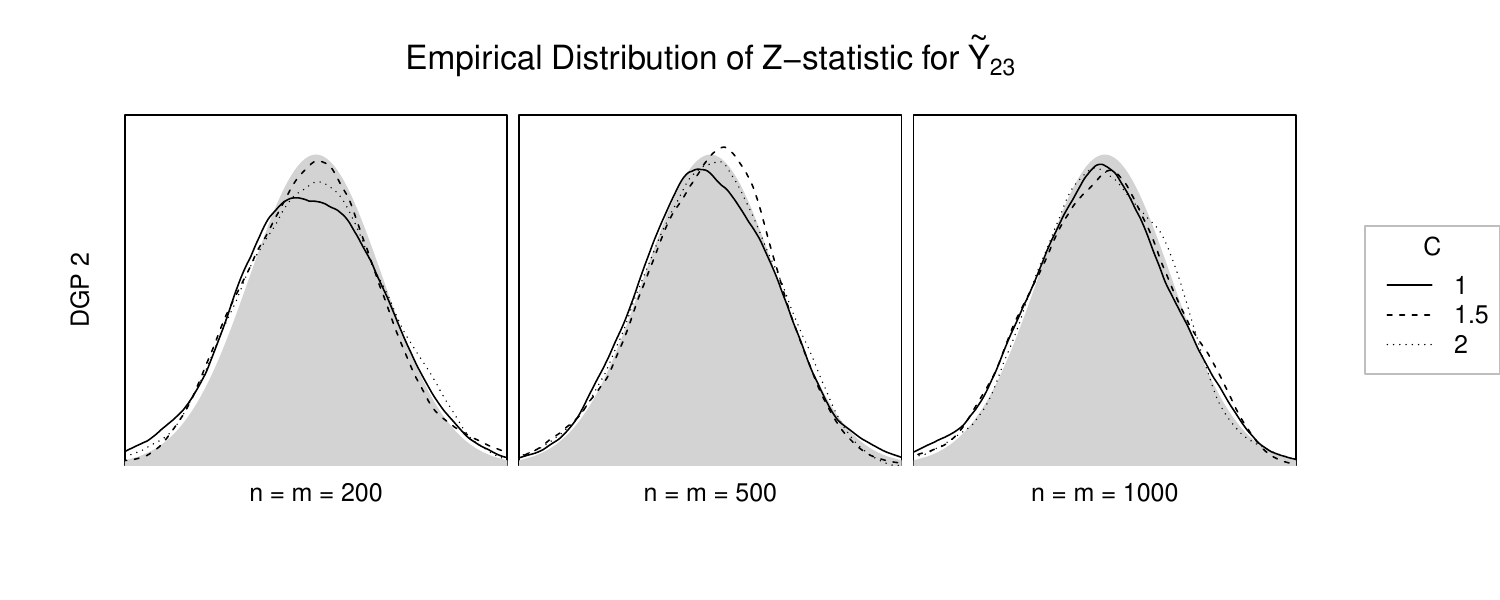}\vspace{-20pt}
        \captionx[Simulation - empirical distribution]{The empirical distribution of $\frac{\tY_{23}-\mu_{23}}{\hsigma_{n,m}(\tY_{23})}$ in different simulation settings. The shaded area is the density of standard normal distribution.}
        \label{fig: asymp dist ey}
    \end{figure}

    Table \ref{tab: bias and coverage dgp2} shows the biases and the empirical coverage rates of 95\% CI of three arbitrarily chosen estimators obtained at $g=3$. Each value is calculated based on 500 simulation replicates. We observe that all the biases are very small, and the coverage rates are close to the nominal value except for the cases with a very small effective size ($n=m=200,\ C=1$). The empirical distributions of the $Z$-statistics of $\tY_{i,j}$ for $(i,j)=(2,3)$ are shown in Figure \ref{fig: asymp dist ey}. We can see that the distributions are close to the standard normal (shaded area) in those cases with larger effective sample sizes. Results for settings in DGP 1 are similar and provided in the Supplementary Materials\footnotemark[1].

    \renewcommand{\arraystretch}{0.9}
    \begin{table}[ht]
    \tabfnsymbol \centering
        \captionx[Simulation - performance comparison in MSE]{The $\MSE$ of different estimators in DGP 1.}
        \label{tab: mse dgp1}
        \vspace{12pt}{\footnotesize
        \begin{tabularx}{\textwidth}{lLRRRRRRRRRR}\toprule
            \multicolumn{2}{c}{DGP 1} & \multicolumn{2}{c}{initial} & \multicolumn{4}{c}{iterative PCA} & \multicolumn{4}{c}{iterative LS}\\
            \cmidrule(lr){1-2}\cmidrule(lr){3-4}\cmidrule(lr){5-8}\cmidrule(lr){9-12}
 $n,m $  &  $\pi$  &  $\hbeta$  &  $\hGamma$  &  $\tbeta_{pca}^{(3)}$  &  $\tbeta_{pca}^{(c)}$  &  $\tGamma_{pca}^{(3)}$  &  $\tGamma_{pca}^{(c)}$  &  $\tbeta_{ls}^{(3)}$  &  $\tbeta_{ls}^{(c)}$  &  $\tGamma_{ls}^{(3)}$  &  $\tGamma_{ls}^{(c)}$\\
\hline
 $200 $  &  $0.2$  &  $ 0.614$  &  $ 7.469$  &  $ 0.258$  &  $ 0.157$  &  $ 3.075$  &  $ 0.419$  &  $ 0.197$  &  $ 0.176$  &  $ 0.631$  &  $ 0.457$\\
 $    $  &  $0.5$  &  $ 0.230$  &  $ 1.121$  &  $ 0.126$  &  $ 0.125$  &  $ 0.321$  &  $ 0.283$  &  $ 0.125$  &  $ 0.125$  &  $ 0.285$  &  $ 0.285$\\
 $    $  &  $0.8$  &  $ 0.145$  &  $ 0.430$  &  $ 0.119$  &  $ 0.119$  &  $ 0.258$  &  $ 0.258$  &  $ 0.119$  &  $ 0.119$  &  $ 0.258$  &  $ 0.258$\\
 $    $  &  $1  $  &  $ 0.117$  &  $ 0.250$  &  $     -$  &  $     -$  &  $     -$  &  $     -$  &  $     -$  &  $     -$  &  $     -$  &  $     -$\\
\hline
 $500 $  &  $0.2$  &  $ 0.224$  &  $ 2.354$  &  $ 0.068$  &  $ 0.057$  &  $ 0.835$  &  $ 0.150$  &  $ 0.058$  &  $ 0.058$  &  $ 0.152$  &  $ 0.152$\\
 $    $  &  $0.5$  &  $ 0.087$  &  $ 0.399$  &  $ 0.047$  &  $ 0.047$  &  $ 0.116$  &  $ 0.108$  &  $ 0.047$  &  $ 0.047$  &  $ 0.108$  &  $ 0.108$\\
 $    $  &  $0.8$  &  $ 0.054$  &  $ 0.167$  &  $ 0.044$  &  $ 0.044$  &  $ 0.099$  &  $ 0.099$  &  $ 0.044$  &  $ 0.044$  &  $ 0.099$  &  $ 0.099$\\
 $    $  &  $1  $  &  $ 0.043$  &  $ 0.096$  &  $     -$  &  $     -$  &  $     -$  &  $     -$  &  $     -$  &  $     -$  &  $     -$  &  $     -$\\
\hline
 $1000$  &  $0.2$  &  $ 0.110$  &  $ 0.726$  &  $ 0.030$  &  $ 0.029$  &  $ 0.246$  &  $ 0.074$  &  $ 0.029$  &  $ 0.029$  &  $ 0.074$  &  $ 0.074$\\
 $    $  &  $0.5$  &  $ 0.044$  &  $ 0.195$  &  $ 0.024$  &  $ 0.024$  &  $ 0.057$  &  $ 0.055$  &  $ 0.024$  &  $ 0.024$  &  $ 0.055$  &  $ 0.055$\\
 $    $  &  $0.8$  &  $ 0.028$  &  $ 0.084$  &  $ 0.023$  &  $ 0.023$  &  $ 0.050$  &  $ 0.050$  &  $ 0.023$  &  $ 0.023$  &  $ 0.050$  &  $ 0.050$\\
 $    $  &  $1  $  &  $ 0.023$  &  $ 0.048$  &  $     -$  &  $     -$  &  $     -$  &  $     -$  &  $     -$  &  $     -$  &  $     -$  &  $     -$
\\
        \bottomrule \\[-7pt]
        \end{tabularx}
        }
    \end{table}
\vspace{-0.5cm}
    \renewcommand{\arraystretch}{0.9}
    \begin{table}[ht]
    \tabfnsymbol \centering
        \captionx[Simulation - computing time]{Computing time in seconds\footnotemark[1] in each setting.} \label{tab: cputime dgp1}
        \vspace{12pt}{\footnotesize
        \begin{tabularx}{\textwidth}{llRRRRRRR}\toprule
            \multicolumn{2}{c}{} & \multicolumn{3}{c}{Time in sec} & \multicolumn{2}{c}{Number} & \multicolumn{2}{c}{Ave. time} \\[-3pt]
            \multicolumn{2}{c}{DGP 1} & \multicolumn{3}{c}{to get estimators} & \multicolumn{2}{c}{of iterations} & \multicolumn{2}{c}{for 1 iteration}\\
            \cmidrule(lr){1-2}\cmidrule(lr){3-5}\cmidrule(lr){6-7}\cmidrule(lr){8-9}
$n,m $  &  $\pi$  &  $\hGamma$  &  $\tGamma^{(c)}_{ls}$  &  $\tGamma^{(c)}_{pca}$  &  $\tGamma^{(c)}_{ls}$  &  $\tGamma^{(c)}_{pca}$  &  $     ls$  &  $    pca$\\
\cline{1-9}  
$200 $  &  $0.8$  &  $   0.04$  &  $   0.16$  &  $   0.38$  &  $    4.0$  &  $    9.3$  &  $  0.041$  &  $  0.041$\\
$    $  &  $0.4$  &  $   0.05$  &  $   0.28$  &  $   1.37$  &  $    7.0$  &  $   33.6$  &  $  0.040$  &  $  0.041$\\
$    $  &  $0.2$  &  $   0.04$  &  $   0.60$  &  $   4.01$  &  $   15.7$  &  $   99.9$  &  $  0.038$  &  $  0.040$\\
\cline{1-9}  
$500 $  &  $0.8$  &  $   0.41$  &  $   0.44$  &  $   3.18$  &  $    3.6$  &  $    7.5$  &  $  0.124$  &  $  0.423$\\
$    $  &  $0.4$  &  $   0.42$  &  $   0.58$  &  $  10.06$  &  $    5.0$  &  $   23.5$  &  $  0.115$  &  $  0.429$\\
$    $  &  $0.2$  &  $   0.41$  &  $   0.80$  &  $  27.54$  &  $    7.4$  &  $   64.5$  &  $  0.108$  &  $  0.427$\\
\cline{1-9}  
$1000$  &  $0.8$  &  $   3.43$  &  $   1.05$  &  $  24.24$  &  $    3.0$  &  $    6.9$  &  $  0.351$  &  $  3.514$\\
$    $  &  $0.4$  &  $   3.03$  &  $   1.16$  &  $  62.22$  &  $    4.0$  &  $   19.9$  &  $  0.287$  &  $  3.130$\\
$    $  &  $0.2$  &  $   2.96$  &  $   1.39$  &  $ 149.12$  &  $    5.5$  &  $   48.8$  &  $  0.252$  &  $  3.056$
\\
            \bottomrule \\[-7pt]
            \tabfootnote{1}{The values are calculated based on 100 simulation replicates}{9}{.95}
        \end{tabularx}%
        }
    \end{table}

    \renewcommand{\arraystretch}{0.9}
    \begin{table}[ht]
    \tabfnsymbol \centering
        \captionx[Simulation - performance comparison in MSE]{The $\MSE$ of different estimators in DGP 2.}
        \label{tab: mse dgp2}
        \vspace{12pt}{\footnotesize
        \begin{tabularx}{\textwidth}{lLRRRRRRR}\toprule
            \multicolumn{2}{c}{DGP 2} & \multicolumn{2}{c}{initial} & \multicolumn{5}{c}{iterative LS}\\
            \cmidrule(lr){1-2}\cmidrule(lr){3-4}\cmidrule(lr){5-9}
$n,m $  &  $C  $  &  $\hbeta$  &  $\hGamma$  &  $\tbeta_{ls}^{(3)}$  &  $\tbeta_{ls}^{(c)}$  &  $\tGamma_{ls}^{(3)}$  &  $\tGamma_{ls}^{(c)}$  &  $N^{(c)}_{ls}$\footnotemark[2]\\
\cline{1-9}
$200 $  &  $1  $  &  $ 0.420$  &  $ 5.993$  &  $ 0.183$  &  $ 0.178$  &  $ 0.516$  &  $ 0.456$  &  $ 12.7$\\
$    $  &  $1.5$  &  $ 0.318$  &  $ 3.612$  &  $ 0.155$  &  $ 0.154$  &  $ 0.379$  &  $ 0.375$  &  $  9.7$\\
$    $  &  $2  $  &  $ 0.268$  &  $ 2.315$  &  $ 0.142$  &  $ 0.141$  &  $ 0.337$  &  $ 0.334$  &  $  8.4$\\
\cline{1-9}
$500 $  &  $1  $  &  $ 0.192$  &  $ 3.271$  &  $ 0.067$  &  $ 0.066$  &  $ 0.185$  &  $ 0.182$  &  $  8.9$\\
$    $  &  $1.5$  &  $ 0.143$  &  $ 1.472$  &  $ 0.057$  &  $ 0.057$  &  $ 0.147$  &  $ 0.147$  &  $  7.2$\\
$    $  &  $2  $  &  $ 0.119$  &  $ 0.888$  &  $ 0.053$  &  $ 0.053$  &  $ 0.132$  &  $ 0.132$  &  $  6.3$\\
\cline{1-9}
$1000$  &  $1  $  &  $ 0.115$  &  $ 1.535$  &  $ 0.034$  &  $ 0.034$  &  $ 0.094$  &  $ 0.094$  &  $  7.3$\\
$    $  &  $1.5$  &  $ 0.085$  &  $ 0.721$  &  $ 0.030$  &  $ 0.030$  &  $ 0.077$  &  $ 0.077$  &  $  6.1$\\
$    $  &  $2  $  &  $ 0.070$  &  $ 0.482$  &  $ 0.028$  &  $ 0.028$  &  $ 0.071$  &  $ 0.071$  &  $  5.6$
\\
        \bottomrule \\[-7pt]
        \tabfootnote{1}{The average number of complete iterations to get converged results.}{9}{.95}
        \end{tabularx}
        }
    \end{table}

    \begin{table}[ht]
    	\tabfnsymbol \centering
    	\captionx[Simulation - bias and CI coverage rate]{Average bias and 95\% CI coverage rate for some estimators\footnotemark[1].}\label{tab: bias and coverage dgp2}
    	\vspace{12pt}{\footnotesize
    	\begin{tabularx}{\textwidth}{llRRRRRRRRRRRR}\toprule
    		\multicolumn{2}{c}{DGP 2}  & \multicolumn{6}{c}{Bias ($10^{-2}$)} & \multicolumn{6}{c}{95\% CI coverage rate} \\
    		\cmidrule(lr){1-2}\cmidrule(lr){3-8}\cmidrule(lr){9-14}
$n,m $  &  $C $  &  $\tGamma_{11}$  &  $\tGamma_{23}$  &  $\tGamma_{35}$  &  $\tY_{11}$  &  $\tY_{23}$  &  $\tY_{35}$  &  $\tGamma_{11}$  &  $\tGamma_{23}$  &  $\tGamma_{35}$  &  $\tY_{11}$  &  $\tY_{23}$  &  $\tY_{35}$\\
\cline{1-14}
$200 $  &  $1  $  &  $   0.3$  &  $  -1.9$  &  $  -1.9$  &  $  -0.5$  &  $  -2.6$  &  $  -1.7$  &  $  0.93$  &  $  0.90$  &  $  0.92$  &  $  0.92$  &  $  0.92$  &  $  0.92$\\
$    $  &  $1.5$  &  $  -1.0$  &  $  -0.1$  &  $  -1.1$  &  $  -0.5$  &  $  -0.4$  &  $   0.5$  &  $  0.92$  &  $  0.92$  &  $  0.94$  &  $  0.95$  &  $  0.95$  &  $  0.96$\\
$    $  &  $2  $  &  $  -0.1$  &  $  -0.0$  &  $  -2.0$  &  $  -0.0$  &  $   0.2$  &  $  -0.4$  &  $  0.92$  &  $  0.92$  &  $  0.94$  &  $  0.93$  &  $  0.95$  &  $  0.95$\\
\cline{1-14}
$500 $  &  $1  $  &  $   0.7$  &  $  -0.4$  &  $  -0.4$  &  $   0.3$  &  $  -0.3$  &  $   2.0$  &  $  0.92$  &  $  0.93$  &  $  0.93$  &  $  0.94$  &  $  0.95$  &  $  0.93$\\
$    $  &  $1.5$  &  $   0.5$  &  $  -0.4$  &  $   0.6$  &  $   0.2$  &  $  -0.4$  &  $   1.9$  &  $  0.94$  &  $  0.94$  &  $  0.93$  &  $  0.95$  &  $  0.96$  &  $  0.95$\\
$    $  &  $2  $  &  $   0.1$  &  $  -0.8$  &  $   0.5$  &  $  -0.1$  &  $  -0.4$  &  $   1.4$  &  $  0.95$  &  $  0.95$  &  $  0.93$  &  $  0.95$  &  $  0.96$  &  $  0.95$\\
\cline{1-14}
$1000$  &  $1  $  &  $   0.3$  &  $  -2.1$  &  $   1.3$  &  $  -0.9$  &  $  -1.0$  &  $  -1.1$  &  $  0.94$  &  $  0.92$  &  $  0.94$  &  $  0.94$  &  $  0.94$  &  $  0.96$\\
$    $  &  $1.5$  &  $   0.7$  &  $  -2.4$  &  $   1.8$  &  $  -0.3$  &  $  -0.7$  &  $   0.0$  &  $  0.95$  &  $  0.92$  &  $  0.96$  &  $  0.93$  &  $  0.96$  &  $  0.95$\\
$    $  &  $2  $  &  $   0.9$  &  $  -1.9$  &  $   1.3$  &  $   0.0$  &  $  -0.4$  &  $   0.0$  &  $  0.96$  &  $  0.93$  &  $  0.95$  &  $  0.93$  &  $  0.95$  &  $  0.95$
\\
    	\bottomrule \\[-7pt]
    	\tabfootnote{1}{$\tGamma_{i,j} = \tGamma^{(3)}_{i,j}$ and $\tY_{i,j}=X'_i\tbeta^{(3)}_j + \tGamma^{(3)}_{i,j}$.}{14}{.9}
    	\end{tabularx}
    	}
    \end{table}

\vspace{-1cm}
    \subsection{Rank Estimation}\label{sec: sim rank est}
    {\color{black}
   The number of factors $r$ is often unknown in practice. Section \ref{sec: rank est} introduces the estimator $\hr^{\eIC}$ to estimate the unknown rank $r$. With large $n,m$, Theorem \ref{thm: r est eIC} shows that this estimator can find the correct rank with a high probability if the penalty $h(n,m)$ satisfies the stated condition. We let
   \vspace{-0.5cm}
    \begin{equation}
 \setlength{\belowdisplayskip}{0pt}
 \setlength{\abovedisplayskip}{0pt}
        h(n,m) = C_hn^{\delta_h} \sqrt{(m+n)/(mn\hat\alpha_n)} , \label{eq: pen h}
           \vspace{-0.3cm}
    \end{equation}
    with $C_h=0.9$ and $\delta_h=0.1$, where $\hat\alpha_n=\overline \pi = n^{-1}\sumi \hat \pi_i$ for DGP1 and $\hat\alpha_n=e^{\hat\gamma_{0,n}}$ for DGP2, so that $h(n,m)$ satisfies the condition given in Theorem \ref{thm: r est eIC}. To see the performance of the $\eIC$ method for rank estimation, we use the $500$ simulation replicates in each setting of DGP 1 and DGP 2, and estimate the rank using the $\eIC$ criterion given in (\ref{eq: est r eIC}).
    We set $g=3$ for the $\eIC$ method because our iterative LS estimates perform well with three iterations as shown in Section \ref{sec: sim performance}.

   We report the accuracy (the percentage of obtaining the true rank) along with the average of the rank estimates based on $500$ simulation replicates for all settings in Table \ref{tab: rank est dgp1 dgp2}. Note that cases with $\pi=1$ in DGP 1 are omitted since we aim to find a method that can accurately estimate the rank when the data have missing entries. We see that our proposed $\eIC$ method performs well in all settings for both DGP 1 and DGP 2. Even with a relatively small sample size ($n=m=200$) and low observation rate ($\pi=0.2$ or $C=1.0$), the $\eIC$ method can correctly estimate the true rank with high probability. Its performance further improves as the sample size becomes larger.
    }

    \begin{table}[ht]
    \tabfnsymbol \centering
    \captionx[Simulation - rank estimation]{The rank estimation results of $\hr^{\eIC(3)}$ based on 500 simulations in each setting with the true rank $r=3$.}
    \label{tab: rank est dgp1 dgp2}
    \vspace{12pt}{\footnotesize
    \begin{minipage}{.6\linewidth}
    \begin{tabularx}{\textwidth}{lCRRCRR}\toprule
    \multicolumn{1}{c}{} & \multicolumn{3}{c}{DGP 1} & \multicolumn{3}{c}{DGP 2}\\
    \cmidrule(lr){2-4} \cmidrule(lr){5-7}
$n,m $  &  $\pi$  &  $\mbox{Acc.}\footnotemark[1]$  &  $\mbox{Ave.}\footnotemark[2]$  &  $\mbox{C}$  &  $\mbox{Acc.}\footnotemark[1]$  &  $\mbox{Ave.}\footnotemark[2]$\\
\cline{1-7}
$200 $  &  $0.2$  &  $  93.8$  &  $  2.95$  &  $1.0$  &  $  97.8$  &  $  3.02$\\
$    $  &  $0.5$  &  $ 100.0$  &  $  3.00$  &  $1.5$  &  $  99.8$  &  $  3.00$\\
$    $  &  $0.8$  &  $ 100.0$  &  $  3.00$  &  $2.0$  &  $  99.8$  &  $  3.00$\\
\cline{1-7}
$500 $  &  $0.2$  &  $ 100.0$  &  $  3.00$  &  $1.0$  &  $ 100.0$  &  $  3.00$\\
$    $  &  $0.5$  &  $ 100.0$  &  $  3.00$  &  $1.5$  &  $ 100.0$  &  $  3.00$\\
$    $  &  $0.8$  &  $ 100.0$  &  $  3.00$  &  $2.0$  &  $ 100.0$  &  $  3.00$\\
\cline{1-7}
$1000$  &  $0.2$  &  $ 100.0$  &  $  3.00$  &  $1.0$  &  $ 100.0$  &  $  3.00$\\
$    $  &  $0.5$  &  $ 100.0$  &  $  3.00$  &  $1.5$  &  $ 100.0$  &  $  3.00$\\
$    $  &  $0.8$  &  $ 100.0$  &  $  3.00$  &  $2.0$  &  $ 100.0$  &  $  3.00$
\\
    \bottomrule
    \tabfootnote{1}{The percentage of $\hr^{\eIC(3)}=r$.}{7}{.9}\\
    \tabfootnote{2}{The average of $\hr^{\eIC(3)}$.}{7}{.9}\\
    \end{tabularx}
    \end{minipage}\hspace{12pt}
    }
    \end{table}
\vspace{-0.5cm}

    \subsection{Simultaneous Inference for The Coefficients}\label{sec: sim hypo}
    In this section, we conduct hypothesis tests on $H_{0}:A_j\cdot \beta_j=a^0_j$ for $j\in \Gcal$ at the significant level $\alpha=0.05$ by the multiplier bootstrap method given in Section~\ref{sec: bootstrap inference of beta}. We consider the following hypotheses:
    (i) $H_{0}:\beta_{j,p}=0\quad \forall \ j,p$; 
    (ii) $H_{0}:\beta_{j,p_0}=0\quad \forall \ j$. 

    \noindent Note that in $\textii$, $p_0$ is a fixed value (could be $1, 2$ or $3$ in our DGPs).
    \begin{rmk}
    To follow the notation in Section $\ref{sec: bootstrap inference of beta}$, $A_j=I$ in (i), and $A_j=(1,0,0)$, $(0,1,0)$, $(0,0,1)$ in (ii) for $p_0=1,2,3$ respectively, and $\Gcal=\{1, \cdots, m\}$ in all the tests.
    \end{rmk}
    To see the performance of the testing procedure under null and different alternative hypotheses, we generate our $\beta$ from $N(0, 4\rho^2 I)$ in DGPs 1 and 2. We run 500 simulation replications in each setting with  $\rho=0$, $e^{-3}$, $e^{-2.5}$, $e^{-2}$, $e^{-1.5}$ and $e^{-1}$, respectively. Note that the null hypothesis $H_0$ is true when $\rho=0$. For each setting and $\rho $ value, we compute the empirical rejection rate of each test based on the 500 simulation replicates.

\begin{figure}[ht]
    \tabfnsymbol \centering
    \includegraphics[width=\linewidth]{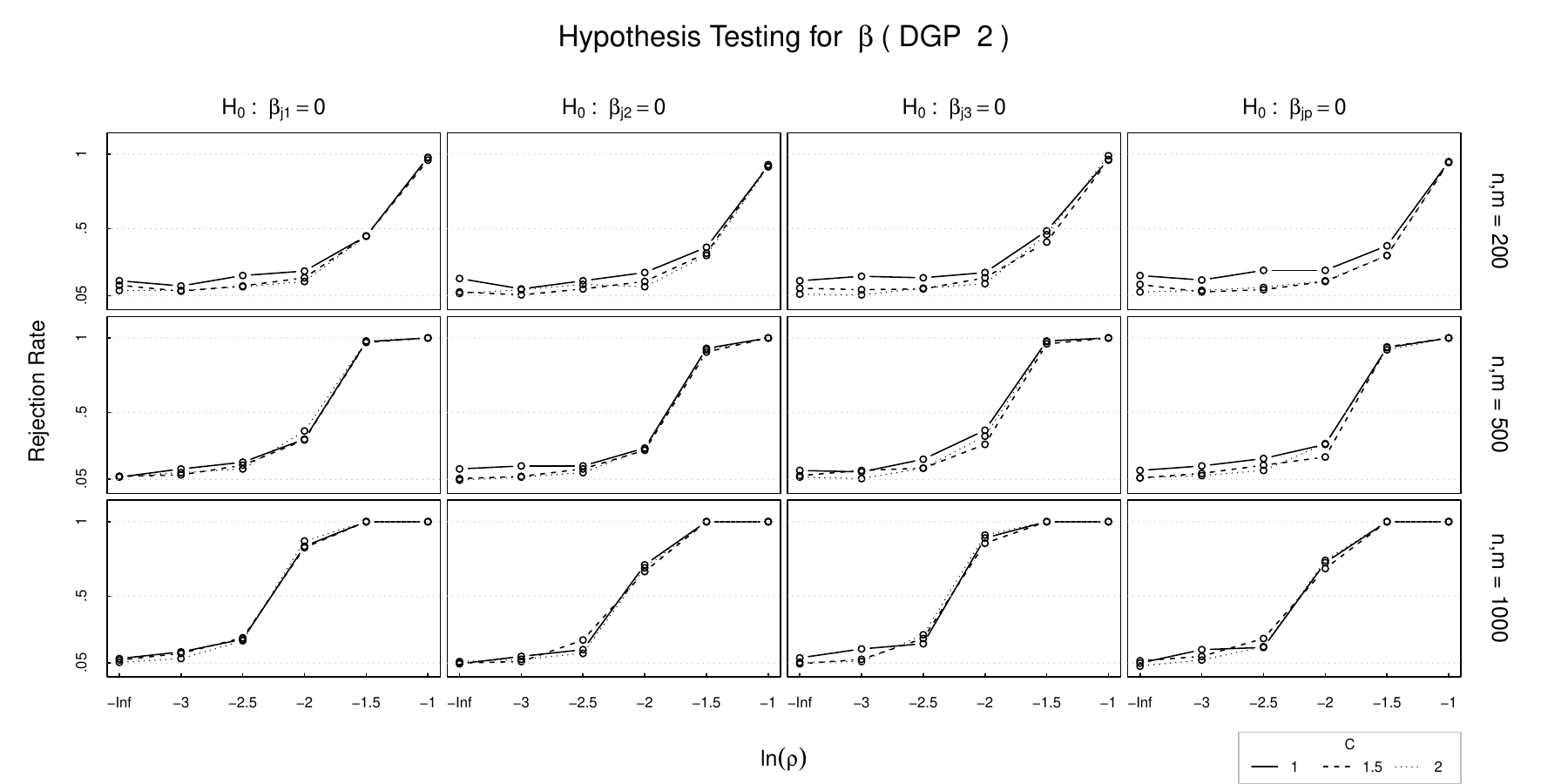}\vspace{-5pt}
    \captionx[Simulation - empirical rejection rates of simultaneous testing]{Empirical rejection rates at level $\alpha = 0.05$. Each column represents a hypothesis, and each row represents a sample size. When $x = \ln(\rho)=-{\rm Inf}$, the null hypothesis is true. }
    \label{fig: hypothesis test for beta dgp 2 rank 3}
    \end{figure}

    The results for DGP 2, $r=3$ are presented in Figure \ref{fig: hypothesis test for beta dgp 2 rank 3}. The numerical results of all scenarios are relegated to the Supplementary Materials\footnotemark[1].  We observe that except for the case with a small sample size $n=m=200$ and $\pi =0.2$, the rejection rate is very close to the significant level $0.05$ under the null hypothesis ($\rho =0$). The power approaches $1$ quickly as the $\rho $ value becomes larger or the sample size increases. This corroborates our theoretical results for the proposed simultaneous testing method given in Section \ref{sec: bootstrap inference of beta}. The rejection rate for the case with $n=m=200$ and $\pi =0.2$ is slightly larger due to the small effective sample size.

\section{\label{sec: application}Application}
\vspace{-0.2cm}
In this section, we apply the proposed method to the MovieLens 1M dataset\footnote{website: https://grouplens.org/datasets/movielens/}. MovieLens is a website where people can sign up and rate movies in their database, and it is run by a lab at the University of Minnesota called GroupLens. They provide movie recommendations to the users based on their rating history. The 1M dataset contains $1,000,209$ ratings on $3,952$ movies from $6,040$ users. Some demographic information of users is provided using an assigned ID, including the user's gender, age, occupation, and zip code. Each rating is a number between $0.5$ and $5$ with $0.5$ gaps between two ratings, linked to a user and a movie. The timestamp at which a rating was given was also recorded. In the dataset, each user has rated at least 20 movies. To provide appropriate recommendations to users, our goal is to : (i) estimate the ratings based on the proposed low-rank model with covariates given in Model (\ref{eq: model Y}) through our iterative LS procedure, (ii) conduct pointwise inference for each rating based on the established asymptotic distribution, and (iii) conduct simultaneous inference for the coefficients of the covariates based on our bootstrap procedure.

    \vspace{-0.5cm}
    \subsection{\label{sec: app how}Application of The Proposed Method to MovieLens 1M}
    \vspace{-0.2cm}
    To apply our method to the MovieLens 1M dataset, we use $Y$ to represent the rating matrix in which the $i$th row and $j$th column correspond to the $i$th user and $j$th movie, respectively. As a result, the dimension of $Y$ should be $6,040\times 3,952$. 
    We consider gender and age as the covariates in Model (\ref{eq: model Y}); both of them may have effects on the movie ratings and the missingness of the ratings. Then, in the covariate matrix, the gender is encoded as "0" (female) and "1" (male); the age is factorized into $4$ groups: "0-24", "25-34", "35-49", "50+", and it is represented by $3$ dummy variables. We also include the interactive terms between gender and age groups in the covariate matrix $X$, and then the dimension of $X$ including the intercept is $m=6040$ by $d=8$. The dataset is split into a training set and a test set, and the test set contains $60,400$ ratings with $10$ ratings from each user.

   Let $G_i$ be the indicator of the gender for the $i^{\rm th}$ user, and $A_{i,k}$, $1\leq k\leq 3$, be the indicator of age groups "25-34", "35-49", "50+", respectively. Both gender and age may affect the missingness of movie ratings, so we fit a logistic model for $\pi_i=P(\xi_{i,j}=1|X_i)$:
        \begin{equation}
        \setlength{\belowdisplayskip}{0.5pt}
        \setlength{\abovedisplayskip}{0.5pt}
        \mbox{logit}(\pi_{i}) = \gamma_{0} + G_i\gamma_{1} + A_{i,1}\gamma_{2}+ A_{i,2}\gamma_{3} + A_{i,3}\gamma_{4} +
     (G_i\cdot A_{i,1})\gamma_{5}+ (G_i\cdot A_{i,2})\gamma_{6} + (G_i\cdot A_{i,3})\gamma_{7}, \label{eq: ml model of pi}
    \end{equation}
    where $\mbox{logit}(x)=\log(x/(1-x))$. Moreover, we fit the following model for the responses:
    \begin{align}\label{eq: ml model of y}
        Y_{i,j} =\,& \beta_{j,0} + G_i\beta_{j,1} + A_{i,1}\beta_{j,2}+ A_{i,2}\beta_{j,3} + A_{i,3}\beta_{j,4} + \nonumber\\
        &(G_i\cdot A_{i,1})\beta_{j,5}+ (G_i\cdot A_{i,2})\beta_{j,6} + (G_i\cdot A_{i,3})\beta_{j,7} + L'_iF_j +\varepsilon_{i,j}.
    \end{align}
    We also consider the sub-model with only the main effects of $G_{\cdot}$ and $A_{\cdot,k}$ as well as other sub-models which include only partial interactions between gender and age groups to see which model has the best prediction. We use the $\eIC$ method with the penalty given in \eqref{eq: pen h} with  $C_h=0.2$, $\delta_h=0.1$ to obtain the estimated rank $\hr = \hr^{\eIC(c)}$. Tables \ref{tab: cputime dgp1} and \ref{tab: mse dgp2} in Section \ref{sec: sim performance} show that the iterative LS algorithm in general only needs a few iterations to converge. Then, in the real data analysis,
    we estimate $\beta_{j}$ and $\Gamma_{i,j}$ by running the iterative LS algorithm until convergence or stopped at step=30. With the estimated rank, we then obtain the estimated ratings in the training set and the predicted ratings in the test set by $\tY_{i,j} = X'_i\tbeta_{j} + \tGamma_{i,j}$ where $\tbeta=\tbeta^{(c)}$ and $\tGamma_{i,j}=\tL^{(c)'}_i \tF^{(c)}_j$. In addition, since the rating is limited to be between $0.5$ and $5$, we define the adjusted estimated rating as $\tY^{\rm adj}_{i,j} = (\tY_{i,j}\vee 0.5)\wedge 5$, so as to enable the estimated ratings to have values between $(0.5, 5)$.
\vspace{-0.5cm}
    \subsection{The Fitting Results}\label{sec: app fitting result}
    The fitting result of the logistic model for $\pi_i$ given in \eqref{eq: ml model of pi} is shown in Table \ref{tab: ml logistic result for pi}. It shows the estimate and the standard error of each coefficient, and the $p$-values for testing whether each coefficient is zero or not. Table \ref{tab: ml logistic result for pi} shows that the  $p$-values are all close to zero, indicating that all the coefficients for both the main and interaction effects are significantly different from zero. This result further demonstrates that the two baseline covariates, gender and age, and their interactions should have significant effects on the missing pattern of the movie rates. We will use the full model \eqref{eq: ml model of pi} for $\pi_i$ in the follow-up analysis.

    \begin{table}[ht]
    \tabfnsymbol \centering
        \captionx[logistic regression on pi for ML data]{The fitting result fors the logistic model for $\pi_i$.}\label{tab: ml logistic result for pi}
        \vspace{12pt}{\footnotesize
        \begin{tabularx}{\textwidth}{lRRRRRRRR}\toprule
                      &   (Intercept)   &        $G$   &       $A_1$   &       $A_2$   &       $A_3$   &     $G\cdot A_1$   &     $G\cdot A_2$   &     $G\cdot A_3$ \\
\hline
 Estimate     &  $-3.347$  &  $ 0.148$  &  $ 0.140$  &  $-0.037$  &  $-0.304$  &  $ 0.061$  &  $ 0.083$  &  $ 0.072$\\
 Std. Error ($10^{-3}$)   &  $  4.53$  &  $  5.25$  &  $  5.71$  &  $  5.97$  &  $  7.87$  &  $  6.58$  &  $  6.93$  &  $  9.07$\\
 $p$-value    &  \footnote[1]  &  \footnote[1]  &  \footnote[1]  &  $10^{-9}$  &  \footnote[1]  &  \footnote[1]  &  \footnote[1]  &  $10^{-14}$\\
        \bottomrule \\[-6pt]
        \tabfootnotex{1}{Value $<10^{-15}$}{9}{.95}
        \end{tabularx}
        }
    \end{table}
    Next, we calculate the root mean square error (RMSE) of the estimated ratings, where
     ${\rm RMSE} = \left[\,(n_S)^{-1}\sum\nolimits_{(i,j)\in S}(Y_{i,j}-\tY_{i,j})^2\,\right]^{1/2}$
    respectively, for the training and test datasets, to check the prediction performance of each model. In the above formula, the $S$ is the set of observed indices in the training set or the indices in the test set, and $n_S=|S|$ is the number of elements in $S$. The RMSEs for the training and test datasets are provided in Table \ref{tab: ml RMSE}. Note that in this table, we also calculate the RMSE using the adjusted rating $\tY^{adj}_{i,j}$.
    \begin{table}[ht]
        \tabfnsymbol \centering
        \captionx[RMSE of different models for ML data]{RMSE of different models in training and test set.}\label{tab: ml RMSE}
        \vspace{12pt} {\footnotesize
        \begin{tabularx}{\textwidth}{lrlRRRR}\toprule
        \multicolumn{3}{l}{Model} & \multicolumn{2}{c}{RMSE} & \multicolumn{2}{c}{adj. RMSE\footnotemark[2]}\\[-3pt]
 Covariate(s)                                                     &       $p$   &     $\hr$   &   training   &    tested   &   training   &    tested\\
\cmidrule(lr){1-3}\cmidrule(lr){4-5}\cmidrule(lr){6-7}
$1 + G + A_1 + A_2 + A_3 + G\cdot A_1 + G\cdot A_2 + G\cdot A_3$  &  $      8$  &  $      2$  &  $ 0.8381$  &  $ 0.9067$  &  $ 0.8379$  &  $ 0.8993$\\
$1 + G + A_1 + A_2 + A_3 + G\cdot A_1 + G\cdot A_2             $  &  $      7$  &  $      2$  &  $ 0.8395$  &  $ 0.9046$  &  $ 0.8393$  &  $ 0.8982$\\
$1 + G + A_1 + A_2 + A_3 + G\cdot A_1 + G\cdot A_3             $  &  $      7$  &  $      2$  &  $ 0.8398$  &  $ 0.9057$  &  $ 0.8396$  &  $ 0.8980$\\
$1 + G + A_1 + A_2 + A_3 + G\cdot A_2 + G\cdot A_3             $  &  $      7$  &  $      2$  &  $ 0.8398$  &  $ 0.9064$  &  $ 0.8396$  &  $ 0.8982$\\
$1 + G + A_1 + A_2 + A_3 + G\cdot A_1                          $  &  $      6$  &  $      2$  &  $ 0.8411$  &  $ 0.9032$  &  $ 0.8409$  &  $ 0.8962$\\
$1 + G + A_1 + A_2 + A_3 + G\cdot A_2                          $  &  $      6$  &  $      2$  &  $ 0.8411$  &  $ 0.9043$  &  $ 0.8409$  &  $ 0.8966$\\
$1 + G + A_1 + A_2 + A_3 + G\cdot A_3                          $  &  $      6$  &  $      2$  &  $ 0.8415$  &  $ 0.9044$  &  $ 0.8413$  &  $ 0.8968$\\
$1 + G + A_1 + A_2 + A_3                                       $  &  $      5$  &  $      2$  &  $ 0.8428$  &  $ 0.9020$  &  $ 0.8426$  &  $ 0.8949$\\
$1                                                             $  &  $      1$  &  $      2$  &  $ 0.8582$  &  $ 0.9813$  &  $ 0.8581$  &  $ 0.9010$
\\
        \bottomrule\\[-6pt]
        \tabfootnotex{1}{adj. RMSE uses $\tY^{adj}_{i,j}$ in stead of $\tY_{i,j}$ in the RMSE formula.}{7}{.95}
        \end{tabularx}
        }
    \end{table}
    According to Table \ref{tab: ml RMSE}, we can see that all the estimated rank $\hr$ by the $\eIC$ method is $2$ for all cases, and the best prediction which gives the lowest $\rmse$ in the test set is the model with only the main effects, no matter we use the original estimators or the adjusted ones. As a result, we will use the model with only the main effects in the follow-up analysis, and it is formulated as
    \begin{equation}
        \setlength{\belowdisplayskip}{0pt}
        \setlength{\abovedisplayskip}{0.5pt}
         Y_{i,j} =\beta_{j,0} + G_i\beta_{j,1} + A_{i,1}\beta_{j,2}+ A_{i,2}\beta_{j,3} + A_{i,3}\beta_{j,4} + L'_iF_j +\varepsilon_{i,j}  \label{eq: ml selected model}.
    \end{equation}

    \vspace{-0.5cm}
    \subsection{Insight into MovieLens 1M}\label{sec: app insight}
\begin{figure}[ht]
    \tabfnsymbol \centering
    \includegraphics[width=\linewidth]{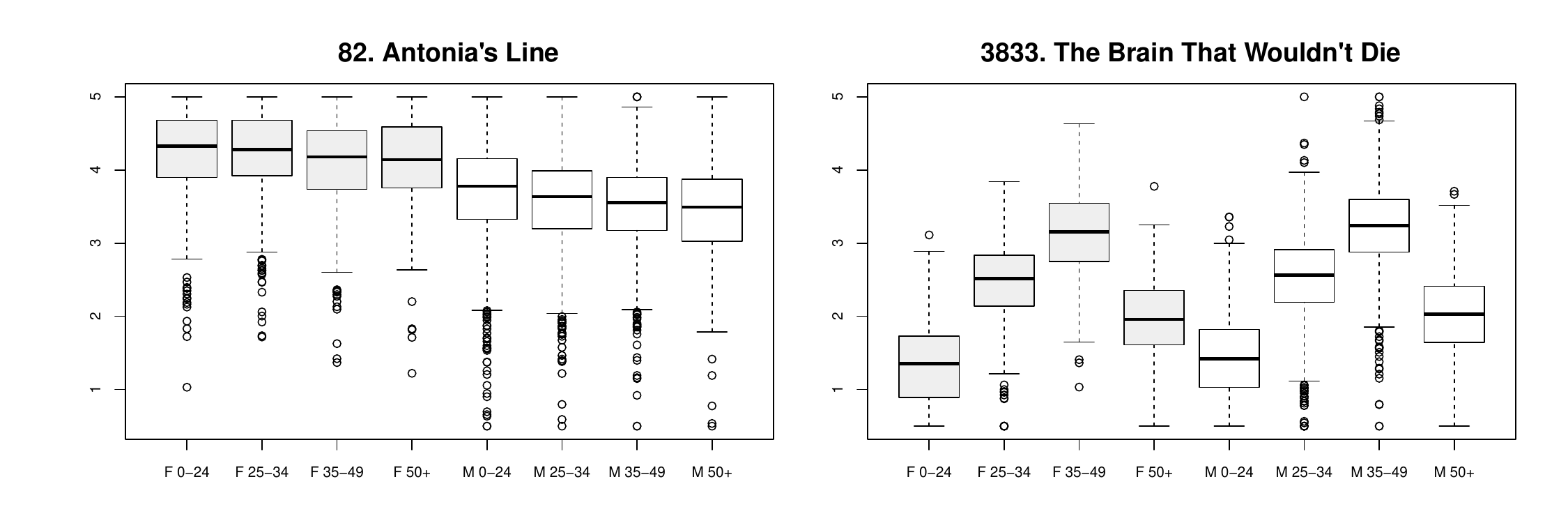}
    \captionx[ Boxplot examples of ML data]{{\small Boxplots of the estimated ratings in different gender and age groups for some movies.}}\label{fig: ml boxplots}
    \end{figure}

    With the selected model \eqref{eq: ml selected model}, we can run contrast tests on the coefficient matrix $\beta$ to see if any category in the covariates is unnecessary or if any two (or more) of them can be combined. Since the tests concern the high dimensional coefficient matrix $\beta$, we use the multiplier bootstrap method provided in Section \ref{sec: bootstrap inference of beta} to conduct simultaneous inference, and the results are presented in Table \ref{tab: ml hypo test}.
    \begin{table}[ht]
    \tabfnsymbol \centering
        \captionx[Simultaneous testing results of ML data]{The hypothesis testing results for all contrasts.}\label{tab: ml hypo test}
        \vspace{12pt}{\footnotesize
        \begin{tabularx}{\textwidth}{llRR}\toprule
 $H_0$                                           &   meaning                                          &   test statistic   &   $p$-value\footnotemark[1] \\
\cmidrule(lr){1-1}\cmidrule(lr){2-2}\cmidrule(lr){3-3}\cmidrule(lr){4-4} 
 $\beta_{j1} = 0\quad \forall \ j$               &   No difference in gender                          &  $ 5.25$  &  $<0.001$\\
 $\beta_{j2} = 0\quad \forall \ j$               &   No difference in age group (-24) and (25-34)     &  $ 5.50$  &  $<0.001$\\
 $\beta_{j3} = 0\quad \forall \ j$               &   No difference in age group (-24) and (35-49)     &  $10.75$  &  $<0.001$\\
 $\beta_{j4} = 0\quad \forall \ j$               &   No difference in age group (-24) and (50+)       &  $ 6.15$  &  $<0.001$\\
 $\beta_{j2} - \beta_{j3} = 0\quad \forall\ j$   &   No difference in age group (25-34) and (35-49)   &  $10.75$  &  $<0.001$\\
 $\beta_{j2} - \beta_{j4} = 0\quad \forall\ j$   &   No difference in age group (25-34) and (50+)     &  $ 7.81$  &  $<0.001$\\
 $\beta_{j3} - \beta_{j4} = 0\quad \forall\ j$   &   No difference in age group (35-49) and (50+)     &  $10.75$  &  $<0.001$
\\
        \bottomrule \\[-6pt]
        \tabfootnotex{1}{Each $p$-value is calculated through $1000$ bootstrap results.}{4}{.95}
        \end{tabularx}
        }
    \end{table}
    All the p-values in Table \ref{tab: ml hypo test} are very small, indicating that the different age and gender groups have significant effects on the prediction of movie ratings. For further illustration, Figure \ref{fig: ml boxplots} shows the box plots of the estimated movie ratings in different gender and age groups for some movies. We can see that for the movie ``Antonia's Line'', the ratings are very different between genders, but they are similar among different age groups of the same gender. However, for some other movies such as ``The Brain That Wouldn't Die'', in which both age and gender significantly affect the movie ratings;  see more boxplot examples in the supplementary materials\footnotemark[1].

\begin{figure}[ht]
    \tabfnsymbol \centering
    \includegraphics[width = \linewidth]{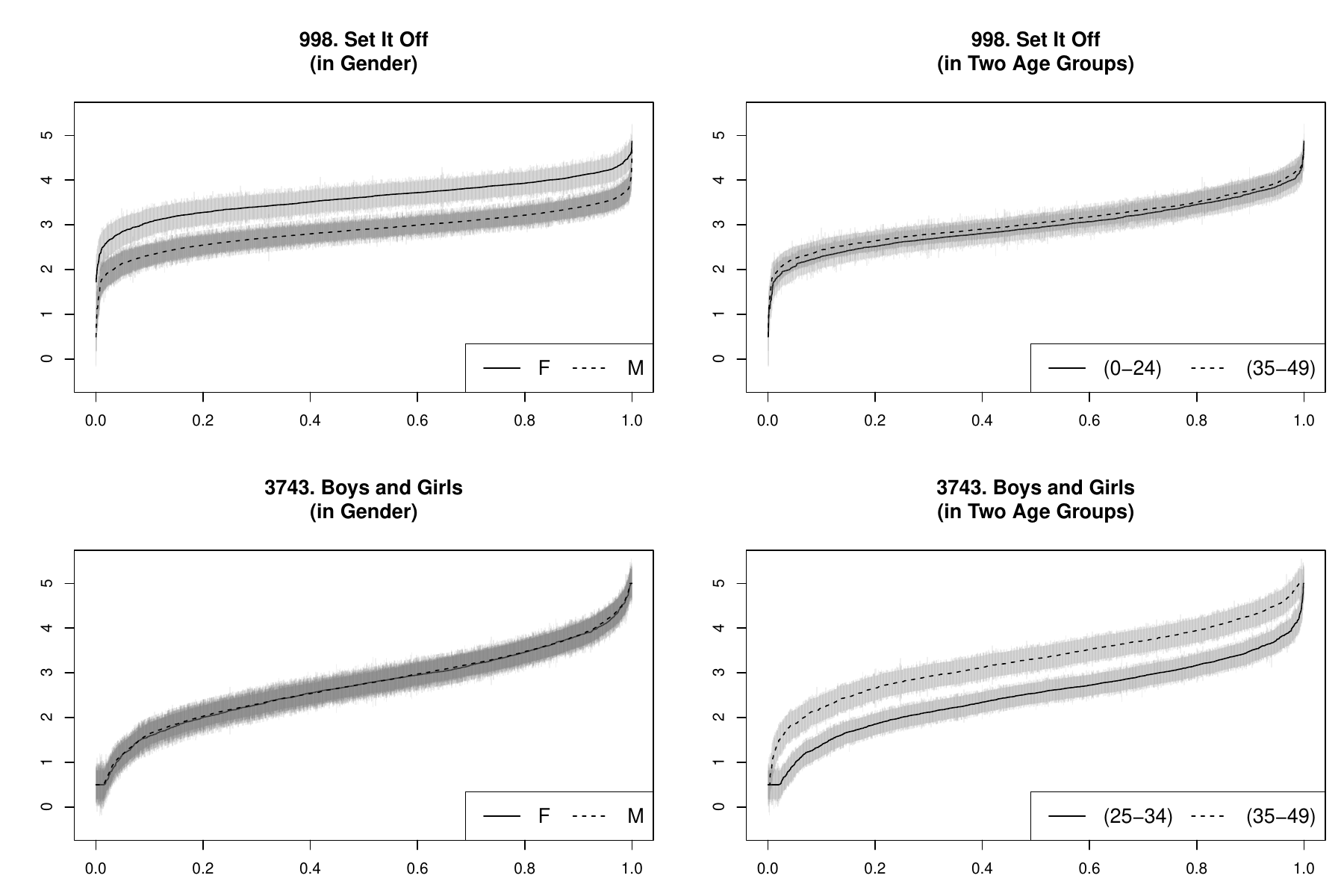}
    \captionx[Example of estimated ratings and 90\% point-wise CI for ML data]{{\small Estimated ratings and 90\% point-wise confidence intervals in different groups. The $y$-axis is the rating and the $x$-axis is the percentile. Ratings are grouped by gender or age.}}\label{fig: ml rating CI}
    \end{figure}

    While the overall effects of different covariates on the ratings can be investigated through Table \ref{tab: ml hypo test}, examples in Figure \ref{fig: ml boxplots} motivate us to perform individual tests on each movie, so that we can understand the effect of covariate on each movie. A $z$-test is then conducted for each movie based on the asymptotic result in Theorem \ref{thm: beta est}. Table S5 in Section H.1 of the Supplementary Materials shows the top 10 movies with the smallest p-values in each test. All the $p$-values shown in Table S5 are significant after a Bonferroni adjustment. In Figure \ref{fig: ml rating CI}, we select two movies in which either gender or age has a significant effect and draw a quantile plot with  $90\%$ point-wise confidence intervals (CI) to further illustrate the effects. The movie `` Set It Off'' is the one in which the ratings are significantly different in gender (non-overlapping CI bands), but not at all in age, and ``Boys and Girls'' shows the other way. Note that only the most significant pair of age groups are shown in this figure. We refer to the supplementary materials for the numerical results of more movie examples.
    \vspace{-1cm}
    \renewcommand{\arraystretch}{0.4}

\vspace{-0.5cm}
\section{Conclusion}\label{sec: conclusion}
\vspace{-0.2cm}
This paper studies statistical inference for noisy matrix completion with auxiliary information when the missing pattern of the responses depends on baseline covariates and the observed rates can go to zero as the sample size increases. We show that the iterative LS method has a computational advantage over the iterative PCA method, and it is supported by reliable statistical properties for inference.  With only a finite number of iterations, the resulting estimators of the latent low-rank matrix and the coefficient matrix for the observed covariates are asymptotically unbiased and guaranteed to have asymptotic normality under mild conditions.  A new information criterion $\eIC$ method based on the iterative LS estimation is proposed for rank estimation. It is supported by the consistency property and is demonstrated to have better numerical performance than the widely used $\IC$ criterion method based on the singular value estimation.

Moreover, we propose a simultaneous testing method for the high dimensional coefficient matrix $\beta$ via a Gaussian multiplier bootstrap procedure. This inferential procedure can help us investigate the effects (or contrast effects) of the auxiliary covariates for the prediction of the missing entries. We have discussed in Section \ref{sec: sim performance} and have shown in the real data application Section \ref{sec: application} that the use of the observed covariates in matrix completion does help the prediction and improves the prediction accuracy. Our proposed method has immediate applications in collaborative filtering, biological and social network recovery,  recommender systems, and so forth. The semi-supervised model considered in our paper makes use of row-feature information such as the user's demographic information to help the prediction of movie ratings. It is worth noting that \cite{ZhuShenYe2016} have considered a different model that incorporates user-specific and content-specific predictors by letting their coefficients be the same across all $j$ and $i$, respectively. As an extended work, we can also consider incorporating the column-feature information into our proposed framework. Moreover, the development of the asymptotic distributions of the iterative LS estimators in the setting with the growing number of factors \citep{Mao2019} or $\Gamma$ with high rank is also an interesting future research topic to explore.

\vspace{-0.5cm}

\section{Appendix}

In this document, Section \ref{sec:pihat} derives the statistical properties of $\hat \pi_i$ which will be used later in the proofs of the main results, Sections \ref{app: proof of beta est g=1}-\ref{app: inference beta} provide the technical proofs for Theorems \ref{thm: beta est}-\ref{thm: beta hypothesis}, Section \ref{sec: sim misspecify rank} provides an argument about the MSE pattern when $\eIC$ is constructed based on the initial estimates,
and Section \ref{sec: add num} contains additional numerical results of the simulation studies and the real data application.

\section{Properties of $\hat \pi_i$}
\label{sec:pihat}
\begin{prop}
Denote the true value $\gamma = (\gamma_{0,n},\gamma_1)$. Then, for any constant $q>0$, we have
\begin{align*}
\hat{\gamma} - \gamma = \frac{1}{nm\alpha_n}\sum_{i=1}^n \sum_{j=1}^m H_0^{-1}(\xi_{i,j} - \eta(\gamma_{0,n}+ \tilde{X}_i'  \gamma_1))X_i + o_p((nm\alpha_n)^{-1/2}),
\end{align*}
\begin{align}
& \hat{\gamma} = \gamma + O_p((nm\alpha_n)^{-1/2}),
\label{eq:gammahat}
\end{align}
\begin{align*}
& \max_{1 \leq i \leq n} \pi_i^{-1} \lessp \alpha_n^{-1} n^{1/q}, \quad  \max_{1 \leq i \leq n}|\hat{\pi}_i^{-1}\pi_i-1|\lessp (nm\alpha_n)^{-1/2}  n^{1/q} = o_P(1),
\end{align*}
and
\begin{align*}
\max_{1 \leq i \leq n}\left|\hat \pi_i^{-1} - \pi_i^{-1}  + \frac{(1-\pi_i)X_i'(\hat{\gamma} - \gamma)}{\pi_i}\right| & \lessp \alpha_n^{-2} n^{1/q} (nm)^{-1},
\end{align*}

\label{prop:pihat}
\end{prop}

\begin{proof}
The first two results can be established by the usual analysis for logistic regressions.

For the third result, note that $\max_{1 \leq i \leq n}(|\exp(\gamma_0 + \tilde{X}_i' \gamma_1)| + |\exp(-(\gamma_0 + \tilde{X}_i' \gamma_1))|) \lessp n^{1/q}$ for any $q>0$ due to the sub-Gaussianity of $||\tilde{X}_i||$. This implies $\max_{1 \leq i\leq n}\pi_{i}^{-1} \lessp \alpha_n^{-1} n^{1/q}$. In addition, we note that  there exists a constant $C>0$ such that for any $q>0$,
\begin{align*}
\max_{1 \leq i \leq n}|\hat \pi_i - \pi_i| \lessp \max_{1 \leq i \leq n}\exp(C||X_i||) ||X_i|| \alpha_n (nm\alpha_n)^{-1/2} \lessp  \alpha_n (nm\alpha_n)^{-1/2} n^{1/q}.
\end{align*}
Therefore, we have
\begin{align*}
\max_{1 \leq i \leq n} |\hat \pi_i\pi_i^{-1} - 1| \lessp  \max_{1 \leq i \leq n}|\hat \pi_i - \pi_i| \alpha_n^{-1} n^{1/q}  \lessp (nm\alpha_n)^{-1/2} n^{2/q} = o_P(1),
\end{align*}
as $q$ can be arbitrarily large. This further implies
\begin{align*}
\max_{1 \leq i \leq n} |\hat \pi_i^{-1}\pi_i - 1| \lesssim (nm\alpha_n)^{-1/2}  n^{1/q} = o_P(1).
\end{align*}

For the last result, we note that
\begin{align*}
\hat \pi_i - \pi_i = \pi_i(1-\pi_i) X_i'(\hat \gamma - \gamma) + \frac{1}{2}  \tilde \pi_i(1-\tilde \pi_i)(1-2\tilde \pi_i)(X_i'(\hat \gamma - \gamma))^2,
\end{align*}
where $\pi_i = \eta(\gamma_{0,n} + \tilde{X}_i'\gamma_1)$, $\tilde \pi_i = \eta(\gamma_{0,n} + \tilde r_0  + \tilde{X}_i'(\gamma_1+\tilde r_1))$, $\tilde r_0$ is between $0$ and $\hat \gamma_0 - \gamma_{0,n}$, and $\tilde r_1$ is between $0$ and $\hat \gamma_1 - \gamma_{1}$. This implies that for any $e>0$, there exists a constant $C>0$ such that with probability greater than $1-e$,
\begin{align*}
\max_{1 \leq i \leq n} \left|\hat \pi_i - \pi_i - \pi_i(1-\pi_i) X_i'(\hat \gamma - \gamma)\right| \leq \max_{1 \leq i \leq n} \alpha_n | C\exp(C||X_i||)(X_i'(\hat{\gamma} - \gamma))^2 \lessp n^{1/q} (nm)^{-1}.
\end{align*}

In addition, we have
\begin{align*}
\hat \pi_i^{-1} - \pi_i^{-1} = - \frac{\hat \pi_i - \pi_i}{\pi_i^2} + \frac{(\hat \pi_i - \pi_i)^2}{\pi_i^2 \hat \pi_i},
\end{align*}
which implies
\begin{align*}
\max_{1 \leq i \leq n}\left|\hat \pi_i^{-1} - \pi_i^{-1}  + \frac{(1-\pi_i)X_i'(\hat{\gamma} - \gamma)}{\pi_i}\right| & \lessp \max_{1 \leq i \leq n} \frac{(\hat \pi_i - \pi_i)^2}{\hat \pi_i  \pi_i^2} + \alpha_n^{-2} n^{3/q} (nm)^{-1} \\
& \lessp\alpha_n^{-2} n^{5/q} (nm)^{-1}.
\end{align*}
As $q$ is an arbitrary positive constant, we obtain the desired result.
\end{proof}

\section[Proof of Theorem \ref{thm: beta est} at first iteration]{Auxiliary Results for Proof of Theorem \ref{thm: beta est} at $g=1$}\label{app: proof of beta est g=1}

\subsection{Notations}
For the sake of simplicity, we define the operator $E_n(\cdot) = n^{-1}\sumi $. Since we only focus on $g=1$ in this section, we denote $\tbeta = \tbeta^{(1)}$, $\tW = \tW^{(1)}$. $\tF = \tF^{(1)}$, $\tL = \tL^{(1)}$ and omit the iteration counter. Let $\Delta=\hGamma-\Gamma$ be the difference between the true value and the initial estimator of the hidden matrix. Then
\begin{align}
    \tbeta_{j}
    & =(E_{n}X_{i}X_{i}'\xi_{i,j})^{-1}E_{n}(X_{i}(X_{i}'\beta_{j}+\Gamma_{i,j}+\varepsilon_{i,j}-\hGamma_{i,j})\xi_{i,j}).\nonumber \\
    & =\beta_{j}+(E_{n}X_{i}X_{i}'\xi_{i,j})^{-1}E_{n}(X_{i}\varepsilon_{i,j}\xi_{i,j})-(E_{n}X_{i}X_{i}'\xi_{i,j})^{-1}E_{n}(X_{i}\Delta_{i,j}\xi_{i,j}).\label{eq: tbeta decomp g=1}
\end{align}
Let $\delta_{j}=\hbeta_{j}-\beta_{j}$,
\begin{align*}
    u_{i,j} &= Y_{i,j} - L_i' F_j - X_i' \hbeta_j = \varepsilon_{i,j} - X_i'\delta_j,
\end{align*}
$\pi = (\pi_1,\cdots,\pi_n)'$, and $\hat \pi = (\hat \pi_1,\cdots,\hat \pi_n)'$.
Recall that
\begin{align*}
    W_{i,j} & =\hpi_i^{-1}\xi_{i,j}(Y_{i,j}-X_{i}'\hbeta_{j})=\hpi_i^{-1}\xi_{i,j}(L_i' F_j + u_{i,j}),
\end{align*}
and we can define
\begin{align}
    e_{i,j} & = W_{i,j} - L_{i}'F_{j} \notag \\
            & = \hpi_i^{-1}\xi_{i,j}L_{i}'F_{j}+\hpi_i^{-1}\xi_{i,j}u_{i,j} - L_i' F_j \notag  \\
            & = L_{i}'F_{j}(\hpi_i^{-1}\xi_{i,j}-1)+\hpi_i^{-1}\xi_{i,j}u_{i,j} \notag \\
            & = \hpi_i^{-1}\cdot(\xi_{i,j}-\pi_i)L_{i}'F_{j}+\hpi_i^{-1}\cdot(\pi_i-\hpi_i)L_{i}'F_{j}+\hpi_i^{-1}\xi_{i,j}\varepsilon_{i,j}-\hpi_i^{-1}\xi_{i,j}X_{i}'\delta_{j} \notag  \\
            & = \xi_{i,j} (\hat{\pi}_i^{-1} - \pi_i^{-1})L_i'F_j + \pi_i^{-1}(\xi_{i,j}- \pi_i)L_i'F_j + \hat{\pi}_i^{-1} \xi_{i,j}\eps_{i,j} - \hat{\pi}_i^{-1} \xi_{i,j}X_i'\delta_j.
            \label{eq:e}
\end{align}
In matrix notation, we can write
\begin{align*}
    W & = \diag(\hpi)^{-1}(Y-X\hbeta)\circ\Xi = LF'+e.
\end{align*}

Let $W= U D V'$ be the conventional SVD representation of $W$, and $ D_r\in\RR^{r\times r},  U_r\in\RR^{n\times r},  V_r\in\RR^{m\times r}$ be matrices of the largest $r$ singular values and the corresponding singular vectors. By definition,
\begin{align*}
    \hL =\,&\sqrt{n} U_r\\
    \hF =\,& n^{-1}W'\hL = \frac{1}{\sqrt{n}} V_r D_r
\end{align*}
Let $\hOme_r= D_r(nm)^{-1/2}$. By the definition of $\hL, \hF$, we have $W\hF=n^{-1}\hL D_r^{2}$
and thus
\begin{align*}
    \hL
        &= nW\hF D_r^{-2}=n(LF'+e)\hF D_r^{-2}\\
        &= L\left(m^{-1}F'\hF\hOme_r^{-2}\right)+m^{-1}e\hF\hOme_r^{-2}\\
        &= L H+\Delta_{L},
\end{align*}
where $H=\left(m^{-1}F'\hF\hOme_r^{-2}\right)$ and $\Delta_{L}=m^{-1}e\hF\hOme_r^{-2}$. Since $\hF=n^{-1}(FL'+e')\hL$, we have $\hL\hF'=n^{-1}\hL\hL'(LF'+e)$ and thus
\begin{align*}
\Delta & =(n^{-1}\hL\hL'-I_{n})LF'+n^{-1}\hL\hL'e\\
 & =(n^{-1}\hL\hL'-I_{n})(\hL-\Delta_{L}) H^{-1}F'+n^{-1}\hL\hL'e\\
 & =(I_{n}-n^{-1}\hL\hL')\Delta_{L} H^{-1}F'+n^{-1}\hL\hL'e.
\end{align*}
This means that
\begin{equation}
\Delta_{i,j}=\Delta_{L,i}'H^{-1}F_{j}-n^{-1}\hL_{i}'\hL'\Delta_{L}H^{-1}F_{j}+n^{-1}\hL_{i}'\hL'e_{\jj},\label{eq: PCA 1}
\end{equation}
where $e_{\jj}\in\RR^{n}$ is the $j$-th column of $e$. We also write $e_{\ii}\in\RR^m$ to be the $i$-th row of $e$. Thus,
\begin{align*}
e=\begin{pmatrix}e_{1,\cdot}'\\
\vdots\\
e_{n,\cdot}'
\end{pmatrix}=\begin{pmatrix}e_{\cdot, 1} & \cdots & e_{\cdot, m}\end{pmatrix}\in\RR^{n\times m}.
\end{align*}
Based on (\ref{eq: tbeta decomp g=1}), we only need to show
\begin{equation*}
E_{n}(X_{i}\Delta_{i,j}\xi_{i,j})=-E_{n}(\xi_{i,j}X_{i}L_{i}')F_{j}+o_{P}(n^{-1/2}).
\end{equation*}
To show this, we observe that (\ref{eq: PCA 1}) implies
\begin{align}
 &\ E_{n}(X_{i}\Delta_{i,j}\xi_{i,j})\nonumber\\
=&\ E_{n}\xi_{i,j}X_{i}\Delta_{L,i}'H^{-1}F_{j}-n^{-1}E_{n}\xi_{i,j}X_{i}\hL_{i}'\hL'\Delta_{L}H^{-1}F_{j}+n^{-1}E_{n}\xi_{i,j}X_{i}\hL_{i}'\hL'e_{\jj}.\label{eq: key decomposition}
\end{align}
This is the key decomposition in the proof, and we will revisit it later. To begin with, we will establish some lemmas that give the asymptotic properties we need. We also define \begin{align*}\delta=\begin{pmatrix}\delta_{1}'\\
\vdots\\
\delta_{m}'
\end{pmatrix}=\hbeta -\beta\in\RR^{m\times k}.\end{align*}
We use $X_{n,m}\lessp a_{n,m}$ to denote $X_{n,m}=O_{P}(a_{n,m})$ as $n\wedge m \rightarrow \infty$. Throughout the supplement, we use $q$ to denote a positive constant that can be arbitrarily large.

\subsection{Auxiliary results}
\begin{lem}\label{lem: general asymp prop}Let Assumption \ref{assu: moment cond} hold.
Then we have the following asymptotic properties.
	\begin{enumerate}
		\item[(1)]
		\begin{align}
		&  \max_{1 \leq j \leq m} ||(E_n X_iX_i' \xi_{i,j})^{-1} - (E X_iX_i' \xi_{i,j})^{-1}|| \lessp \alpha_n^{-1} \sqrt{n^{-1} \alpha_n^{-1} \log n}, \label{eq:lem1_1}\\
		&  \max_{1 \leq i \leq n} ||\sumj (\xi_{i,j} - \pi_i)F_jF_j'|| \lessp \sqrt{m \alpha_n \log n}, \label{eq:lem1_1'} \\
		&  \max_{1 \leq j \leq m}||E_n \pi_i^{-1/2}X_i L_i'\xi_{i,j}|| \lessp \sqrt{n^{-1}\log n}, \label{eq:lem1_3}\\
		&  \max_{1 \leq j \leq m}||E_n X_i \eps_{i,j}\xi_{i,j}|| \lessp \sqrt{n^{-1} \alpha_n \log n}, \label{eq:lem1_5}\\
		&  \max_{1 \leq s \leq m}\left\Vert \sum_{i=1}^n \sum_{j=1}^m \pi_i^{-1}X_i F_j'\xi_{i,j}\xi_{i,s}\eps_{i,j}\right\Vert \lessp  \sqrt{nm \log n}, \label{eq:lem1_7} \\
		& \max_{1 \leq s \leq m, 1\leq r \leq R} \left\Vert \sum_{i=1}^n \sum_{j=1}^m \xi_{i,s} \xi_{i,j} \eps_{i,j}  \frac{1-\pi_i}{\pi_i}F_{j,r}X_iX_i'\right\Vert \lessp  \sqrt{nm \log n}, \label{eq:lem1_8}
		\end{align}
			\item[(2)]
		\begin{align}
		\max_{1 \leq j \leq m}\left\Vert\delta_j - (EX_iX_i'\xi_{i,j})^{-1}(E_nX_i \eps_{i,j}\xi_{i,j} + (E_n X_i L_i'\xi_{i,j})F_j) \right\Vert \lessp n^{-1}\alpha_n^{-1} \log^{3/2} n \label{eq:lem1_2_8},
		\end{align}	
	\item[(3)]
	\begin{align}
		\max_{1 \leq j \leq m} ||\delta_j||_2 \lessp \sqrt{n^{-1}\alpha_n^{-1} \log^{3/2} n}, \label{eq:lem1_9}
	\end{align}
	\item[(4)]
	\begin{align}
	||\delta||_F^2 \lessp m n^{-1} \alpha_n^{-1} \log n, \label{eq:lem1_4_1}
	\end{align}
	\item[(5)]
	\begin{align}
&  \max_{1 \leq j \leq m}\left\Vert \sum_{i=1}^n L_i \xi_{i,j} \eps_{i,j} \right\Vert_2 \lessp \sqrt{n \alpha_n \log n}, \label{eq:lem1_5_5} \\
&  \max_{1 \leq s \leq m}\left\Vert \sum_{i=1}^n \sum_{1 \leq j \leq m} \pi_i^{-1}L_i F_j' \xi_{i,s} \xi_{i,j} \eps_{i,j}   \right\Vert \lessp \sqrt{n m  \log n}, \label{eq:lem1_5_9} \\
&   \max_{1 \leq s \leq m}\left\Vert \sum_{i=1}^n \sum_{1 \leq j \leq m} L_i L_i' F_j F_j' \xi_{i,s} \pi_i^{-1}(\xi_{i,j}-\pi_i)  \right\Vert \lessp \sqrt{n m \log n} + n \log n,  \label{eq:lem1_5_10}
	\end{align}
\item[(6)]
	\begin{align}
	 &  \max_{1 \leq j \leq m}\left\Vert\sum_{i=1}^n  \frac{(1-\pi_i)L_i X_i'}{\pi_i} \xi_{i,j} \eps_{i,j} \right\Vert \lessp \sqrt{ n \alpha_n^{-1} \log n}, \label{eq:ex_lem_1} \\
&  \max_{1\leq j \leq m, 1 \leq r \leq R, 1 \leq k \leq K}\left\Vert \sum_{i=1}^n \frac{1-\pi_i}{\pi_i} L_{i,r} \xi_{i,j} X_{i,k}X_i\right\Vert_2 \lessp  \sqrt{ n \alpha_n^{-1} \log n},
\label{eq:422} \\
	 &  \sum_{j=1}^m \left\Vert \sum_{i=1}^n \frac{1-\pi_i}{\pi_i} L_i \xi_{i,j} X_i'\delta_jX_i'\right\Vert^2 \lessp m \alpha_n^{-2} \log^2 n, \label{eq:ex_lem_2} \\
	 &  \max_{1 \leq j \leq m} \left\Vert \sum_{i=1}^n \pi_i^{-1} L_i \xi_{i,j} X_i'\right\Vert \lessp \sqrt{ n \alpha_n^{-1} \log n}, \label{eq:ex_lem_3} \\
  & \max_{1 \leq j \leq m} \left\Vert \sum_{i=1}^n \hat \pi_i^{-1} L_i \xi_{i,j} X_i'\right\Vert \lessp \sqrt{ n \alpha_n^{-1} \log n}, \label{eq:ex_lem_3'} \\
	 & \max_{1 \leq j \leq n} \left\Vert \sum_{i=1}^n \frac{1-\pi_i}{\pi_i}L_i L_i'F_j\xi_{i,j}X_i'\right\Vert \lessp n \log^{1/2} n, \label{eq:ex_lem_4}  \\
	 & \max_{1\leq j \leq m} \left\Vert \sum_{i=1}^n L_i(L_i'F_j)( \xi_{i,j}\pi_i^{-1} - 1) \right\Vert_2 \lessp \sqrt{n \alpha_n^{-1} \log^2 n}, \label{eq:ex_lem_5} \\
	 &  \max_{1\leq j\leq m} \left\Vert \sum_{i=1}^n \pi_i^{-1} L_i \xi_{i,j} \eps_{i,j}\right\Vert_2 \lessp \sqrt{n \alpha_n^{-1} \log n}, \label{eq:ex_lem_6} \\
	 & \max_{1 \leq j \leq m} \left\Vert E_n \hat \pi_i^{-1}  \xi_{i,j} X_i
	 X_i' - \Sigma_X\right\Vert \lessp \sqrt{n \alpha_n^{-1} \log n},  \label{eq:ex_lem_7} \\
	 &  \max_{1\leq s\leq m} \sum_{i=1}^n \sum_{j=1}^m \xi_{i,s} ||L_i||_2^2 ||F_j||_2^2 \xi_{i,j} \lessp nm \alpha_n^2, \label{eq:ex_lem_8} \\
	 & \max_{1\leq s\leq m}\left\Vert\sum_{i=1}^n \sum_{j=1}^m \xi_{i,s}L_i L_i'F_j\xi_{i,j}\left(\frac{1-\pi_i}{\pi_i}X_i'(\hat{\gamma} - \gamma)\right) F_j' \right\Vert  \lessp \sqrt{nm\alpha_n}, \label{eq:ex_lem_9} \\
  	 & \max_{1\leq s\leq m}\left\Vert\sum_{i=1}^n \sum_{j=1}^m (\xi_{i,s}-\pi_i)L_i L_i'F_j\xi_{i,j}\left(\frac{1-\pi_i}{\pi_i}X_i'(\hat{\gamma} - \gamma)\right) F_j' \right\Vert \notag \\
    & \lessp  (m \log n)^{1/2} + n^{1/2} \log^2 n (m \alpha_n)^{-1/2}, \label{eq:ex_lem_9'} \\
	 &  \max_{1\leq s\leq m}\left\Vert\sum_{i=1}^n \sum_{j=1}^m \xi_{i,s}L_i \left(\frac{1-\pi_i}{\pi_i} \right)X_i'(\hat \gamma - \gamma) \xi_{i,j}\eps_{i,j} F_j' \right\Vert \lessp \sqrt{\alpha_n^{-1} \log n}, \label{eq:ex_lem_10} \\
	& \max_{1 \leq s \leq m} \left\Vert \sum_{1\leq j \leq m,j\neq s}\sum_{i=1}^n \eps_{i,s} \xi_{i,s} \xi_{i,j}\left(\frac{(1-\pi_i) }{\pi_i}\right)L_i'F_j F_jX_i'(\hat\gamma - \gamma) \right\Vert_2 \lessp (m \log n)^{1/2},  \label{eq:ex_lem_10'} \\
& \max_{1\leq s \leq m}\left\Vert \sum_{1\leq j \leq m, j\neq s}\sum_{i=1}^n \eps_{i,s} \xi_{i,s} \pi_i^{-1}(\xi_{i,j} - \pi_i)L_i'F_j F_j \right\Vert_2 \lessp (mn \log^2 n)^{1/2},  \label{eq:ex_lem_10''} \\
& \max_{1\leq s \leq m}\left\Vert \sum_{1\leq j \leq m, j \neq s}\sum_{i=1}^n \eps_{i,s} \xi_{i,s} \pi_i^{-1} \xi_{i,j} \eps_{i,j}F_j\right\Vert_2 \lessp (m n \log^2 n)^{1/2}, \label{eq:ex_lem_10'''} \\
&\max_{1\leq s \leq m}\left\Vert \sum_{1\leq j \leq m, j \neq s}\sum_{i=1}^n \eps_{i,s} \xi_{i,s} \frac{1-\pi_i}{\pi_i}X_i'(\hat \gamma - \gamma) \xi_{i,j} \eps_{i,j}F_j\right\Vert_2 \lessp \alpha_n^{-1/2} \log n,  \label{eq:ex_lem_10''''} \\
	 & \max_{1\leq j,s \leq m, j \neq s}\left\Vert \sum_{i=1}^n L_i \xi_{i,s}L_i'F_j \pi_i^{-1} (\xi_{i,j} - \pi_i)\right\Vert_2 \lessp \sqrt{n} \log n + \alpha_n^{-1} n^{1/q} \log n, \label{eq:ex_lem_11}\\
& \max_{1\leq j,s \leq m, j \neq s}\left\Vert \sum_{i=1}^n L_i \xi_{i,s}L_i'F_j \pi_i^{-1} \xi_{i,j} \eps_{i,j}\right\Vert_2 \lessp \sqrt{n} \log n + \alpha_n^{-1} n^{1/q} \log n \label{eq:ex_lem_12} \\
	 &  \max_{1\leq j,s \leq m, j \neq s}\left\Vert \sum_{i=1}^n L_i \xi_{i,s}L_i'F_j \left(\frac{1-\pi_i}{\pi_i}\right)X_i' \xi_{i,j} \eps_{i,j}\right\Vert  \lessp \sqrt{n} \log n + \alpha_n^{-1} n^{1/q} \log n, \label{eq:ex_lem_13} \\
	 &   \max_{1\leq j,s \leq m, j \neq s} \left\Vert \sum_{i=1}^n L_i \xi_{i,s} \pi_i^{-1} \xi_{i,j} X_i'\right\Vert \lessp \sqrt{n} \log n + \alpha_n^{-1} n^{1/q},  \label{eq:ex_lem_16}\\
	 & \max_{1\leq j,s \leq m, j \neq s} \sum_{i=1}^n ||L_i||_2 ||X_i||_2 \pi_i^{-1} \xi_{i,s} \xi_{i,j} \lessp  n \alpha_n n^{1/q}, \label{eq:ex_lem_18} \\
  	 & \max_{1\leq j,s \leq m, j \neq s} \sum_{i=1}^n ||L_i||_2 ||X_i||_2 \pi_i^{-2} \xi_{i,s} \xi_{i,j} \lessp  n^{1+1/q}, \label{eq:ex_lem_18'} \\
	 &  \max_{1\leq j,s \leq m, j \neq s} \left\| \sum_{i=1}^n \eps_{i,s}\xi_{i,s}\xi_{i,j}\frac{1-\pi_i}{\pi_i}X_i L_i'F_j \right\| \lessp \sqrt{n \log n} + \alpha_n^{-1} n^{1/q}, \label{eq:ex_lem_19} \\
&  \max_{1\leq s,j \leq m, s \neq j}\left\|\sum_{i=1}^n \eps_{i,s}\xi_{i,s}\xi_{i,j}\hat \pi_i^{-1}\pi_i^{-1}(1-\pi_i)X_i L_i'F_j\right\| \lessp (n\alpha_n^{-2} \log^2 n)^{1/2}, \label{eq:ex_lem_19'} \\
&\max_{1\leq s \leq m}||\sum_{i=1}^n L_{i} \xi_{i,s}L_i' \pi_i^{-1} (\xi_{i,j} - \pi_i)F|| \lessp  (mn \log n)^{1/2},
	\label{eq:ex_lem_14} \\
 &	\max_{1\leq s \leq m} ||     \sum_{i=1}^n  L_i \pi_i^{-1} \xi_{i,s}L_i'F_j \pi_i^{-1} (\xi_{i,j} - \pi_i)|| \lessp (mn \log n)^{1/2} \alpha_n^{-1},
 \label{eq:ex_lem_14'} \\
 & 	\max_{1\leq s \leq m}||\sum_{i=1}^n L_i \xi_{i,s} \pi_i^{-1} \xi_{i,j} \eps_{i,j}|| \lessp  (mn \log n)^{1/2},  \label{eq:ex_lem_15} \\
& \max_{1\leq s \leq m}||    \sum_{i=1}^n L_i \pi_i^{-1} \xi_{i,s} \pi_i^{-1} \xi_{i,j} \eps_{i,j} || \lessp  (mn \log n)^{1/2} \alpha_n^{-1} , \label{eq:ex_lem_15'} \\
& 	\max_{1\leq s \leq m}||\sum_{i=1}^n L_i \xi_{i,s} \pi_i^{-1} \xi_{i,j} X_i'\delta_j || \lessp  m^{1/2}\alpha_n^{-1/2} \log^{5/4} n, \label{eq:ex_lem_17}\\
& 	\max_{1\leq s \leq m}||\sum_{i=1}^n \pi_i^{-1} L_i \xi_{i,s} \pi_i^{-1} \xi_{i,j} X_i'\delta_j || \lessp  m^{1/2}\alpha_n^{-3/2} \log^{5/4}. \label{eq:ex_lem_17'}
\end{align}
 \end{enumerate}
\end{lem}

\begin{proof}[\textbf{Proof of Lemma \ref{lem: general asymp prop}}]
For \eqref{eq:lem1_1}, we note that $E_n X_iX_i' \xi_{ij} -E X_iX_i' \xi_{i,j} = E_n X_iX_i' (\xi_{ij}-\pi_i) + (E_n - E)\pi_iX_iX_i'$. For the first term on the RHS of the above display, by Bernstein inequality, we have
\begin{align*}
& P\left(\max_{1\leq k_1,k_2 \leq K, 1\leq j \leq m}\left|\sum_{i=1}^n X_{i,k_1}X_{i,k_2}(\xi_{i,j}-\pi_i)  \right| \geq t\bigg|X \right) \\
& \leq \sum_{1\leq k_1,k_2 \leq K, 1\leq j \leq m}P\left(\left|\sum_{i=1}^n X_{i,k_1}X_{i,k_2}(\xi_{i,j}-\pi_i)  \right| \geq t|X \right) \\
& \leq \sum_{1\leq k_1,k_2 \leq K, 1\leq j \leq m} 2\exp\left(-\frac{t^2/2 }{\sum_{i=1}^nX_{i,k_1}^2X_{i,k_2}^2 \pi_{i}  + M_n t/3}\right),
\end{align*}
where $M_n = \max_{1\leq k_1,k_2 \leq K, 1\leq i \leq n}|X_{i,k_1}X_{i,k_2}|$. We have $M_n \leq c \log n$ w.p.a.1. and \newline $\sum_{i=1}^nX_{i,k_1}^2X_{i,k_2}^2 \pi_{i} \lessp n \alpha_n$. Then, by taking $t = C n\alpha_n \log n$ for sufficiently large $C$, we have
\begin{align*}
\max_{1\leq k_1,k_2 \leq K, 1\leq j \leq m}\left|\sum_{i=1}^n X_{i,k_1}X_{i,k_2}(\xi_{i,j}-\pi_i)  \right| \lessp \sqrt{n\alpha_n \log n}.
\end{align*}
In addition, we have
\begin{align*}
\max_{1\leq k_1,k_2 \leq K}\left|\sum_{i=1}^n X_{i,k_1}X_{i,k_2}\pi_i - EX_{i,k_1}X_{i,k_2}\pi_i  \right| \lessp \alpha_n \sqrt{n}.
\end{align*}
This implies
\begin{align*}
\max_{1 \leq j \leq m}||\alpha_n^{-1} E_n X_iX_i' \xi_{ij} - \alpha_n^{-1}E X_iX_i' \xi_{ij} ||\lessp \sqrt{n^{-1} \alpha_n^{-1} \log n}.
\end{align*}
As $\sigma_{K}(\alpha_n^{-1}E X_iX_i' \xi_{ij} ) >0$ and $\log n = o(n \alpha_n)$, we have
\begin{align*}
||(\alpha_n^{-1} E_n X_iX_i' \xi_{ij})^{-1} - (\alpha_n^{-1}E X_iX_i' \xi_{ij})^{-1}|| \lessp \sqrt{n^{-1} \alpha_n^{-1} \log n}.
\end{align*}

We can establish \eqref{eq:lem1_1'} in the same manner.


We can establish \eqref{eq:lem1_3} in part (1) by the same argument used in \eqref{eq:lem1_1} and noticing that $E(\xi_{i,j}|X_i,L_i) = E(\xi_{i,j}|X_i) = \pi_i$ and $E(L_i|X_i) = 0$.


For \eqref{eq:lem1_5} in Part (1), because  $E(\xi_{i,j}|X_i,\eps_{i,j}) = E(\xi_{i,j}|X_i) = \pi_i$, we have
\begin{align*}
E_nX_i \eps_{i,j}\xi_{i,j} = E_nX_i \eps_{i,j}(\xi_{i,j} - \pi_i) + E_n X_i \eps_{i,j}\pi_i.
\end{align*}
Following the argument in the proof of \eqref{eq:lem1_1}, we have
\begin{align*}
\max_{1 \leq j \leq m}||E_nX_i \eps_{i,j}(\xi_{i,j} - \pi_i)|| \lessp \sqrt{n^{-1}\alpha_n \log n}.
\end{align*}
In addition, let $\{e_i\}_{1\leq i \leq n}$ be a sequence of i.i.d. Rademacher random variable that is also independent of the data and $\mathcal{A}_n(C) = \{\max_{1 \leq j \leq n}\sum_{i=1}^n X_{i,k}^2 \eps_{i,j}^2 \pi_i^2 \leq C n \alpha_n^2 \log n\}$. Then, for any $\eta>0$, we can choose sufficiently large constants $C'$ and $C$ such that
\begin{align*}
P(\mathcal{A}_n^c(C)) \leq P(C'\sum_{i=1}^n X_{i,k}^2  \pi_i^2 \geq C n \alpha_n^2) + P( \max_{1 \leq j \leq m, 1\leq i \leq n}|\eps_{i,j}^2| \geq C' \log n ) \leq \eta.
\end{align*}
Then, by van der Vaart (1996, Lemma 2.3.7) we have
\begin{align*}
& \left(1- \max_{1 \leq j \leq m, 1 \leq k_1 \leq K} P\left( \left|\sum_{i=1}^n X_{i,k} \eps_{i,j}\pi_i\right| \geq t/2\bigg|X\right)\right)P\left(\max_{1 \leq j \leq m, 1 \leq k_1 \leq K} \left|\sum_{i=1}^n X_{i,k} \eps_{i,j}\pi_i\right| \geq t \bigg|X\right) \\
& \leq 2P\left(\max_{1 \leq j \leq m, 1 \leq k_1 \leq K} 4 \left|\sum_{i=1}^n e_iX_{i,k} \eps_{i,j}\pi_i\right| \geq t \bigg|X\right) \\
& \leq 2E\left[P\left(\max_{1 \leq j \leq m, 1 \leq k_1 \leq K} 4 \left|\sum_{i=1}^n e_iX_{i,k} \eps_{i,j}\pi_i\right| \geq t \bigg|X, \eps\right)1\{\mathcal{A}_n(C) \}\bigg|X\right] + \eta \\
& \leq 2 \sum_{{1 \leq j \leq m, 1 \leq k_1 \leq K} }E\left[P\left( 4 \left|\sum_{i=1}^n e_iX_{i,k} \eps_{i,j}\pi_i\right| \geq t \bigg|X, \eps\right)1\{\mathcal{A}_n(C)  \}\bigg|X\right] + \eta \\
& \leq 4 \sum_{{1 \leq j \leq m, 1 \leq k_1 \leq K} } \exp\left(- \frac{t^2}{\sum_{i=1}^n X_{i,k}^2 \eps_{i,j}^2 \pi_i^2}\right)1\{\mathcal{A}_n(C)  \} + \eta\\
& \lesssim \exp\left(\log (mK) - \frac{  t^2}{C n \alpha_n^2 \log n}\right) + \eta.
\end{align*}
Let $t = C \sqrt{n \alpha_n^2 \log^2 n }$ with a sufficiently large $C$. Then, by letting $n,m$ diverge to infinity first followed by $\eta$ converging to zero, the RHS of the above display will converge to zero. In addition, we have
\begin{align*}
\max_{1 \leq j \leq m, 1 \leq k_1 \leq K} P\left( \left|\sum_{i=1}^n X_{i,k} \eps_{i,j}\pi_i\right| \geq C \sqrt{n \alpha_n^2 \log^2 n }/2\bigg|X\right) \lessp \frac{n \alpha_n^2}{n \alpha_n^2 \log^2 n} \convP 0.
\end{align*}
This implies
\begin{align}
\max_{1 \leq j \leq m, 1 \leq k_1 \leq K} \left|\sum_{i=1}^n X_{i,k} \eps_{i,j}\pi_i\right| \lessp \sqrt{n \alpha_n^2 \log^2 n }.
\label{eq:eg}
\end{align}
Combining the above two results, we have
\begin{align*}
\max_{1 \leq j \leq m}||E_nX_i \eps_{i,j}\xi_{i,j}|| \lessp \sqrt{n^{-1}\alpha_n \log n}.
\end{align*}


For \eqref{eq:lem1_7} in Part (1), we have
\begin{align}
\max_{1 \leq s \leq m}\left\Vert \sum_{i=1}^n \sum_{j=1}^m \pi_i^{-1}X_i F_j'\xi_{i,j}\xi_{i,s}\eps_{i,j}\right\Vert \leq \max_{1 \leq s \leq m}\left\Vert \sum_{i=1}^n \sum_{j \neq s} \pi_i^{-1}X_i F_j'\xi_{i,j}\xi_{i,s}\eps_{i,j}\right\Vert + \max_{1 \leq s \leq m}\left\Vert \sum_{i=1}^n \pi_i^{-1}X_i F_s'\xi_{i,s}\eps_{i,s}\right\Vert.
\label{eq:lem1_7'}
\end{align}
For the first term of \eqref{eq:lem1_7'}, we note that
\begin{align*}
\max_{1 \leq s \leq m} \sum_{i=1}^n \sum_{j \neq s}Var(\pi_i^{-1}X_{i,k}F_{j,r}\xi_{i,j}\xi_{i,s}\eps_{i,j}|X,F,\eps) \lessp nm
\end{align*}
and
\begin{align*}
\max_{1 \leq s \leq m}\sum_{i=1}^n \sum_{j \neq s} X_{i,k}^2 F_{j,r}^2 \pi_i^2 \eps_{i,j}^2 \lessp n m \alpha_n^2.
\end{align*}
Therefore, we have
\begin{align*}
& \max_{1 \leq s \leq m,  1\leq k \leq K, 1 \leq r \leq R}\left\Vert \sum_{i=1}^n \sum_{j \neq s} \pi_i^{-1}X_{i,k} F_{j,r}'\xi_{i,j}\xi_{i,s}\eps_{i,j}\right\Vert \\
& \leq \max_{1 \leq s \leq m,  1\leq k \leq K, 1 \leq r \leq R}\left\Vert \sum_{i=1}^n \sum_{j \neq s}\pi_i^{-1} X_{i,k} F_{j,r}(\xi_{i,j}\xi_{i,s} - \pi_i^2)\eps_{i,j}\right\Vert + \max_{1 \leq s \leq m,  1\leq k \leq K, 1 \leq r \leq R}\left\Vert \sum_{i=1}^n \sum_{j \neq s} X_{i,k} F_{j,r}\pi_i\eps_{i,j}\right\Vert \\
& \lessp \sqrt{nm \log n} + \log^{5/2} n + \sqrt{n m \alpha_n^2 \log n} \\
& \lessp \sqrt{nm \log n},
\end{align*}
where the second last inequality is due to the Bernstein's inequality.

For the second term on the RHS of \eqref{eq:lem1_7'}, we have
\begin{align*}
\max_{1 \leq s \leq m, 1\leq r_1 \leq r, 1\leq k \leq K} |F_{s,r}| \left\Vert \sum_{i=1}^n \pi_i^{-1} X_{i,k}\xi_{i,s}\eps_{i,s}\right\Vert \lessp \sqrt{ n \alpha_n^{-1} \log^2 n},
\end{align*}
where we use \eqref{eq:lem1_5} and the fact that $\max_{1 \leq s \leq m, 1\leq r_1 \leq r}|F_{r,s}| \lessp \sqrt{\log n}$. Combining the two results, we have established \eqref{eq:lem1_7}. \eqref{eq:lem1_8} can be established in the same manner.

For \eqref{eq:lem1_2_8} is Part (2), we have
\begin{align*}
\delta_j & = (\alpha_n^{-1}E_nX_iX_i'\xi_{i,j})^{-1} \alpha_n^{-1} E_nX_i(L_i'F_j + \eps_{i,j})\xi_{i,j}\\
& = \left[(\alpha_n^{-1}EX_iX_i'\xi_{i,j})^{-1}+O_p(\sqrt{n^{-1}\alpha_n^{-1}\log n})\right] \alpha_n^{-1} E_nX_i(L_i'F_j + \eps_{i,j})\xi_{i,j} \\
& = (\alpha_n^{-1}EX_iX_i'\xi_{i,j})^{-1} \alpha_n^{-1} E_nX_i(L_i'F_j + \eps_{i,j})\xi_{i,j} +  O_p(n^{-1}\alpha_n^{-1}\log^{3/2} n ),
\end{align*}
where we have used \eqref{eq:lem1_1}, \eqref{eq:lem1_3}, \eqref{eq:lem1_5} and the fact that $F_j$ is sub-Gaussian.

Part (3) directly follows Part (2).

For part (4), we have
\begin{align*}
\sum_{j=1}||\delta_j||_2^2 & \lesssim \sum_{j=1}^m ||(\alpha_n^{-1}EX_iX_i'\xi_{i,j})^{-1} \alpha_n^{-1} E_nX_i(L_i'F_j + \eps_{i,j})\xi_{i,j}||_2^2 + O_p( m n^{-2}\alpha_n^{-2}\log^{3} n ) \\
& \lesssim \sum_{j=1}^m ||\alpha_n^{-1} E_nX_i(L_i'F_j + \eps_{i,j})\xi_{i,j}||_2^2 + O_p( m n^{-2}\alpha_n^{-2}\log^{3} n ) \\
& = \alpha_n^{-2}trace\left(\sum_{j=1}^m  (E_nX_iL_i'\xi_{i,j})'(E_nX_iL_i'\xi_{i,j}) F_jF_j'\right) + \alpha_n^{-2} \sum_{j=1}^m (E_n X_i'\eps_{i,j}\xi_{i,j})(E_n X_i\eps_{i,j}\xi_{i,j}) \\
& + O_p( m n^{-2}\alpha_n^{-2}\log^{3} n ) \\
& \lessp m n^{-1} \alpha_n^{-1} \log n,
\end{align*}
where the last inequality is by \eqref{eq:lem1_3} and \eqref{eq:lem1_5}.



\eqref{eq:lem1_5_5} in Part (5) is straightforward. For \eqref{eq:lem1_5_9}, we note that
\begin{align*}
& \max_{1 \leq s \leq m}\left\Vert \sum_{i=1}^n \sum_{1 \leq j \leq m} \pi_i^{-1}L_i F_j' \xi_{i,s} \xi_{i,j} \eps_{i,j}   \right\Vert \\
& \leq \max_{1 \leq s \leq m}\left\Vert \sum_{i=1}^n \pi_i^{-1}L_i F_j' \xi_{i,s} \eps_{i,s}   \right\Vert  + \max_{1 \leq s \leq m}\left\Vert \sum_{i=1}^n \sum_{1 \leq j \leq m, j\neq s} \pi_i^{-1}L_i F_j' \xi_{i,s} \xi_{i,j} \eps_{i,j}   \right\Vert \\
& \lessp \sqrt{n \alpha_n^{-1} \log n} +  \max_{1 \leq s \leq m}\left\Vert \sum_{i=1}^n \sum_{1 \leq j \leq m, j\neq s} \pi_i^{-1}L_i F_j' \xi_{i,s} (\xi_{i,j}-\pi_i) \eps_{i,j}   \right\Vert \\
& +  \max_{1 \leq s \leq m}\left\Vert \sum_{i=1}^n \sum_{j=1}^m L_i F_j' \xi_{i,s} \eps_{i,j}   \right\Vert + \max_{1 \leq j \leq m}\left\Vert \sum_{i=1}^n L_i \xi_{i,j} \eps_{i,j}   \right\Vert_2 ||F_j||_2\\
& \lessp \sqrt{n \alpha_n^{-1} \log n} +  \max_{1 \leq s \leq m}\left\Vert \sum_{i=1}^n \sum_{1 \leq j \leq m, j\neq s} \pi_i^{-1}L_i F_j' \xi_{i,s} (\xi_{i,j}-\pi_i) \eps_{i,j}   \right\Vert \\
& +  \max_{1 \leq s \leq m}\left\Vert \sum_{i=1}^n \sum_{j=1}^m L_i F_j' \xi_{i,s} \eps_{i,j}   \right\Vert.
\end{align*}

For the second term on the RHS of the above display, we note that $\{\xi_{i,j} - \pi_i\}_{1\leq i \leq n, 1\leq j \leq m, j \neq s}$ is independent conditional on the sigma field generated by $(L, F, X, \xi_{\cdot,s}, \eps)$,
\begin{align*}
& \max_{1\leq s \leq m} \sum_{i=1}^n \sum_{j \neq s} \pi_i^{-1} ||L_i||_2^2 ||F_j||_2^2 \eps_{i,j}^2 \xi_{i,s} \lessp \max_{1\leq s \leq m} \left(\sum_{i=1}^n  \pi_i^{-1} \xi_{i,s} ||L_i||_2^2 \right)\left( \max_{1\leq i \leq n} \sum_{j \neq s}||F_j||_2^2 \eps_{i,j}^2 \right) \lessp mn \\
& \max_{1\leq i \leq n, 1\leq j,s \leq m, j \neq s} \pi_i^{-1} |\xi_{i,j}-\pi_i| ||L_i||_2 ||F_j||_2 |\eps_{i,j}| \xi_{i,s} \leq \alpha_n^{-1} n^{1/q}.
\end{align*}
Therefore, by the same argument leads to \eqref{eq:lem6_2}, we have
\begin{align*}
\max_{1 \leq s \leq m}\left\Vert \sum_{i=1}^n \sum_{1 \leq j \leq m, j\neq s} \pi_i^{-1}L_i F_j' \xi_{i,s} (\xi_{i,j}-\pi_i) \eps_{i,j}   \right\Vert  \lessp \sqrt{ nm \log n}.
\end{align*}

In addition, we note that $\{\eps_{i,j}\}_{1\leq i \leq n, 1\leq j \leq m}$ is independent conditionally on $(L, F, X, \xi)$,
\begin{align*}
& \max_{1 \leq s \leq m} \sum_{i=1}^n \sum_{j=1}^m ||L_i||_2^2 ||F_j||_2^2 \xi_{i,s} E(\eps_{i,j}^2|L, F, X, \xi) \lessp nm\alpha_n, \\
& \max_{1 \leq i \leq n, 1 \leq s,j \leq m} ||L_i||_2 ||F_j||_2 \xi_{i,s} |\eps_{i,j}| \lessp \log^{3/2} n.
\end{align*}
Therefore, by the Bernstein's inequality conditional on $(L, F, X, \xi)$, we have
\begin{align*}
\max_{1 \leq s \leq m}\left\Vert \sum_{i=1}^n \sum_{j=1}^m L_i F_j' \xi_{i,s} \eps_{i,j}   \right\Vert \lessp \sqrt{ n m \alpha_n \log n},
\end{align*}
which implies the desired result in \eqref{eq:lem1_5_9}.

For \eqref{eq:lem1_5_10}, we note that
\begin{align*}
& \max_{1 \leq s \leq m}\left\Vert \sum_{i=1}^n \sum_{1 \leq j \leq m} L_i L_i' F_j F_j' \xi_{i,s} \pi_i^{-1}(\xi_{i,j}-\pi_i)  \right\Vert \\
& \lessp\max_{1 \leq s \leq m}\left\Vert \sum_{i=1}^n \sum_{1 \leq j \leq m, j\neq s} L_i L_i' F_j F_j' \xi_{i,s} \pi_i^{-1}(\xi_{i,j}-\pi_i)  \right\Vert + \max_{1 \leq s \leq m}\left\Vert \sum_{i=1}^n  L_i L_i' F_s F_s' \xi_{i,s} \pi_i^{-1}(1-\pi_i)  \right\Vert \\
& \lessp \sqrt{nm \log n} + n \log n,
\end{align*}
where the last inequality is by the Bernstein's inequality conditional on $L, F, X, \{\xi_{i,s}\}_{1\leq i\leq n}$.

The rest of the inequalities in Part (5) can be established in the same manner.

For \eqref{eq:ex_lem_1} in Part (6), we have
\begin{align*}
& \max_{1 \leq j \leq m}\left\Vert\sum_{i=1}^n  \frac{(1-\pi_i)L_i X_i'}{\pi_i} \xi_{i,j} \eps_{i,j} \right\Vert \\
& \leq \max_{1 \leq j \leq m}\left\Vert\sum_{i=1}^n  \frac{(1-\pi_i)L_i X_i'}{\pi_i} (\xi_{i,j} - \pi_i) \eps_{i,j} \right\Vert + \max_{1 \leq j \leq m}\left\Vert\sum_{i=1}^n  (1-\pi_i)L_i X_i' \eps_{i,j} \right\Vert.
\end{align*}
For the first term on the RHS, we note that
\begin{align*}
& \max_{1\leq j\leq m}\sum_{i=1}^n E\left( \left(\frac{(1-\pi_i)L_{i,r} X_{i,k}}{\pi_i} (\xi_{i,j} - \pi_i) \eps_{i,j}\right)^2|X,L,\eps\right) \lessp \alpha_n^{-1} n, \\
& \max_{1\leq j\leq m, 1 \leq i \leq n}\left|\frac{(1-\pi_i)L_{i,r} X_{i,k}}{\pi_i} (\xi_{i,j} - \pi_i) \eps_{i,j}\right| \lessp \alpha_n^{-1} n^{1/q}.
\end{align*}
Therefore, by the Bernstein's inequality conditional on $X,L,\eps$, we have
\begin{align*}
\max_{1 \leq j \leq m}\left\Vert\sum_{i=1}^n  \frac{(1-\pi_i)L_i X_i'}{\pi_i} (\xi_{i,j} - \pi_i) \eps_{i,j} \right\Vert \lessp  \sqrt{ n \alpha_n^{-1} \log n},
\end{align*}
where we use the fact that $q$ is arbitrary and $n\alpha_n$ diverges to infinity in a polynomial rate in $n$.

In addition, we have $\max_{1\leq j\leq m} \sum_{i=1}^n (1-\pi_i)^2 L_{i,r}^2 X_{i,k}^2 \eps_{i,j}^2 \lesssim n $. Following the same argument in the proof of \eqref{eq:eg}, we have
\begin{align*}
 \max_{1 \leq j \leq m}\left\Vert\sum_{i=1}^n  (1-\pi_i)L_i X_i' \eps_{i,j} \right\Vert \lessp \sqrt{ n \log n},
\end{align*}
which leads to the desired result.

Following the proof of \eqref{eq:ex_lem_1}, we can show \eqref{eq:422}.

For \eqref{eq:ex_lem_2}, we have
\begin{align*}
& \sum_{j=1}^m \left\Vert \sum_{i=1}^n \frac{1-\pi_i}{\pi_i} L_i \xi_{i,j} X_i'\delta_jX_i'\right\Vert^2 \leq \sum_{j=1}^m \sum_{1 \leq r \leq R, 1 \leq k \leq K}\left\Vert \sum_{i=1}^n \frac{1-\pi_i}{\pi_i} L_{i,r} \xi_{i,j} X_{i,k}X_i\right\Vert_2^2 ||\delta_j||_2^2.
\end{align*}
Note that $E(L_i|X_i) = 0$.

Then, by \eqref{eq:422}, we have
\begin{align*}
\sum_{j=1}^m \left\Vert \sum_{i=1}^n \frac{1-\pi_i}{\pi_i} L_i \xi_{i,j} X_i'\delta_jX_i'\right\Vert^2 & \lessp n \alpha_n^{-1} \log n \sum_{j=1}^m ||\delta_j||_2^2  \lessp m \alpha_n^{-2} \log^2 n.
\end{align*}

For \eqref{eq:ex_lem_3}, we have
\begin{align*}
\max_{1 \leq j \leq m} \left\Vert \sum_{i=1}^n \pi_i^{-1} L_i \xi_{i,j} X_i'\right\Vert & \leq \max_{1 \leq j \leq m} \left\Vert \sum_{i=1}^n \pi_i^{-1} L_i (\xi_{i,j} - \pi_i) X_i'\right\Vert + \left\Vert \sum_{i=1}^n  L_i X_i'\right\Vert \\
& \lessp \max_{1 \leq j \leq m} \left\Vert \sum_{i=1}^n \pi_i^{-1} L_i (\xi_{i,j} - \pi_i) X_i'\right\Vert + n^{-1/2}.
\end{align*}
In addition, by the Bernstein's inequality conditional on $(L,X)$, we have
\begin{align*}
\max_{1 \leq j \leq m} \left\Vert \sum_{i=1}^n \pi_i^{-1} L_i (\xi_{i,j} - \pi_i) X_i'\right\Vert \lessp \sqrt{ n \alpha_n^{-1} \log n}.
\end{align*}
This leads to the desired result.

For \eqref{eq:ex_lem_3'}, we have
\begin{align*}
& \max_{1 \leq j \leq m} \left\Vert \sum_{i=1}^n \hat \pi_i^{-1} L_i \xi_{i,j} X_i'\right\Vert \\
&  \lessp \sqrt{ n \alpha_n^{-1} \log n} + \max_{1 \leq j \leq m} \left\Vert \sum_{i=1}^n (\hat \pi_i^{-1} - \pi_i^{-1}) L_i \xi_{i,j} X_i'\right\Vert \\
& \lessp \sqrt{ n \alpha_n^{-1} \log n} + \alpha_n^2(nm)^{-1} n^{1/q} \max_{1 \leq j \leq m} \sum_{i=1}^n \| L_i\| \xi_{i,j} \|X_i\| \\
& + \max_{1 \leq j \leq m} \left\Vert \sum_{i=1}^n \frac{1-\pi_i}{\pi_i} L_i \xi_{i,j} X_i X_i' (\hat \gamma - \gamma)\right\Vert \\
& \lessp \sqrt{ n \alpha_n^{-1} \log n}.
\end{align*}

For \eqref{eq:ex_lem_4}, we note that
\begin{align*}
& \max_{1 \leq j \leq n} \left\Vert \sum_{i=1}^n \frac{1-\pi_i}{\pi_i}L_i L_i'F_j\xi_{i,j}X_i'\right\Vert \\
& \leq \max_{1 \leq j \leq m} \left|\sum_{i=1}^n \frac{1 - \pi_i}{\pi_i} ||L_i||_2^2 ||X_i||_2 ||F_j||_2 (\xi_{i,j} - \pi_i)\right| + \max_{1 \leq j \leq m} \sum_{i=1}^n ||L_i||_2^2 ||X_i||_2 ||F_j||_2 \\
& \lessp \sqrt{n \alpha_n^{-1} \log^2 n} + n \log^{1/2} n \lessp n \log^{1/2} n.
\end{align*}

We can establish \eqref{eq:ex_lem_5} by the Bernstein's inequality conditional on $(L,F,X)$. \eqref{eq:ex_lem_6} can be established in the same manner as \eqref{eq:ex_lem_1}.

For \eqref{eq:ex_lem_7}, we have
\begin{align*}
& \max_{1 \leq j \leq m} \left\Vert E_n \hat \pi_i^{-1}  \xi_{i,j} X_i
X_i' - \Sigma_X\right\Vert \\
& \leq \max_{1 \leq j \leq m} \left\Vert E_n (\hat \pi_i^{-1} - \pi_i^{-1})  \xi_{i,j} X_i
X_i'\right\Vert + \max_{1 \leq j \leq m} \left\Vert E_n \pi_i^{-1}  \xi_{i,j} X_i
X_i' - \Sigma_X\right\Vert \\
& \lessp \alpha_n^{-2} n^{1/q} (nm)^{-1} \max_{1 \leq j \leq m} E_n \xi_{i,j}||X_i||_2^2 + \max_{1 \leq j \leq m} \left\Vert E_n \left(\frac{1 - \pi_i}{\pi_i}\right)X_i'(\hat{\gamma} - \gamma) \xi_{i,j} X_i
X_i'\right\Vert + \sqrt{n^{-1} \alpha_n^{-1} \log n} \\
& \lessp\sqrt{n^{-1} \alpha_n^{-1} \log n},
\end{align*}
where we use the facts that
\begin{align*}
&  \max_{1 \leq j \leq m} \left\Vert E_n \left(\frac{1 - \pi_i}{\pi_i}\right)X_i'(\hat{\gamma} - \gamma) \xi_{i,j} X_i
X_i'\right\Vert \leq  \max_{1 \leq j \leq m} E_n \pi_i^{-1}\xi_{i,j} ||X_i||_2^3 ||\hat \gamma - \gamma||_2 \lessp (nm\alpha_n)^{-1/2}, \\
& \max_{1 \leq j \leq m} \left\Vert E_n \pi_i^{-1}  \xi_{i,j} X_i
X_i' - \Sigma_X\right\Vert \lessp\sqrt{n^{-1} \alpha_n^{-1} \log n}.
\end{align*}

For \eqref{eq:ex_lem_8}, we note that
\begin{align*}
& \max_{1\leq s\leq m} \sum_{i=1}^n \sum_{j=1}^m \xi_{i,s} ||L_i||_2^2 ||F_j||_2^2 \xi_{i,j}\\ & \lessp 	\max_{1\leq s\leq m} \sum_{i=1}^n \sum_{j \neq s} \xi_{i,s} ||L_i||_2^2 ||F_j||_2^2 \xi_{i,j} + n\alpha_n \\
&\lessp 	\max_{1\leq s\leq m} \left\vert \sum_{i=1}^n \sum_{j \neq s} \xi_{i,s} ||L_i||_2^2 ||F_j||_2^2 (\xi_{i,j} - \pi_i)\right\vert + \max_{1 \leq j \leq m} \left\vert \sum_{i=1}^n \xi_{i,j} ||L_i||_2^2  \pi_{i} \right\vert (\sum_{j =1}^m ||F_j||_2^2 ) + n\alpha_n \\
& \lessp 	\max_{1\leq s\leq m} \left\vert \sum_{i=1}^n \sum_{j \neq s} \xi_{i,s} ||L_i||_2^2 ||F_j||_2^2 (\xi_{i,j} - \pi_i)\right\vert+ nm\alpha_n^2.
\end{align*}
We note that $\{\xi_{i,j}\}_{i=1,\cdots,n, 1\leq j \leq m, j \neq s}$ is independent conditional on $\mathcal{F}_s'$, where $\mathcal{F}_s'$ is the sigma field generated by $(L,X,F,\{\xi_{i,s}\}_{1\leq i \leq n})$. In addition, we have
\begin{align*}
& \max_{1\leq s \leq m}\sum_{i=1}^n \sum_{j \neq s} \xi_{i,s} \pi_i ||L_i||_2^4 ||F_j||_2^4 \leq \sum_{i=1}^n \sum_{j =1}^m  \xi_{i,j} \pi_i ||L_i||_2^4 ||F_j||_2^4 \lessp \alpha_n^2 nm
\end{align*}
and
\begin{align*}
& \max_{1\leq i\leq n, 1\leq j,s \leq m, j \neq s} |\xi_{i,j} - \pi_i| \xi_{i,s} ||L_i||_2^4 ||F_j||_2^4 \leq \max_{1\leq i \leq n, 1\leq j \leq m}  ||L_i||_2^4 ||F_j||_2^4 \lessp  n^{1/q}.
\end{align*}
Then, following the argument that leads to \eqref{eq:lem6_2} and the Bernstein's inequality, we have
\begin{align*}
	\max_{1\leq s\leq m} \left\vert \sum_{i=1}^n \sum_{j \neq s} \xi_{i,s} ||L_i||_2^2 ||F_j||_2^2 (\xi_{i,j} - \pi_i)\right\vert \lessp \sqrt{nm \alpha_n^2 \log n},
\end{align*}
which leads to the desired result.

For \eqref{eq:ex_lem_9}, following the similar argument in \eqref{eq:ex_lem_8} and by Proposition \ref{prop:pihat}, we have
\begin{align*}
& \max_{1\leq s\leq m}\left\Vert\sum_{i=1}^n \sum_{j=1}^m \xi_{i,s}L_i L_i'F_j\xi_{i,j}\left(\frac{1-\pi_i}{\pi_i}X_i'(\hat{\gamma} - \gamma)\right) F_j' \right\Vert \\
&\lessp \max_{1\leq s\leq m}\sum_{i=1}^n \sum_{j=1}^m \xi_{i,s}||L_i||_2^2 ||F_j||_2^2 \xi_{i,j}\pi_i^{-1}||X_i||_2||\hat{\gamma} - \gamma||_2 \\
& \lessp \sqrt{nm\alpha_n}.
\end{align*}

For \eqref{eq:ex_lem_9'}, we have
\begin{align*}
& \max_{1\leq s\leq m}\left\Vert\sum_{i=1}^n \sum_{j=1}^m (\xi_{i,s} - \pi_i)L_i L_i'F_j\xi_{i,j}\left(\frac{1-\pi_i}{\pi_i}X_i'(\hat{\gamma} - \gamma)\right) F_j' \right\Vert \\
&\lesssim \max_{1\leq s\leq m}\left\Vert\sum_{i=1}^n \sum_{j \neq s} (\xi_{i,s} - \pi_i)L_i L_i'F_j\xi_{i,j}\left(\frac{1-\pi_i}{\pi_i}X_i'(\hat{\gamma} - \gamma)\right) F_j' \right\Vert\\
& + \max_{1\leq s\leq m}\left\Vert\sum_{i=1}^n \xi_{i,s}( 1- \pi_i)L_i L_i'F_s\left(\frac{1-\pi_i}{\pi_i}X_i'(\hat{\gamma} - \gamma)\right) F_s' \right\Vert \\
& \lessp \max_{1\leq s\leq m}\left\Vert\sum_{i=1}^n \sum_{j \neq s} (\xi_{i,s} - \pi_i)L_i L_i'F_j(\xi_{i,j} - \pi_i)\left(\frac{1-\pi_i}{\pi_i}X_i'(\hat{\gamma} - \gamma)\right) F_j' \right\Vert\\
& + \max_{1\leq s\leq m}\left\Vert\sum_{i=1}^n \sum_{j \neq s} (\xi_{i,s} - \pi_i)L_i L_i'F_j\left((1-\pi_i)X_i'(\hat{\gamma} - \gamma)\right) F_j' \right\Vert\\
& + \max_{1\leq s\leq m}\left\Vert\sum_{i=1}^n \xi_{i,s}( 1- \pi_i)L_i L_i'F_s\left(\frac{1-\pi_i}{\pi_i}X_i'(\hat{\gamma} - \gamma)\right) F_s' \right\Vert \\
& \lessp ((mn)^{1/2} + (n \alpha_n \log n)^{1/2} m + n \log^2 n)(mn\alpha_n)^{-1/2} \\
& \lessp (m \log n)^{1/2} + n^{1/2} \log^2 n (m \alpha_n)^{-1/2}.
\end{align*}

For \eqref{eq:ex_lem_10}, we have
\begin{align}
& \max_{1\leq s\leq m}\left\Vert\sum_{i=1}^n \sum_{j=1}^m \xi_{i,s}L_i \left(\frac{1-\pi_i}{\pi_i} \right)X_i'(\hat \gamma - \gamma) \xi_{i,j}\eps_{i,j} F_j' \right\Vert \notag \\
& \lessp \max_{1\leq s\leq m, 1\leq r_1,r_2 \leq R} \left\Vert\sum_{i=1}^n \sum_{j=1}^m \xi_{i,s}L_{i,r_1} \left(\frac{1-\pi_i}{\pi_i} \right)X_i \xi_{i,j}\eps_{i,j} F_{j,r_2} \right\Vert_2 ||\hat \gamma - \gamma||_2 \notag \\
& \lessp (nm\alpha_n)^{-1/2} \left[\max_{1\leq s\leq m, 1\leq r_1,r_2 \leq R} \left\Vert\sum_{i=1}^n \sum_{j \neq s} \xi_{i,s}L_{i,r_1} \left(\frac{1-\pi_i}{\pi_i} \right)X_i \xi_{i,j}\eps_{i,j} F_{j,r_2} \right\Vert_2\right] \notag \\
& + (nm\alpha_n)^{-1/2} \left[\max_{1\leq s\leq m, 1\leq r_1,r_2 \leq R} \left\Vert\sum_{i=1}^n \xi_{i,s}L_{i,r_1} \left(\frac{1-\pi_i}{\pi_i} \right)X_i\eps_{i,s} F_{s,r_2} \right\Vert_2 \right] \notag \\
& \lessp (nm\alpha_n)^{-1/2} \left[\max_{1\leq s\leq m, 1\leq r_1,r_2 \leq R} \left\Vert\sum_{i=1}^n \sum_{j \neq s} \xi_{i,s}L_{i,r_1} \left(\frac{1-\pi_i}{\pi_i} \right)X_i \xi_{i,j}\eps_{i,j} F_{j,r_2} \right\Vert_2\right] + m^{-1/2} \alpha_n^{-1} \log n.
\label{eq:exlem_10_1}
\end{align}
In addition, we have
\begin{align*}
& \max_{1\leq s\leq m, 1\leq r_1,r_2 \leq R} \left\Vert\sum_{i=1}^n \sum_{j \neq s} \xi_{i,s}L_{i,r_1} \left(\frac{1-\pi_i}{\pi_i} \right)X_i \xi_{i,j}\eps_{i,j} F_{j,r_2} \right\Vert_2 \\
& \lessp \max_{1\leq s\leq m, 1\leq r_1,r_2 \leq R} \left\Vert\sum_{i=1}^n \sum_{j \neq s} \xi_{i,s}L_{i,r_1} \left(\frac{1-\pi_i}{\pi_i} \right)X_i (\xi_{i,j} - \pi_i)\eps_{i,j}F_{j,r_2} \right\Vert_2 \\
& + \max_{1\leq s\leq m, 1\leq r_1,r_2 \leq R} \left\Vert\sum_{i=1}^n \sum_{j \neq s} \xi_{i,s}L_{i,r_1} \left(1-\pi_i \right)X_i \eps_{i,j} F_{j,r_2} \right\Vert_2.
\end{align*}
We note that
\begin{align*}
& \max_{1\leq s\leq m, 1\leq r_1,r_2 \leq R} \sum_{i=1}^n \sum_{j \neq s} \xi_{i,s} ||L_{i}||_2^2 \left(\frac{1-\pi_i}{\pi_i} \right)^2 ||X_i||_2^2 \pi_i \eps^2_{i,j} ||F_{j}||_2^2 \lessp mn \\
& \max_{1\leq i \leq n, 1\leq s,j\leq m,j\neq s, 1\leq r_1,r_2 \leq R}  \xi_{i,s} ||L_{i}||_2 \left(\frac{1-\pi_i}{\pi_i} \right) ||X_i||_2 |\xi_{i,j}- \pi_i| |\eps_{i,j}| ||F_{j}||_2 \lessp \alpha_n^{-1} n^{1/q}.
\end{align*}
Therefore, by the same argument that leads to \eqref{eq:lem6_2} and the Bernstein's inequality, we have
\begin{align*}
\max_{1\leq s\leq m, 1\leq r_1,r_2 \leq R} \left\Vert\sum_{i=1}^n \sum_{j \neq s} \xi_{i,s}L_{i,r_1} \left(\frac{1-\pi_i}{\pi_i} \right)X_i (\xi_{i,j} - \pi_i)\eps_{i,j}F_{j,r_2} \right\Vert_2 \lessp \sqrt{nm \log n}.
\end{align*}

Similarly, we note that $\{\eps_{i,j}\}_{1 \leq i\leq n, 1 \leq j \leq m, j \neq s}$ is independent conditionally on the sigma field generated by $(L,F,X,\xi)$,
\begin{align*}
& \max_{1\leq s\leq m, 1\leq r_1,r_2 \leq R} \sum_{i=1}^n \sum_{j \neq s} \xi_{i,s} ||L_{i}||_2^2 ||X_i||_2^2 E(\eps^2_{i,j}|L,F,X,\xi) ||F_{j}||_2 \lessp  n m \alpha_n, \\
& \max_{1\leq s\leq m, 1\leq r_1,r_2 \leq R}  \xi_{i,s} ||L_{i}||_2 ||X_i||_2 |\eps_{i,j}| ||F_{j}|| \lessp  \log^2 n.
\end{align*}

Therefore, by the same argument that leads to \eqref{eq:lem6_2} and the Bernstein's inequality, we have
\begin{align*}
\max_{1\leq s\leq m, 1\leq r_1,r_2 \leq R} \left\Vert\sum_{i=1}^n \sum_{j \neq s} \xi_{i,s}L_{i,r_1} \left(1-\pi_i \right)X_i \eps_{i,j} F_{j,r_2} \right\Vert_2 \lessp \sqrt{nm \alpha_n \log n},
\end{align*}
which implies
\begin{align*}
\max_{1\leq s\leq m, 1\leq r_1,r_2 \leq R} \left\Vert\sum_{i=1}^n \sum_{j \neq s} \xi_{i,s}L_{i,r_1} \left(\frac{1-\pi_i}{\pi_i} \right)X_i \xi_{i,j}\eps_{i,j} F_{j,r_2} \right\Vert_2 \lessp \sqrt{nm \log n}.
\end{align*}
Combining this with \eqref{eq:exlem_10_1}, we obtain the desired result.

For \eqref{eq:ex_lem_10'}, we note that
\begin{align*}
& \max_{1 \leq s \leq m} \left\Vert \sum_{1\leq j \leq m,j\neq s}\sum_{i=1}^n \eps_{i,s} \xi_{i,s} \xi_{i,j}\left(\frac{(1-\pi_i) }{\pi_i}\right)L_i'F_j F_jX_i'(\hat\gamma - \gamma) \right\Vert_2\\	
& \leq \max_{1 \leq s \leq m} \left\Vert \sum_{1\leq j \leq m,j\neq s}\sum_{i=1}^n \eps_{i,s} \xi_{i,s} \xi_{i,j}\left(\frac{(1-\pi_i) }{\pi_i}\right)L_i'F_j F_jX_i'\right\Vert ||(\hat\gamma - \gamma) ||_2 \\
& \lessp \max_{1 \leq s \leq m, 1\leq r_1,r_2 \leq r, 1\leq d_1 \leq d} \left\vert \sum_{i=1}^n \eps_{i,s} \xi_{i,s} \left(\frac{1-\pi_i}{\pi_i}\right)L_{i,r_1}X_{i,d_1} (\sum_{1\leq j \leq m,j\neq s} \xi_{i,j}F_{j,r_2} F_{j,r_1})\right\vert ||(\hat\gamma - \gamma) ||_2 \\
& \lessp (m \log n)^{1/2},
\end{align*}
where the last inequality is by the Bernstein's inequality and the fact that
$$\max_{i,s,r_1,r_2}|\sum_{1\leq j \leq m,j\neq s} \xi_{i,j}F_{j,r_1} F_{j,r_2}| \lessp \alpha_n m.$$

For \eqref{eq:ex_lem_10''}, we note that
\begin{align*}
 \max_{1\leq s \leq m, 1\leq r_1,r_2 \leq r} |\sum_{1\leq j \leq m, j\neq s} (\xi_{i,j} - \pi_i)F_{j,r_1}F_{j,r_2} | \lessp (m \alpha_n \log n)^{1/2}.
\end{align*}
Then, by the Bernstein's inequality conditional on $(X,L,F)$ and $\{\xi_{i,j}\}_{ j \neq s}$, we have
\begin{align*}
& \max_{1\leq s \leq m}\left\Vert \sum_{1\leq j \leq m, j\neq s}\sum_{i=1}^n \eps_{i,s} \xi_{i,s} \pi_i^{-1}(\xi_{i,j} - \pi_i)L_i'F_j F_j \right\Vert_2 \\
& \lesssim  \max_{1\leq s \leq m,1\leq r_1,r_2,r_3 \leq r}\left\Vert \sum_{i=1}^n \eps_{i,s} \xi_{i,s} \pi_i^{-1}L_{i,r_3} \left[\sum_{1\leq j \leq m, j\neq s} (\xi_{i,j} - \pi_i) F_{j,r_1} F_{j,r_2} \right]\right\Vert_2\lessp (mn \log^2 n)^{1/2}.
\end{align*}

We can establish \eqref{eq:ex_lem_10'''} and \eqref{eq:ex_lem_10''''} in the same manner by noticing that
\begin{align*}
    \max_{1\leq i \leq n, 1\leq s \leq m, 1\leq r_1 \leq r} ||\sum_{j \neq s} \xi_{i,j}\eps_{i,j}F_{j,r_1}|| \lessp (m \alpha_n \log n)^{1/2}.
\end{align*}

For \eqref{eq:ex_lem_11}, we note that $\{\xi_{i,j} - \pi_i\}_{1\leq i \leq n}$ is independent conditional on $(X,F,L,\xi_{i,s})$ when $s\neq j$. In addition, we have
\begin{align*}
& \max_{1\leq s \leq n}\sum_{i=1}^{n} ||L_i||_2^2\pi_i^{-1}  \xi_{i,s} \lessp n  \\
& \max_{1\leq j,s \leq n, s \neq j, 1\leq 1\leq n}||L_i||_2\pi_i^{-1}  \xi_{i,s}|\xi_{i,s} - \pi_i| \lessp \alpha_n^{-1} n^{1/q}.
\end{align*}
Then, by the conditional Bernstein's inequality, we have
\begin{align*}
& \max_{1\leq j,s \leq m, j \neq s}\left\Vert \sum_{i=1}^n L_i \xi_{i,s}L_i'F_j \pi_i^{-1} (\xi_{i,j} - \pi_i)\right\Vert_2 \\
& \lessp \max_{1\leq j,s \leq m, j \neq s}\left\Vert \sum_{i=1}^n L_i \xi_{i,s}L_i' \pi_i^{-1} (\xi_{i,j} - \pi_i)\right\Vert \log^{1/2} n \\
& \lessp \sqrt{n} \log n + \alpha_n^{-1} n^{1/q} \log n
\end{align*}

For \eqref{eq:ex_lem_12}, we note that
\begin{align*}
& \max_{1\leq j,s \leq m, j \neq s}\left\Vert \sum_{i=1}^n L_i \xi_{i,s}L_i'F_j \pi_i^{-1} \xi_{i,j} \eps_{i,j}\right\Vert_2 \\
& \leq \max_{1\leq j,s \leq m, j \neq s}\left\Vert \sum_{i=1}^n L_i (\xi_{i,s}  \xi_{i,j} - \pi_i^2)L_i'F_j \pi_i^{-1} \eps_{i,j}\right\Vert_2 + \max_{1\leq j,s \leq m, j \neq s}\left\Vert \sum_{i=1}^n L_i L_i'F_j \pi_i \eps_{i,j}\right\Vert_2 \\
& \lessp \sqrt{n} \log n + \alpha_n^{-1} n^{1/q} \log n + \max_{1\leq j,s \leq m, j \neq s}\left\Vert \sum_{i=1}^n L_i L_i'F_j \pi_i \eps_{i,j}\right\Vert_2\\
& \lessp \sqrt{n} \log n + \alpha_n^{-1} n^{1/q} \log n,
\end{align*}
where the second inequality follows the same argument in the proof of \eqref{eq:ex_lem_11} and the last inequality is by the Bernstein's inequality conditional on $(L,F,X)$.

We note \eqref{eq:ex_lem_13} can be established in the same manner above.

To see \eqref{eq:ex_lem_16}, we note
\begin{align*}
\max_{1\leq j,s \leq m, j \neq s} \left\Vert \sum_{i=1}^n L_i \xi_{i,s} \pi_i^{-1} \xi_{i,j} X_i'\right\Vert & \leq \max_{1\leq j,s \leq m, j \neq s} \left\Vert \sum_{i=1}^n L_i \xi_{i,s} \pi_i^{-1} (\xi_{i,j}- \pi_i) X_i'\right\Vert + \max_{1\leq s \leq m} \left\Vert \sum_{i=1}^n L_i \xi_{i,s} X_i'\right\Vert \\
& \lessp \max_{1\leq j,s \leq m, j \neq s} \left\Vert \sum_{i=1}^n L_i \xi_{i,s} \pi_i^{-1} (\xi_{i,j}- \pi_i) X_i'\right\Vert + \sqrt{ n \alpha_n \log n} \\
& \lessp \sqrt{n} \log n + \alpha_n^{-1} n^{1/q} \log n,
\end{align*}
where the last inequality follows the same argument in the proof of Lemma \eqref{eq:ex_lem_12}.

To see \eqref{eq:ex_lem_18}, we note that
\begin{align*}
& \max_{1\leq j,s \leq m, j \neq s} \sum_{i=1}^n ||L_i||_2 ||X_i||_2 \pi_i^{-1} \xi_{i,s} \xi_{i,j}\\
& \leq \max_{1\leq j,s \leq m, j \neq s} \left\vert \sum_{i=1}^n ||L_i||_2 ||X_i||_2 \pi_i^{-1} (\xi_{i,s} \xi_{i,j}- \pi_i^2)\right\vert + \sum_{i=1}^n ||L_i||_2 ||X_i||_2 \pi_i \\
& \lessp  \sqrt{n \log n} + \alpha_n^{-1} n^{1/q}  + n\alpha_n \lessp   n \alpha_n n^{1/q}.
\end{align*}

We can establish \eqref{eq:ex_lem_18'} in the same manner.

\eqref{eq:ex_lem_19} can be established in the same manner of \eqref{eq:ex_lem_12}.

For \eqref{eq:ex_lem_19'}, we have
\begin{align*}
& \max_{1\leq s,j \leq m, s \neq j}\left\|\sum_{i=1}^n \eps_{i,s}\xi_{i,s}\xi_{i,j}\hat \pi_i^{-1}\pi_i^{-1}(1-\pi_i)X_i L_i'F_j\right\| \\
& \leq \max_{1\leq s,j \leq m, s \neq j}\left\|\sum_{i=1}^n \eps_{i,s}\xi_{i,s}\xi_{i,j}\pi_i^{-1}\pi_i^{-1}(1-\pi_i)X_i L_i'F_j\right\| \\
& + \alpha_n^{-2} (nm)^{-1} n^{1/q} \max_{1\leq s,j \leq m, s \neq j}\sum_{i=1}^n |\eps_{i,s}| \xi_{i,s}\xi_{i,j} \pi_i^{-1}(1-\pi_i) ||X_i|| ||L_i||||F_j|| \\
& + \max_{1\leq s,j \leq m, s \neq j}\left\|\sum_{i=1}^n \eps_{i,s}\xi_{i,s}\xi_{i,j}\pi_i^{-2}(1-\pi_i)^2X_i X_i'(\hat \gamma - \gamma) L_i'F_j\right\| \\
& \lessp (n\alpha_n^{-2} \log^2 n)^{1/2}.
\end{align*}

For \eqref{eq:ex_lem_14}, let $A^{(s)}$ be a matrix with its typical entry $\sumi L_{i,r} \xi_{i,s}L_i'F_j \pi_i^{-1} (\xi_{i,j} - \pi_i)$. we first define $A^{(s,1)}$ as the same as $A^{(s)}$ except its $s$th column, which is instead just zero. We further define $A^{(s,0)} = A^{(s)} - A^{(s,1)}$ which contains all zero entries except its $s$th column. Then, we have
\begin{align}
\max_{1\leq s \leq m}||A^{(s)}|| \leq \max_{1\leq s \leq m} ||A^{(s,1)}|| + \max_{1\leq s \leq m} ||A^{(s,0)}||_F.
\label{eq:As}
\end{align}
We first bound $ \max_{1\leq s \leq m} ||A^{(s,0)}||_F$. Note that
\begin{align}
\max_{1\leq s \leq m} ||A^{(s,0)}||_F = \max_{1\leq s \leq m}\left\Vert \sum_{i=1}^n L_i \xi_{i,s}L_i'F_s \pi_i^{-1} (\xi_{i,s} - \pi_i)\right\Vert_2 \lessp n \log^{1/2} n.
\label{eq:As0}
\end{align}
Next, we bound $||A^{(s,1)}||$. Note that $A^{(s,1)} = \sum_{i=1}^{n} A_{i}^{(s,1)}$ where $A_{i}^{(s,1)}$ is a $R \times m$ matrix with its $j$th column being $L_i \xi_{i,s}L_i'F_j \pi_i^{-1} (\xi_{i,j} - \pi_i)$ if $j \neq s$ and $0_{m}$ otherwise where $0_m$ is an $m\times 1$ vector of zeros. Then, we see that $A_{i}^{(s,1)}$ is independent across $i$ and mean-zero conditional on $(L,F,X)$. In addition, we have
\begin{align*}
& \max_{1\leq i \leq n, 1\leq s\leq m} ||A_i^{(s,1)}|| \leq  \max_{1\leq i \leq n, 1\leq s\leq m} ||A_i^{(s,1)}||_F \lessp \sqrt{m \alpha_n^{-1}} n^{1/q}, \\
& \max_{1\leq s \leq m} \left\Vert \sum_{i=1}^{n} E (A_iA_i'|L,F,X )\right\Vert \leq \max_{1\leq s\leq m} \left\Vert \sum_{i=1}^{n} \sum_{j \neq s}  E\left( L_iL_i' \xi_{i,s}^2 (L_i'F_j)^2 \pi_i^{-2}(\xi_{i,j} - \pi_i)^2|L,F,X\right)\right\Vert \lessp nm,\\
& \max_{1\leq s \leq m} \left\Vert \sum_{i=1}^{n} E (A_i'A_i|L,F,X)\right\Vert \leq \max_{1\leq j,s \leq m, j \neq s}\left\Vert \sum_{i=1}^{n}E\left( ||L_i||_2^2 \xi_{i,s}^2 (L_i'F_j)^2 \pi_i^{-2}(\xi_{i,j} - \pi_i)^2|L,F,X\right)\right\Vert \lessp n.
\end{align*}
Therefore, by the matrix Bernstein inequality (\citet[Theorem 1.6]{T12}), we have
\begin{align*}
\max_{1\leq s\leq m}||A^{(s,1)}|| \lesssim  (mn \log n)^{1/2}.
\end{align*}
This leads to the desired result.

We can establish \eqref{eq:ex_lem_14'} in the same manner.

For \eqref{eq:ex_lem_15}, we note that
\begin{align*}
 \sum_{i=1}^n L_i \xi_{i,s} \pi_i^{-1} \xi_{i,j} \eps_{i,j}  =   \sum_{i=1}^n L_i \xi_{i,s} \pi_i^{-1} (\xi_{i,j}-\pi_i) \eps_{i,j} + \sum_{i=1}^n L_i \xi_{i,s} \eps_{i,j}.
\end{align*}
We note that $\max_{1\leq s,j\leq m}||\sum_{i=1}^n L_i \xi_{i,s} \eps_{i,j}||_2 \lessp \sqrt{ n \alpha_n \log n}$ so that
\begin{align*}
\max_{1\leq s \leq m}\sum_{1\leq j \leq m}||\sum_{i=1}^n L_i \xi_{i,s} \eps_{i,j}||_2^2 \lesssim m n \alpha_n \log n.
\end{align*}
In addition, let $A^{(s)} \in \Re^{R \times m}$ be a matrix with its $j$th column being $\sum_{i=1}^n L_i \xi_{i,s} \pi_i^{-1} (\xi_{i,j}-\pi_i) \eps_{i,j}$. Then, following the same argument in the proof of \eqref{eq:ex_lem_14}, we have
\begin{align*}
\max_{1\leq s \leq m}||A^{(s)}|| \lessp (mn \log n)^{1/2},
\end{align*}
which implies the desired result.

We can establish \eqref{eq:ex_lem_15'} in the same manner.

For \eqref{eq:ex_lem_17}, recall $B_{\cdot,j}^{(s,4,1)} =  \sum_{i=1}^n L_i \xi_{i,s} \pi_i^{-1} \xi_{i,j} X_i'\delta_j $. Then, we note that, for $t=1,\cdots,d$,
\begin{align*}
B_{\cdot,j}^{(s,4,1)} & = \sum_{t=1}^d \sum_{i=1}^n \left(L_i \xi_{i,s} \pi_i^{-1}(\xi_{i,j} - \pi_i) X_{i,t}\right)\delta_{j,t} + \sum_{i=1}^n L_i \xi_{i,s} X_i'\delta_j \\
& \equiv \sum_{t=1}^d B_{\cdot,j}^{(s,4,1,t)} \delta_{j,t} + B_{\cdot,j}^{(s,4,1,0)}.
\end{align*}
Define $B^{(s,4,1,t)} = ( B_{\cdot,1}^{(s,4,1,t)},\cdots, B_{\cdot,m}^{(s,4,1,t)})$ for $t =0,\cdots,d$. Following the above argument, we can show
\begin{align*}
& \max_{1\leq s \leq m, 1\leq t\leq d}||B^{(s,4,1,t)}|| \lessp (nm \log n)^{1/2}.
\end{align*}
In addition, by \eqref{eq:lem1_9}, we have
\begin{align*}
& \max_{1\leq s \leq m} \left\Vert \sum_{i=1}^n L_i \xi_{i,s} X_i'\delta_j \right\Vert_2 \leq \max_{1\leq s \leq m} \left\Vert \sum_{i=1}^n L_i \xi_{i,s} X_i' \right\Vert \max_{1\leq j \leq m}||\delta_j||_2 \lessp \log^{5/4} n.
\end{align*}
This implies
\begin{align*}
\max_{1\leq s \leq m} ||B^{(s,4,1,0)}|| \leq \max_{1\leq s \leq m} ||B^{(s,4,1,0)}||_F \lessp m^{1/2} \log^{5/4} n,
\end{align*}
and thus,
\begin{align*}
\max_{1\leq s \leq m} ||B^{(s,4,1)}|| \leq \sum_{t=1}^d ||B^{(s,4,1,d)}|| \max_{1\leq  j \leq m} ||\delta_j||_2 + \max_{1\leq s \leq m} ||B^{(s,4,1,0)}|| \lessp m^{1/2}\alpha_n^{-1/2} \log^{5/4}.
\end{align*}

We can establish \eqref{eq:ex_lem_17'} in the same manner.

\end{proof}

\begin{lem}
\label{lem: part 2}\label{lem: order of e}Let Assumption \ref{assu: moment cond} hold. Then
$||e||\lessp \sqrt{\alpha_n^{-1} (m+n)}n^{1/q}$ for any $q>0$ and $||\Xi \circ \eps|| \lessp \sqrt{(m+n)\alpha_n} n^{1/q}$.
\end{lem}
\begin{proof}[\textbf{Proof of Lemma \ref{lem: part 2}}]

Recall $e_{i,j}$ in \eqref{eq:e}. In matrix notation, we can write
\begin{align*}
e & = \diag(\hat \pi^{-1}) ((\Xi - \pi 1_m') \circ \Gamma) + \diag(\hat \pi_1^{-1}(\hat \pi_1 - \pi_1),\cdots,\hat \pi_n^{-1}(\hat \pi_n - \pi_n) ) \Gamma \\
& + \diag(\hat \pi^{-1}) ( \Xi \circ \eps) - \diag(\hat \pi^{-1} ) ( \Xi \circ (X\delta)').
\end{align*}

Note that $(\Xi - \pi 1_m') \circ \Gamma$ are independent conditional on $(\Gamma, X)$ with mean zero. In addition, we have
\begin{align*}
& \max_{1 \leq i \leq n} E (\sum_{j=1}^m(\xi_{i,j}-\pi_i)^2 \Gamma_{i,j}^2|\Gamma,X) \lessp m \alpha_n \log n , \\
& \max_{1 \leq j \leq m} E (\sum_{i=1}^n(\xi_{i,j}-\pi_i)^2 \Gamma_{i,j}^2|\Gamma,X) \lessp n \alpha_n \log n, \\
& \max_{1 \leq i,j \leq m} |\Gamma_{i,j}| \lessp  \log n.
\end{align*}
Therefore, by \citet[Corollary 3.12 and Remark 3.13]{BV16}, there exist universal constants $C$ and $c$ that we have
\begin{align*}
P\left(||(\Xi - \pi 1_m') \circ \Gamma || \geq C \sqrt{(m+n)\alpha_n  \log n} + t |\Gamma,X\right) \leq (n+m)\exp\left(-\frac{t^2}{c \log^2 n} \right).
\end{align*}
By letting $t = C \log^{3/2} n$ for some sufficiently large constant $C$, we have
\begin{align*}
\left\Vert \diag(\hat \pi^{-1})((\Xi - \pi 1_m') \circ \Gamma)\right\Vert & \leq \left\Vert \diag(\hat \pi^{-1} \pi)\right\Vert \left\Vert \diag(\pi^{-1})\right\Vert \left\Vert((\Xi - \pi 1_m') \circ \Gamma)\right\Vert \\
& \lessp \alpha_n^{-1}n^{1/q}\sqrt{(m+n)\alpha_n}.
\end{align*}

Similarly, we can show that
\begin{align*}
\left\Vert\diag(\hat \pi^{-1})(\Xi - \pi 1_m') \circ \eps\right\Vert \lessp \alpha_n^{-1} n^{1/q}\sqrt{(m+n)\alpha_n}.
\end{align*}
In addition, we have
\begin{align*}
\left\Vert\diag(\hat \pi^{-1})(\pi 1_m') \circ \eps\right\Vert \lessp \max_{i \in [n]}(\hat \pi_i^{-1} \pi_i ) ||\eps|| \lessp \sqrt{(m+n)}.
\end{align*}

Next, by Proposition \ref{prop:pihat}, we have
\begin{align*}
||\diag(\hat \pi_1^{-1}(\hat \pi_1 - \pi_1),\cdots,\hat \pi_n^{-1}(\hat \pi_n - \pi_n) ) \Gamma|| \lesssim (nm\alpha_n)^{-1/2}  n^{1/q} ||\Gamma|| \lesssim \alpha_n^{-1/2} n^{1/q},
\end{align*}
where we use the fact that $||\Gamma||^2 \leq ||\Gamma||_F^2 \leq ||L||_F^2 ||F||_F^2 \lessp mn.$

Last,  by \eqref{eq:lem1_9} and \eqref{eq:lem1_4_1}, we have
\begin{align*}
& \max_{1\leq j\leq m}\sum_{i=1}^n \hat \pi_i^{-1} \xi_{i,j} |X_i'\delta_j| \leq \max_{1\leq j\leq m}\sum_{i=1}^n \hat \pi_i^{-1} \xi_{i,j} ||X_i||_2 ||\delta_j||_2 \lessp \sqrt{ n \alpha_n^{-1} \log^{3/2} n}, \\
& \max_{1\leq i\leq n}\sum_{j=1}^m \hat \pi_i^{-1} \xi_{i,j} |X_i'\delta_j| \leq \max_{1\leq i\leq n}||X_i||_2 \hat  \pi_i^{-1}  \sum_{j=1}^m  \xi_{i,j} ||\delta_j||_2 \lessp m \sqrt{ n^{-1} \alpha_n^{-1} \log^{5/2} n} n^{1/q}.
\end{align*}

Then, by \citet[Corollary 2.3.2]{GV16}, we have
\begin{align*}
& ||( (\diag(\hat \pi^{-1})\Xi) \circ (X\delta)')||\\
& \leq \left[\max_{1\leq j\leq m}\sum_{i=1}^n \hat \pi_i^{-1} \xi_{i,j} |X_i'\delta_j|  \right]^{1/2} \left[\max_{1\leq i\leq n}\sum_{j=1}^m \hat \pi_i^{-1} \xi_{i,j} |X_i'\delta_j|  \right]^{1/2} \\
& \lessp (m \alpha_n^{-1})^{1/2} n^{1/q},
\end{align*}
which leads to the first desired result.

For the second result, we have
\begin{align*}
||\Xi \circ \eps|| \leq ||\Xi || \max_i \pi_i + ||\Xi \circ (\eps - \pi 1_m')|| \lessp \sqrt{(m+n)\alpha_n} n^{1/q}.
\end{align*}

\end{proof}

\begin{lem}
\label{lem: part 3}\label{lem: order of L'e, F'e'}Let Assumption \ref{assu: moment cond} hold.
Then
\begin{enumerate}
	\item $||L'e||_F = O_P(\sqrt{\alpha_n^{-1}mn \log n})$,
	\item $\max_{1 \leq j \leq m}||L'e_{\cdot,j}||_2 = O_P(\sqrt{ n \alpha_n^{-1} \log n}. )$,
\item 	 $||F'e'||_F = O_P(\sqrt{m (m +n)\alpha_n^{-1} \log^{3/2} n})$.
\end{enumerate}
\end{lem}
\begin{proof}[\textbf{Proof of Lemma \ref{lem: part 3}}]
For the first result, by \eqref{eq:e} in Lemma \ref{lem: part 3}, we have
\begin{align}
L'e_{\cdot,j} & = \sum_{i=1}^n L_i \pi_i^{-1} \xi_{i,j}\eps_{i,j} + \sum_{i=1}^n L_i \pi_i^{-1}(\xi_{i,j}- \pi_i)L_i'F_j + \sum_{i=1}^n  L_i \xi_{i,j} (\hat{\pi}_i^{-1} - \pi_i^{-1})L_i'F_j  \notag \\
&+ \sum_{i=1}^n L_i (\hat{\pi}_i^{-1} - \pi_i^{-1}) \xi_{i,j}\eps_{i,j} - \sum_{i=1}^{n} L_i \hat{\pi}_i^{-1} \xi_{i,j}X_i'\delta_j .
\label{eq:Le}
\end{align}

For the third term on the RHS of \eqref{eq:Le}, we have
\begin{align*}
& \max_{1\leq j \leq m} \left\Vert \sum_{i=1}^n(\hat \pi_i^{-1} - \pi_i^{-1})  L_i \xi_{i,j} L_i'F_j\right\Vert_2 \notag \\
& \lessp   \alpha_n^{-2} n^{1/q} (nm)^{-1} \max_{1\leq j \leq m} \left[\sum_{i=1}^n \left\Vert L_i \xi_{i,j} L_i'F_j \right\Vert_2\right] +  \max_{1\leq j \leq m} \left[\left\Vert\sum_{i=1}^n  \frac{(1-\pi_i)L_i X_i'}{\pi_i} \xi_{i,j} L_i'F_j \right\Vert\right] ||\hat{\gamma} - \gamma||_2\\
& \lessp m^{-1}\alpha_n^{-1} n^{1/q} + n^{1/2} (m\alpha_n)^{-1/2},
\end{align*}
where the first inequality is by Proposition \ref{prop:pihat} and the second inequality holds because of \eqref{eq:gammahat}, \eqref{eq:ex_lem_4}, and the fact that $\max_{1\leq j \leq m} \left[\sum_{i=1}^n \left\Vert L_i \xi_{i,j} L_i'F_j \right\Vert_2\right] \lessp \alpha_n n$.

For the fourth term on the RHS of \eqref{eq:Le}, we have
\begin{align*}
& \max_{1\leq j \leq m} \left\Vert \sum_{i=1}^n L_i (\hat{\pi}_i^{-1} - \pi_i^{-1}) \xi_{i,j}\eps_{i,j} \right\Vert_2 \notag \\
& \lessp   \alpha_n^{-2} n^{1/q} (nm)^{-1} \max_{1\leq j \leq m} \left[\sum_{i=1}^n \left\Vert L_i \xi_{i,j} \eps_{i,j} \right\Vert_2\right] +  \max_{1\leq j \leq m} \left[\left\Vert\sum_{i=1}^n  \frac{(1-\pi_i)L_i X_i'}{\pi_i} \xi_{i,j} \eps_{i,j} \right\Vert\right] ||\hat{\gamma} - \gamma||_2\\
& \lessp m^{-1}\alpha_n^{-1} n^{1/q},
\end{align*}
where the first inequality is by Proposition \ref{prop:pihat} and the second inequality is by \eqref{eq:gammahat}, \eqref{eq:ex_lem_1}, and the fact that $\max_{1\leq j \leq m} \left[\sum_{i=1}^n \left\Vert L_i \xi_{i,j} \eps_{i,j} \right\Vert_2\right] \lessp \alpha_n n$.

For the fifth term, by Proposition \ref{prop:pihat}, we have
\begin{align}
& \max_{1\leq j \leq m}  \left\Vert \sum_{i=1}^n \hat{\pi}_i^{-1} L_i \xi_{i,j} X_i'\delta_j\right\Vert_2 \notag \\
& \lessp \alpha_n^{-2} n^{1/q} (nm)^{-1} \max_{1\leq j \leq m}  \left[\sum_{i=1}^n ||L_i \xi_{i,j} X_i'\delta_j||_2\right] \notag \\
& + \max_{1\leq j \leq m}  \left\Vert \sum_{i=1}^n \frac{1-\pi_i}{\pi_i} L_i \xi_{i,j} X_i'\delta_jX_i'\right\Vert \left\Vert \hat{\gamma} - \gamma \right\Vert_2 +  \max_{1\leq j \leq m}  \left[\left\Vert \sum_{i=1}^n \pi_i^{-1} L_i \xi_{i,j} X_i'\delta_j\right\Vert_2\right]
\label{eq:lem3_2'}.
\end{align}

For the first term on the RHS of \eqref{eq:lem3_2'}, we have
\begin{align*}
\alpha_n^{-2} n^{1/q} (nm)^{-1} \max_{1\leq j \leq m}  \left[\sum_{i=1}^n ||L_i \xi_{i,j} X_i'\delta_j||_2\right] & \lessp \alpha_n^{-2} n^{1/q} (nm)^{-1}\max_{1\leq j \leq m}  ||\delta_j||_2 \left[\sum_{i=1}^n ||L_i \xi_{i,j} X_i'||_F\right] \\
& \lessp  (m \alpha_n)^{-1} (n \alpha_n)^{-1/2} n^{1/q},
\end{align*}
where we use \eqref{eq:lem1_9} and the fact that $ \max_{1\leq j \leq m}  \left[\sum_{i=1}^n ||L_i \xi_{i,j} X_i'\delta_j||_2\right]  \lessp \alpha_n n.$

For the second term on the RHS of \eqref{eq:lem3_2'}, by \eqref{eq:gammahat}, \eqref{eq:422}, and \eqref{eq:ex_lem_2}, we have
\begin{align*}
& \max_{1\leq j \leq m} \left\Vert \sum_{i=1}^n \frac{1-\pi_i}{\pi_i} L_i \xi_{i,j} X_i'\delta_jX_i'\right\Vert \left\Vert \hat{\gamma} - \gamma \right\Vert_2 \lessp (mn)^{-1/2}\alpha_n^{-3/2} \log^{1/2} n.
\end{align*}

For the third term on the RHS of \eqref{eq:lem3_2'},  we have
\begin{align*}
\max_{1\leq  j \leq m} \left[\left\Vert \sum_{i=1}^n \pi_i^{-1} L_i \xi_{i,j} X_i'\delta_j\right\Vert_2\right] & \leq \max_{1\leq  j \leq m}  \left[\left\Vert \sum_{i=1}^n \pi_i^{-1} L_i \xi_{i,j} X_i'\right\Vert\right] \max_{1\leq  j \leq m} ||\delta_j||_2  \lessp \alpha_n^{-1}  \log^{5/4} n,
\end{align*}
where the last inequality is by \eqref{eq:lem1_9} and \eqref{eq:ex_lem_3}.

Combining the above three bounds, we have
\begin{align*}
\max_{1\leq j \leq m} \left\Vert \sum_{i=1}^n \hat{\pi}_i^{-1} L_i \xi_{i,j} X_i'\delta_j\right\Vert_2 \lessp \alpha_n^{-1}  \log^{5/4} n.
\end{align*}

Therefore, we have
\begin{align}
\max_{1\leq  j \leq m}\left\Vert L' e_{\cdot,j} - \sum_{i=1}^n L_i \pi_i^{-1} \xi_{i,j}\eps_{i,j} - \sum_{i=1}^n L_i \pi_i^{-1}(\xi_{i,j}- \pi_i)L_i'F_j\right\Vert_2 \lessp \alpha_n^{-1} \log ^{5/4} n + n^{1/2} (m \alpha_n)^{-1/2}.
\label{eq:Le1}
\end{align}

For the first term on the RHS of \eqref{eq:Le}, we define $A_{i} \in \Re^{R \times m}$ with its $(r,j)$th entry $A_{i,r,j} = L_{i,r} \pi_i^{-1} \xi_{i,j}\eps_{i,j}$. In addition, we have
\begin{align*}
& ||A_i|| \leq ||A_i||_F \lessp m^{1/2} \alpha_n^{-1/2} n^{1/q}, \\
& ||E (\sum_{i=1}^{n} A_i A_i'|L,X)|| = mn \alpha_n^{-1}, \\
& ||E (\sum_{i=1}^{n} A_i' A_i|L,X)|| \leq \max_{1\leq j \leq m} \sum_{i=1}^{n} ||L_i||_2^2 \pi_i^{-1} E(\eps_{i,j}^2 |X,L) = n \alpha_n^{-1}.
\end{align*}
Then, by \citet[Theorem 1.6]{T12}, we have
\begin{align*}
\left\| \sum_{i=1}^n A_i \right\| \lessp \left(nm\alpha_n^{-1} \log n\right)^{1/2}.
\end{align*}

Similarly, for the second term on the RHS of \eqref{eq:Le}, we have
\begin{align*}
\left\Vert (\sum_{i=1}^n L_i \pi_i^{-1}(\xi_{i,1}- \pi_i)L_i'F_1,\cdots,\sum_{i=1}^n L_i \pi_i^{-1}(\xi_{i,m}- \pi_i)L_i'F_m )  \right\Vert \lessp \left(nm\alpha_n^{-1} \log n\right)^{1/2}.
\end{align*}

Therefore, we have $||L'e|| \lessp \left(nm\alpha_n^{-1} \log n\right)^{1/2}.$

For the second result in Lemma \ref{lem: part 3}, we have
\begin{align*}
& \max_{1\leq j\leq m}|| \sum_{i=1}^n L_i \pi_i^{-1} \xi_{i,j}\eps_{i,j}||_2 \lessp \sqrt{n \alpha_n^{-1} \log n}, \\
& \max_{1\leq j\leq m} \left\Vert \sum_{i=1}^n L_i \pi_i^{-1}(\xi_{i,j}- \pi_i)L_i'F_j \right\Vert_2 \lessp \sqrt{n \alpha_n^{-1} \log n},
\end{align*}
as shown in \eqref{eq:ex_lem_5} and \eqref{eq:ex_lem_6}. Combining the above bounds with \eqref{eq:Le1}, we obtained the desired result.

For the third result in Lemma \ref{lem: part 3}, by \eqref{eq:e}, we have
\begin{align*}
& \max_{1\leq i \leq n} \left\Vert \sum_{j=1}^m F_j e_{i,j} \right\Vert_2 \\
& \leq \max_{1\leq i \leq n} \left\Vert \sum_{j=1}^m F_j \xi_{i,j} (\hat{\pi}_i^{-1} - \pi_i^{-1})L_i'F_j \right\Vert_2 +  \max_{1\leq i \leq n} \left\Vert \sum_{j=1}^m F_j \pi_i^{-1}(\xi_{i,j}- \pi_i)L_i'F_j  \right\Vert_2 \\
& + \max_{1\leq i \leq n} \left\Vert \sum_{j=1}^m F_j \hat{\pi}_i^{-1} \xi_{i,j}\eps_{i,j} \right\Vert_2 +  \max_{1\leq i \leq n} \left\Vert \sum_{j=1}^m F_j (\hat{\pi}_i^{-1} - \pi_i^{-1}) \xi_{i,j}\eps_{i,j} \right\Vert_2 \\
& + \max_{1\leq i \leq n} \left\Vert \sum_{j=1}^m F_j \pi_i^{-1} \xi_{i,j}X_i'\delta_j \right\Vert_2 + \max_{1\leq i \leq n} \left\Vert \sum_{j=1}^m F_j (\hat{\pi}_i^{-1} - \pi_i^{-1}) \xi_{i,j}X_i'\delta_j \right\Vert_2.
\end{align*}
We can further show
\begin{align*}
& \max_{1 \leq i \leq n}\left\Vert \sum_{j=1}^m F_j F_j'\xi_{i,j} \right\Vert \lessp m\alpha_n, \\
& \max_{1 \leq i \leq n} \left\Vert \sum_{j=1}^m F_jF_j'\pi_i^{-1}(\xi_{i,j} - \pi_i )  \right\Vert \lessp \sqrt{m \alpha_n^{-1} \log n}, \\
& \max_{1 \leq i \leq n} \left\Vert\sum_{j=1}^m F_j
\pi_i^{-1}\xi_{i,j}\eps_{i,j}  \right\Vert_2 \lessp\sqrt{m \alpha_n^{-1} \log n}, \\
& \max_{1\leq i \leq n}\left\Vert \sum_{j=1}^m \delta_jF_j \pi_i^{-1}\xi_{i,j}  \right\Vert_2 \leq \max_{1\leq j \leq m}||\delta_j||_2 \left(\sum_{j=1}^m ||F_j||_2 \pi_i^{-1}\xi_{i,j}\right)\lessp m \sqrt{ n^{-1} \alpha_n^{-1} \log^{3/2} n}.
\end{align*}
Following the previous argument, these bounds imply that
\begin{align}
\max_{1\leq i \leq n}||F'e_i||_2 \lessp  \sqrt{m(1+m/n) \alpha_n^{-1} \log^{3/2} n},
\label{eq:Fei}
\end{align}
which leads to the desired result.

\end{proof}

\begin{lem}
\label{lem: part 4}\label{lem: order of Omega H k=r}Let Assumption \ref{assu: moment cond} hold.
Then $||\hat{\Omega}_r^{-1}|| \lessp 1$ and $||H|| \lessp 1$.
\end{lem}
\begin{proof}[\textbf{Proof of Lemma \ref{lem: part 4}}]

	Note that $\sigma_r(W) \geq \sigma_r(LF') - ||e||$ and $P((nm)^{-1/2} \sigma_r(LF') > b^2) \rightarrow 1$ as shown in the paper. Therefore, we have
\begin{align*}
P((nm)^{-1/2} \sigma_r(W) + (nm)^{-1/2}||e||) \geq P((nm)^{-1/2} \sigma_r(LF') > b^2) \rightarrow 1.
\end{align*}	
Since 	$(nm)^{-1/2}||e|| = o_P(1)$ by Lemma \ref{lem: part 2}. Note $||\hat \Omega_r^{-2}|| = nm \sigma_r^{-2}(W)$. Therefore, $||\hat \Omega_r^{-2}||$ is bounded above by $4/b^2$ with probability approaching one, i.e.,  $||\hat \Omega_r^{-1}|| = O_P(1)$. Also note that
\begin{align*}
||H|| = ||F'W' \hat{L} \hat \Omega_r^{-2} (nm)^{-1}|| \leq ||F|| ||W|| ||\hat{L}|| ||\hat \Omega_r^{-2}||/(nm) = O_P(1).
\end{align*}
\end{proof}

\begin{lem}
\label{lem: part 5}Let Assumption \ref{assu: moment cond} hold. Then, we have
	\begin{enumerate}
		\item $||\Delta_L|| \lessp \sqrt{\alpha_n^{-1}(1 + n/m)}n^{1/q}$,
	\item $\max_{1\leq j \leq m} ||\hat L'e_{\cdot,j}||_2=O_P(\sqrt{n(1+n/m)\alpha_n^{-2}} n^{1/q})$,
	\item $||\hat L'e|| =O_P(\sqrt{n(n+m)\alpha_n^{-1}} n^{1/q})$.
\end{enumerate}	
\end{lem}
\begin{proof}[\textbf{Proof of Lemma \ref{lem: part 5}}]
We have
\begin{align*}
||\Delta_L|| = ||eW'\hat L \hat \Omega_r^{-2}(nm)^{-1}||\lessp ||e|| ||W|| ||\hat L|| (nm)^{-1} \lessp \sqrt{(1+n/m)\alpha_n^{-1}} n^{1/q}.
\end{align*}

By \eqref{eq:e}, we have
\begin{align*}
    e_{i,j} = \sum_{l=1}^4 A_{l,i,j},
\end{align*}
where
\begin{align*}
  A_{1,i,j} & = \xi_{i,j} (\hat \pi_i^{-1} - \pi_i^{-1})L_i'F_j, \\
  A_{2,i,j} & = \pi_i^{-1}(\xi_{i,j}-\pi)L_i'F_j, \\
  A_{3,i,j} & = \hat \pi_i^{-1} \xi_{i,j}\eps_{i,j}, \\
  A_{4,i,j} & = - \hat \pi_i^{-1} \xi_{i,j} X_i'\delta_j.
\end{align*}

Then, we have
\begin{align*}
& \max_j    \sumi A_{1,i,j}^2 \leq (m \alpha_n^2)^{-1} n^{1/q}, \\
& \max_j    \sumi A_{2,i,j}^2 \leq n \alpha_n^{-1} \log n, \\
& \max_j    \sumi A_{2,i,j}^2 \leq n \alpha_n^{-1}, \\
& \max_j    \sumi A_{2,i,j}^2 \leq n \alpha_n^{-1}.
\end{align*}
This implies
\begin{align*}
    \max_{j} ||e_{\cdot,j}|| \lessp \sqrt{n \alpha_n^{-1} \log n}.
\end{align*}

By Lemmas \ref{lem: part 2} and \ref{lem: part 3}, we have
\begin{align*}
\max_{1 \leq j \leq m} ||\hat L' e_{\cdot,j}||_2 & \leq \max_{1 \leq j \leq m} ||H' L' e_{\cdot,j}||_2 + \max_{1 \leq j \leq m} ||\Delta_L' e_{\cdot,j}||_2 \\
& \leq ||H|| \max_{1 \leq j \leq m} ||L' e_{\cdot,j}||_2 + ||\Delta_L|| \max_{1 \leq j \leq m}||e_{\cdot,j}|| \\
& \lessp \sqrt{ n \alpha_n^{-1} \log n} + \sqrt{(1+n/m)\alpha_n^{-1}} n^{1/q} \sqrt{n \alpha_n^{-1} \log n} \\
& \lessp \sqrt{n(1+n/m)\alpha_n^{-2}} n^{1/q}.
\end{align*}

In addition, we have
\begin{align*}
||\hat L'e|| & \leq ||\hat L|| ||e|| \lessp \sqrt{n(n+m)\alpha_n^{-1}} n^{1/q}.
\end{align*}

\end{proof}

\subsection{More auxiliary results: bound on $E_{n}\xi_{i,j}X_{i}\Delta_{L,i}'$}
\begin{lem}[bound on $E_{n}\xi_{i,j}X_{i}\Delta_{L,i}'$]
\label{lem: part 6}Let Assumption \ref{assu: moment cond} hold.
Then, uniformly over $j$
\begin{align*}
E_n \xi_{i,j} X_i \Delta_{L,i}' = - (nm)^{-1}(E_n \pi_iX_iL_i') F'FL'\hat L \hat \Omega_r^{-2}+ O_P(n^{-1}\log^2 n + (nm)^{-1/2}\log n),
\end{align*}	
and
\begin{align*}
\max_{1 \leq j \leq m} ||E_n \xi_{i,j} X_i \Delta_{L,i}'||\lessp \alpha_n n^{-1/2} + n^{-1}\log^2 n + (nm)^{-1/2}\log n.
\end{align*}

\end{lem}
\begin{proof}[\textbf{Proof of Lemma \ref{lem: part 6}}]
By the definition of $\Delta_{L,i}$, we have \begin{align*}
\Delta_{L,i}' = (nm)^{-1}e_i'FL'\hat L \hat \Omega_r^{-2} + (nm)^{-1}e_i'e'\hat L \hat \Omega_r^{-2}.
\end{align*}
This implies
\begin{align*}
E_n \xi_{i,j} X_i \Delta_{L,i}' & = n^{-2} m^{-1} \sum_{i=1}^n  \xi_{i,j} X_i e_i'FL' \hat L \hat \Omega_r^{-2} + n^{-2} m^{-1} \sum_{i=1}^n  \xi_{i,j} X_i e_i'e' \hat L \hat \Omega_r^{-2} \\
& \equiv Q_{1,j} + Q_{2,j}.
\end{align*}

\textbf{Step 1: bound $\max_{1 \leq j \leq m}||Q_{2,j}||_2$. } We have
\begin{align*}
\max_{1 \leq j \leq m} \left\Vert  \sum_{i=1}^n  \xi_{i,j} X_i e_i'e' \hat L \hat \Omega_r^{-2} \right\Vert & \leq \sqrt{ \left(\max_{1 \leq j \leq m} \sum_{i=1}^n \xi_{i,j} ||X_i||_2^2 \right) \times \left(\sum_{i=1}^n ||e_i'e' \hat L \hat \Omega_r^{-2}||_2^2 \right)} \\
& \lessp \sqrt{n \alpha_n  trace\left(\sum_{i=1}^n \hat \Omega_r^{-2} \hat L' e e_i e_i' e' \hat L \hat \Omega_r^{-2} \right) } \\
& \lessp \sqrt{n \alpha_n  trace\left( \hat \Omega_r^{-2} \hat L' e e'e e' \hat L \hat \Omega_r^{-2} \right) } \\
& \lessp \sqrt{n\alpha_n} ||\hat \Omega_r^{-2}|| ||\hat L' e|| ||e|| \\
& \lessp   n(n+m)\alpha_n^{-1/2} n^{1/q}.
\end{align*}
This implies $\max_{1 \leq j \leq m}||Q_{2,j}|| \lessp n^{-1}\alpha_n^{-1/2} (1 + n/m) n^{1/q}$.

\textbf{Step 2: bound $\max_{1 \leq j \leq m} ||Q_{1,j}||_2$.} By the definition of $e_{i,j}$ and $u_{i,j}$, we observe that
\begin{align}
& \sum_{i=1}^n \xi_{i,j} X_i e_i'F \notag \\
& = \sum_{i=1}^n \xi_{i,j} X_i \left( \sum_{s=1}^m e_{i,s} F_s'\right) \notag \\
& = \sum_{i=1}^n \xi_{i,j} X_i \left[\sum_{s=1}^m \left(L_i'F_s(\xi_{i,s}\hat{\pi}_i^{-1} - 1) + \hat
\pi_i^{-1} \xi_{i,s} u_{i,s} \right) F_s'\right] \notag \\
& = \sum_{i=1}^n \xi_{i,j} X_i \left[ \sum_{s=1}^m \left(L_i'F_s\xi_{i,s}(\hat{\pi}_i^{-1} - \pi_i^{-1})F_s'\right) \right] + \sum_{i=1}^n \xi_{i,j} X_i  \left[ \sum_{s=1}^m \left(L_i'F_s\pi_i^{-1}(\xi_{i,s}-\pi_i)F_s'\right) \right]\notag \\
& + \sum_{i=1}^n \xi_{i,j} X_i \hat
\pi_i^{-1} \left[ \sum_{s=1}^m \xi_{i,s} \eps_{i,s}  F_s'\right] - \sum_{i=1}^n \xi_{i,j} X_i \left[ \sum_{s=1}^m \left(\hat
\pi_i^{-1} \xi_{i,s} X_i'\delta_s \right) F_s'\right].
\label{eq:lem6_step2_1}
\end{align}

For the first term on the RHS of \eqref{eq:lem6_step2_1}, we have
\begin{align*}
& \max_{1 \leq j \leq m} \left\Vert \sum_{i=1}^n \xi_{i,j} X_i \left[\sum_{s=1}^m \left(L_i'F_s\xi_{i,s}(\hat{\pi}_i^{-1} - \pi_i^{-1})F_s'\right)\right] \right\Vert \\
& \lessp \alpha_n^{-2} n^{1/q}(nm)^{-1} \max_{1\leq j \leq m} \sum_{i=1}^n \sum_{s=1}^m \xi_{i,j} ||X_i||_2 ||L_i'F_s F_s||_2 \xi_{i,s} \\
& + \max_{1 \leq j \leq m} \sum_{i=1}^n \xi_{i,j} ||X_i||_2 ||X_i L_i'|| \left\Vert \sum_{s=1}^m \left(\xi_{i,s}\left(\frac{1 - \pi_i}{\pi_i} \right)F_sF_s'\right) \right\Vert ||(\hat{\gamma} - \gamma)||_2 \\
& \lessp n^{1/q} + (nm \alpha_n)^{1/2} \lessp \sqrt{nm \alpha_n},
\end{align*}
where we rely on \eqref{eq:gammahat} and the facts that
\begin{align*}
& \max_{1\leq j \leq m} \sum_{i=1}^n \sum_{s=1}^m \xi_{i,j} ||X_i||_2 ||L_i'F_s F_s||_2 \xi_{i,s}  \lessp n m\alpha_n^2, \\
& \max_{1 \leq i \leq n} \left\Vert \sum_{s=1}^m \left(\xi_{i,s}\left(\frac{1 - \pi_i}{\pi_i} \right)F_sF_s'\right) \right\Vert \lessp m, \quad \text{and} \\
& \max_{1 \leq j \leq m}\sum_{i = 1}^n \xi_{i,j} ||X_i||_2||X_i L_i'|| \lessp \alpha_n n.
\end{align*}

For the second term on the RHS of \eqref{eq:lem6_step2_1}, we have
\begin{align*}
&\max_{1\leq j \leq m} \left\Vert \sum_{i=1}^n \xi_{i,j} X_i \left[ \sum_{s=1}^m \left(L_i'F_s\pi_i^{-1}(\xi_{i,s}-\pi_i)F_s'\right) \right] \right\Vert \\
& \leq \max_{1\leq j \leq m} \left\Vert \sum_{i=1}^n \pi_i^{-1/2}\xi_{i,j} X_iL_i' \right\Vert \max_{1 \leq i \leq n} \left\Vert\sum_{s=1}^m \left(\pi_i^{-1/2}(\xi_{i,s}-\pi_i)F_sF_s'\right) \right\Vert \\
& \lessp \sqrt{n \log n} \sqrt{m  \log n} \lesssim \sqrt{nm \log^2 n},
\end{align*}
where the last inequality is by \eqref{eq:lem1_3} and the fact that
\begin{align*}
\max_{1 \leq i \leq n} \left\Vert\sum_{s=1}^m \left(\pi_i^{-1/2}(\xi_{i,s}-\pi_i)F_sF_s'\right) \right\Vert \lessp \sqrt{ m \log n}.
\end{align*}

For the third term on the RHS of \eqref{eq:lem6_step2_1}, by \eqref{eq:lem1_7} and \eqref{eq:lem1_8}, we have
\begin{align*}
& \max_{1 \leq j \leq m} \left\Vert \sum_{i=1}^n \xi_{i,j} X_i \hat \pi_i^{-1} \left[\sum_{s=1}^m
 \xi_{i,s} \eps_{i,s}  F_s'\right] \right\Vert \\
 & \lessp \max_{1 \leq j \leq m} \left\Vert \sum_{i=1}^n \xi_{i,j} X_i \pi_i^{-1} \sum_{s=1}^m
 \xi_{i,s} \eps_{i,s}  F_s' \right\Vert\\
 & + \alpha_n^{-2} n^{1/q} (nm)^{-1}\max_{1 \leq j \leq m} \left\Vert \sum_{i=1}^n \xi_{i,j} ||X_i||_2  \left(\sum_{s=1}^m
 \xi_{i,s} |\eps_{i,s}| || F_s||_2\right) \right\Vert \\
 & + \max_{1 \leq j \leq m} \left\Vert \sum_{i=1}^n \sum_{s=1}^m\xi_{i,j} X_i \frac{1-\pi_i}{\pi_i}X_i'(\hat \gamma - \gamma)
 \xi_{i,s} \eps_{i,s}  F_s' \right\Vert \\
 & \lessp \sqrt{nm \log n} +  \max_{1 \leq j \leq m, 1\leq r \leq R} \left\Vert \sum_{i=1}^n \sum_{s=1}^m \xi_{i,j} \xi_{i,s} \eps_{i,s}  \frac{1-\pi_i}{\pi_i}F_{s,r}X_iX_i'\right\Vert ||\hat \gamma - \gamma|| \\
 & \lessp \sqrt{nm \log n}.
\end{align*}


For the last term on the RHS of \eqref{eq:lem6_step2_1}, we have
\begin{align}
\sum_{i=1}^n \xi_{i,j} X_i \sum_{s=1}^m \left(\hat
\pi_i^{-1} \xi_{i,s} X_i'\delta_s \right) F_s' & = \sum_{i=1}^n \xi_{i,j} X_i \sum_{s \neq j} \left(\hat
\pi_i^{-1} \xi_{i,s} X_i'\delta_s \right) F_s' + n \left(E_n \hat \pi_i^{-1}  \xi_{i,j} X_i
X_i'\right) \delta_j  F_j' .
\label{eq:Q1j_1}
\end{align}
For the first term on the RHS of \eqref{eq:Q1j_1}, we have
\begin{align}
& \sum_{i=1}^n \xi_{i,j} X_i \sum_{s \neq j} \left(\hat
\pi_i^{-1} \xi_{i,s} X_i'\delta_s \right) F_s' \notag \\
& = \sum_{i=1}^n \xi_{i,j}\pi_i^{-1} X_i \sum_{s \neq j} \left(
 \xi_{i,s} X_i'\delta_s \right) F_s'+  \sum_{s \neq j}\sum_{i=1}^n \xi_{i,j} X_i \left((\hat
\pi_i^{-1} - \pi_i^{-1})\xi_{i,s} X_i'\delta_s \right) F_s' \notag \\
& = \sum_{s \neq j} \left(\sum_{i=1}^n \xi_{i,s} X_i
 X_i'\right)\delta_s F_s'+ \sum_{i=1}^n (\xi_{i,j}-\pi_i)\pi_i^{-1} X_iX_i' \left[\sum_{s \neq j} \left(
\xi_{i,s} \delta_s \right) F_s'\right] \notag \\
& + \sum_{s \neq j}\sum_{i=1}^n \xi_{i,j} X_i \left((\hat
\pi_i^{-1} - \pi_i^{-1})\xi_{i,s} X_i'\delta_s \right) F_s'.
\label{eq:lem6_1}
\end{align}

For the first term on the RHS of \eqref{eq:lem6_1}, we have
\begin{align*}
& \sum_{s \neq j} \left(\sum_{i=1}^n \xi_{i,s} X_i
X_i'\right)\delta_s F_s' \\
& = \sum_{s \neq j} \left(\sum_{i=1}^n \xi_{i,s} X_i
X_i'\right)\left[
(EX_iX_i'\pi_{i})^{-1}(E_nX_i \eps_{i,s}\xi_{i,s} + (E_n X_i L_i'\xi_{i,s})F_s) \right] F_s' \\
& + \sum_{s \neq j} \left(\sum_{i=1}^n \xi_{i,s} X_i
X_i'\right)\left[ \delta_s-
(EX_iX_i'\pi_{i})^{-1}(E_nX_i \eps_{i,s}\xi_{i,s} + (E_n X_i L_i'\xi_{i,s})F_s) \right] F_s' \\
& = n \sum_{s \neq j}\left[E_nX_i \eps_{i,s}\xi_{i,s} + (E_n X_i L_i'\xi_{i,s}F_s) \right] F_s' + O_P(m \log^{3/2} n),
\end{align*}
where the last $O_P(\cdot)$ term holds uniformly over $j$ and the last equality holds by \eqref{eq:lem1_2_8} and the facts that
\begin{align*}
& \max_{1 \leq s \leq m} ||(E_n  \xi_{i,s} X_iX_i')(E \xi_{i,s} X_iX_i')^{-1} -I_K|| \lessp \sqrt{n^{-1}\alpha_n^{-1} \log n},\\
& \max_{1\leq j \leq m} ||\sum_{s \neq j} (E_n X_i\eps_{i,s}\xi_{i,s} + E_nX_iL_i'\xi_{i,s}F_s)F_s'|| \lessp m\sqrt{n^{-1} \alpha_n \log^2 n}.
\end{align*}

For the second term on the RHS of \eqref{eq:lem6_1}, we note that when $s \neq j$, $\delta_s = \hat \beta_s - \beta_s$ is $\mathcal{F}_j$-measurable where $\mathcal{F}_j$ is the sigma field generated by $(L,F,X,\{\epsilon_{i,s},\xi_{i,s}\}_{1 \leq i \leq n, 1\leq s\leq m, s\neq j})$. In addition, conditionally on $\mathcal{F}_j$, $\{\xi_{i,j}\}_{1\leq i \leq n}$ is independent and
\begin{align*}
& \max_{1\leq j \leq m} E\sum_{i=1}^n\left\{(\xi_{i,j}-\pi_i)\pi_i^{-1} X_{i,k}X_i' \left[\sum_{s \neq j} \left(
\xi_{i,s} \delta_s \right) F_{s,r}\right] \bigg| \mathcal{F}_j\right\}^2\\
& \leq  \max_{1\leq j \leq m}\sum_{i=1}^n \pi_i^{-1} X_{i,k}^2 (X_{i}' \sum_{s \neq j} \xi_{i,s}\delta_sF_{s,r})^2 \\
& \lessp\max_{1\leq j \leq m, 1\leq k \leq d, 1\leq r\leq R}\sum_{i=1}^n \pi_i^{-1} X_{i,k}^2 ||X_{i}||_2^2 (\sum_{s \neq j} \xi_{i,s}||\delta_s||_2 |F_{s,r}|)^2 \\
& \lessp\max_{1\leq k \leq d, 1\leq r\leq R}\sum_{i=1}^n \pi_i^{-1} X_{i,k}^2 ||X_{i}||_2^2 (\sum_{s =1}^m \xi_{i,s}||\delta_s||_2 |F_{s,r}|)^2 \\
& \lessp m^2 \log^{3/2} n.
\end{align*}
We also have
\begin{align*}
& \max_{1\leq i\leq n, 1\leq j \leq m, 1\leq k \leq d, 1\leq r \leq R}\left\vert (\xi_{i,j}-\pi_i)\pi_i^{-1} X_{i,k}X_i' \left[\sum_{s \neq j} \left(
\xi_{i,s} \delta_s \right) F_{s,r}\right] \right\vert \\
& \leq \max_{1\leq i\leq n}\pi_i^{-1} ||X_i||_2^2 \sum_{s = 1}^m \xi_{i,s} ||\delta_s||_2 ||F_s||_2  \\
& \lessp m n^{-1/2} \alpha_n^{-1/2} n^{1/q}.
\end{align*}

 For any $\eta>0$, we can choose a sufficiently large constant $C_1$ such that
\begin{align*}
P(\mathcal{E}_n^c) \leq \eta, \quad \mathcal{E}_n = \begin{Bmatrix}
& \max_{1\leq k \leq d, 1\leq r\leq R}\sum_{i=1}^n \pi_i^{-1} X_{i,k}^2 ||X_{i}||_2^2 (\sum_{s =1}^m \xi_{i,s}||\delta_s||_2 |F_{s,r}|)^2 \leq C_1 m^2 \log^{3/2} n \\
& \max_{1\leq i\leq n}\pi_i^{-1} ||X_i||_2^2 \sum_{s = 1}^m \xi_{i,s} ||\delta_s||_2 ||F_s||_2  \leq C_1 m n^{-1/2} \alpha_n^{-1/2} n^{1/q}
\end{Bmatrix}.
\end{align*}
Further define
\begin{align*}
\mathcal{E}_{j,n} = \begin{Bmatrix}
\max_{1\leq k \leq d, 1\leq r\leq R}\sum_{i=1}^n \pi_i^{-1} X_{i,k}^2 ||X_{i}||_2^2 (\sum_{s \neq j} \xi_{i,s}||\delta_s||_2 |F_{s,r}|)^2 \leq C_1 m^2 \log^{3/2}\\
\max_{1\leq i\leq n, 1\leq k \leq d, 1\leq r \leq R}\left\vert (\xi_{i,j}-\pi_i)\pi_i^{-1} X_{i,k}X_i' \left[\sum_{s \neq j} \left(
\xi_{i,s} \delta_s \right) F_{s,r}\right] \right\vert \leq C_1 m n^{-1/2} \alpha_n^{-1/2} n^{1/q}
\end{Bmatrix}
\end{align*}
so that $\mathcal{E}_n \subset \mathcal{E}_{j,n}$ and $\mathcal{E}_{j,n} \in \mathcal{F}_j$.

Therefore, we have
\begin{align*}
& P\left(\max_{1\leq j \leq m}\left\Vert \sum_{i=1}^n (\xi_{i,j}-\pi_i)\pi_i^{-1} X_iX_i' \left[\sum_{s \neq j} \left(
\xi_{i,s} \delta_s \right) F_s'\right]\right\Vert \geq C m \log^{2} n\right) \\
& \leq P\left(\max_{1\leq j \leq m}\left\Vert \sum_{i=1}^n (\xi_{i,j}-\pi_i)\pi_i^{-1} X_iX_i' \left[\sum_{s \neq j} \left(
\xi_{i,s} \delta_s \right) F_s'\right]\right\Vert \geq C m \log^{2} n, \mathcal{E}_n\right) + \eta \\
& \leq \sum_{1 \leq j \leq m} P\left(\left\Vert \sum_{i=1}^n (\xi_{i,j}-\pi_i)\pi_i^{-1} X_iX_i' \left[\sum_{s \neq j} \left(
\xi_{i,s} \delta_s \right) F_s'\right]\right\Vert \geq C m \log^{2} n, \mathcal{E}_n\right) + \eta\\
& \leq \sum_{1 \leq j \leq m} P\left(\left\Vert \sum_{i=1}^n (\xi_{i,j}-\pi_i)\pi_i^{-1} X_iX_i' \left[\sum_{s \neq j} \left(
\xi_{i,s} \delta_s \right) F_s'\right]\right\Vert \geq C m \log^{2} n, \mathcal{E}_{j,n}\right) + \eta \\
& \leq \sum_{1 \leq j \leq m} E P\left(\left\Vert \sum_{i=1}^n (\xi_{i,j}-\pi_i)\pi_i^{-1} X_iX_i' \left[\sum_{s \neq j} \left(
\xi_{i,s} \delta_s \right) F_s'\right]\right\Vert \geq C m \log^{2} n \bigg| \mathcal{F}_j\right)1\{\mathcal{E}_{j,n}\} + \eta \\
& \leq \sum_{1 \leq j \leq m, 1\leq k \leq d, 1\leq r \leq R} E P\left(\left\vert \sum_{i=1}^n (\xi_{i,j}-\pi_i)\pi_i^{-1} X_{i,k}X_i' \left[\sum_{s \neq j} \left(
\xi_{i,s} \delta_s \right) F_{s,r}\right]\right\vert \geq C m \log^{2} n \bigg| \mathcal{F}_j\right)1\{\mathcal{E}_{j,n}\} + \eta \\
& \lesssim \exp\left(\log (dRm) - \frac{\frac{1}{2}C^2 m^2 \log^4 n }{ C_1 m^2 \log^{3/2} n + \frac{1}{3} C_1 C m^2 (n \alpha_n)^{-1/2}  n^{1/q} m \log^{2} n}\right) + \eta,
\end{align*}
where the last inequality is by Bernstein's inequality and the definition of $\mathcal{E}_{j,n}$. By choosing a sufficiently large $C$ and letting $m \rightarrow \infty$ followed by $\eta \downarrow 0$, the RHS of the above display vanishes, which implies
\begin{align}
\max_{1\leq j \leq m}\left\Vert \sum_{i=1}^n (\xi_{i,j}-\pi_i)\pi_i^{-1} X_iX_i' \left[\sum_{s \neq j} \left(
\xi_{i,s} \delta_s \right) F_s'\right]\right\Vert \lessp m \log^{2} n
\label{eq:lem6_2}
\end{align}

For the third term on the RHS of  \eqref{eq:lem6_1}, we have
\begin{align*}
& \max_{1 \leq j \leq m}|| \sum_{s \neq j}\sum_{i=1}^n \xi_{i,j} X_i \left((\hat
\pi_i^{-1} - \pi_i^{-1})\xi_{i,s} X_i'\delta_s \right) F_s'|| \\
& \lessp \alpha_n^{-2} n^{1/q} (nm)^{-1} \sum_{s=1}^m \sum_{i=1}^n \xi_{i,j} ||X_i||_2^2 \xi_{i,s} ||\delta_s||_2||F_s||_2 \\
& + \max_{1 \leq j \leq m}\left\Vert \sum_{s \neq j}\sum_{i=1}^n \xi_{i,j} X_i \left(\frac{1-\pi_i}{\pi_i}X_i'(\hat \gamma - \gamma)\xi_{i,s} X_i'\delta_s \right) F_s'\right\Vert \\
& \lessp n^{1/q} \sqrt{n^{-1}\alpha_n^{-1} \log^{5/2} n} + \max_{1 \leq j \leq m, 1\leq k \leq K, 1\leq r\leq R}\left\Vert \sum_{s \neq j}\sum_{i=1}^n \xi_{i,j} X_{i,k} \left(\frac{1-\pi_i}{\pi_i}\xi_{i,s} X_iX_i' ||\hat \gamma - \gamma||_2||\delta_s||_2 \right) F_{s,r}\right\Vert\\
& \lessp \sqrt{m \log^{3/2} n}.
\end{align*}

This implies
\begin{align*}
\sum_{i=1}^n \xi_{i,j} X_i \sum_{s \neq j} \left(\hat
\pi_i^{-1} \xi_{i,s} X_i'\delta_s \right) F_s' = n \sum_{s \neq j}\left[E_nX_i \eps_{i,s}\xi_{i,s} + (E_n X_i L_i'\xi_{i,s}F_s) \right] F_s' + O_P(m \log^{2} n),
\end{align*}
where the $O_P(\cdot)$ term holds uniformly over $j$.

For the second term on the RHS of \eqref{eq:Q1j_1}, by \eqref{eq:ex_lem_7}, we have
\begin{align*}
\max_{1 \leq j \leq m} n \left\Vert \left(E_n \hat \pi_i^{-1}  \xi_{i,j} X_i
X_i' - \Sigma_X\right) \delta_j  F_j' \right\Vert & \lessp \sqrt{n \alpha_n^{-1} \log n} \max_{1 \leq j \leq m} ||\delta_j||_2 ||F_j||  \lessp \alpha_n^{-1} \log^{7/4} n.
\end{align*}

In addition, by \eqref{eq:lem1_2_8}, we have
\begin{align*}
\max_{1 \leq j \leq m} n \left\Vert \Sigma_X \left[\delta_j - (EX_iX_i'\xi_{i,j})^{-1}(E_nX_i \eps_{i,j}\xi_{i,j} + (E_n X_i L_i'\xi_{i,j})F_j)\right]  F_j' \right\Vert \lessp \alpha_n^{-1} \log^{5/2} n.
\end{align*}
This implies
\begin{align*}
n \left(E_n \hat \pi_i^{-1}  \xi_{i,j} X_i
X_i'\right) \delta_j  F_j' & = n \Sigma_X (EX_iX_i'\xi_{i,j})^{-1}(E_nX_i \eps_{i,j}\xi_{i,j} + (E_n X_i L_i'\xi_{i,j})F_j)  F_j' \\
& + O_P(\alpha_n^{-1} \log^{5/2} n) \\
& = O_P( \sqrt{n \alpha_n^{-1}\log^3 n})
\end{align*}
where the $O_P(\cdot)$ term holds uniformly over $j$.

Therefore, \eqref{eq:Q1j_1} implies
\begin{align*}
& \sum_{i=1}^n \xi_{i,j} X_i \sum_{s=1}^m \left(\hat
\pi_i^{-1} \xi_{i,s} X_i'\delta_s \right) F_s' \\
& = n \sum_{s \neq j} (E_nX_i \eps_{i,s}\xi_{i,s} + (E_n X_i L_i'\xi_{i,s})F_s)  F_s' + O_P( \sqrt{n \alpha_n^{-1}\log^3 n}) + O_P(m \log^2 n) \\
& = n \sum_{s =1}^m (E_nX_i \eps_{i,s}\xi_{i,s} + (E_n X_i L_i'\xi_{i,s})F_s)  F_s' + O_P( \sqrt{n \alpha_n^{-1}\log^3 n}) + O_P(m \log^2 n),
\end{align*}
where  $O_P(\cdot)$ term holds uniformly over $j$. Furthermore, \eqref{eq:lem6_step2_1} implies
\begin{align*}
\sum_{i=1}^n \xi_{i,j} X_i e_i'F = - n \sum_{s =1}^m (E_nX_i \eps_{i,s}\xi_{i,s} + (E_n X_i L_i'\xi_{i,s})F_s)  F_s' +  O_P(m \log^2 n + (nm)^{1/2} \log n) ,
\end{align*}
where  $O_P(\cdot)$ term holds uniformly over $j$, which further implies
\begin{align*}
& \max_{1 \leq j \leq m}||Q_{1,j} + (nm)^{-1} \sum_{s =1}^m (E_nX_i \eps_{i,s}\xi_{i,s} + (E_n X_i L_i'\xi_{i,s})F_s)  F_s'L' \hat L \hat \Omega_r^{-2}|| \\
& \lessp (n^{-2}m^{-1})(m \log^2 n + (nm)^{1/2} \log n) ||L' \hat L|| \lessp n^{-1}\log^2 n + (nm)^{-1/2}\log n.
\end{align*}

We also note that
\begin{align*}
(nm)^{-1} \left\Vert \sum_{s =1}^m E_nX_i \eps_{i,s}\xi_{i,s}  F_s'L' \hat L \hat \Omega_r^{-2} \right\Vert \lessp (nm)^{-1}\left\Vert \sum_{s =1}^m \sum_{i=1}^n X_i \eps_{i,s}\xi_{i,s}  F_s'\right\Vert \lessp (nm)^{-1/2} \alpha_n^{1/2}
\end{align*}
and
\begin{align*}
\left\Vert (nm)^{-1} \sum_{s =1}^m (E_n X_i L_i'(\xi_{i,s} - \pi_i))F_s  F_s'L' \hat L \hat \Omega_r^{-2}\right\Vert  \lessp  (nm)^{-1/2} \alpha_n^{1/2}.
\end{align*}

This implies that
\begin{align*}
\max_{1 \leq j \leq m}\left\Vert Q_{1,j} + (nm)^{-1} \sum_{s =1}^m  (E_n X_i L_i'\pi_{i})F_s  F_s'L' \hat L \hat \Omega_r^{-2}\right\Vert \lessp n^{-1}\log^2 n + (nm)^{-1/2}\log n.
\end{align*}
Given the bound for $Q_{2,j}$ derived in Step 1, we obtain the desired result.

\end{proof}

\subsection{More auxiliary results: bound on $E_{n}\xi_{i,j}X_{i}\protect\hL_{i}'$ and $\protect\hL'\Delta_{L}$}
\begin{lem}[bound on $E_{n}\xi_{i,j}X_{i}\protect\hL_{i}'$]\label{lem: part 7}\label{lem: order of Delta_L}
Let Assumption \ref{assu: moment cond} hold. Then, we have
	\begin{enumerate}
		\item $||L' e F|| \lessp (nm\alpha_n^{-1})^{1/2} +  m \alpha_n^{-1} \log^{5/4} n$,
		\item 	\begin{align*}
		\max_{1\leq s\leq m}\left\Vert\sum_{i=1}^n \sum_{j=1}^m \xi_{i,s}L_i e_{i,j} F_j' \right\Vert \lessp \sqrt{nm \log n} + m \alpha_n^{-1/2} \log^{5/4} n.
		\end{align*}
  \item $||L' \diag(\pi) e F|| \lessp (nm\alpha_n)^{1/2} +  m \log^{5/4} n$
	\end{enumerate}

\end{lem}
\begin{proof}[\textbf{Proof of Lemma \ref{lem: part 7}}]

We have
\begin{align}
L'eF & = \sum_{i=1}^n \sum_{j=1}^m L_i e_{i,j}F_j' \notag \\
& = \sum_{i=1}^n \sum_{j=1}^m L_i L_i'F_j \xi_{i,j}(\hat \pi_i^{-1} - \pi_i^{-1})F_j' + \sum_{i=1}^n \sum_{j=1}^m L_i L_i'F_j\pi_i^{-1} (\xi_{i,j} - \pi_i)F_j' \notag \\
& + \sum_{i=1}^n \sum_{j=1}^m L_i \hat \pi_i^{-1} \xi_{i,j}\eps_{i,j}F_j' - \sum_{i=1}^n \sum_{j=1}^m L_i \hat \pi_i^{-1} \xi_{i,j} X_i'\delta_jF_j'.
\label{eq:lem7_1}
\end{align}
For the first term on the RHS of \eqref{eq:lem7_1}, we have
\begin{align*}
\left\Vert \sum_{i=1}^n \sum_{j=1}^m L_i L_i'F_j \xi_{i,j}(\hat \pi_i^{-1} - \pi_i^{-1})F_j' \right\Vert & \lessp \alpha_n^{-2}(nm)^{-1} n^{1/q}\sum_{i=1}^n \sum_{j=1}^m \xi_{i,j} ||L_i||_2^2||F_j||_2^2 \\
& + \left\Vert \sum_{i=1}^n \sum_{j=1}^m \xi_{i,j} \frac{1-\pi_i}{\pi_i}X_i'(\hat{\gamma} - \gamma) L_iL_i'F_jF_j'\right\Vert \\
& \lessp \alpha_n^{-1} n^{1/q} + (nm)^{1/2} \alpha_n^{-1/2} \lessp (nm)^{1/2} \alpha_n^{-1/2}.
\end{align*}

For the second term on the  RHS of \eqref{eq:lem7_1}, by the Markov's inequality, we have
\begin{align*}
\left\Vert \sum_{i=1}^n \sum_{j=1}^m L_i L_i'F_j\pi_i^{-1} (\xi_{i,j} - \pi_i)F_j' \right\Vert \lessp (nm)^{1/2} \alpha_n^{-1/2}.
\end{align*}

For the third term on the  RHS of \eqref{eq:lem7_1}, we have
\begin{align*}
\left\Vert \sum_{i=1}^n \sum_{j=1}^m L_i \hat \pi_i^{-1} \xi_{i,j} \eps_{i,j}F_j' \right\Vert & \lessp \left\Vert \sum_{i=1}^n \sum_{j=1}^m L_i \pi_i^{-1} \eps_{i,j} \xi_{i,j}F_j' \right\Vert + \left\Vert \sum_{i=1}^n \sum_{j=1}^m L_i \frac{1-\pi_i}{\pi_i}X_i'(\hat \gamma - \gamma) \eps_{i,j}\xi_{i,j}F_j' \right\Vert \\
& + \alpha_n^{-2}(nm)^{-1}n^{1/q} \sum_{i=1}^n \sum_{j=1}^m ||L_i||_2  |\eps_{i,j}| \xi_{i,j}||F_j||_2 \\
& \lessp (nm\alpha_n^{-1})^{1/2} + \alpha_n^{-1} n^{1/q} \lessp (nm)^{1/2} \alpha_n^{-1/2}.
\end{align*}

For the last term  on the  RHS of \eqref{eq:lem7_1}, we have
\begin{align*}
\left\Vert \sum_{i=1}^n \sum_{j=1}^m L_i \hat \pi_i^{-1} \xi_{i,j} X_i'\delta_jF_j'\right\Vert & \lessp  \max_{1\leq j \leq m}\left\Vert \sum_{i=1}^n L_i \pi_i^{-1} \xi_{i,j} X_i'\right\Vert \sum_{j=1}^m ||\delta_j||_2 ||F_j||_2 \\
& + \max_{1 \leq j \leq m} \left\Vert \sum_{i=1}^n L_i \frac{1-\pi_i}{\pi_i}X_i'(\hat{\gamma} - \gamma) \xi_{i,j} X_i' \right\Vert  \sum_{j=1}^m ||\delta_j||_2||F_j||_2\\
& + \alpha_n^{-2}(nm)^{-1}n^{1/q} \sum_{i=1}^n \sum_{j=1}^m ||L_i||_2 \xi_{i,j} ||X_i||_2 ||\delta_j||_2 ||F_j||_2 \\
& \lessp m \alpha_n^{-1} \log^{5/4} n,
\end{align*}
where the last inequality holds by \eqref{eq:lem1_9}, \eqref{eq:ex_lem_3}, and the fact that
\begin{align*}
    \max_{1 \leq j \leq  m}\left\Vert \sum_{i=1}^n L_i \frac{1-\pi_i}{\pi_i}X_i'\xi_{i,j} X_i' \right\Vert \lessp \sqrt{n \alpha_n^{-1}} n^{1/q}
\end{align*}

This implies
\begin{align*}
||L'eF|| \lessp (nm\alpha_n^{-1})^{1/2} +  m \alpha_n^{-1} \log^{5/4} n.
\end{align*}

For the second result of Lemma \ref{lem: part 7}, note that
\begin{align}
	\max_{1\leq s\leq m}\left\Vert\sum_{i=1}^n \sum_{j=1}^m \xi_{i,s}L_i e_{i,j} F_j' \right\Vert & \lessp 	\max_{1\leq s\leq m}\left\Vert\sum_{i=1}^n \sum_{j=1}^m \xi_{i,s}L_i L_i'F_j\xi_{i,j}(\hat \pi_i^{-1} - \pi_i^{-1}) F_j' \right\Vert \notag \\
	& + \max_{1\leq s\leq m}\left\Vert\sum_{i=1}^n \sum_{j=1}^m \xi_{i,s}L_i L_i'F_j \pi_i^{-1}(\xi_{i,j} - \pi_i) F_j' \right\Vert \notag \\
	& + \max_{1\leq s\leq m}\left\Vert\sum_{i=1}^n \sum_{j=1}^m \xi_{i,s}L_i \hat \pi_i^{-1} \xi_{i,j}\eps_{i,j} F_j' \right\Vert \notag \\
	& + \max_{1\leq s\leq m}\left\Vert\sum_{i=1}^n \sum_{j=1}^m \xi_{i,s}L_i \hat \pi_i^{-1} \xi_{i,j}X_i'\delta_j F_j' \right\Vert.
	\label{eq:lem7_3}
\end{align}
For the first term on the RHS of \eqref{eq:lem7_3}, by \eqref{eq:ex_lem_8} and \eqref{eq:ex_lem_9}, we have
\begin{align*}
& 	\max_{1\leq s\leq m}\left\Vert\sum_{i=1}^n \sum_{j=1}^m \xi_{i,s}L_i L_i'F_j\xi_{i,j}(\hat \pi_i^{-1} - \pi_i^{-1}) F_j' \right\Vert \\
& \lessp \alpha_n^{-2} (nm)^{-1}n^{1/q} \max_{1\leq s\leq m} \sum_{i=1}^n \sum_{j=1}^m \xi_{i,s} ||L_i||_2^2 ||F_j||_2^2 \xi_{i,j} \\
 	& +  \max_{1\leq s\leq m}\left\Vert\sum_{i=1}^n \sum_{j=1}^m \xi_{i,s}L_i L_i'F_j\xi_{i,j}\left(\frac{1-\pi_i}{\pi_i}X_i'(\hat{\gamma} - \gamma)\right) F_j' \right\Vert\\
 	& \lessp  n^{1/q} + \sqrt{nm\alpha_n} \lessp \sqrt{nm\alpha_n}.
\end{align*}

For the second term on the RHS of \eqref{eq:lem7_3}, by  \eqref{eq:lem1_5_10}, we have
\begin{align*}
\max_{1\leq s\leq m}\left\Vert\sum_{i=1}^n \sum_{j=1}^m \xi_{i,s}L_i L_i'F_j \pi_i^{-1}(\xi_{i,j} - \pi_i) F_j' \right\Vert \lessp \sqrt{mn \log n}.
\end{align*}

For the third term on the RHS of \eqref{eq:lem7_3}, we have
\begin{align*}
\max_{1\leq s\leq m}\left\Vert\sum_{i=1}^n \sum_{j=1}^m \xi_{i,s}L_i \hat \pi_i^{-1} \xi_{i,j}\eps_{i,j} F_j' \right\Vert & \lessp \max_{1\leq s\leq m}\left\Vert\sum_{i=1}^n \sum_{j=1}^m \xi_{i,s}L_i \pi_i^{-1} \xi_{i,j}\eps_{i,j} F_j' \right\Vert \\
& + \max_{1\leq s\leq m} \alpha_n^{-2} (mn)^{-1} n^{1/q} \sum_{i=1}^n \sum_{j=1}^m \xi_{i,s} ||L_i||_2  \xi_{i,j}|\eps_{i,j}| ||F_j||_2 \\
& + \max_{1\leq s\leq m}\left\Vert\sum_{i=1}^n \sum_{j=1}^m \xi_{i,s}L_i \left(\frac{1-\pi_i}{\pi_i} \right)X_i'(\hat \gamma - \gamma) \xi_{i,j}\eps_{i,j} F_j' \right\Vert \\
& \lessp \sqrt{nm \log n},
\end{align*}
where the second inequality is by \eqref{eq:lem1_5_9}, \eqref{eq:ex_lem_10}, and the fact that
\begin{align*}
\max_{1\leq s\leq m} \sum_{i=1}^n \sum_{j=1}^m \xi_{i,s} ||L_i||_2 ||F_j||_2 |\eps_{i,j}| \xi_{i,j} \lessp nm \alpha_n^2.
\end{align*}

For the last term on the RHS of \eqref{eq:lem7_3}, we have
\begin{align*}
& \max_{1\leq s\leq m}\left\Vert\sum_{i=1}^n \sum_{j=1}^m \xi_{i,s}L_i \hat \pi_i^{-1} \xi_{i,j}X_i'\delta_j F_j' \right\Vert \\
& \lessp  \max_{1\leq s\leq m}\left\Vert\sum_{i=1}^n \sum_{j=1}^m \xi_{i,s}L_i \pi_i^{-1} \xi_{i,j}X_i'\delta_j F_j' \right\Vert \\
 & +  \alpha_n^{-2}(nm)^{-1}n^{1/q} \max_{1\leq s\leq m}\sum_{i=1}^n \sum_{j=1}^m \xi_{i,s} ||L_i||_2 \xi_{i,j} ||X_i||_2 ||\delta_j||_2 ||F_j||_2 \\
 & +  \max_{1\leq s\leq m} \sum_{i=1}^n \sum_{j=1}^m \xi_{i,s} ||L_i||_2 \left(\frac{1-\pi_i}{\pi_i}\right) ||X_i||_2^2 ||\hat \gamma - \gamma||_2 \xi_{i,j} ||\delta_j||_2 || F_j||_2 \\
 & \lessp \left(\sum_{j =1}^m ||\delta_j||_2 ||F_j||_2\right) \max_{1\leq j,s\leq m, j\neq s} \left\Vert\sum_{i=1}^n \xi_{i,s}\xi_{i,j}L_i X_i'\pi_i^{-1}\right\Vert\\
 & + \max_{1\leq j\leq m} \left\Vert \sum_{i=1}^n \xi_{i,j}L_iX_i' \pi_i^{-1}\right\Vert ||\delta_j||_2 ||F_j||_2 + (n \alpha_n)^{-1/2} n^{1/q} + \sqrt{m \log^{3/2} n} \\
 & \lessp m \alpha_n^{-1/2} n^{1/q},
\end{align*}
where the last inequality is by \eqref{eq:lem1_9}, \eqref{eq:ex_lem_3}, and \eqref{eq:ex_lem_16}.

This implies
\begin{align*}
\max_{1\leq s\leq m}\left\Vert\sum_{i=1}^n \sum_{j=1}^m \xi_{i,s}L_i e_{i,j} F_j' \right\Vert \lessp \sqrt{nm \log n} + m \alpha_n^{-1/2} n^{1/q}.
\end{align*}

For the last result of Lemma \ref{lem: part 7}, we have
\begin{align}
L'\diag(\pi)eF & = \sum_{i=1}^n \sum_{j=1}^m L_i \pi_i e_{i,j}F_j' \notag \\
& = \sum_{i=1}^n \sum_{j=1}^m L_i \pi_i L_i'F_j \xi_{i,j}(\hat \pi_i^{-1} - \pi_i^{-1})F_j' + \sum_{i=1}^n \sum_{j=1}^m L_i L_i'F_j (\xi_{i,j} - \pi_i)F_j' \notag \\
& + \sum_{i=1}^n \sum_{j=1}^m L_i \pi_i \hat \pi_i^{-1} \xi_{i,j}\eps_{i,j}F_j' - \sum_{i=1}^n \sum_{j=1}^m L_i \pi_i \hat \pi_i^{-1} \xi_{i,j} X_i'\delta_jF_j'.
\label{eq:lem7_4}
\end{align}
For the first term on the RHS of \eqref{eq:lem7_4}, we have
\begin{align*}
\left\Vert \sum_{i=1}^n \sum_{j=1}^m L_i \pi_i L_i'F_j \xi_{i,j}(\hat \pi_i^{-1} - \pi_i^{-1})F_j' \right\Vert & \lessp \alpha_n^{-2}(nm)^{-1} n^{1/q}\sum_{i=1}^n \sum_{j=1}^m \xi_{i,j} \pi_i ||L_i||_2^2||F_j||_2^2 \\
& + \left\Vert \sum_{i=1}^n \sum_{j=1}^m \xi_{i,j} (1-\pi_i)X_i'(\hat{\gamma} - \gamma) L_iL_i'F_jF_j'\right\Vert \\
& \lessp n^{1/q} + (nm)^{1/2} \alpha_n^{-1/2} \lessp (nm \alpha_n)^{1/2}.
\end{align*}

For the second term on the  RHS of \eqref{eq:lem7_4}, by the Markov's inequality, we have
\begin{align*}
\left\Vert \sum_{i=1}^n \sum_{j=1}^m L_i L_i'F_j(\xi_{i,j} - \pi_i)F_j' \right\Vert \lessp (nm \alpha_n)^{1/2}.
\end{align*}

For the third term on the  RHS of \eqref{eq:lem7_4}, we have
\begin{align*}
\left\Vert \sum_{i=1}^n \sum_{j=1}^m L_i \pi_i \hat \pi_i^{-1} \xi_{i,j} \eps_{i,j}F_j' \right\Vert & \lessp \left\Vert \sum_{i=1}^n \sum_{j=1}^m L_i \eps_{i,j} \xi_{i,j}F_j' \right\Vert + \left\Vert \sum_{i=1}^n \sum_{j=1}^m L_i (1-\pi_i)X_i'(\hat \gamma - \gamma) \eps_{i,j}\xi_{i,j}F_j' \right\Vert \\
& + \alpha_n^{-2}(nm)^{-1}n^{1/q} \sum_{i=1}^n \sum_{j=1}^m ||L_i||_2  |\eps_{i,j}| \pi_i \xi_{i,j}||F_j||_2 \\
& \lessp (nm\alpha_n)^{1/2} + n^{1/q} \lessp (nm \alpha_n)^{1/2}.
\end{align*}

For the last term  on the  RHS of \eqref{eq:lem7_4}, we have
\begin{align*}
\left\Vert \sum_{i=1}^n \sum_{j=1}^m L_i \pi_i \hat \pi_i^{-1} \xi_{i,j} X_i'\delta_jF_j'\right\Vert & \lessp  \max_{1\leq j \leq m}\left\Vert \sum_{i=1}^n L_i \xi_{i,j} X_i'\right\Vert \sum_{j=1}^m ||\delta_j||_2 ||F_j||_2 \\
& + \max_{1 \leq j \leq m} \left\Vert \sum_{i=1}^n L_i (1-\pi_i)X_i'(\hat{\gamma} - \gamma) \xi_{i,j} X_i' \right\Vert  \sum_{j=1}^m ||\delta_j||_2||F_j||_2\\
& + \alpha_n^{-2}(nm)^{-1}n^{1/q} \sum_{i=1}^n \sum_{j=1}^m \pi_i ||L_i||_2 \xi_{i,j} ||X_i||_2 ||\delta_j||_2 ||F_j||_2 \\
& \lessp m \log^{5/4} n,
\end{align*}
where the last inequality holds by \eqref{eq:lem1_9} and the facts that
\begin{align*}
& \max_{1\leq j \leq m}\left\Vert \sum_{i=1}^n L_i \xi_{i,j} X_i'\right\Vert \lessp (n \alpha_n \log n)^{1/2}, \\
 &   \max_{1 \leq j \leq  m}\left\Vert \sum_{i=1}^n L_i (1-\pi_i)X_i'\xi_{i,j} X_i' \right\Vert \lessp \sqrt{n \alpha_n} n^{1/q}.
\end{align*}

This implies
\begin{align*}
||L'\diag(\pi) eF|| \lessp (nm\alpha_n)^{1/2} +  m \log^{5/4} n.
\end{align*}

\end{proof}

\begin{lem}[bound on $\protect\hL'\Delta_{L}$]
\label{lem: part 8}Let Assumption \ref{assu: moment cond} hold.
Then, we have
\begin{align*}
& \max_{1 \leq j \leq m} ||E_n \xi_{i,j} X_i \hat L_i'|| \lessp \sqrt{n^{-1} \alpha_n \log n}, \\
& ||L'\Delta_L|| \lessp  \alpha_n^{-1}(1+n/m)n^{1/q},\\
& ||\hat L'\Delta_L|| \lessp  \alpha_n^{-1}(1+n/m)n^{1/q},\\
& ||H^{-1}|| \lessp 1, \\
& H^{-1} = \hat \Omega_r^{2}\left(n^{-1}L' \hat L\right)^{-1}\Sigma_F^{-1} + O_P((n\alpha_n)^{-1} \log^{5/4} n + (m \alpha_n)^{-1}  (1+m/n)^{1/2}n^{1/q}),\\
& H^{-1}\Sigma_L^{-1} = H' + O_P( (n\alpha_n)^{-1} (1+n/m)n^{1/q})\\
& ||L'\diag(\pi)\Delta_L|| \lessp  (1+n/m)n^{1/q}.
\end{align*}
\end{lem}
\begin{proof}[\textbf{Proof of Lemma \ref{lem: part 8}}]

Because $\hat L = LH + \Delta_L$, we have $\hat L_i' = L_i' H + \Delta_{L,i}'$, and thus,
\begin{align*}
E_n \xi_{i,j} X_i \hat L_i' = E_n \xi_{i,j} X_i L_i' H + E_n \xi_{i,j} X_i \Delta_{L,i}'.
\end{align*}
By Lemma \ref{lem: part 6}, we have $\max_{1 \leq j \leq m}||E_n \xi_{i,j} X_i \Delta_{L,i}'|| \lessp \alpha_n n^{-1/2} + n^{-1}\log^2 n + (nm)^{-1/2}\log n$. In addition, we have $\max_{1 \leq j \leq m} ||E_n \xi_{i,j} X_i L_i' || \lessp \sqrt{n^{-1} \alpha_n \log n}$, which implies
\begin{align*}
\max_{1 \leq j \leq m}||E_n \xi_{i,j} X_i \hat L_i' || \lessp \sqrt{n^{-1} \alpha_n \log n}.
\end{align*}

By the definition of $\Delta_L$ and Lemmas \ref{lem: part 2} and \ref{lem: part 7}, we have
\begin{align*}
||L'\Delta_L|| & = ||L'e W' \hat L \hat \Omega_r^{-2}||(nm)^{-1} \\
& \leq ||L'e FL' \hat L \hat \Omega_r^{-2}||(nm)^{-1} + ||L'e e' \hat L \hat \Omega_r^{-2}||(nm)^{-1}\\
& \lessp ||L'eF|| m^{-1} + ||L'e|| ||\hat L' e|| (nm)^{-1} \\
& \lessp ||L'eF|| m^{-1} + || e||^2 m^{-1} \\
& \lessp (n m^{-1}\alpha_n^{-1})^{1/2} + n^{1/q} (1+n/m)  \alpha_n^{-1}\\
& \lessp n^{1/q} (1+n/m)  \alpha_n^{-1}.
\end{align*}

This further implies
\begin{align*}
||\hat L'\Delta_L|| \leq ||\Delta_L'\Delta_L|| + ||H' L'\Delta_L|| \lessp  \alpha_n^{-1}(1+n/m)n^{1/q},
\end{align*}
where the second inequality is by Lemmas \ref{lem: part 4} and \ref{lem: part 5}.
Next, note that $I = \hat L' \hat L/n = (H'L'+\Delta_L')(HL+\Delta_L)/n$. This means
\begin{align}
H'\Sigma_L H & = I - \Delta_L' LH/n - H'L'\Delta_L/n - \Delta_L'\Delta_L/n \notag \\
& = I - O_P( (n\alpha_n)^{-1} (1+n/m)n^{1/q})  = I - o_P(1).
\label{eq:HSigmaLH}
\end{align}
This implies $\Sigma_L = (H^{-1})'(I - o_P(1))H^{-1}$, and thus, $H^{-1} = O_P(1)$. It also implies that
\begin{align*}
H^{-1}\Sigma_L^{-1} = H' + O_P((n\alpha_n)^{-1} (1+n/m)n^{1/q}).
\end{align*}

Last, we note that
\begin{align*}
H = F'W' \hat L \hat \Omega_r^{-2}(nm)^{-1} = \Sigma_F L' \hat L \hat \Omega_r^{-2} n^{-1} + F'e' \hat L \hat \Omega_r^{-2} (nm)^{-1}
\end{align*}
and by Lemmas \ref{lem: part 3}, \ref{lem: part 4}, \ref{lem: part 5}, and \ref{lem: part 7},
\begin{align*}
||\hat L' e F|| \leq ||H|| ||L'e F|| + ||\Delta_L|| ||F'e'|| \lessp m \alpha_n^{-1} \log^{5/4} n + \alpha_n^{-1} n^{1/q} \sqrt{n(m+n)}.
\end{align*}
This implies
\begin{align*}
||H - \Sigma_F L' \hat L \hat \Omega_r^{-2} n^{-1}|| \lessp (n\alpha_n)^{-1} \log^{5/4} n + (m \alpha_n)^{-1}  (1+n/m)^{1/2}n^{1/q},
\end{align*}
and thus,
\begin{align*}
||H^{-1} -   \hat \Omega_r^{2} (n^{-1}L' \hat L)^{-1}\Sigma_F^{-1}|| \lessp (n\alpha_n)^{-1} \log^{5/4} n + (m \alpha_n)^{-1}  (1+m/n)^{1/2}n^{1/q},
\end{align*}
given $H^{-1} = O_P(1)$.

Last, we have
\begin{align*}
||L'\diag(\pi)\Delta_L||& = ||L'\diag(\pi)e W' \hat L \hat \Omega_r^{-2}||(nm)^{-1} \\
& \leq ||L'\diag(\pi)e FL' \hat L \hat \Omega_r^{-2}||(nm)^{-1} + ||L'\diag(\pi)e e' \hat L \hat \Omega_r^{-2}||(nm)^{-1}\\
& \lessp ||L'\diag(\pi)eF|| m^{-1} + ||L'\diag(\pi)e|| ||\hat L' e|| (nm)^{-1} \\
& \lessp ||L'\diag(\pi)eF|| m^{-1} + \alpha_n n^{1/q} || e||^2 m^{-1} \\
& \lessp (n m^{-1}\alpha_n)^{1/2} + n^{1/q} (1+n/m)\\
& \lessp n^{1/q} (1+n/m).
\end{align*}

\end{proof}

\section{Proof of Theorem \ref{thm: beta est} at $g=1$} \label{app: sub proof of beta est g=1}
We aim to show that
\begin{align}
\max_{1 \leq j \leq m} \left\Vert \tilde \beta^{(1)}_j - \beta_j - (EX_iX_i'\pi_i)^{-1} \left[E_n(X_i \eps_{i,j} \xi_{i,j} + \pi_i X_iL_i'F_j) \right]\right\Vert_2 =  O_P(\kappa_n).  \label{eq: betatilda}
\end{align}

Recall  $\Delta_{i,j} = \hat \Gamma_{i,j} - \Gamma_{i,j} = \Delta'_{L,i}H^{-1}F_j - n^{-1}\hat L_i' \hat L' \Delta_L H^{-1}F_j+n^{-1}\hat L_i \hat L'e_{\cdot,j}$. Then, we have
\begin{align*}
\tilde \beta_j^{(1)} - \beta_j = (E_n \xi_{i,j}X_iX_i')^{-1} E_n(X_i \xi_{i,j}  (\eps_{i,j} - \Delta_{i,j}))
\end{align*}
and
\begin{align*}
E_n (\xi_{i,j}X_i \Delta_{i,j} ) = E_n (\xi_{i,j}X_i \Delta_{L,i}H^{-1}F_j) - n^{-1}E_n (\xi_{i,j}X_i \hat L_i' \hat L' \Delta_L H^{-1}F_j) + n^{-1}E_n (\xi_{i,j}X_i \hat L_i \hat L'e_{\cdot,j}).
\end{align*}

We first notice that uniformly over $j$,
\begin{align*}
& E_n (\xi_{i,j}X_i \Delta_{L,i}H^{-1}F_j)\\
& = - (nm)^{-1}(E_n \pi_iX_iL_i')F'FL'\hat L \hat \Omega_r^{-2}H^{-1}F_j + O_P((n^{-1}+(nm)^{-1/2}) \log^2 n)\\
& = - (nm)^{-1}(E_n \pi_iX_iL_i')F'FL'\hat L \hat \Omega_r^{-2}(\hat \Omega_r^2(n^{-1}L'\hat L)^{-1}\Sigma_F^{-1} + R_{1,n})F_j + O_P((n^{-1}+(nm)^{-1/2}) \log^2 n)\\
& =  -E_n \pi_iX_iL_i'F_j +  O_P((n^{-1}+(nm)^{-1/2}) \log^2 n)
\end{align*}
where the first equality is by Lemma \ref{lem: part 6}, $R_{1,n}$ in the second equality satisfies $$||R_{1,n}|| \lessp (n\alpha_n)^{-1} \log^{5/4} n + (m \alpha_n)^{-1}  (1+m/n)^{1/2}n^{1/q} \log n$$ by Lemma \ref{lem: part 8}, and the last inequality is by the fact that
\begin{align*}
||E_n \pi_iX_iL_i|| \lessp \alpha_n \sqrt{n^{-1} \log n}.
\end{align*}

Next, we have, uniformly over $j$, that
\begin{align*}
& n^{-1}|E_n (\xi_{i,j}X_i \hat L_i' \hat L' \Delta_L H^{-1}F_j)| \\
& \leq n^{-1}|E_n (\xi_{i,j}X_i L_i'H \hat L' \Delta_L H^{-1}F_j)| + n^{-1}|E_n (\xi_{i,j}X_i \Delta_{L,i}' \hat L' \Delta_L H^{-1}F_j)|\\
& \lessp n^{-1}||E_n\xi_{i,j}X_i L_i'|| ||H|| ||\hat L' \Delta_L|| ||H^{-1}|| ||F_j||_2 + n^{-1}\left\Vert E_n \xi_{i,j}X_i\Delta_{L,i}'\right\Vert || \hat L' \Delta_L|| ||H^{-1} || ||F_j||_2\\
& \lessp  (n\alpha_n)^{-1/2} (n^{-1} + m^{-1}) n^{1/q}
\end{align*}
where the third inequality is by Lemmas \ref{lem: part 6} and \ref{lem: part 8} and the fact that $\max_{1 \leq j \leq m}||E_n \xi_{i,j} X_i L_i'||\lessp \sqrt{n^{-1} \alpha_n \log n}$.

Finally, we observe that, by Lemma \ref{lem: part 3},
\begin{align*}
n^{-1} ||E_n\xi_{i,j}X_i \hat L_i \hat L'e_{\cdot,j} ||_2 &  \lessp n^{-1} ||E_n \xi_{i,j} X_i \hat L_i'|| ||\hat L' e_{\cdot,j}||_2 \\
& \lessp n^{-1} (||E_n \xi_{i,j} X_i L_i'|| + ||E_n \xi_{i,j} X_i \Delta_{L,i}'||) ||\hat L' e_{\cdot,j}||_2 \\
& \lessp n^{-1} (\sqrt{n^{-1} \alpha_n \log n}) \sqrt{n(1+n/m)\alpha_n^{-2}} n^{1/q} \\
& \lessp (1/n+1/m) \alpha_n^{-1/2} n^{1/q}.
\end{align*}

This implies that, uniformly over $j$,
\begin{align*}
E_n X_i \Delta_{i,j}\xi_{i,j} = - E_n \pi_iX_i L_i'F_j + O_P((1/n+1/m) \alpha_n^{-1/2} n^{1/q}).
\end{align*}
In addition, we have
\begin{align*}
\max_{1 \leq j \leq m} ||E_n \pi_iX_i L_i'F_j|| \lessp \alpha_n n^{-1/2}\log^{1/2} n \quad \text{and} \quad (EX_iX_i' \xi_{i,j})^{-1} = (EX_iX_i' \pi_i)^{-1} \lesssim \alpha_n^{-1}.
\end{align*}

By \eqref{eq:lem1_1}, we have that, uniformly over $j$
\begin{align*}
& (E_nX_iX_i' \xi_{i,j})^{-1}E_n X_i \Delta_{i,j}\xi_{i,j} \\
& = \left[(EX_iX_i' \xi_{i,j})^{-1}+O_P(\alpha_n^{-2} \sqrt{n^{-1}\alpha_n^{-1} \log n})\right]\left[- E_n \pi_iX_i L_i'F_j + O_P((1/n+1/m) \alpha_n^{-1/2} n^{1/q})\right] \\
& = -(EX_iX_i' \xi_{i,j})^{-1} E_n \pi_iX_i L_i'F_j +O_P((1/n+1/m) \alpha_n^{-3/2} n^{1/q}).
\end{align*}
Similarly, we have that, uniformly over $j$,
\begin{align*}
(E_nX_iX_i' \xi_{i,j})^{-1}E_n X_i \eps_{i,j}\xi_{i,j} = (EX_iX_i' \xi_{i,j})^{-1}E_n X_i \eps_{i,j}\xi_{i,j} + O_P((1/n+1/m) \alpha_n^{-3/2} n^{1/q}).
\end{align*}
This leads to the desired result.

\section{Proofs of Theorems \ref{thm: beta est} and \ref{thm: Gamma est} for  $g\geq 1$}{Proofs of Theorems \ref{thm: beta est} and \ref{thm: Gamma est} for any finite $g$ satisfying $g\geq 1$}\label{app: proof of beta and gamma est}

\subsection{Preliminary results}
\begin{lem}
\label{lem: part 9} Let Assumption \ref{assu: moment cond} hold. Then
\begin{align*}
\max_{1 \leq s \leq m}\left\Vert \sum_{i=1}^n L_i \xi_{i,s}e_i'\right\Vert \lessp (mn )^{1/2} n^{1/q},
\end{align*}	
\begin{align*}
\max_{1 \leq s \leq m}\left\Vert\sum_{i=1}^n L_i \xi_{i,s}\Delta_{L,i}'\right\Vert \lessp (1+n/m)\alpha_n^{-1/2} n^{1/q},
\end{align*}	
\begin{align*}
\max_{1 \leq s \leq m}\left\Vert \sum_{i=1}^n L_i (\xi_{i,s} - \pi_i)e_i'\right\Vert \lessp (mn )^{1/2} n^{1/q},
\end{align*}
\begin{align*}
 \max_{1 \leq s \leq m}\left\Vert\sum_{i=1}^n L_i (\xi_{i,s} - \pi_i)\Delta_{L,i}'\right\Vert \lessp (1+n/m)\alpha_n^{-1/2} n^{1/q}.
\end{align*}

\end{lem}
\begin{proof}[\textbf{Proof of Lemma \ref{lem: part 9}}]

For the first result of Lemma \ref{lem: part 9}, let
\begin{align*}
& B_{\cdot,j}^{(s,1)} = \sum_{i=1}^n L_i \xi_{i,s}L_i'F_j (\hat \pi_i^{-1} - \pi_i^{-1})\xi_{i,j} \\
& B_{\cdot,j}^{(s,2)} = \sum_{i=1}^n  L_i \xi_{i,s}L_i'F_j \pi_i^{-1} (\xi_{i,j} - \pi_i) \\
& B_{\cdot,j}^{(s,3)} =  \sum_{i=1}^n L_i \xi_{i,s}\hat \pi_i^{-1} \xi_{i,j} \eps_{i,j} \\
& B_{\cdot,j}^{(s,4)} =  \sum_{i=1}^n L_i \xi_{i,s}\hat \pi_i^{-1} \xi_{i,j} X_i'\delta_j
\end{align*}
and $B^{(s,d)} = (B_{\cdot,1}^{(s,d)}, \cdots, B_{\cdot,m}^{(s,d)})$ for $d =1,\cdots,4$. Then, by \eqref{eq:e}, we have
\begin{align*}
\max_{1 \leq s \leq m}||\sum_{i=1}^n L_i \xi_{i,s}e_i'||  \leq \sum_{d=1}^4\max_{1 \leq s \leq m}||B^{(s,d)}||.
\end{align*}


For $B^{(s,1)}$, by Proposition \ref{prop:pihat}, we have
\begin{align*}
& \max_{1\leq s,j \leq m, s\neq j}\left\Vert \sum_{i=1}^n L_i \xi_{i,s}L_i'F_j (\hat \pi_i^{-1} - \pi_i^{-1})\xi_{i,j} \right\Vert_2 \\
& \lessp \alpha_n^{-2} (nm)^{-1} n^{1/q} \max_{1\leq s,j \leq m, s\neq j} \sum_{i=1}^n ||L_i||_2^2 \xi_{i,s} \xi_{i,j} ||F_j||_2 \\
& + \max_{1\leq s,j \leq m, s\neq j}\sum_{i=1}^n ||L_i||_2^2 \xi_{i,s} \xi_{i,j} ||F_j||_2 \pi_i^{-1} ||X_i||_2 ||\hat \gamma - \gamma||  \\
& \lessp m^{-1} n^{1/q} + (n\alpha_n)^{1/2} m^{-1/2} \log^{1/2} n
\end{align*}
and
\begin{align*}
& \max_{1\leq s,j \leq m, s = j}\left\Vert \sum_{i=1}^n L_i \xi_{i,s}L_i'F_j (\hat \pi_i^{-1} - \pi_i^{-1})\xi_{i,j} \right\Vert_2 \\
& \lessp \max_{1\leq j\leq m}\alpha_n^{-2} (nm)^{-1} n^{1/q} \sum_{i=1}^n ||L_i||_2^2 \xi_{i,j} ||F_j||_2 + \max_{1\leq j\leq m} \sum_{i=1}^n ||L_i||_2^2 \xi_{i,j} ||F_j||_2 \pi_i^{-1} ||X_i||_2 ||\hat \gamma - \gamma||  \\
& \lessp (m \alpha_n)^{-1} n^{1/q} + n^{1/2} (m \alpha_n)^{-1/2} \log^{1/2} n.
\end{align*}
This implies
\begin{align*}
\max_{1\leq s \leq m}||B^{(s,1)}|| \leq \max_{1\leq s \leq m}||B^{(s,1)}||_F \leq \left[\max_{1\leq s \leq m} \sum_{j = 1}^m \left\Vert \sum_{i=1}^n L_i \xi_{i,s}L_i'F_j (\hat \pi_i^{-1} - \pi_i^{-1})\xi_{i,j} \right\Vert_2^2\right]^{1/2} \lessp (n \alpha_n \log n)^{1/2}.
\end{align*}

For $B^{(s,2)}$, by \eqref{eq:ex_lem_14}, we have
\begin{align*}
\max_{1\leq s \leq m}||B^{(s,2)}|| \lessp (mn \log n)^{1/2}.
\end{align*}


For $B^{(s,3)}$, we define $B_{\cdot,j}^{(s,3,1)} = \sum_{i=1}^n L_i \xi_{i,s} \pi_i^{-1} \xi_{i,j} \eps_{i,j}$ and $B_{\cdot,j}^{(s,3,2)} = B_{\cdot,j}^{(s,3)} - B_{\cdot,j}^{(s,3,1)} = \sum_{i=1}^n L_i \xi_{i,s}(\hat \pi_i^{-1} - \pi_i^{-1})\xi_{i,j} \eps_{i,j}$, $B^{(s,3,1)} = (B_{\cdot,1}^{(s,3,1)},\cdots,B_{\cdot,m}^{(s,3,1)})$, and $B^{(s,3,2)} = (B_{\cdot,1}^{(s,3,2)},\cdots,B_{\cdot,m}^{(s,3,2)})$. Note that, by \eqref{eq:ex_lem_15}, we have
\begin{align*}
\max_{1\leq s \leq m} ||B^{(s,3,1)}|| \lessp (mn \log n)^{1/2}.
\end{align*}
We next bound $||B^{(s,3,2)}||$. Note that
\begin{align*}
& \max_{1\leq j,s \leq m, j \neq s}\left\Vert \sum_{i=1}^n L_i \xi_{i,s} (\hat\pi_i^{-1} - \pi_i^{-1})\xi_{i,j} \eps_{i,j}\right\Vert_2 \\
& \leq  \max_{1\leq j,s \leq m, j \neq s} \alpha_n^{-2} (nm)^{-1} n^{1/q} \sum_{i=1}^n ||L_i||_2 \xi_{i,s}\xi_{i,j} |\eps_{i,j}| \\
& + \max_{1\leq j,s \leq m, j \neq s}\left\Vert \sum_{i=1}^n L_i \xi_{i,s} \left(\frac{1-\pi_i}{\pi_i}\right)X_i' \xi_{i,j} \eps_{i,j}\right\Vert ||(\hat \gamma -\gamma)||_2 \\
& \lessp m^{-1} n^{1/q} + (m\alpha_n)^{-1/2} \log^{1/2} n,
\end{align*}
where the last inequality is by \eqref{eq:gammahat} and the facts that
\begin{align*}
\max_{1\leq j,s \leq m, j \neq s}  \sum_{i=1}^n ||L_i||_2 \xi_{i,s}\xi_{i,j} |\eps_{i,j}| \lessp n \alpha_n^2
\end{align*}
and
\begin{align*}
\max_{1\leq j,s \leq m, j \neq s}\left\Vert \sum_{i=1}^n L_i \xi_{i,s} \left(\frac{1-\pi_i}{\pi_i}\right)X_i' \xi_{i,j} \eps_{i,j}\right\Vert \lessp (n \log n)^{1/2}.
\end{align*}

In addition, we have
\begin{align*}
& \max_{1\leq j,s \leq m, j = s}\left\Vert \sum_{i=1}^n L_i \xi_{i,s}(\hat \pi_i^{-1} - \pi_i^{-1})\xi_{i,j} \eps_{i,j}\right\Vert_2 \\
& \leq  \max_{1\leq j \leq m} \alpha_n^{-2} (nm)^{-1} n^{1/q} \sum_{i=1}^n ||L_i||_2\xi_{i,j} |\eps_{i,j}|  + \max_{1\leq j \leq m}\left\Vert \sum_{i=1}^n  \left(\frac{1-\pi_i}{\pi_i}\right)X_i' \xi_{i,j} \eps_{i,j}\right\Vert ||(\hat \gamma -\gamma)||_2 \\
& \lessp \alpha_n^{-1} m ^{-1/2} \log^{1/2} n,
\end{align*}
which implies
\begin{align*}
\max_{1\leq s \leq m}||B^{(s,3,2)}|| & \leq \max_{1\leq s \leq m}||B^{(s,3,2)}||_F \\
& = \left[\max_{1\leq s \leq m}\sum_{1 \leq j \leq m}\left\Vert \sum_{i=1}^n L_i \xi_{i,s}(\hat \pi_i^{-1} - \pi_i^{-1}) \xi_{i,j} \eps_{i,j}\right\Vert_2^2\right]^{1/2} \\
& \lessp m^{-1/2} \alpha_n^{-1} \log^{1/2} n + \alpha_n^{-1/2} \log^{1/2} n,
\end{align*}
and thus,
\begin{align*}
\max_{1\leq s \leq m}||B^{(s,3)}|| \lessp (mn)^{1/2} n^{1/q} .
\end{align*}

For $B^{(s,4)}$, we define $B_{\cdot,j}^{(s,4,1)} =  \sum_{i=1}^n L_i \xi_{i,s} \pi_i^{-1} \xi_{i,j} X_i'\delta_j $, $B_{\cdot,j}^{(s,4,2)} = \sum_{i=1}^n L_i \xi_{i,s} (\hat \pi_i^{-1} - \pi_i^{-1}) \xi_{i,j} X_i'\delta_j $, $B^{(s,4,1)} = (B_{\cdot,1}^{(s,4,1)},\cdots,B_{\cdot,m}^{(s,4,1)})$, and $B^{(s,4,2)} = (B_{\cdot,1}^{(s,4,2)},\cdots,B_{\cdot,m}^{(s,4,2)})$. Then, by \eqref{eq:ex_lem_17}, we have
\begin{align*}
\max_{1\leq s \leq m} ||B^{(s,4,1)}|| \lessp  m^{1/2}\alpha_n^{-1/2} \log^{5/4}.
\end{align*}
In addition, we have
\begin{align*}
& \max_{1\leq j,s \leq m, j \neq s} \left\Vert \sum_{i=1}^n L_i \xi_{i,s} (\hat \pi_i^{-1} - \pi_i^{-1}) \xi_{i,j} X_i'\delta_j\right\Vert_2 \\
& \lessp   \max_{1\leq j,s \leq m, j \neq s} \left\Vert \sum_{i=1}^n L_i \xi_{i,s} (\hat \pi_i^{-1} - \pi_i^{-1})\xi_{i,j} X_i'\right\Vert \max_{1\leq j \leq m}||\delta_j||_2
\end{align*}
and by \eqref{eq:ex_lem_18}
\begin{align*}
& \max_{1\leq j,s \leq m, j \neq s} \left\Vert \sum_{i=1}^n L_i \xi_{i,s} (\hat \pi_i^{-1} - \pi_i^{-1})\xi_{i,j} X_i'\right\Vert \\
& \lessp   \max_{1\leq j,s \leq m, j \neq s} \alpha_n^{-2} (nm)^{-1} n^{1/q} \sum_{i=1}^n ||L_i||_2 ||X_i||_2  \xi_{i,s} \xi_{i,j} \\
& + \max_{1\leq j,s \leq m, j \neq s} \sum_{i=1}^n ||L_i||_2 ||X_i||_2 \pi_i^{-1} \xi_{i,s} \xi_{i,j} ||\hat \gamma - \gamma||_2 \\
& \lessp \alpha_n^{-2} (mn)^{-1} n^{1/q} (n \alpha_n^2 + \log^2 n ) +  (mn\alpha_n)^{-1/2} ( n \alpha_n n^{1/q}) \\
& \lessp m^{-1/2}(n\alpha_n)^{1/2} n^{1/q}.
\end{align*}
This implies
\begin{align*}
& \max_{1\leq j,s \leq m, j \neq s} \left\Vert \sum_{i=1}^n L_i \xi_{i,s} (\hat \pi_i^{-1} - \pi_i^{-1})\xi_{i,j} X_i'\delta_j\right\Vert_2 \lessp m^{-1/2} n^{1/q}.
\end{align*}

Similarly, we can show that
\begin{align*}
& \max_{1\leq j \leq m} \left\Vert \sum_{i=1}^n L_i (\hat \pi_i^{-1} - \pi_i^{-1}) \xi_{i,j} X_i'\delta_j\right\Vert_2 \\
& \lessp  \alpha_n^{-2} (mn)^{-1} n^{1/q} \max_{1\leq j \leq m}  \sum_{i=1}^n ||L_i||_2  \xi_{i,j} ||X_i||_2 ||\delta_j||_2\\
& + \max_{1\leq j \leq m} (mn\alpha_n)^{-1/2} \sum_{i=1}^n ||L_i||_2 \pi_i^{-1} \xi_{i,j} ||X_i||_2 ||\delta_j||_2 \\
& \lessp m^{-1}\alpha_n^{-1} \log^{3/4} n.
\end{align*}
This implies
\begin{align*}
\max_{1\leq s \leq m}||B^{(s,4,2)}|| \leq \max_{1\leq s \leq m}||B^{(s,4,2)}||_F \lessp n^{1/q},
\end{align*}
and thus,
\begin{align*}
& \max_{1\leq s \leq m}||B^{(s,4)}|| \lessp  m^{1/2} \alpha_n^{-1/2} \log^{5/4} n.
\end{align*}

Combining the four bounds for $(B^{(s,1)},\cdots,B^{(s,4)})$ above, we obtain that
\begin{align*}
\max_{1 \leq s \leq m}||\sum_{i=1}^n L_i \xi_{i,s}e_i'|| \lessp (mn)^{1/2} n^{1/q}.
\end{align*}

For the second result in Lemma \ref{lem: part 9}, we note that
\begin{align*}
& \max_{1\leq s \leq m}||\sum_{i=1}^n L_i \xi_{i,s}\Delta_{L,i}'|| \\
& = \max_{1\leq s \leq m}\left\Vert \sum_{i=1}^n L_i \xi_{i,s}\left(e_i'e'\hat L + \sum_{j = 1}^m e_{i,j}F_j' L' \hat L \right) \hat \Omega_r^{-2} (mn)^{-1} \right\Vert \\
& \lessp (mn)^{-1}\max_{1\leq s \leq m} \left\Vert \sum_{i=1}^{n}L_i \xi_{i,s} e_i' \right\Vert ||\hat L' e|| + \max_{1\leq s \leq m} \left\Vert \sum_{i=1}^n \sum_{j = 1}^m L_i \xi_{i,s} e_{i,j}F_j' \right\Vert ||L' \hat L \hat \Omega_r^{-2}|| (nm)^{-1} \\
& \lessp (1+n/m)\alpha_n^{-1/2} n^{1/q} + \max_{1\leq s \leq m} \left\Vert \sum_{i=1}^n \sum_{j = 1}^m L_i \xi_{i,s} e_{i,j}F_j' \right\Vert ||L' \hat L \hat \Omega_r^{-2}|| (nm)^{-1} \\
& \lessp (1+n/m)\alpha_n^{-1/2} n^{1/q} + (n/m)^{1/2}\alpha_n^{-1/2} \log^{5/4} n \lessp (1+n/m)^{1/2}\alpha_n^{-1/2} n^{1/q}.
\end{align*}
where the second inequality is by Lemma \ref{lem: part 5} and the third inequality is by Lemma \ref{lem: part 7} and the first result in Lemma \ref{lem: part 9} shown above.

The third and fourth results can be established in the same manner.

 \end{proof}

\begin{lem}
\label{lem: general asymp prop0}\label{lem: order of hL'XihL-I}Let Assumption \ref{assu: moment cond} hold.
Then
\begin{align*}
\max_{1\leq j \leq m} \left\Vert \left(E_n \hat L_i \hat L_i' \xi_{i,j} \right)^{-1} -  \left(H'E (L_i L_i' \pi_i) H\right)^{-1} \right\Vert \lessp n^{-1}(1+n/m)\alpha_n^{-3} n^{1/q} + n^{-1/2}\alpha_n^{-1}.
\end{align*}

\end{lem}

\begin{proof}[\textbf{Proof of Lemma \ref{lem: general asymp prop0}}]

We note that
\begin{align*}
& \left\Vert E_n \hat L_i \hat L_i' (\xi_{i,j} - \pi_i)\right\Vert \\
& = \left\Vert E_n (H' L_i + \Delta_{L,i}) (L_i'H + \Delta_{L,i}') (\xi_{i,j} - \pi_i)\right\Vert \\
& \leq \left\Vert H'(E_n L_iL_i'(\xi_{i,j} - \pi_i))H \right\Vert + 2||H|| ||E_n L_i\Delta_{L,i}'(\xi_{i,j} - \pi_i)|| + ||E_n \Delta_{L,i} \Delta_{L,i}' (\xi_{i,j} - \pi_i)||.
\end{align*}
For the first term, we have
\begin{align*}
\max_{1\leq j \leq m}\left\Vert H' E_n  L_iL_i' (\xi_{i,j} - \pi_i) H\right\Vert  \lessp (n^{-1}\alpha_n \log n)^{1/2}.
\end{align*}

For the second term, by Lemma \ref{lem: part 9}, we have
\begin{align*}
\max_{1\leq j \leq m} ||H|| ||E_n L_i\Delta_{L,i}'(\xi_{i,j} - \pi_i)|| \lessp n^{-1}(1+n/m)\alpha_n^{-1/2} n^{1/q}.
\end{align*}

For the third term, by Lemma \ref{lem: part 5}, we have
\begin{align*}
\max_{1 \leq j \leq m} ||E_n \Delta_{L,i} \Delta_{L,i}' (\xi_{i,j} - \pi_i)|| \leq n^{-1} ||\Delta_{L}||_F^2 \lessp (n\alpha_n)^{-1} (1+n/m) n^{1/q}.
\end{align*}

This implies
\begin{align*}
\max_{1\leq j \leq m}\left\Vert E_n \hat L_i \hat L_i' (\xi_{i,j} - \pi_i)\right\Vert \lessp (n\alpha_n)^{-1} (1+n/m) n^{1/q}.
\end{align*}

In addition, we have
\begin{align*}
	E_n \hat L_i \hat L_i' \pi_i & = 	E_n (H'  L_i + \Delta_{L,i}) (H'  L_i + \Delta_{L,i})' \pi_i \\
	& = H' E_n L_i L_i' \pi_i H + H' E_n  L_i \Delta_{L,i}' \pi_i + (E_n  L_i \Delta_{L,i}' \pi_i)' H + H' E_n \Delta_{L,i}\Delta_{L,i}' \pi_i H,
\end{align*}
\begin{align*}
||H' E_n  L_i \Delta_{L,i}' \pi_i || \lessp ||H|| n^{-1} ||L'\diag(\pi) \Delta_L|| \lessp  n^{1/q} n^{-1}(1+n/m),
\end{align*}
and
\begin{align*}
	H' E_n \Delta_{L,i}\Delta_{L,i}' \pi_i H \lessp \alpha_n n^{1/q} n^{-1} ||\Delta_L||^2 \lessp  n^{1/q} n^{-1}(1+n/m).
\end{align*}

Last, we have
\begin{align*}
	||E_n L_iL_i'\pi_i - EL_iL_i'\pi_i|| \lessp n^{-1/2} \alpha_n.
\end{align*}
Therefore, we have
\begin{align*}
\max_{1\leq j \leq m} \left\Vert \left(E_n \hat L_i \hat L_i'\xi_{i,j} \right) -  \left(H'E (L_i L_i' \pi_i) H\right) \right\Vert \lessp (n\alpha_n)^{-1}(1+n/m) n^{1/q} + n^{-1/2} \alpha_n.
\end{align*}
which implies
\begin{align*}
	\max_{1\leq j \leq m} \left\Vert \left(E_n \hat L_i \hat L_i'\xi_{i,j} \right)^{-1} -  \left(H'E (L_i L_i' \pi_i) H\right)^{-1} \right\Vert \lessp n^{-1}(1+n/m)\alpha_n^{-3} n^{1/q} + n^{-1/2}\alpha_n^{-1}.
\end{align*}

\end{proof}

\begin{lem}
\label{lem: general asymp prop1}Let Assumption \ref{assu: moment cond} hold.
Then
	\begin{align*}
	\max_{1 \leq s \leq m}\left\Vert n^{-1} \sum_{i=1}^n \hat L_i u_{i,s} \xi_{i,s} - n^{-1} H' \sum_{i=1}^n L_i \eps_{i,s} \xi_{i,s} \right\Vert \lessp (n^{-1}+m^{-1})\alpha_n^{-1/2} n^{1/q},
	\end{align*}
 and
	\begin{align*}
	\max_{1 \leq s \leq m}\left\Vert n^{-1} \sum_{i=1}^n  \hat L_i \tilde u_{i,s} \xi_{i,s} - n^{-1} H' \sum_{i=1}^n L_i \eps_{i,s} \xi_{i,s} \right\Vert \lessp (n^{-1}+m^{-1})\alpha_n^{-1/2} n^{1/q}
	\end{align*}
where $u_{i,s} = \eps_{i,s} - X_i'\delta_s$, $\tilde u_{i,s} = \eps_{i,s} - X_i'\tilde \delta_s$, $\delta_s = \hat \beta_s - \beta_s$, and $\tilde \delta_s = \tilde \beta_s^{(1)} - \beta_s$.
\end{lem}
\begin{proof}[\textbf{Proof of Lemma \ref{lem: general asymp prop1}}]
\textbf{The first result of Lemma \ref{lem: general asymp prop1}.} We have
\begin{align*}
& n^{-1} \sum_{i=1}^n \hat L_i u_{i,s} \xi_{i,s}	- n^{-1} H' \sum_{i=1}^n L_i \eps_{i,s} \xi_{i,s} \\
& = n^{-1} \sum_{i=1}^n (H' L_i + \Delta_{L,i}) (\eps_{i,s} - X_i'\delta_s) \xi_{i,s}	- n^{-1} H' \sum_{i=1}^n L_i \eps_{i,s} \xi_{i,s} \\
& = n^{-1} \sum_{i=1}^n \Delta_{L,i} \eps_{i,s} \xi_{i,s}	- H' E_n L_i X_i'\delta_s \xi_{i,s} - E_n \Delta_{L,i} X_i'\delta_s \xi_{i,s}.
\end{align*}
In addition, by \eqref{eq:lem1_9} and Lemma \ref{lem: part 5}, we have
\begin{align*}
\max_{1\leq s \leq m}||E_n L_i \xi_{i,s} X_i'\delta_s||_2 \leq \max_{1\leq s \leq m}||E_n L_i \xi_{i,s} X_i'||_2 \max_{1\leq s \leq m} ||\delta_s||_2 \lessp n^{-1}\log^{5/4} n
\end{align*}
 and
\begin{align*}
\max_{1\leq s \leq m} ||E_n \Delta_{L,i} X_i'\delta_s \xi_{i,s} ||_2 & \leq n^{-1} \left[\sum_{i=1}^n ||\Delta_{L,i}||_2^2 \times \max_{1\leq s \leq m} \sum_{i =1}^n \xi_{i,s}(X_i'\delta_s)^2 \right]^{1/2} \\
& \lessp n^{-1} \sqrt{\alpha_n^{-1} (1+n/m)} n^{1/q}.
\end{align*}

Next, we turn to the first term. Note $\Delta_{L,i}' = (nm)^{-1} e_i' W' \hat L \hat \Omega_r^{-2} = (nm)^{-1} e_i' FL' \hat L \hat \Omega_r^{-2} + (nm)^{-1} e_i' e' \hat L \hat \Omega_r^{-2}$. The rest of the proof proceeds in three steps.

\textbf{Step 1: bound $\sum_{i=1}^n \eps_{i,s} \xi_{i,s} e_i' e \hat L$.}
Note that
\begin{align*}
\sum_{i=1}^n \eps_{i,s} \xi_{i,s} e_{i,j} & = \sum_{i=1}^n \eps_{i,s} \xi_{i,s}\xi_{i,j}(\hat \pi_i^{-1} - \pi_i^{-1}) L_i'F_j + \sum_{i=1}^n \eps_{i,s} \xi_{i,s}\pi_i^{-1}(\xi_{i,j} - \pi_i)L_i'F_j \\
& +  	\sum_{i=1}^n \eps_{i,s} \xi_{i,s}\hat \pi_i^{-1}\xi_{i,j}\eps_{i,j}- \sum_{i=1}^n \eps_{i,s} \xi_{i,s}\hat \pi_i^{-1}\xi_{i,j}X_i'\delta_j \\
& \equiv \sum_{l=1}^4 A_{l,s,j}.
\end{align*}
For $A_{1,s,j}$, by \eqref{eq:ex_lem_19}, we have
\begin{align*}
& \max_{1\leq s,j \leq m, s \neq j}|A_{1,s,j}| \\
& \leq \alpha_n^{-2}n^{1/q}(nm)^{-1}  \max_{1\leq s,j \leq m, s \neq j}\sum_{i =1}^n |\eps_{i,s}|
\xi_{i,s}\xi_{i,j} ||L_i||_2||F_j||_2 \\
& +   \max_{1\leq s,j \leq m, s \neq j}|\sum_{i=1}^n \eps_{i,s}\xi_{i,s}\xi_{i,j}\pi_i^{-1}(1-\pi_i)X_i'(\hat \gamma - \gamma) L_i'F_j| \\
& \lessp m^{-1} n^{1/q} + \left(\sqrt{n}\log n + \alpha_n^{-1} n^{1/q} \log n \right) (nm\alpha_n)^{-1/2} \\
& \lessp (m \alpha_n)^{-1/2} \log n
\end{align*}
and
\begin{align*}
	\max_{1\leq s\leq m}|A_{1,s,s}| & = \max_{1\leq s\leq m}|\sum_{i=1}^n \eps_{i,s} \xi_{i,s}(\hat \pi_i^{-1} - \pi_i^{-1}) L_i'F_s|\\
	& \leq \alpha_n^{-2}n^{1/q}(nm)^{-1} \max_{1\leq s\leq m} \sum_{i =1}^n |\eps_{i,s}|
	\xi_{i,s}||L_i||_2||F_s||_2 \\
 & +  \max_{1\leq s\leq m}|\sum_{i=1}^n \eps_{i,s}\xi_{i,s}\pi_i^{-1}(1-\pi_i)X_i'(\hat \gamma - \gamma) L_i'F_s| \\
	& \lessp \alpha_n^{-1}m^{-1} n^{1/q} + \left(\sqrt{n \alpha_n^{-1}}n^{1/q}\right) (nm\alpha_n)^{-1/2}.
\end{align*}

This implies
\begin{align*}
  \max_{1\leq s \leq m}  ||A_{1,s}|| \leq   \max_{1\leq s \leq m}  ||A_{1,s}||_F \lessp \alpha_n^{-1/2} \log n.
\end{align*}

For $A_{2,s,j}$, we have
\begin{align*}
\max_{1\leq s,j \leq m, s \neq j}|A_{2,s,j}| \lessp n^{1/2+1/q}
\end{align*}
and
\begin{align*}
	\max_{1\leq s \leq m}|A_{2,s,s}| \lessp \alpha_n^{-1/2}n^{1/2+1/q}.
\end{align*}

This implies
\begin{align*}
  \max_{1\leq s \leq m}  ||A_{2,s}|| \lessp (mn \log n)^{1/2} n^{1/q}.
\end{align*}

For $A_{3,s,j}$, we have

\begin{align*}
	\max_{1\leq s,j \leq m, s \neq j}|A_{3,s,j}| & = \max_{1\leq s,j \leq m, s \neq j}|\sum_{i=1}^n \eps_{i,s} \xi_{i,s}\hat \pi_i^{-1}\xi_{i,j}\eps_{i,j}|\\
 & \lessp \max_{1\leq s,j \leq m, s \neq j}|\sum_{i=1}^n \eps_{i,s} \xi_{i,s} \pi_i^{-1}\xi_{i,j}\eps_{i,j}| \\
 & + \alpha_n^{-2}n^{1/q}(nm)^{-1} 	\max_{1\leq s,j \leq m, s \neq j} \sum_{i=1}^n |\eps_{i,s}||\eps_{i,j}|\xi_{i,s} \xi_{i,j} \\
 & + 	\max_{1\leq s,j \leq m, s \neq j} \left\vert \sum_{i=1}^n \eps_{i,s}\eps_{i,j}\xi_{i,s} \xi_{i,j}\frac{1-\pi_i}{\pi_i}X_i'(\gamma - \gamma) \right\vert \\
	 & \lessp (n \log n)^{1/2} + n^{1/q}m^{-1} + \left(\sqrt{n}\log n + \alpha_n^{-1} n^{1/q} \log n \right) (nm\alpha_n)^{-1/2} \\
  & \lessp (n \log n)^{1/2}
\end{align*}
and
\begin{align*}
	\max_{1\leq s \leq m}|A_{3,s,s}| & = \max_{1\leq s \leq m}|\sum_{i=1}^n \xi_{i,s}\eps_{i,s}^2 \hat \pi_i^{-1}| \\
 & \lessp \max_{1\leq s \leq m}|\sum_{i=1}^n \xi_{i,s}\eps_{i,s}^2 \pi_i^{-1}| \\
	& + \max_{1\leq s \leq m}\alpha_n^{-2}n^{1/q}(nm)^{-1} \sum_{i=1}^n \eps_{i,s}^2\xi_{i,s} + \max_{1\leq s \leq m}\left\vert \sum_{i=1}^n \eps_{i,s}^2\xi_{i,s}\frac{1-\pi_i}{\pi_i}X_i'(\gamma - \gamma) \right\vert \\
	& \lessp n.
\end{align*}

This implies
\begin{align*}
  \max_{1\leq s \leq m}  ||A_{3,s}|| \lessp (mn \log n)^{1/2} + n .
\end{align*}

For $A_{4,s,j}$, we have
\begin{align*}
	\max_{1\leq s,j\leq m, s\neq j}|A_{4,s,j}| & = 	\max_{1\leq s,j\leq m, s\neq j} |\sum_{i=1}^n \eps_{i,s} \xi_{i,s}\hat \pi_i^{-1}\xi_{i,j}X_i'\delta_j|\\
 & \lessp 	\max_{1\leq s,j\leq m, s\neq j} |\sum_{i=1}^n \eps_{i,s} \xi_{i,s}\pi_i^{-1}\xi_{i,j}X_i'\delta_j|\\
& + \alpha_n^{-2}n^{1/q}(nm)^{-1} \max_{1\leq s,j\leq m, s\neq j} \sum_{i=1}^n |\eps_{i,s}||X_i'\delta_j|\xi_{i,s}\xi_{i,j} \\
 & + \max_{1\leq s,j\leq m, s\neq j}\left\vert \sum_{i=1}^n \eps_{i,s}\xi_{i,s}\xi_{i,j} X_i'\delta_j \frac{1-\pi_i}{\pi_i}X_i'(\gamma - \gamma) \right\vert \\
	& \lessp \alpha_n^{-1/2} \log^{5/4} n
\end{align*}
and
\begin{align*}
	\max_{1\leq s\leq m}|A_{4,s,s}| & \lessp \max_{1\leq s,j\leq m, s\neq j} |\sum_{i=1}^n \eps_{i,s}^2 \xi_{i,s}\pi_i^{-1}X_i'\delta_s| + \alpha_n^{-2}n^{1/q}(nm)^{-1} \max_{1\leq s\leq m}\sum_{i=1}^n |\eps_{i,s}||X_i'\delta_s|\xi_{i,s} \\
 & + \max_{1\leq s\leq m}\left\vert \sum_{i=1}^n \eps_{i,s}\xi_{i,s} X_i'\delta_s \frac{1-\pi_i}{\pi_i}X_i'(\gamma - \gamma) \right\vert \\
	& \lessp n^{1/2} \alpha_n^{-1/2} \log^{3/4} n.
\end{align*}

This implies
\begin{align*}
  \max_{1\leq s \leq m}  ||A_{4,s}|| \lessp (m^{1/2}+n^{1/2}) \alpha_n^{-1/2} \log^{5/4} n.
\end{align*}

Therefore, by Lemma \ref{lem: part 5}, we have
\begin{align*}
\max_{1\leq s \leq m}	||\sum_{i=1}^n \eps_{i,s} \xi_{i,s} e_i' e \hat L|| & \leq \max_{1\leq s \leq m} ||\sum_{i=1}^n \eps_{i,s} \xi_{i,s} e_i'||_2 ||e \hat L|| \\
& \leq \max_{1\leq s \leq m} \left[\sum_{1 \leq j \leq m}\left(\sum_{i=1}^n \eps_{i,s} \xi_{i,s} e_{i,j}\right)^2\right]^{1/2} ||e \hat L|| \\
& \lessp n (n+m)\alpha_n^{-1/2} n^{1/q}.
\end{align*}

\textbf{Step 2: bound $\sum_{i=1}^n \eps_{i,s} \xi_{i,s} e_i'F$.}  We have
\begin{align*}
\sum_{i=1}^n \eps_{i,s} \xi_{i,s} e_i'F & = \sum_{j=1}^m\sum_{i=1}^n \eps_{i,s} \xi_{i,s} e_{i,j}F_j \\
& = \sum_{j=1}^m\sum_{i=1}^n \eps_{i,s} \xi_{i,s} \xi_{i,j}(\hat \pi_i^{-1} - \pi_i^{-1})L_i'F_j F_j + \sum_{j=1}^m\sum_{i=1}^n \eps_{i,s} \xi_{i,s} \pi_i^{-1}(\xi_{i,j} - \pi_i)L_i'F_j F_j \\
& + \sum_{j=1}^m\sum_{i=1}^n \eps_{i,s} \xi_{i,s} \hat \pi_i^{-1} \xi_{i,j} \eps_{i,j}F_j - \sum_{j=1}^m\sum_{i=1}^n \eps_{i,s} \xi_{i,s} \hat \pi_i^{-1} \xi_{i,j}X_i'\delta_j F_j \\
& \equiv \sum_{l=1}^4 B_{l,s}.
\end{align*}

For $B_{1,s}$, we have
\begin{align*}
\max_{1 \leq s \leq m}||B_{1,s}||_2 \leq & \alpha_n^{-2} n^{1/q} (nm)^{-1} \max_{1 \leq s \leq m}\sum_{j = 1}^m \sum_{i =1}^n |\eps_{i,s}| \xi_{i,j}\xi_{i,s} |L_i'F_j| ||F_j||_2 \\
& + \max_{1 \leq s \leq m} \left\Vert \sum_{1\leq j \leq m,j\neq s}\sum_{i=1}^n \eps_{i,s} \xi_{i,s} \xi_{i,j}\left(\frac{(1-\pi_i) }{\pi_i}\right)L_i'F_j F_jX_i'(\hat\gamma - \gamma) \right\Vert_2 \\
& + \max_{1 \leq s \leq m} \left\Vert \sum_{i=1}^n \eps_{i,s} \xi_{i,s}\left(\frac{(1-\pi_i) X_i'(\hat\gamma - \gamma)}{\pi_i}\right)L_i'F_s F_s \right\Vert_2 \\
& \lessp n^{1/q} + (m \log n)^{1/2} + \alpha_n^{-1} m^{-1/2} n^{1/q},
\end{align*}
where the last inequality is by \eqref{eq:ex_lem_10'}.

For $B_{2,s}$, we have
\begin{align*}
& \max_{1\leq s \leq m}\left\Vert \sum_{j=1}^m\sum_{i=1}^n \eps_{i,s} \xi_{i,s} \pi_i^{-1}(\xi_{i,j} - \pi_i)L_i'F_j F_j \right\Vert_2 \\
& \leq \max_{1\leq s \leq m}\left\Vert \sum_{1\leq j \leq m, j\neq s}\sum_{i=1}^n \eps_{i,s} \xi_{i,s} \pi_i^{-1}(\xi_{i,j} - \pi_i)L_i'F_j F_j \right\Vert_2 + \max_{1\leq s \leq m}\left\Vert \sum_{i=1}^n \eps_{i,s} \xi_{i,s} \pi_i^{-1}(\xi_{i,s} - \pi_i)L_i'F_s F_s \right\Vert_2\\
& \lessp (mn\log^2 n)^{1/2},
\end{align*}
where by \eqref{eq:ex_lem_10''}, we have
\begin{align*}
\max_{1\leq s \leq m}\left\Vert \sum_{1\leq j \leq m, j\neq s}\sum_{i=1}^n \eps_{i,s} \xi_{i,s} \pi_i^{-1}(\xi_{i,j} - \pi_i)L_i'F_j F_j \right\Vert_2 \lessp (mn \log^2 n)^{1/2}
\end{align*}

For $B_{3,s}$, we have
\begin{align*}
& \max_{1\leq s \leq m}\left\Vert \sum_{j=1}^m\sum_{i=1}^n \eps_{i,s} \xi_{i,s} \hat \pi_i^{-1} \xi_{i,j} \eps_{i,j}F_j\right\Vert_2 \\
& \lessp \max_{1\leq s \leq m}\left\Vert \sum_{j=1}^m\sum_{i=1}^n \eps_{i,s} \xi_{i,s} \pi_i^{-1} \xi_{i,j} \eps_{i,j}F_j\right\Vert_2 + \alpha_n^{-2}(nm)^{-1} n^{1/q}\max_{1\leq s \leq m} \sum_{j=1}^m\sum_{i=1}^n |\eps_{i,s}| \xi_{i,s}  \xi_{i,j} |\eps_{i,j}| ||F_j||_2 \\
& + \max_{1\leq s \leq m}\left\Vert \sum_{j=1}^m\sum_{i=1}^n \eps_{i,s} \xi_{i,s} \frac{1-\pi_i}{\pi_i}X_i'(\hat \gamma - \gamma) \xi_{i,j} \eps_{i,j}F_j\right\Vert_2 \\
& \lessp \max_{1\leq s \leq m}\left\Vert \sum_{1\leq j \leq m, j \neq s}\sum_{i=1}^n \eps_{i,s} \xi_{i,s} \pi_i^{-1} \xi_{i,j} \eps_{i,j}F_j\right\Vert_2 + \max_{1\leq s \leq m}\left\Vert \sum_{i=1}^n \eps_{i,s}^2 \xi_{i,s} \pi_i^{-1} F_s\right\Vert_2  + n^{1/q}\\
& + \max_{1\leq s \leq m}\left\Vert \sum_{1\leq j \leq m, j \neq s}\sum_{i=1}^n \eps_{i,s} \xi_{i,s} \frac{1-\pi_i}{\pi_i}X_i'(\hat \gamma - \gamma) \xi_{i,j} \eps_{i,j}F_j\right\Vert_2 + \max_{1\leq s \leq m}\left\Vert \sum_{i=1}^n \eps_{i,s}^2 \xi_{i,s} \frac{1-\pi_i}{\pi_i}X_i'(\hat \gamma - \gamma) F_s\right\Vert_2\\
& \lessp (mn \log^2 n)^{1/2} + n \log n,
\end{align*}
where by \eqref{eq:ex_lem_10'''} and \eqref{eq:ex_lem_10''''}, we have
\begin{align*}
\max_{1\leq s \leq m}\left\Vert \sum_{1\leq j \leq m, j \neq s}\sum_{i=1}^n \eps_{i,s} \xi_{i,s} \pi_i^{-1} \xi_{i,j} \eps_{i,j}F_j\right\Vert_2 \lessp (m n \log^2 n)^{1/2}
\end{align*}
and
\begin{align*}
\max_{1\leq s \leq m}\left\Vert \sum_{1\leq j \leq m, j \neq s}\sum_{i=1}^n \eps_{i,s} \xi_{i,s} \frac{1-\pi_i}{\pi_i}X_i'(\hat \gamma - \gamma) \xi_{i,j} \eps_{i,j}F_j\right\Vert_2 \lessp \alpha_n^{-1/2} \log n.
\end{align*}

For $B_{4,s}$, we have
\begin{align*}
& \max_{1\leq s \leq m}\left\Vert \sum_{j=1}^m\sum_{i=1}^n \eps_{i,s} \xi_{i,s} \hat \pi_i^{-1} \xi_{i,j}X_i'\delta_j F_j \right\Vert_2 \\
& \lessp \max_{1\leq s \leq m}\left\Vert \sum_{j=1}^m\sum_{i=1}^n \eps_{i,s} \xi_{i,s} \pi_i^{-1} \xi_{i,j}X_i'\delta_j F_j \right\Vert_2  + \alpha_n^{-2}(nm)^{-1} n^{1/q}\max_{1\leq s \leq m} \sum_{j=1}^m\sum_{i=1}^n |\eps_{i,s}| \xi_{i,s} \xi_{i,j}|X_i'\delta_j| ||F_j||_2 \\
& + \max_{1\leq s \leq m}\left\Vert \sum_{j=1}^m\sum_{i=1}^n \eps_{i,s} \xi_{i,s} \frac{1-\pi_i}{\pi_i}X_i'(\hat \gamma-\gamma) \xi_{i,j}X_i'\delta_j F_j \right\Vert_2\\
& \lessp  \sum_{j=1}^m \max_{1\leq j,s \leq m, j \neq s}\left\Vert \sum_{i=1}^n \eps_{i,s} \xi_{i,s} \pi_i^{-1} \xi_{i,j} X_i\right\Vert_2 ||\delta_j||_2 || F_j||_2  + \max_{1\leq s \leq m}\left\Vert \sum_{i=1}^n \eps_{i,s} \xi_{i,s} \pi_i^{-1}X_i'\delta_s F_s \right\Vert_2\\
& + \sum_{j=1}^m \max_{1\leq j,s\leq m, j\neq s}\left\Vert \sum_{i=1}^n \eps_{i,s} \xi_{i,s} \frac{1-\pi_i}{\pi_i}X_i'(\hat \gamma-\gamma) \xi_{i,j}X_i\right\Vert_2 ||\delta_j||_2 ||F_j||_2 \\
& + \max_{1\leq s \leq m}\left\Vert \sum_{i=1}^n \eps_{i,s} \xi_{i,s} \frac{1-\pi_i}{\pi_i}X_i'(\hat \gamma-\gamma)X_i'\delta_s F_s \right\Vert_2+ (n\alpha_n)^{-1/2}n^{1/q} \\
& \lessp m \alpha_n^{-1/2} \log^{5/4} n + \alpha_n^{-1} \log^{9/4} n,
\end{align*}
where we use the facts that
\begin{align*}
& \max_{1\leq j,s \leq m, j \neq s}\left\Vert \sum_{i=1}^n \eps_{i,s} \xi_{i,s} \pi_i^{-1} \xi_{i,j} X_i\right\Vert_2 \lessp (n \log n)^{1/2}, \\
& \max_{1\leq s \leq m}\left\Vert \sum_{i=1}^n \eps_{i,s} \xi_{i,s} \pi_i^{-1}X_i \right\Vert_2 \lessp (n \alpha_n^{-1} \log n)^{1/2}, \\
& \max_{1\leq j,s\leq m, j\neq s}\left\Vert \sum_{i=1}^n \eps_{i,s} \xi_{i,s} \frac{1-\pi_i}{\pi_i} \xi_{i,j}X_iX_i'\right\Vert \lessp  (n \log n )^{1/2}, \\
& \max_{1\leq s \leq m}\left\Vert \sum_{i=1}^n \eps_{i,s} \xi_{i,s} \frac{1-\pi_i}{\pi_i}X_iX_i'\right\Vert \lessp (n \alpha_n^{-1} \log n)^{1/2}.
\end{align*}

Combining all bounds, we have
\begin{align*}
\max_{1\leq s \leq m}\left\Vert \sum_{i=1}^n \eps_{i,s} \xi_{i,s} e_i'F\right\Vert_2 \lessp n \log n +  m \alpha_n^{-1/2} \log^{5/4} n .
\end{align*}

\textbf{Step 3: derive the final result.} Combining the results in Steps 1 and 2, we have
\begin{align*}
& \max_{1\leq s \leq m}||E_n \Delta_{L,i}'\eps_{i,s}\xi_{i,s}||_2 \\
& \leq  n^{-2} m^{-1} \max_{1\leq s \leq m}||\sum_{i=1}^n \eps_{i,s}\xi_{i,s}  e_i' FL'\hat L \hat \Omega_r^{-2}||_2 + n^{-2} m^{-1}  \max_{1\leq s \leq m}||\sum_{i=1}^n \eps_{i,s}\xi_{i,s}  e_i' e'\hat L \hat \Omega_r^{-2}||_2 \\
& \leq  n^{-2} m^{-1} \max_{1\leq s \leq m}||\sum_{i=1}^n \eps_{i,s}\xi_{i,s}  e_i'F||_2 ||L'\hat L|| ||\hat \Omega_r^{-2}||+ n^{-2} m^{-1}  \max_{1\leq s \leq m}||\sum_{i=1}^n \eps_{i,s}\xi_{i,s}  e_i' e'\hat L||_2 ||\hat \Omega_r^{-2}|| \\
& \lessp  m^{-1} \log n + n^{-1} \alpha_n^{-1/2} \log ^{5/4} n + (1+m/n)^{1/2} (nm\alpha_n)^{-1/2} n^{1/q} \\
& \lessp (n^{-1} + m^{-1}) \alpha_n^{-1/2} n^{1/q}.
\end{align*}

\textbf{The second result of Lemma \ref{lem: general asymp prop1}.} This can be proved in the same manner as the first result by noticing that $\max_{1\leq j \leq m}\|\tdel_j\| \lessp \sqrt{ n^{-1} \alpha_n^{-1} \log^{3/2} n}$ due to \eqref{eq: betatilda}.
\end{proof}

\subsection{Asymptotics for $\tF^{(1)}_j$}\label{sec: asymp tF g=1}

\begin{lem}
\label{lem: asymp tF g=1}Let Assumption \ref{lem: general asymp prop} hold. Then
\begin{align*}
\max_{1\leq j\leq m}\left\| \tF^{(1)}_j-H^{-1}F_{j}-n^{-1} \left(H'E (L_i L_i' \pi_i) H\right)^{-1} H'\sumi L_{i} \xi_{i,j}\varepsilon_{i,j}\right\|_{2}\lessp (1/n + 1/m)\alpha_n^{-3/2} n^{1/q}.
\end{align*}
\end{lem}

\begin{proof}[\textbf{Proof of Lemma \ref{lem: asymp tF g=1}}]
For the sake of simplicity of notations, we denote $\tF^{(1)}$ as $\tF$. Since $\hL_{i}'=L_{i}'H+\Delta_{L,i}$, we notice that
\begin{align}
 &\quad \tF_j-H^{-1}F_{j}\nonumber \\
 & =\left(\sumi \hL_{i}\hL_{i}' \xi_{i,j}\right)^{-1}\left(\sumi \hL_{i}(L_{i}'F_{j}+\tu_{i,j}) \xi_{i,j}-\hL_{i}\hL_{i}' \xi_{i,j}H^{-1}F_{j}\right)\nonumber \\
 & =\left(\sumi \hL_{i}\hL_{i}' \xi_{i,j}\right)^{-1}\left(\sumi \hL_{i}\tu_{i,j}  \xi_{i,j}\right)\nonumber\\
 & +\left(\sumi \hL_{i}\hL_{i}'  \xi_{i,j}\right)^{-1}\left(\sumi \hL_{i}L_{i}'F_{j}  \xi_{i,j}-\hL_{i}\hL_{i}'\xi_{i,j}H^{-1}F_{j}\right)\nonumber \\
 & =\left(\sumi \hL_{i}\hL_{i}' \xi_{i,j}\right)^{-1}\left(\sumi \hL_{i}\tu_{i,j}\xi_{i,j}\right)-\left(\sumi \hL_{i}\hL_{i}'\xi_{i,j}\right)^{-1}\left(\sumi \hL_{i}\Delta_{L,i}' \xi_{i,j}\right)H^{-1}F_{j}.\label{eq: F est eq 1}
\end{align}
Now we bound $\sumi \hL_{i}\Delta_{L,i}'\xi_{i,j}$. Note
that
\begin{align*}
\sumi \hL_{i}\Delta_{L,i}' \xi_{i,j}=H'\sumi L_{i}\Delta_{L,i}'\xi_{i,j}+\sumi \Delta_{L,i}\Delta_{L,i}' \xi_{i,j}.
\end{align*}
To bound the second term, we notice that
\begin{align*}
\max_{1\leq j\leq m}\left\| \sumi \Delta_{L,i}\Delta_{L,i}'\xi_{i,j}\right\| \lessp  \sumi \|\Delta_{L,i}\|_{2}^{2}\leq r\|\Delta_{L}\|^{2}\overset{\texti}{\lessp } \alpha_n^{-1} (1+n/m) n^{1/q},
\end{align*}
where (i) follows by Lemma \ref{lem: part 5}. To bound the first
term, we notice that since $\|H\|=O_{P}(1)$ (due to Lemma \ref{lem: part 4}). Then, by Lemma \ref{lem: part 9}, we have
\begin{align*}
\max_{1\leq j\leq m}\left\| \sumi \hL_{i}\Delta_{L,i}' \xi_{i,j}\right\| \lessp (1+n/m)\alpha_n^{-1} n^{1/q}.
\end{align*}
Since $\|H^{-1}\|=O_{P}(1)$ (Lemma \ref{lem: part 8}) and $\max_{j}\|F_{j}\|_{2}\lessp \sqrt{\log m}$,
Lemma \ref{lem: general asymp prop0} and the above display imply
\begin{align*}
&\quad \max_{1\leq j\leq m}\left\| \left(\sumi \hL_{i}\hL_{i}'  \xi_{i,j}\right)^{-1}\left(\sumi \hL_{i}\Delta_{L,i}'  \xi_{i,j}\right)H^{-1}F_{j}\right\|_{2}\\
& =O_{P}((\alpha_n n)^{-1})\cdot O_{P}\left((1+n/m)\alpha_n^{-1} n^{1/q}\right).
\end{align*}
By Lemmas \ref{lem: general asymp prop0} and \ref{lem: general asymp prop1}, we have that uniformly in $j$,
\begin{align*}
 &\quad \left(n^{-1}\sumi \hL_{i}\hL_{i}'  \xi_{i,j}\right)^{-1}\left(n^{-1}\sumi \hL_{i}\tu_{i,j} \xi_{i,j}\right)\\
 & =\left(\left(H'E (L_i L_i' \pi_i) H\right)^{-1} +O_{P}(n^{-1}(1+n/m)\alpha_n^{-3} n^{1/q} + n^{-1/2}\alpha_n^{-1})\right) \\
 & \cdot\left(n^{-1}H'\sumi L_{i}\xi_{i,j}\varepsilon_{i,j}+O_{P}\left((n^{-1} + m^{-1}) \alpha_n^{-1/2} n^{1/q}.\right)\right)\\
 & \overset{\texti}{=} n^{-1} \left(H'E (L_i L_i' \pi_i) H\right)^{-1} H'\sumi L_{i}\xi_{i,j}\varepsilon_{i,j}+O_{P}\left((1/n + 1/m)\alpha_n^{-3/2} n^{1/q}\right),
\end{align*}
where (i) follows by $\|H\|=O_{P}(1)$ (Lemma \ref{lem: part 4})
and \eqref{eq:lem1_5_5} in Lemma \ref{lem: general asymp prop}. We now combine (\ref{eq: F est eq 1}) with the above two displays, obtaining
\begin{align*}
\max_{1\leq s\leq m}\left\| \tF_j-H^{-1}F_{j}-n^{-1} \left(H'E (L_i L_i' \pi_i) H\right)^{-1} H'\sumi L_{i}\xi_{i,j}\varepsilon_{i,j}\right\|_{2}\lessp (1/n + 1/m)\alpha_n^{-3/2} n^{1/q}.
\end{align*}
The proof is complete.
\end{proof}

\subsection{Auxiliary results for iterations}
We would like to prove Theorem \ref{thm: beta est} and \ref{thm: Gamma est}  for any $g\geq 1$ by induction.
We already proved the result of Theorem \ref{thm: beta est} at $g=1$ in \ref{app: proof of beta est g=1}, so now it is sufficed by proving that
\begin{itemize}
    \item[(1)] for any $g\geq 1$, the representation of $\tbeta^{(g)}$ implies the representation of $\tGamma^{(g)}$, and
    \item[(2)] for any $g\geq 1$, the representation of $\tGamma^{(g)}$ implies the representation of $\tbeta^{(g+1)}$
\end{itemize}
 Note that with $\tbeta^{(g)}$, the estimator $\tGamma^{(g)}$ is totally determined by $\tF^{(g)}$ because $\tL^{(g)}$ is actually a function of $\tbeta^{(g)}$, $\tF^{(g)}$ and the given data. Since We already have the asymptotic representation of $\tF^{(1)}$ in \ref{sec: asymp tF g=1}, the proof can also be completed by proving that
\begin{itemize}
    \item[(1)] the representations of $\tF^{(g)}$ and $\tbeta^{(g)}$ imply the representation of $\tGamma^{(g)}$,
    \item[(2)] the representation of $\tGamma^{(g)}$ implies the representation of $\tbeta^{(g+1)}$, and
    \item[(3)] the representations of $\tbeta^{(g+1)}$ and $\tF^{(g)}$ imply the representation of $\tF^{(g+1)}$.
\end{itemize}
Now we make a supporting assumption that gives us the representation for $\tF^{(g)}$, and later we will prove that it is automatically satisfied under Assumption \ref{assu: moment cond}.
\begin{ass}
\label{assu: asymp tF} Suppose that $\tF^{(g)}_j=H^{-1}F_{j}+a_{j}+b^{(g)}_{j}$, where
\begin{align*}
    &a_{j}=n^{-1} \left(H'E (L_i L_i' \pi_i) H\right)^{-1} H'\sumi L_{i}\xi_{i,j}\varepsilon_{i,j}\\
    & \max_{1\leq j\leq m}\|b^{(g)}_{j}\|_{2} =O_P\left( \kappa_{n}\right)
\end{align*}
where $\kappa_{n} =  (1/n + 1/m)\alpha_n^{-3/2} n^{1/q}$.
\end{ass}


\begin{lem}\label{lem: induction for asymp tL}
Under Assumptions \ref{assu: moment cond}, if $\tF^{(g)}$ satisfies Assumption \ref{assu: asymp tF} and $\tbeta^{(g)}$ satisfies the representation in Theorem \ref{thm: beta est}, then
\begin{align*}
 \tL^{(g)}_{i}-H'L_{i}
 &= H'\pi_i^{-1}\Sigma_F^{-1}m^{-1}\sumj F_j\xi_{i,j}\varepsilon_{i,j} - n^{-1} H' \sum_{k=1}^n \pi_k L_k X_k' (EX_iX_i'\pi_i)^{-1}   X_i + d^{(g)}_i,\\
 &\equiv c_{1,i} + c_{2,i} + d^{(g)}_i\\[6pt]
\max_i \|d^{(g)}_i\|_2 &= O_P\left(\kappa_n\right).
\end{align*}
In particular, $\tL^{(1)}$ satisfies the above representation.
\end{lem}

\begin{proof}[\textbf{Proof of Lemma \ref{lem: induction for asymp tL}}]
For notation simplicity, we denote $\tF^{(g)}, \tL^{(g)}, \tbeta^{(g)}$, $\tW^{(g)}$, $b^{(g)}_j$ by $\tF, \tL, \tbeta, \tW$ and $b_j$. By the definition of $\tL$ and Assumption \ref{assu: asymp tF},
\begin{align}
    \tL_i &= \left(\sumj \tF_j\tF'_j\xi_{i,j}\right)^{-1}\sumj \tF_j\xi_{i,j}\tW_{i,j}\nonumber\\
    &= \left(\sumj \tF_j\tF'_j\xi_{i,j}\right)^{-1}\sumj \tF_j\xi_{i,j}\left(F'_jL_i + \tu_{i,j}\right)\nonumber\\
    &= \left(\sumj \tF_j\tF'_j\xi_{i,j}\right)^{-1}\sumj \tF_j\xi_{i,j}\left(\tF_j-a_j-b_j\right)'H'L_i +\left(\sumj \tF_j\tF'_j\xi_{i,j}\right)^{-1}\sumj \tF_j\xi_{i,j}\tu_{i,j}\nonumber\\
    &= H'L_i - \left(\sumj \tF_j\tF'_j\xi_{i,j}\right)^{-1}\sumj \tF_j\xi_{i,j}\left(a_j+b_j\right)'H'L_i\nonumber \\
    & \hspace{7cm} +\left(\sumj \tF_j\tF'_j\xi_{i,j}\right)^{-1}\sumj \tF_j\xi_{i,j}\tu_{i,j}\label{eq: tL decomp 1}
\end{align}
\textbf{Step 1:} Asymptotic of $\left(\sum_j \tF_j\tF'_j\xi_{i,j}\right)^{-1}$. Observe that, by Assumption \ref{assu: asymp tF},
\begin{align*}
    \sumj \tF_j\tF'_j\xi_{i,j}
    =\ & \sumj (H^{-1}F_j + a_j + b_j)(H^{-1}F_j + a_j + b_j)'\xi_{i,j}\\
    =\ & H^{-1} \left(\sumj F_jF'_j\xi_{i,j}\right)H^{-1} + (H')^{-1} \left(\sumj F_j(a_j + b_j)'\xi_{i,j}\right)\\
       & \qquad +  \left(\sumj (a_j + b_j)F'_j\xi_{i,j}\right)(H')^{-1} + \sumj (a_j + b_j)(a_j + b_j)'\xi_{i,j}.
\end{align*}
Note that by \eqref{eq:lem1_1'} in Lemma \ref{lem: general asymp prop} and the fact that $\|H^{-1}\|=O_P(1)$, we have
\begin{align*}
    \max_i \left\|m^{-1}H^{-1} \left(\sumj F_jF'_j\xi_{i,j}\right)(H')^{-1} - \pi_i H^{-1} \Sigma_F (H')^{-1}\right\| = O_P(\sqrt{m^{-1} \alpha_n\log n}).
\end{align*}
On the other hand, by Lemma \ref{lem: general asymp prop5} below, we observe that
\begin{align*}
    &\quad \max_i \left\|m^{-1}\sumj F_j(a_j + b_j)'\xi_{i,j}\right\| \\
    & \lessp \max_i \left\|m^{-1}\sumj F_ja_j'\xi_{i,j}\right\| +  \max_i \left\|m^{-1}\sumj ||F_j||_2 \xi_{i,j}\right\| \kappa_n \\
    & \lessp (mn)^{-1/2} \log n + \alpha_n \kappa_n
\end{align*}
and
\begin{align*}
    & \max_i \left\|m^{-1}\sumj (a_j + b_j)(a_j + b_j)'\xi_{i,j}\right\| \leq m^{-1} \max_i \sumj \|a_j + b_j\|_2^2\xi_{i,j}=O_P\left(n^{-1} \log m + \alpha_n \kappa_n^2\right).
\end{align*}

Therefore, we have
\begin{align}
    &\max_i \left\|m^{-1}\sumj \tF_j\tF'_j\xi_{i,j} - \pi_i H^{-1}\Sigma_F (H')^{-1}\right\|=O_P(\sqrt{m^{-1} \alpha_n \log n} + \sqrt{n^{-1}\alpha_n \log m}) \quad \text{and}\nonumber\\
    &\max_i \left\|\left(m^{-1}\sumj \tF_j\tF'_j\xi_{i,j}\right)^{-1} - \pi_i^{-1} H'\Sigma^{-1}_F H\right\|=O_P\left(\left(\sqrt{m^{-1} \alpha_n^{-3}} + \sqrt{n^{-1}\alpha_n^{-3} }\right)n^{1/q}\right)\label{eq: tL decomp 2}
\end{align}
\textbf{Step 2:} Asymptotic of $\sum_j \tF_j\xi_{i,j}\left(a_j+b_j\right)'H'L_i$. Following the above argument, we have
\begin{align*}
    &m^{-1}\sumj \tF_j\xi_{i,j}\left(a_j+b_j\right)'H'L_i \\
    =\ &m^{-1}H^{-1}\sumj F_j\xi_{i,j}\left(a_j+b_j\right)'H'L_i + m^{-1}\sumj (a_j+b_j)(a_j+b_j)'\xi_{i,j}H'L_i\\
    =\ &m^{-1}H^{-1}\sumj F_j\xi_{i,j}a'_j H'L_i+ m^{-1}H^{-1}\sumj F_j\xi_{i,j}b'_j H'L_i \\
    &  + m^{-1}\sumj (a_j+b_j)(a_j+b_j)'\xi_{i,j}H'L_i
\end{align*}
\begin{align}
    \max_i \left\|m^{-1}\sumj \tF_j\xi_{i,j}\left(a_j+b_j\right)'H'L_i \right\|_2 \lessp \alpha_n \kappa_n \label{eq: tL decomp 3}
\end{align}
\textbf{Step 3:} Asymptotic of $\sum_j \tF_j\xi_{i,j}\tu_{i,j}$. Recall that $\tu_{i,j}=\varepsilon_{i,j} - X'_i(\tbeta_j - \beta_j)$. Let $\tdel_j = \tbeta_j - \beta_j$. We have
\begin{align*}
    \sumj \tF_j\xi_{i,j}\tu_{i,j} = H^{-1}\sumj F_j\xi_{i,j}\varepsilon_{i,j} - H^{-1}\sumj F_j\xi_{i,j}X'_i\tdel_j + \sumj a_j\xi_{i,j}\tu_{i,j} + \sumj b_j\xi_{i,j}\tu_{i,j}
\end{align*}
Since $\tbeta$ satisfies the representation in Theorem \ref{thm: beta est}, and $\max_i\|X_i\|_2=O_P(\sqrt{\log n})$, we have
\begin{align*}
    &\max_j \|\tdel_j\|_2=O_P(\sqrt{n^{-1} \alpha_n^{-1} \log m}\ ),\\
    &\max_{i,j} |\tu_{i,j}|=O_P( \sqrt{\log(nm)} +\sqrt{n^{-1}\alpha_n^{-1}\log m \log n}\ )=O_P\left(\sqrt{\log n} + \sqrt{\log m}\right),
\end{align*}
\begin{align*}
    \max_i \left\| m^{-1}\sumj b_j\xi_{i,j}\tu_{i,j}\right\|_2=O_P\left( \alpha_n \kappa_n (\sqrt{\log n} + \sqrt{\log m}\ ) \right),
\end{align*}
and
\begin{align*}
    \max_i \left\| m^{-1}\sumj a_j\xi_{i,j}\tu_{i,j}\right\|_2
    &\leq \max_i\left\| m^{-1}\sumj a_j\xi_{i,j}\varepsilon_{i,j}\right\|_2 + \max_i\left\| m^{-1}\sumj a_j\xi_{i,j}X'_i\tdel_j\right\|_2\\[6pt]
    & \lessp \alpha_n \kappa_n + n^{-1} \log n\log m \\
    &\lessp \alpha_n \kappa_n,
\end{align*}
where we use Lemma \ref{lem: general asymp prop5}.

By the representation of $\tdel_j$ in Theorem \ref{thm: beta est},
\begin{align*}
    m^{-1}\sumj F_j\xi_{i,j}X'_i\tdel_j &= m^{-1}\sumj F_j\xi_{i,j}\tdel'_jX_i \\
    &= m^{-1}\sumj F_j\xi_{i,j} \left( n^{-1}\left( \sum_{k=1}^n(X_k \eps_{k,j} \xi_{k,j} + \pi_k X_k L_k'F_j) \right)'(EX_iX_i'\pi_i)^{-1}  X_i + s_j\right),
\end{align*}
where $\max_j \|s_j\| \lessp \kappa_n$. Observe that
\begin{align*}
    & \max_i\left\|(nm)^{-1}\sumj\sum_{k=1}^n F_j\xi_{i,j}\xi_{k,j}\varepsilon_{k,j}X'_k (EX_iX_i'\pi_i)^{-1}  X_i \right\|_2=O_P\left((nm)^{-1/2} \log n\right),\\
    & \max_i\left\|m^{-1}\sumj F_j\xi_{i,j}s_j\right\|_2 = O_P\left( \alpha_n \kappa_n\right),
\end{align*}
and
\begin{align*}
     &\max_i \left\|n^{-1}\left(m^{-1}\sumj F_j\xi_{i,j}F'_j\right) \sum_{k=1}^n \pi_k L_k X_k' (EX_iX_i'\pi_i)^{-1}   X_i  - n^{-1} \left(\pi_i \Sigma_F\right) \sum_{k=1}^n \pi_k L_k X_k' (EX_iX_i'\pi_i)^{-1}   X_i\right\|_2 \\
     =&\,  O_P\left( \sqrt{m^{-1} \alpha_n \log n}\ n^{-1/2} \sqrt{\log n}\right).
\end{align*}
Therefore
\begin{align}
    &m^{-1} \sumj \tF_j\xi_{i,j}\tu_{i,j} = m^{-1} H^{-1}\sumj F_j\xi_{i,j}\varepsilon_{i,j} -  H^{-1}n^{-1} \left(\pi_i \Sigma_F\right) \sum_{k=1}^n \pi_k L_k X_k' (EX_iX_i'\pi_i)^{-1}   X_i + \tau_{1,i}\label{eq: tL decomp 4},
\end{align}
where $  \max_i \|\tau_{1,i}\|_2=O_P\left(\alpha_n \kappa_n \right).$

\textbf{Step 4:} Combining \eqref{eq: tL decomp 1}, \eqref{eq: tL decomp 3}, \eqref{eq: tL decomp 4}, we have
\begin{align*}
    &\quad \tL_i - H'L_i \\
    & = \left(m^{-1}\sumj \tF_j\tF'_j\xi_{i,j}\right)^{-1}\left(H^{-1} m^{-1}\sumj F_j\xi_{i,j}\varepsilon_{i,j} - H^{-1}n^{-1} \left(\pi_i \Sigma_F\right) \sum_{k=1}^n \pi_k L_k X_k' (EX_iX_i'\pi_i)^{-1}   X_i\right) + \tau_{2,i},
\end{align*}
where $ \max_i \|\tau_{2,i}\|_2 =O_P\left( \kappa_n \log n\right).$ Then by \eqref{eq: tL decomp 2} and the fact that $\max_i\|m^{-1}\sumj F_j\xi_{i,j}\varepsilon_{i,j}\|_2=O_P(\sqrt{m^{-1} \alpha_n \log n})$ and $\max_i\| H^{-1}n^{-1} \left(\pi_i \Sigma_F\right) \sum_{k=1}^n \pi_k L_k X_k' (EX_iX_i'\pi_i)^{-1}   X_i \|_2=O_P(n^{-1/2} \alpha_n \sqrt{\log n} )$, we have
\begin{align*}
    &\quad \tL_i - H'L_i \\
    & = H'\pi_i^{-1}\Sigma_F^{-1}H\left(H^{-1}m^{-1}\sumj F_j\xi_{i,j}\varepsilon_{i,j} - H^{-1}n^{-1} \left(\pi_i \Sigma_F\right) \sum_{k=1}^n \pi_k L_k X_k' (EX_iX_i'\pi_i)^{-1}   X_i \right) + \tau_{3,i}\\
    &= H'\pi_i^{-1}\Sigma_F^{-1}m^{-1}\sumj F_j\xi_{i,j}\varepsilon_{i,j} - n^{-1} H' \sum_{k=1}^n \pi_k L_k X_k' (EX_iX_i'\pi_i)^{-1}   X_i +\tau_{3,i}\\
    &\max_i \|\tau_{3,i}\|_2 =O_P\left( \alpha_n^{1/2} \kappa_n +\kappa_n\right) =O_P \left(  \kappa_n \right).
\end{align*}
This concludes the proof.
\end{proof}

\begin{lem}\label{lem: induction for asymp tGamma}
Under the same assumption in Lemma \ref{lem: induction for asymp tL}, the estimator $\tGamma^{(g)}$ satisfies the representation in Theorem \ref{thm: Gamma est}. That is
\begin{align*}
    \tGamma^{(g)}_{i,j} - \Gamma_{i,j} - \Delta_{i,j} = t^{(g)}_{i,j}
\end{align*}
where
\begin{align*}
& \Delta_{i,j} = \frac{1}{n} L'_i (EL_iL_i \pi_i)^{-1} \sum_{k=1}^n L_k \xi_{k,j}\varepsilon_{k,j} + \frac{1}{m \pi_i}\sum_{t=1}^m F'_t\xi_{i,t}\varepsilon_{i,t} \Sigma^{-1}_F F_j -  \frac{1}{n}X'_i E(X_iX_i'\pi_i)^{-1} \sum_{k=1}^n \pi_k X_k L_k' F_j\\
    &\max_{i,j}|t^{(g)}_{i,j}| = O_P\left(\kappa_n\right)
\end{align*}
\end{lem}

\begin{proof}[\textbf{Proof of Lemma \ref{lem: induction for asymp tGamma}}] Again, we denote $b_j^{(g)}$, $d_i^{(g)}$, $\tF^{(g)}$, $\tL^{(g)}$ and $\tGamma^{(g)}$ by $b_j$, $d_i$, $\tF$, $\tL$ and $\tGamma$. By Assumption \ref{assu: asymp tF} and Lemma \ref{lem: induction for asymp tL}, we have
\begin{align*}
    &\tF_j = H^{-1}F_j + a_j + b_j\\
    &\tL_i = H' L_i + c_{1,i} + c_{2,i} + d_i
\end{align*}
and note that
\begin{align*}
    &\max_j \|a_j + b_j\|_2 = O_P\left(\sqrt{n^{-1}\alpha_n^{-1} \log m}\right),\\
    &\max_j \|b_j\|_2 = O_P\left(\kappa_n\right),\\
    &\max_j \|c_{1,i} + c_{2,i} + d_i\|_2 = O_P\left(\sqrt{m^{-1} \alpha_n^{-1}}n^{1/q} + n^{-1/2} \log n\right),\\
    &\max_j \|d_i\|_2 = O_P\left(\kappa_n\right).
\end{align*}
Thus, we have
\begin{align*}
    &\tL'_i\tF_j- L'_i F_j = L'_i H a_j + (c_{1,i}+c_{2,i})'H^{-1} F_j + \tau_{1,i,j},
\end{align*}
where
\begin{align*}
 \max_{i,j} |  \tau_{1,i,j} | & =  \max_{i,j} | (c_{1,i} + c_{2,i}+d_i)'(a_j + b_j) + L_i'H b_j + d_i'H^{-1} F_j|  \lessp \kappa_n.
\end{align*}

By Assumption \ref{assu: asymp tF}, we have
\begin{align*}
    L'_iH a_j &= n^{-1} L'_i H (H' EL_iL_i \pi_i H)^{-1} H'\sum_{k=1}^n L_k\xi_{k,j}\varepsilon_{k,j}\\
              &= n^{-1} L'_i (EL_iL_i \pi_i)^{-1} \sum_{k=1}^n L_k \xi_{k,j}\varepsilon_{k,j}.
\end{align*}
By Lemma \ref{lem: induction for asymp tL},
\begin{align*}
    (c_{1,i} + c_{2,i})'H^{-1}F_j & = m^{-1}\pi_i^{-1}\sum_{t=1}^m F'_t\xi_{i,t}\varepsilon_{i,t} \Sigma^{-1}_F F_j -  n^{-1}X'_i E(X_iX_i'\pi_i)^{-1} \sum_{k=1}^n \pi_k X_k L_k' F_j
\end{align*}
and therefore
\begin{align*}
    \tGamma_{i,j} - \Gamma_{i,j} = \tL'_i\tF_j - L'_iF_j = \tDel_{i,j} + \tau_{1,i,j}
\end{align*}
where
\begin{align*}
    &\quad \max_{i,j}|\tau_{1,i,j} |  \lessp \kappa_n.
\end{align*}
\end{proof}

\begin{lem}\label{lem: induction tgamma to tbeta}
Under the same assumption in Lemma \ref{lem: induction for asymp tL}, $\tbeta^{(g+1)}$ has the following representation.
\begin{align*}
    &\tbeta^{(g+1)}_j - \beta_j = n^{-1} E(X_iX_i'\pi_i)^{-1}  \left(\sumi X_i\varepsilon_{i,j}\xi_{i,j} + \pi_i X_i L_i' F_j\right)  + s^{(g)}_j\\
    &\max_j \|s^{(g)}_j\| = O_P\left( \kappa_n \right).
\end{align*}
\end{lem}
\begin{proof}[\textbf{Proof of Lemma \ref{lem: induction tgamma to tbeta}}]
For the sake of simplicity of notations, we denote $\tbeta^{(g+1)}$, $t_{i,j}^{(g)}$, and $\tGamma^{(g)}$ as $\tbeta$, $t_{i,j}$, and $\tGamma$, respectively. By Lemma \ref{lem: induction for asymp tGamma}, we have
\begin{align}
\label{eq: tbeta decomp}
    \tbeta_j - \beta_j
    & = \left(n^{-1}\sumi X_iX'_i\xi_{i,j}\right)^{-1}\left(n^{-1}\sumi X_i\varepsilon_{i,j}\xi_{i,j}\right) \nonumber\\
    &\qquad - \left(n^{-1}\sumi X_iX'_i\xi_{i,j}\right)^{-1}\left(n^{-1}\sumi X_i(\Delta_{i,j} + t_{i,j})\xi_{i,j}\right)
\end{align}
where
\begin{align*}
\Delta_{i,j} = \frac{1}{n} L'_i (EL_iL_i \pi_i)^{-1} \sum_{k=1}^n L_k \xi_{k,j}\varepsilon_{k,j} + \frac{1}{m \pi_i}\sum_{t=1}^m F'_t\xi_{i,t}\varepsilon_{i,t} \Sigma^{-1}_F F_j -  \frac{1}{n}X'_i E(X_iX_i'\pi_i)^{-1} \sum_{k=1}^n \pi_k X_k L_k' F_j.
\end{align*}

Observe that
\begin{align*}
    &\max_j \left\|n^{-2}\sumi (X_i\xi_{i,j}L'_i) (EL_iL_i'\pi_i)^{-1}\sum_{s=1}^n (L_s\xi_{s,j}\varepsilon_{s,j})\right\|_2
    = O_P\left(n^{-1}\log m\right),\\
    &\max_j \left\|(nm)^{-1}\sumi\sum_{t=1}^m \pi_i^{-1}(X_i\xi_{i,j}\varepsilon_{i,t}\xi_{i,t}F'_t)\Sigma^{-1}_F F_j\right\|_2
    =O_P\left((nm)^{-1/2}\log m\right),\\
    &\max_j \left\|n^{-1}\sumi X_i t_{i,j}\xi_{i,j}\right\|_2=O_P\left( \kappa_ n \alpha_n \right).
\end{align*}
Thus, by \eqref{eq: tbeta decomp}, we have
\begin{align*}
    &\tbeta_j - \beta_j = \left(n^{-1}\sumi X_iX'_i\xi_{i,j}\right)^{-1}\left(n^{-1}\sumi X_i\varepsilon_{i,j}\xi_{i,j}\right) + n^{-1} E(X_iX_i'\pi_i)^{-1} \sum_{k=1}^n \pi_k X_k L_k' F_j + s_{1,j}\\
    &\max_j \|s_{1,j}\| = O_P\left( \kappa_n \right).
\end{align*}
Then, by the facts that
\begin{align*}
 & \max_j\|n^{-1}\sum_i X_i\varepsilon_{i,j}\xi_{i,j}\|_2=O_P(\sqrt{n^{-1} \alpha_n \log m}) \quad \text{and}\\
 & \max_j\|(n^{-1}\sum_i X_iX_i\xi_{i,j})^{-1}-[E(X_iX_i'\pi_i)]^{-1}\|=O_P(\sqrt{n^{-1} \alpha_n^{-3}\log m}),
\end{align*}
we finally get
\begin{align*}
    &\tbeta_j - \beta_j = n^{-1} E(X_iX_i'\pi_i)^{-1}  \left(\sumi X_i\varepsilon_{i,j}\xi_{i,j} + \pi_i X_i L_i' F_j\right) + s_j\\
    &\max_j \|s_j\| = O_P\left(\kappa_n \right).
\end{align*}
\end{proof}
\begin{lem}\label{lem: induction beta to tF}
Under the same assumption in Lemma \ref{lem: induction for asymp tL}, $\tF^{(g+1)}$ also satisfies the decomposition in Assumption \ref{assu: asymp tF} and has order
\begin{align*}
    \max_j \|b^{(g+1)}_j\|_2& \equiv     \max_j \left\|\tF^{(g+1)}_j - H^{-1}F_j - a_j\right\|_2=O_P(\kappa_n).
\end{align*}
\end{lem}
\begin{proof}[\textbf{Proof of Lemma \ref{lem: induction beta to tF}}] For the sake of simplicity of notations, we denote $\tL^{(g)}$, $\tF^{(g+1)}$, $\tbeta^{(g+1)}$, $\tW^{(g+1)}$ by $\tL$, $\tF$, $\tbeta$ and $\tW$. By definition
\begin{align*}
    \tF_j &= \left(\sumi \tL_i\tL'_i\xi_{i,j}\right)^{-1}\sumi \tL_i\xi_{i,j}\tW_{i,j}\nonumber\\
\end{align*}
with the decomposition of $\tL_i$ in Lemma \ref{lem: induction for asymp tL}:
\begin{align*}
    &\tL_i = H' L_i + c_i + d_i\\
    &c_i = H'\pi_i^{-1}\Sigma_F^{-1}m^{-1}\sumj F_j\xi_{i,j}\varepsilon_{i,j} - n^{-1} H' \sum_{k=1}^n \pi_k L_k X_k' (EX_iX_i'\pi_i)^{-1}   X_i\\
    &\max_i \|d_i\|_2 =O_P\left(\kappa_n\right).
\end{align*}
The proof is very similar to the one for Lemma \ref{lem: induction for asymp tL}, so we only sketch it. Recall $\tW_{i,j}=L'_iF_j+\tu_{i,j} = (\tL_i - c_i - d_i)'H^{-1}F_j + \tu_{i,j}$,
\begin{align*}
\tF_j - H^{-1}F_j = \left(\sumi \tL_i\tL'_i\xi_{i,j}\right)^{-1} \sumi \tL_i\xi_{i,j}\left(\tu_{i,j}-( c_i + d_i)'H^{-1}F_j\right).
\end{align*}
Note that
\begin{align*}
    &\quad \max_j\left\|n^{-1}\sumi \tL_i\xi_{i,j}d_i'H^{-1}F_j\right\|_2\\
    &\leq n^{-1} (\max_j \sumi \xi_{i,j}) \max_i \|\tL_i\|_2 \cdot \max_i\|d_i\|_2\cdot\|H\|\cdot\max_j \|F_j\|_2\\
    &\lessp \alpha_n \kappa_n \log n
\end{align*}
and since $\max_i\|c_i\|_2=O_P(\alpha_n^{-1/2}(m^{-1/2} + n^{-1/2})\sqrt{\log n})$, we have
\begin{align*}
    &\, \max_j\left\|n^{-1}\sumi \tL_i\xi_{i,j}c_i'H^{-1}F_j\right\|_2\\
    \leq &\,  \max_j\left\|n^{-1}H'\sumi L_i\xi_{i,j}c_i'H^{-1}F_j\right\|_2 + \max_j\left\|n^{-1}\sumi (c_i + d_i)\xi_{i,j}c_i'H^{-1}F_j\right\|_2\\
    \lessp & \max_j\left\|n^{-1}H'\sumi L_i\xi_{i,j}c_i' \right\| \max_j \left\| F_j\right\|_2 + \max_i (c_i+d_i)^2 \max_j\left(n^{-1}\sumi \xi_{i,j} \right)\max_j ||F_j||_2 \\
    =&\, O_P\left((n^{-1}+m^{-1})\log m \log n\right),
\end{align*}
where we use the fact that
\begin{align*}
\max_s\left\|n^{-1}\sumi L_i\xi_{i,s}c_i' \right\| & \lessp \max_s (mn)^{-1} \left\| \sumi \sumj L_i\xi_{i,s} F_j'\xi_{i,j}\varepsilon_{i,j} \pi_i^{-1} \right\| \\
& + \max_s \left\| n^{-2} (\sumi L_i\xi_{i,s}X_i') (E X_iX_i'\pi_i)^{-1} (\sum_{k=1}^n \pi_k L_k X_k ) \right\| \\
& \lessp (1/n + 1/m) \log^{1/2} n.
\end{align*}

For $\sumi\tL_i\xi_{i,j}\tu_{i,j}=\sumi(H'L_i+c_i+d_i)\xi_{i,j}\tu_{i,j}$, first note that
\begin{align*}
\max_j\left\|n^{-1}\sumi d_i\xi_{i,j}\tu_{i,j}\right\|_2     & \lessp \max_j\left\|n^{-1}\sumi d_i\xi_{i,j}\eps_{i,j}\right\|_2 + \max_j\left\|n^{-1}\sumi d_i\xi_{i,j}X_i' \right\| \|  \tdel_j\|_2 \\
& \lessp \alpha_n \kappa_n \\
    \max_j\left\|n^{-1}\sumi c_i\xi_{i,j}\tu_{i,j}\right\|_2
    &\leq \max_j\left\|n^{-1}\sumi c_i\xi_{i,j}\varepsilon_{i,j}\right\|_2 + \max_j\left\|n^{-1}\sumi c_i\xi_{i,j}X'_i\right\| \|\tdel_j\|_2 \\[6pt]
    & \lessp (1/n + 1/m) \log^{1/2} n,
\end{align*}
where we use the facts that
\begin{align*}
\max_s\left\|n^{-1}\sumi \eps_{i,s}\xi_{i,s}c_i' \right\| & \lessp \max_s (mn)^{-1} \left\| \sumi \sumj \eps_{i,s}\xi_{i,s} F_j'\xi_{i,j}\varepsilon_{i,j} \pi_i^{-1} \right\| \\
& + \max_s \left\| n^{-2} (\sumi \eps_{i,s}\xi_{i,s}X_i') (E X_iX_i'\pi_i)^{-1} (\sum_{k=1}^n \pi_k L_k X_k ) \right\| \\
& \lessp (1/n + 1/m) \log^{1/2} n
\end{align*}
and
\begin{align*}
\max_s\left\|n^{-1}\sumi X_i\xi_{i,s}c_i' \right\| & \lessp \max_s (mn)^{-1} \left\| \sumi \sumj X_i\xi_{i,s} F_j'\xi_{i,j}\varepsilon_{i,j} \pi_i^{-1} \right\| \\
& + \max_s \left\| n^{-2} (\sumi \xi_{i,s} X_iX_i') (E X_iX_i'\pi_i)^{-1} (\sum_{k=1}^n \pi_k L_k X_k ) \right\| \\
& \lessp (1/n + 1/m) \log^{1/2} n + \alpha_n n^{-1/2}.
\end{align*}

In addition, we have
\begin{align*}
    &H'\sumi L_i\xi_{i,j}\tu_{i,j}=H'\sumi L_i\xi_{i,j}\varepsilon_{i,j} - H'\sumi L_i\xi_{i,j} X'_i\tdel_j,\\
    &\max_j\left\|H'\sumi L_i\xi_{i,j} X'_i\tdel_j\right\|_2=O_P\left(\log m\right).
\end{align*}
Last, we have
\begin{align*}
    &\quad \max_j \left\|\left(n^{-1}\sumi \tL_i\tL'_i\xi_{i,j}\right)^{-1} - H^{-1}(E L_iL_i'\pi_i)^{-1} (H')^{-1}\right\|\\
    &=O_P(n^{-1}(1+n/m)\alpha_n^{-3} n^{1/q} + n^{-1/2}\alpha_n^{-1}),\\[6pt]
    &\quad \max_j\left\|H'\sumi L_i\xi_{i,j}\varepsilon_{i,j}\right\|_2=O_P\left(\sqrt{ n \alpha_n \log m}\right).
\end{align*}
Combine all the results, we have
\begin{align*}
    \tF_j - H^{-1}F_j & = \left[ H^{-1}(E L_iL_i'\pi_i)^{-1} + O_P\left(\sqrt{ n \alpha_n \log m}\right) \right]\\
    & \times \left[ \frac{1}{n}  \sumi L_i\xi_{i,j}\varepsilon_{i,j} + O_P(\alpha_n \kappa_n)\right] \\
    = &    \frac{1}{n}H^{-1}(E L_iL_i'\pi_i)^{-1} \sumi L_i\xi_{i,j}\varepsilon_{i,j} + \tau_{1,j}\\
    &\max_j\|\tau_{1,j}\|_2=O_P\left(\kappa_n\right).
\end{align*}

\end{proof}

\begin{lem}\label{lem: assum 1 to 3}
\label{lem: general asymp prop5}Let Assumption \ref{assu: moment cond} hold.
Then
\begin{align*}
 \max_{i}\|m^{-1}\sumj \xi_{i,j}a_{j}\varepsilon_{i,j}\|_{2}\lessp  \alpha_n \kappa_n
\end{align*}
and
\begin{align*}
\max_{i}\|m^{-1}\sumj \xi_{i,j}F_{j}a_{j}'\|\lessp (mn)^{-1/2} \log n,
\end{align*}
where $a_{j}=n^{-1} \left(H'E (L_i L_i' \pi_i) H\right)^{-1} H'\sumi L_{i}\xi_{i,j}\varepsilon_{i,j}$.
\end{lem}
\begin{proof}[\textbf{Proof of Lemma \ref{lem: general asymp prop5}}]
We proceed in two steps.\\[6pt]
\textbf{Step 1:} we bound $\max_{i}\|m^{-1}\sumj \xi_{i,j}a_{j}\varepsilon_{i,j}\|_{2}$.\\
Note that
\begin{align}
    & \max_{1\leq i\leq n}\left\| m^{-1}\sumj \xi_{i,j}a_{j}\varepsilon_{i,j}\right\|_{2} \nonumber\\
    & \lessp \max_{1\leq i\leq n}\left\| (nm\alpha_n)^{-1} \sumj\sum_{h=1}^{n}\xi_{i,j}L_{h}\xi_{h,j}\varepsilon_{h,j}\varepsilon_{i,j}\right\|_{2}\nonumber \\
    & \leq (nm\alpha_n)^{-1} \max_{1\leq i\leq n}\left\| \sumj \sum_{h=1,\ h\neq i}^{n} \xi_{i,j}L_{h}\xi_{h,j}\varepsilon_{h,j}\varepsilon_{i,j}\right\|_{2} + (nm\alpha_n)^{-1}\left\| \sumj L_{i}\xi_{i,j}\varepsilon_{i,j}^{2}\right\|_{2},\label{eq: lem part 15 eq 3}
\end{align}
where (i) follows by $\|H\|=O_{P}(1)$ due to Lemma \ref{lem: part 4}. Let $L_{h,q}$ be the $q$-th entry of $L_{h}$ for $q\in\{1,...,k\}$. Note that conditional on $(\Xi,L)$, the term $\sumj \sum_{h=1,\ h\neq i}^{n}\xi_{i,j}L_{h,q}\xi_{h,j}\varepsilon_{h,j}\varepsilon_{i,j}$ is a summation of $m(n-1)$ independent sub-Gaussian mean-zero variables given $(\Xi,L, \{\eps_{i,j}\}_{j =1}^m)$. Therefore, for any non-random $\lambda>0$,
\begin{align*}
E\left[\left.\exp\left(\lambda\sumj \sum_{h=1,\ h\neq i}^{n} \xi_{i,j}L_{h,q}\xi_{h,j}\varepsilon_{h,j}\varepsilon_{i,j}\right)\right|\, \Xi,L,\{\eps_{i,j}\}_{j =1}^m\right]\leq\exp\left(C_{1}\lambda^{2}\sumj \sum_{h=1,\ h\neq i}^{n}L_{h,q}^{2} \eps_{i,j}^2 \xi_{i,j}\right),
\end{align*}
where $C_{1}>0$ is a constant. In other words,
\begin{align*}
\frac{\sumj \sum_{h=1,\ h\neq i}^{n}\xi_{i,j}L_{h,q}\xi_{h,j}\varepsilon_{h,j}\varepsilon_{i,j}}{\sqrt{\sumj \sum_{h=1,\ h\neq i}^{n}L_{h,q}^{2} \eps_{i,j}^2 \xi_{i,j}}}
\end{align*}
has bounded sub-Gaussian norm. Thus, by the exponential tail of sub-Gaussian
variables and the standard union bound, we have
\begin{align*}
\max_{1\leq i\leq n}\left|\frac{\sumj \sum_{h=1,\ h\neq i}^{n}\xi_{i,j}L_{h,q}\xi_{h,j}\varepsilon_{h,j}\varepsilon_{i,j}}{\sqrt{ \sumj \sum_{h=1,\ h\neq i}^{n}L_{h,q}^{2} \eps_{i,j}^2 \xi_{i,j}}}\right|\lessp \sqrt{\log n}.
\end{align*}
Since we can apply the same analysis for all $q\in\{1,...,k\}$ and
$k$ is bounded, we have
\begin{equation}
\max_{1\leq i\leq n}\left\| \sumj \sum_{h=1,\ h\neq i}^{n}\xi_{i,j}L_{h}\xi_{h,j}\varepsilon_{h,j}\varepsilon_{i,j}\right\|_{2}\lessp \max_i \sqrt{ \sumj \sum_{h=1,\ h\neq i}^{n}L_{h,q}^{2} \eps_{i,j}^2 \xi_{i,j} \log n}\lessp \sqrt{mn \alpha_n \log n}.\label{eq: lem part 15 eq 4}
\end{equation}
We also observe that
\begin{align*}
\max_{1\leq i\leq n}\left\| \sumj L_{i}\xi_{i,j}\varepsilon_{i,j}^{2}\right\|_{2}\lesssim\|L\|_{\infty}\max_{1\leq i\leq n}\sumj \xi_{i,j} \varepsilon_{i,j}^{2}\overset{\texti}{\lessp }m\alpha_n \sqrt{\log n}.
\end{align*}
where (i) follows by $\max_{i}\sumj \xi_{i,j} \varepsilon_{i,j}^{2}\lessp m\alpha_n $
(due to Lemma \ref{lem: general asymp prop}) and $\|L\|_{\infty}\lessp \sqrt{\log n}$. Combining the above two displays with (\ref{eq: lem part 15 eq 3}), we obtain
\begin{align*}
\max_{1\leq i\leq n}\left\| m^{-1}\sumj \xi_{i,j}a_{j}\varepsilon_{i,j}\right\|_{2}\lessp \alpha_n \kappa_n.
\end{align*}
\textbf{Step 2:} we bound $\max_{i}\|m^{-1}\sumj \xi_{i,j}F_{j}a_{j}'\|$.\\
Note that
\begin{align*}
 &\quad \max_{1\leq i\leq n}\left\| m^{-1}\sumj \xi_{i,j}F_{j}a_{j}'H'L_{i}\right\|_{2} \\
 & \lessp (nm\alpha_n)^{-1} \max_{1\leq i\leq n}\left\| \sumj \xi_{i,j}F_{j}\sum_{h=1}^{n}L_{h}'\xi_{h,j}\varepsilon_{h,j}\right\|.
\end{align*}
Again, we observe that conditional on $(\Xi,F,L)$, $\sumj \sum_{h=1}^{n}\xi_{i,j}F_{j}L_{h}'\xi_{h,j}\varepsilon_{h,j}$
is the sum of $mn$ independent sub-Gaussian variables with mean zero.
Therefore, by the same argument as in (\ref{eq: lem part 15 eq 4}),
we have
\begin{align*}
\max_{1\leq i\leq n}\left\| \sumj \sum_{h=1}^{n}\xi_{i,j}F_{j}L_{h}'\xi_{h,j}\varepsilon_{h,j}\right\| & \lessp
 \max_i \sqrt{ \sum_{j=1}^m \sumi \xi_{i,j}\xi_{h,j} ||F_j||_2 ||L_i||_2 \log n} \\
 & \lessp \sqrt{nm\alpha_n^2 \log_n} + \sqrt{ \alpha_n m \log^2 n}.
\end{align*}
The above two displays imply that
\begin{align*}
\max_{1\leq i\leq n}\left\| m^{-1}\sumj \xi_{i,j}F_{j}a_{j}'\right\|_{2}\lessp (mn)^{-1/2}\log n .
\end{align*}
The proof is complete.
\end{proof}

\subsection{Proofs of Theorems \ref{thm: beta est} and \ref{thm: Gamma est}}
\begin{proof}[\textbf{Proof}]\ \\
By Lemma \ref{lem: asymp tF g=1}, we have
\begin{equation}
    \tF_{j}^{(1)}=H^{-1}F_{j}+a_{j}+b^{(1)}_{j},\label{eq: thm Gamma est iter eq 3}
\end{equation}
such that $\max_j \|b^{(1)}_{j}\|_2 = O_P(\kappa_n)$, where
\begin{align*}
    \kappa_n = (n^{-1}+m^{-1})\alpha_n^{-3/2} n^{1/q}
\end{align*}
that satisfies Assumption \ref{assu: asymp tF}. In Section \ref{app: sub proof of beta est g=1}, we have proved that $\tbeta^{(1)}$ satisfies the representation in Theorem \ref{thm: beta est}. Now we have $\tF^{(1)}$ and $\tbeta^{(1)}$ that satisfy Assumption \ref{assu: asymp tF} and Theorem \ref{thm: beta est} respectively. Next, we assume that there exists a positive integer $g$ such that $\tF^{(g)}$ and $\tbeta^{(g)}$ satisfy Assumption \ref{assu: asymp tF} and Theorem \ref{thm: beta est}, respectively. Then by  Lemma \ref{lem: induction tgamma to tbeta}, we have
\begin{align*}
&\Big\|\tbeta^{(g+1)} -\beta -  n^{-1} E(X_iX_i'\pi_i)^{-1}  \left(\sumi X_i\varepsilon_{i,j}\xi_{i,j} + \pi_i X_i L_i' F_j\right) \Big\|_{2,\infty}=O_{P}\left(\kappa_n \right)
\end{align*}
that satisfies Theorem \ref{thm: beta est}. By Lemma \ref{lem: induction beta to tF}, we have
\begin{align*}
&\, \max_j \left\|\tF^{(g+1)}_j - H^{-1}F_j - a_j\right\|_2 = O_{P}\left(\kappa_n \right),
\end{align*}
which again satisfies Assumption \ref{assu: asymp tF}. Thus, the induction for $\tbeta^{(g)}$ and $\tF^{(g)}$ is completed, and hence Theorem \ref{thm: beta est} is proved for all $g$. Lastly, since we also proved that $\tF^{(g)}$ satisfies Assumption \ref{assu: asymp tF} for all $g$, by Lemma \ref{lem: induction for asymp tGamma}, we can get
\begin{align*}
    & \left\|\tGamma^{(g)}-\Gamma- \Delta
        \right\|_\infty   =  O_{P}\left(\kappa_n \right)
        \end{align*}
for all $g$, where $\Delta$ is a $n \times m$ matrix with its $(i,j)$th entry
\begin{align*}
    \Delta_{i,j} =  \frac{1}{n} L'_i (EL_iL_i \pi_i)^{-1} \sum_{k=1}^n L_k \xi_{k,j}\varepsilon_{k,j} + \frac{1}{m \pi_i}\sum_{t=1}^m F'_t\xi_{i,t}\varepsilon_{i,t} \Sigma^{-1}_F F_j -  \frac{1}{n}X'_i E(X_iX_i'\pi_i)^{-1} \sum_{k=1}^n \pi_k X_k L_k' F_j.
\end{align*}
\end{proof}

\section{Proofs of The Theorem of Rank Estimation }\label{app: proof of eIC-star}

{\bf Notations.} We now define the $m$-by-$m$ diagonal matrix $\Xi_{\ii}$ by $\Xi_{\ii}= \diag(\xi_{i,1}, \xi_{i,2}, \cdots, \xi_{i,m})$ $=\diag(\xi_{\ii})$ and similarly $\Xi_{\jj}\in \RR^{n\times n}$. We first prove a special case of Theorem \ref{thm: r est eIC} at $g=1$. Therefore, for the sake of simplicity of notation, we now denote $\tL^{k,(1)}$ and $\tF^{k,(1)}$  by $\tL^k$ and $\tF^k$ and {\bf omit the iteration counter} for all their derivatives. Note that, unlike the estimation, we do not update $\hat \beta$ in each iteration. Again, $W\in\RR^{n\times m}$ and its SVD are defined as follows
\begin{align*}
    W &=\diag(\hpi)^{-1}\Xi\circ\left(LF'-X(\hbeta-\beta)'+\varepsilon\right)\\
      &=UDV'
\end{align*}
Furthermore, the rank-$k$ SVD of $W$ with $k$ largest singular values is denoted by $U_k D_k V'_k$. Also, we define
\begin{align*}
\tW & =  \Xi \circ (Y-X\hbeta') = \Xi \circ (LF'-X\hat \delta' + \varepsilon),\\
e & = \diag(\hpi)^{-1}\tW - LF',
\end{align*}
where both $\tW$ and $e$ do not depend on $k$. We want to use the same strategy that separates the proof into the cases when $k\leq r$ and the cases when $k>r$. Before we start the proof, we need to generalize some of the previous lemmas to the cases when $k$ is not equal to the true rank $r$.

\begin{lem}\label{lem: order of H Delta k<r}
Let $\hL^k$,$\hF^k$ be the estimators defined in Section \ref{sec: rank est}. Under Assumption \ref{assu: moment cond}, for $k\leq r$, we have
\begin{align*}
    \hL^k &= LH^k_L + \Delta^k_L\\
    \hF^k &= FH^k_F + \Delta^k_F
\end{align*}
where $H^k_L=m^{-1}F'\hF^k \hOme_k^{-2}$, $H^k_F=n^{-1}L'\hL^k$, $\hOme_k = \frac{D_k}{\sqrt{nm}}$. Moreover, the matrices $H^k_L$, $H^k_F$, $\Delta^k_L$, $\Delta^k_F$ satisfy
\begin{align*}
    &\left\|H^k_L\right\|=O_P(1), \quad  \left\|H^k_F\right\|=O_P(1), \quad \left\|\Delta^k_L\right\|=O_P\left(\sqrt{\left(1+\frac{n}{m}\right) \alpha_n^{-1}}n^{1/q}\right),\\
    & \left\|\Delta^k_F\right\|=O_P\left(\sqrt{\left(1+\frac{m}{n}\right) \alpha_n^{-1}}n^{1/q}\right).
\end{align*}
\end{lem}
\begin{proof}[\textbf{Proof of Lemma \ref{lem: order of H Delta k<r}}]
    By definition,
\begin{align*}
    \hL^k
    =\, & W\hF^k \cdot n\cdot  D_k^{-2} = (LF'+e)\hF^k n D_k^{-2}\\
    =\, & L \left(m^{-1}F'\hF^k\hOme_k^{-2}\right) + m^{-1}e\hF^k \hOme_k^{-2}\\
    =\, & LH^k_L + \Delta^k_L\\[6pt]
    \hF^k
    = \,& W'\hL^k \cdot n^{-1} = (LF'+e)'\hL^k n^{-1} \\
    =\, & F\left(n^{-1}L'\hL^k \right) + n^{-1}e'\hL^k \\
    =\, & F H^k_F + \Delta^k_F
\end{align*}
Similar arguments as in Lemma \ref{lem: order of Omega H k=r} would give $\|\hOme_k\|=O_P(1)$, $\|\hOme_k^{-1}\|=O_P(1)$ for $k\leq r$. Hence we have
\begin{align*}
    &\left\|H^k_L\right\|\leq \left\|\sqrt{m^{-1}}F\right\|\left\|\sqrt{m^{-1}}\hF^k\right\|\left\|\hOme^{-2}_k\right\|=O_P(1)\\
    &\left\|H^k_F\right\|\leq \left\|\sqrt{n^{-1}}L\right\|\left\|\sqrt{n^{-1}}\hL^k\right\|=O_P(1).
\end{align*}

By Lemma \ref{lem: order of e}, $\|e\|=O_P(\sqrt{(m+n)  \alpha_n^{-1}} n^{1/q})$, and thus we also have
\begin{align*}
    &\left\|\Delta^k_L\right\|=O_P\left(\sqrt{\left(1+\frac{n}{m}\right) \alpha_n^{-1}}n^{1/q}\right)\\
    &\left\|\Delta^k_F\right\|=O_P\left(\sqrt{\left(1+\frac{m}{n}\right) \alpha_n^{-1}}n^{1/q}\right)
\end{align*}

\end{proof}

\begin{lem}\label{lem: order of tDelta_F k<r}
Under Assumption \ref{assu: moment cond}, for $k\leq r$, the estimator $\tF^k=\tF^{k,(1)}\in \RR^{m\times k}$ satisfies
\begin{align*}
    \tF^k = F \bH^k_F  + \tDel^k_F
\end{align*}
where $\tDel^k_F$ satisfies
\begin{align*}
    \left\|\tDel^k_F\right\| = O_P\left(  ((1+m/n)\alpha_n^{-1} )^{1/2} n^{1/q} \right),
\end{align*}
$\bH^k_F= \left( \frac{1}{n}\sumi L_i L_i' H_L^k  \pi_i \right)\left(H_L^{k'} \frac{1}{n} \sumi L_iL_i' \pi_i H_L^k\right)^{-1} $ and $\|\bH^k_F\| = O_P(n^{1/q})$.
\end{lem}
\begin{proof}[\textbf{Proof of Lemma \ref{lem: order of tDelta_F k<r}}]
Let $\hat u_{i,j} = \eps_{i,j} - X_i'\hat \delta_j$. By definition, we have
\begin{align*}
   &  \tF^k_j - \bH^{k'}_F F_j \\
   & =\left(\sumi \hL_{i}^k\hL_{i}^{k'} \xi_{i,j}\right)^{-1}\left(\sumi \hL_{i}^k(L_{i}'F_{j}+\hat u_{i,j}) \xi_{i,j}-\hL_{i}^k\hL_{i}^{k'} \xi_{i,j}\bH^{k'}_FF_{j}\right)\nonumber \\
 & =\left(\sumi \hL_{i}^k\hL_{i}^{k'} \xi_{i,j}\right)^{-1}\left(\sumi \hL_{i}^k\hat u_{i,j}  \xi_{i,j}\right)\nonumber\\
 & +\left(\sumi \hL_{i}^k\hL_{i}^{k'} \xi_{i,j}\right)^{-1}\left(\sumi \hL_{i}^k L_{i}'F_{j}  \xi_{i,j}-\hL_{i}^{k}\hL_{i}^{k'}\xi_{i,j}\bH^{k'}_FF_{j}\right).
\end{align*}

We have
\begin{align*}
& \max_{1\leq j\leq m}\left\|\left(n^{-1}\sumi \Delta^k_{L,i} \Delta^{k'}_{L,i}\xi_{i,j}\right) \right\| \\
& \lessp \max_{1\leq j\leq m}\left\|\left((nm)^{-1}\sumi \Delta^k_{L,i} e_i'(F H_F^k +  \Delta_F^k)) \xi_{i,j}\right) \right\|     \\
& \lessp \max_{1\leq j\leq m}\left\|\left((nm)^{-1}\sumi \Delta^k_{L,i} e_i'F \xi_{i,j}\right) \right\|   + (nm)^{-1} ||\Delta^k_L|| ||e|| ||\Delta^k_F|| \\
& \lessp (nm)^{-1}\max_i ||F'e_i||_2 \sumi || \hat \Delta_{L,i}^{k}||_2 \xi_{i,j} + (1/n+1/m)^{3/2} \alpha_n^{-3/2}n^{1/q} \\
& \lessp (nm)^{-1}\max_i ||F'e_i||_2  ||\hat \Delta_{L}^{k}|| (n\alpha_n)^{1/2} + (1/n+1/m)^{3/2} \alpha_n^{-3/2}n^{1/q} \\
& \lessp (1/n+1/m) \alpha_n^{-1/2}n^{1/q},
\end{align*}
where the last inequality is by \eqref{eq:Fei}. In addition, we have
\begin{align*}
 & \max_{1\leq j\leq m}\left\|\left(n^{-1}\sumi \hL^k_iL_i'\xi_{i,j}\right) - \left(H_L^{k'} \frac{1}{n} \sumi L_iL_i' \pi_i \right)
 \right\|  \\
 & \lessp \max_{1\leq j\leq m}\left\| \left(H_L^{k'} \frac{1}{n} \sumi L_iL_i' (\xi_{i,j} - \pi_i) \right)
 \right\|
  + \max_{1\leq j\leq m}\left\|\left(n^{-1}\sumi \hDel^k_{L,i} L_i'\xi_{i,j}\right) \right\| \\
& \lessp (\alpha_n n^{-1} \log n)^{1/2} + \max_{1\leq j\leq m}\left\|\left((nm)^{-1}\sumi \hF^{k'} e_i'L_i \xi_{i,j}\right) \right\| \\
& \lessp (\alpha_n n^{-1} \log n)^{1/2} + \max_{1\leq j\leq m}\left\|\left((nm)^{-1}\sumi F' e_i'L_i \xi_{i,j}\right) \right\| + (mn)^{-1/2} n^{1/q} (1+m/n)^{1/2}\alpha_n^{-1/2}\\
& \lessp (\alpha_n n^{-1} \log n)^{1/2},
\end{align*}
where the last inequality is due to Lemma \ref{lem: part 7} and the fact that $(n \wedge m) \alpha_n^2 \rightarrow \infty$ in a polynomial rate in $n$.

Therefore, we have
\begin{align*}
 & \max_{1\leq j\leq m}\left\|\left(n^{-1}\sumi \hL^k_i\hL^{k'}_i\xi_{i,j}\right) - \left(H_L^{k'} \frac{1}{n} \sumi L_iL_i' \pi_i H_L^k\right)
 \right\|  \\
 & \lessp \max_{1\leq j\leq m}\left\|\left(n^{-1}\sumi \hL^k_iL_i'\xi_{i,j}\right) - \left(H_L^{k'} \frac{1}{n} \sumi L_iL_i' \pi_i \right) \right\| +\max_{1\leq j\leq m}\left\|\left(n^{-1}\sumi \hDel^k_{L,i} \hDel^{k'}_i\xi_{i,j}\right) \right\| \\
& =O_P\left((1/n+1/m)  \alpha_n^{-1/2} n^{1/q}\right).
\end{align*}

 Last, there exists a constant $c>0$ such that
\begin{align*}
& H_L^{k'}\left( \frac{1}{n} \sumi L_iL_i' \pi_i \right)H_L^{k} \\
& \geq c H_L^{k'}\left( \frac{1}{n} \sumi L_iL_i' \right)H_L^{k} \alpha_n n^{-1/q} \\
& \geq c  \alpha_n n^{-1/q} \left( I_k - \frac{1}{n} \sumi \hL_i^{k'} \hDel_{L,i}^k - \frac{1}{n} \sumi \hDel_{L,i}^{k'} \hL_i^k + \frac{1}{n} \sumi \hDel_{L,i}^{k'} \hDel_{L,i}^{k} \right) \\
& \geq c \alpha_n n^{-1/q} \left( I_k - o_P(1)\right),
\end{align*}
which implies
\begin{align*}
 & \max_{1\leq j\leq m}\left\|\left(n^{-1}\sumi \hL^k_i\hL^{k'}_i\xi_{i,j}\right)^{-1} - \left(H_L^{k'} \frac{1}{n} \sumi L_iL_i' \pi_i H_L^k\right)^{-1}
 \right\| \\
 &=O_P\left((1/n+1/m)  \alpha_n^{-3} n^{1/q} + \alpha_n^{-3/2} n^{-1/2} n^{1/q}\right)
\end{align*}
and
\begin{align*}
& \max_j  \left \|    \left(\sumi \hL_{i}^k\hL_{i}^{k'} \xi_{i,j}\right)^{-1}\left(\sumi \hL_{i}^k L_{i}'F_{j}  \xi_{i,j}-\hL_{i}^{k}\hL_{i}^{k'}\xi_{i,j}\bH^{k'}_FF_{j}\right) \right\| \\
& = \max_j \left\|    \left[\left(\sumi \hL_{i}^k\hL_{i}^{k'} \xi_{i,j}\right)^{-1}\left(\sumi \hL_{i}^k L_{i}' \xi_{i,j}\right)- \bH^{k'}_F \right] F_{j} \right\| \\
& \lessp (1/n+1/m)  \alpha_n^{-3/2} n^{1/q}.
 \end{align*}

In addition, by the same arguments in Lemmas \ref{lem: general asymp prop1} and \ref{lem: asymp tF g=1} respectively, we have
\begin{align*}
\max_j n^{-1} \left\| \sumi \hL_{i}^k\hat u_{i,j}\xi_{i,j}\right\| \lessp \sqrt{ (1/n+1/m) \alpha_n n^{1/q}} .
\end{align*}

Therefore, we have
\begin{align*}
    \left\|\tF^k - F \bH^{k}_F \right\| & \lessp \sqrt{m} \left((1/n+1/m)  \alpha_n^{-3/2} n^{1/q} + \alpha_n^{-1} \sqrt{ (1/n+1/m) \alpha_n n^{1/q}} \right) \\
    & =     O_P\left(  ((1+m/n)\alpha_n^{-1} )^{1/2} n^{1/q} \right).
\end{align*}
\end{proof}

Next, we define the function $\mse(k)$ under rank $k$ estimators as
\begin{align}
    \mse(k)
    &=\frac{1}{nm}\sumi \sumj \left(\tW_{i,j} - \xi_{i,j}\tL^{k'}_i\tF^k_j\right)^2\label{eq: def of mse(k)}\\
    &=\frac{1}{nm}\sumi \left\|\tW_{\ii} - \Xi_{\ii}\tF^k\tL^{k'}_i\right\|^2_2\nonumber\\
    &=\frac{1}{nm}\sumi \left\|\tW_{\ii} - P_{\Xi_{\ii}\tF^k}\tW_{\ii}\right\|^2_2\nonumber\\
    &=\frac{1}{ mn}\sumi \tW^{'}_{\ii}M_{\Xi_{\ii}\tF^k}\tW_{\ii}\nonumber
\end{align}
where $M_{\Xi_{\ii}\tF^k} = I- P_{\Xi_{\ii}\tF^k}$, $P_{\Xi_{\ii}\tF^k}$ is a projection matrix, and we use the fact that $\Xi_{\ii}\tW_{\ii}=\tW_{\ii}$.
By the definition of $\tL^k=\tL^{k,(1)}$, we can write $\eIC^*(k)$ as a function of $\tF^k$
\begin{align*}
    \eIC^*(k)
    & = \frac{1}{nm}\left\|\Xi\circ\left(Y - X\hbeta^{'} - \tL^{k,(1)} \tF^{k,(1)'}\right)\right\|_F^2 + k\cdot h^*(n,m) \\\
    & = \frac{1}{nm}\sumi \left\|\tW_{\ii} - \Xi_{\ii}\tF^k\tL^k_i\right\|^2_2 + k\cdot h^*(n,m) \\
    & = \frac{1}{nm}\sumi \left\|\tW_{\ii} - \Xi_{\ii}\tF^k\left(\tF'^k\Xi_{\ii}\tF^k\right)^{-1}\tF'^k\Xi_{\ii}\tW_{\ii}\right\|^2_2 + k\cdot h^*(n,m)\\
    & = \frac{1}{nm}\sumi \tW^{'}_{\ii} \left( I - P_{\Xi_{\ii}\tF^k}\right)\tW_{\ii} + k\cdot h^*(n,m) \\
    & = \frac{1}{nm}\sumi \tW^{'}_{\ii} M_{\Xi_{\ii}\tF^k} \tW_{\ii} + k\cdot h^*(n,m) \\
    & = \mse(k) + k\cdot h^*(n,m)
\end{align*}
 Therefore, the difference $\eIC^*(r)$ and $\eIC^*(k)$ is
\begin{align*}
    \eIC^*(r) - \eIC^*(k) = \mse(r) - \mse(k) + (r-k)\cdot h^*(n,m)
\end{align*}
where
\begin{align*}
    \mse(r) - \mse(k)
    & = \frac{1}{nm}\left\{\sumi \tW^{'}_{\ii}\left(M_{\Xi_{\ii}\tF^r} - M_{\Xi_{\ii}\tF^k}\right)\tW_{\ii}\right\}\\
    & = \frac{1}{nm}\left\{\sumi \tW^{'}_{\ii}\left(P_{\Xi_{\ii}\tF^k} - P_{\Xi_{\ii}\tF^r}\right)\tW_{\ii}\right\}.
\end{align*}
In order to characterize $\eIC^*(r) - \eIC^*(k)$, we need to study the difference between $P_{\Xi_{\ii}\tF^k}$ and $P_{\Xi_{\ii}\tF^r}$.

\begin{lem}\label{lem: order of inv tF.Xi.tF - H.F.F.H k<r}
Under Assumption \ref{assu: moment cond}, for $k\leq r$, we have
\begin{align*}
    \max_{1\leq i\leq n}\left\| \pi_i\left(\frac{\tF^{k'}\Xi_{\ii}\tF^k}{m}\right)^{-1} - \left( \frac{\bH^{k'}_F F'F \bH^k_F}{m} \right)^{-1} \right\|=O_P\left(\sqrt{(1/n+1/m)\alpha_n^{-2} } n^{1/q}\right).
\end{align*}
\end{lem}

\begin{proof}[\textbf{Proof of Lemma \ref{lem: order of inv tF.Xi.tF - H.F.F.H k<r}}]
Note that
\begin{align*}
     & \left\| \frac{\tF^{k'}\Xi_{\ii}\tF^k}{\pi_i m} -  \frac{\bH^{k'}_F F'F \bH^k_F}{m} \right\|\\
    =\,& \left\| \frac{\left(F\bH^k_F+\tDel^k_F\right)'\Xi_{\ii}\left(F\bH^k_F+\tDel^k_F\right)}{\pi_i m} -  \frac{\bH^{k'}_F F'F \bH^k_F}{m} \right\|\\
\leq & m^{-1}\pi_i^{-1}\left\| \bH^{k'}_F F'\Xi_{\ii}F \bH^k_F - \pi_i \bH^{k'}_F F'F \bH^k_F \right\| \\
 &\, \qquad + 2m^{-1}\pi_i^{-1}\|\bH^{k'}_F F'\Xi_{\ii}\tDel^k_F\| + m^{-1}\pi_i^{-1}\left\|\tDel^{k'}_F\Xi_{\ii}\tDel^k_F \right\|\\
    =\,& I_i + 2II_i +III_i
\end{align*}
For $I_i$, we have
\begin{align*}
    &I_i = \frac{1}{m\pi_i}\left\|\bH^{k'}_F\sumj F_j F'_j\left(\xi_{i,j}-\pi_i\right)\bH^k_F\right\| \quad \text{and}\\
    &\max_{1\leq i\leq n} I_i =O_P\left((m\alpha_n)^{-1/2} n^{1/q}\right)
\end{align*}
by Lemmas \ref{lem: general asymp prop} and \ref{lem: order of H Delta k<r}. For $II_i$ and $III_i$, we have
\begin{align*}
    &\max_{1\leq i\leq n}II_i\leq\frac{1}{m\pi_i}\left\|\bH^k_F\right\|\left\|F'\Xi_{\ii}\right\|\left\|\tDel^k_F\right\|
    =O_P\left(\sqrt{(1/n+1/m)\alpha_n^{-2} } n^{1/q}\right) \quad \text{and}\\
    &\max_{1\leq i\leq n}III_i\leq\frac{1}{m\pi_i}\left\|\tDel^k_F\right\|^2
    =O_P\left( (1/m+1/n)\alpha_n^{-2} n^{1/q}\right)
\end{align*}
by Lemma \ref{lem: order of H Delta k<r} and \ref{lem: order of tDelta_F k<r}. Thus we have
\begin{align*}
    \left\| \frac{\tF^{k'}\Xi_{\ii}\tF^k}{\pi_i m} -  \frac{\bH^{k'}_F F'F \bH^k_F}{m} \right\|=O_P\left(\sqrt{(1/n+1/m)\alpha_n^{-2} } n^{1/q}\right).
\end{align*}
Moreover, since $k\leq r$, $m^{-1}\bH^{k'}_F F'F \bH^k_F$ converges to a full rank matrix whose smallest eigenvalue is bounded below by $c n^{-1/q}$, where the positive constant $c$ depends on the smallest eigenvalue of $\Sigma_F$. This means both $\|(m^{-1}\pi_i^{-1}\tF^{k'}\Xi_{\ii}\tF^k)^{-1}\|$ and $\|(m^{-1}\bH^{k'}_F F'F \bH^k_F)^{-1}\|$ are $O_P(n^{1/q})$. Hence,
\begin{align*}
     &\left\| \left(\frac{\tF^{k'}\Xi_{\ii}\tF^k}{\pi m}\right)^{-1} -  \left(\frac{\bH^{k'}_F F'F \bH^k_F}{m}\right)^{-1} \right\|\\
    =\,&\left\| \left(\frac{\tF^{k'}\Xi_{\ii}\tF^k}{\pi_i m}\right)^{-1}\left(\frac{\tF^{k'}\Xi_{\ii}\tF^k}{\pi_i m} -  \frac{\bH^{k'}_F F'F \bH^k_F}{m}\right)\left(\frac{\bH^{k'}_F F'F \bH^k_F}{m}\right)^{-1} \right\|\\
    =\,& O_P\left(\sqrt{(1/n+1/m)\alpha_n^{-2} } n^{1/q}\right).
\end{align*}
\end{proof}

\begin{lem}\label{lem: order of P_Xi.tF - P_H.F k<r}
Under Assumption \ref{assu: moment cond}, for $k\leq r$, we have
\begin{align*}
    \max_{1\leq i\leq n}\left\|\pi_i \tF^k\left(\tF^{k'}\Xi_{\ii}\tF^k\right)^{-1}\tF^{k'} - P_{F\bH^k_F}\right\|=O_P\left(\sqrt{(1/n+1/m)\alpha_n^{-2} } n^{1/q}\right).
\end{align*}
\end{lem}
\begin{proof}[\textbf{Proof of Lemma \ref{lem: order of P_Xi.tF - P_H.F k<r}}]
Consider
\begin{align*}
    &\pi_i \tF^k\left(\tF^{k'}\Xi_{\ii}\tF^k\right)^{-1}\tF^{k'} - P_{F\bH^k_F}\\
  =\,&m^{-1}\tF^k\left(\frac{\tF^{k'}\Xi_{\ii}\tF^k}{m\pi_i}\right)^{-1}\tF^{k'} -m^{-1}F \bH^k_F\left(\frac{\bH^{k'}_F F'F \bH^k_F}{m}\right)^{-1}\bH^{k'}_F F'
\end{align*}
With the fact $\tF^k=F\bH^k_F +\tDel^k_F$, we can decompose the above difference as $I_i + II_i + III_i$, where
\begin{align*}
I_i    &= m^{-1}F \bH^k_F\left[\left(\frac{\tF^{k'}\Xi_{\ii}\tF^k}{m\pi_i}\right)^{-1} - \left(\frac{\bH^{k'}_F F'F \bH^k_F}{m}\right)^{-1}\right]\bH^{k'}_F F',\\
II_i   &= m^{-1}F \bH^k_F\left(\frac{\tF^{k'}\Xi_{\ii}\tF^k}{m\pi_i}\right)^{-1}\tDel^{k'}_F + m^{-1}\tDel^k_F\left(\frac{\tF^{k'}\Xi_{\ii}\tF^k}{m\pi_i}\right)^{-1}\bH^{k'}_F F', \quad \text{and}\\
III_i  &= m^{-1}\tDel^k_F\left(\frac{\tF^{k'}\Xi_{\ii}\tF^k}{m\pi_i}\right)^{-1}\tDel^{k'}_F.
\end{align*}
By Lemmas \ref{lem: order of H Delta k<r}, \ref{lem: order of Delta_L}, and \ref{lem: order of inv tF.Xi.tF - H.F.F.H k<r}, we have
\begin{align*}
\max_{1\leq i\leq n}\|I_i\|&=O_P\left(\sqrt{(1/n+1/m)\alpha_n^{-2} } n^{1/q}\right),\\
    \max_{1\leq i\leq n}\|II_i\|&=O_P\left(\sqrt{(1/n+1/m)\alpha_n^{-1} } n^{1/q}\right),\\
    \max_{1\leq i\leq n}\|III_i\|&=O_P\left((1/n+1/m)\alpha_n^{-1} n^{1/q}\right).
\end{align*}
The above bounds lead to the desired result.

\end{proof}
\begin{lem}\label{lem: order of mse(FH,k)-mse(F,r) k<r}
    Under Assumption \ref{assu: moment cond}, we have, for $k<r$,
\begin{align*}
  \lim_{n,m\rightarrow \infty}P\left(\frac{1}{nm}\left\{\sumi \tW^{'}_{\ii}\left(P_{\Xi_{\ii}\tF^r}- P_{\Xi_{\ii}\tF^k} \right)\tW_{\ii}\right\} > \alpha_n \tau_k \right) = 1,
\end{align*}
 for some constant $\tau_k>0$.

\end{lem}
\begin{proof}[\textbf{Proof of Lemma \ref{lem: order of mse(FH,k)-mse(F,r) k<r}}]
Let $R_i =  \tF^k\left(\tF^{k'}\Xi_{\ii}\tF^k\right)^{-1}\tF^{k'} - P_{F\bH^k_F}/\pi_i$ and recall that $\diag(\pi^{-1})\tW=LF'+e$.
Then, Lemma \ref{lem: order of e} implies
\begin{align}
\|\diag(\pi^{-1})\tW\|_F^2=\|LF'+e\|_F^2 \lessp mn.
\label{eq:25_1}
\end{align}

Then, we have
\begin{align*}
& \frac{1}{nm }\sumi \tW^{'}_{\ii} P_{F\bH^k_F} \tW_{\ii} \\
& = \frac{1}{nm }\sumi \tW^{'}_{\ii} (P_{F\bH^k_F}/\pi_i + R_i) \tW_{\ii} \\
& = \frac{1}{nm }\sumi (\tW^{'}_{\ii}\pi_i^{-1}) \pi_i (P_{F\bH^k_F} + \pi_i R_i) (\tW_{\ii}\pi_i^{-1}) \\
& \leq \frac{1}{nm }\sumi (\tW^{'}_{\ii}\pi_i^{-1}) \pi_i P_{F\bH^k_F} (\tW_{\ii}\pi_i^{-1}) + \frac{1}{nm } \max_i ||\pi_i^2 R_i|| ||(\tW\diag(\pi^{-1}))||_F^2 \\
& \leq \frac{1}{nm }\sumi (\tW^{'}_{\ii}\pi_i^{-1}) \pi_i P_{F\bH^k_F} (\tW_{\ii}\pi_i^{-1}) + O_P( \alpha_n^2 \sqrt{(1/n+1/m)\alpha_n^{-2} } n^{1/q}),
\end{align*}
where the last inequality is by Lemma \ref{lem: order of P_Xi.tF - P_H.F k<r} and \eqref{eq:25_1}.

Similarly, we have
\begin{align*}
 \frac{1}{nm }\sumi \tW^{'}_{\ii} P_{F\bH^r_F} \tW_{\ii} & \geq \frac{1}{nm }\sumi (\tW^{'}_{\ii}\pi_i^{-1}) \pi_i P_{F\bH^r_F} (\tW_{\ii}\pi_i^{-1}) - O_P( \alpha_n \sqrt{(1/n+1/m)\alpha_n^{-2} } n^{1/q}) \\
 & = \frac{1}{nm }\sumi (\tW^{'}_{\ii}\pi_i^{-1}) \pi_i P_{F} (\tW_{\ii}\pi_i^{-1}) - O_P( \alpha_n^2 \sqrt{(1/n+1/m)\alpha_n^{-2} } n^{1/q}),
\end{align*}
where the equality holds because $\bH_F^r$ is a square matrix.

Next, we have
\begin{align*}
        &\frac{1}{nm }\sumi \tW^{'}_{\ii} \pi_i P_F \tW_{\ii} - \frac{1}{nm }\sumi \tW^{'}_{\ii}  \pi_i P_{F\bH^k_F} \tW_{\ii}\\
        =\,& \frac{1}{nm}\sumi \left(FL_i+e_i\right)' \pi_i\left(P_F - P_{F\bH^k_F}\right) \left(FL_i+e_i\right)\\
        =\,& \frac{1}{nm}\sumi L'_iF' \pi_i \left(P_F - P_{F\bH^k_F}\right) FL_i + 2\frac{1}{nm}\sumi L'_i F' \pi_i \left(P_F - P_{F\bH^k_F}\right) e_i\\
        & \hspace{6cm}+\frac{1}{nm}\sumi e'_i \pi_i \left(P_F - P_{FH^k_F}\right) e_i\\
        =\,& I + 2II + III
    \end{align*}
    Note that, since $k< r$, the column space of $F\bH^k_F$ is a subspace of the column space of $F$, and $P_F - P_{F \bH^k_F}$ is a (p.s.d.) projection matrix. Then it is obvious that $III\geq0$.

For part $II$, we have
    \begin{align*}
        \left|II\right|
        &= \frac{1}{nm}\left|\trace\left\{   F' \left(P_F - P_{F \bH^k_F}\right) \sumi e_i \pi_i L'_i \right\}\right|\\
        &= \frac{1}{nm}\left|\trace\left\{   F' \left(P_F - P_{F \bH^k_F}\right)  e'\diag(\pi) L \right\}\right|\\
        &\leq \frac{r}{nm}\left\|   F' \left(P_F - P_{F \bH^k_F}\right)  e' \diag(\pi)  L \right\|\\
        & \lessp (mn)^{-1} ||F|| ||e|| ||\diag(\pi)|| ||L|| \\
        &=O_P\left((1/n+1/m)^{1/2} \alpha_n^{1/2}n^{1/q}\right) = o_P(\alpha_n).
    \end{align*}
For part $I$, if we write down the explicit form of the two projection matrices, the term can be simplified as
    \begin{align*}
        I&=\frac{1}{nm}\sumi L'_i F' \pi_i \left(F\left(F'F\right)^{-1}F' - F \bH^k_F\left(\bH^{k'}_F F'F \bH^k_F\right)^{-1}\bH^{k'}_F F'\right)FL_i\\
         &=\frac{1}{nm}\trace\left\{ \diag(\pi) L F'\left(F\left(F'F\right)^{-1}F' - F \bH^k_F\left(\bH^{k'}_F F'F \bH^k_F\right)^{-1}\bH^{k'}_F F'\right)FL'\right\}\\
         &=\trace\left\{ \frac{F'F}{m} \frac{L'\diag(\pi)L}{n} - \frac{F'F}{m}\bH^k_F\left(\bH^{k'}_F \frac{F'F}{m} \bH^k_F\right)^{-1}\bH^{k'}_F \frac{F'F}{m} \frac{L'\diag(\pi)L}{n}\right\}\\
         &= \trace\left\{ \left[\Sigma_F - \Sigma_F \bH^k_F\left(\bH^{k'}_F\Sigma_F \bH^k_F\right)^{-1}\bH^{k'}_F\Sigma_F\right] \frac{L'\diag(\pi)L}{n} \right\} \\
& = \trace\left\{ \left[\frac{L'\diag(\pi)L}{n} \right]^{1/2} \Sigma_F^{1/2} \left[I_r - \Sigma_F^{1/2} \bH^k_F\left(\bH^{k'}_F\Sigma_F \bH^k_F\right)^{-1}\bH^{k'}_F\Sigma_F^{1/2}\right]
\Sigma_F^{1/2}\left[\frac{L'\diag(\pi)L}{n} \right]^{1/2}\right\} \\
& =       \trace\left\{ U'
\Sigma_F^{1/2}\left[\frac{L'\diag(\pi)L}{n} \right] \Sigma_F^{1/2} U\right\} \\
& \geq \lambda_{\min}\left( U'
\Sigma_F^{1/2}\left[\frac{L'\diag(\pi)L}{n} \right] \Sigma_F^{1/2} U \right) \\
& \geq \sigma_{r}\left(
\Sigma_F^{1/2}\left[\frac{L'\diag(\pi)L}{n} \right] \Sigma_F^{1/2}\right) \\
& \geq c \alpha_n,
    \end{align*}
for some constant $c>0$, where $U \in \Re^{r \times k}$ such that
\begin{align*}
U'U = I_k \quad \text{and} \quad   UU' = I_r - \Sigma_F^{1/2} \bH^k_F\left(\bH^{k'}_F\Sigma_F \bH^k_F\right)^{-1}\bH^{k'}_F\Sigma_F^{1/2}
\end{align*}
and the last inequality follows Assumption \ref{assu: moment cond}(vi).

    Therefore, we can conclude that $I>\tau_k>0$ for some positive value $\tau_k$. Then the result follows.
\end{proof}

For $k>r$, note that the matrix $D_k$ is asymptotically singular, and thus the $H^k_L$ defined in Lemma \ref{lem: order of H Delta k<r} might behave unpredictably. The following lemma gives another decomposition for $\tL^k$ for $k>r$.

\begin{lem}\label{lem: order of H Delta k>r}
Under Assumption \ref{assu: moment cond}, for $k>r$, we have
\begin{align*}
    \hF^k &= F\bH^k_F + \hat \Delta^k_F
\end{align*}
where $\bH^k_F=n^{-1}L'\hL^k$ and $\hOme_k=(nm)^{-1/2}D_k$. Moreover, the matrices $\bH^k_F$ and $\hat \Delta^k_F$ satisfy
\begin{align*}
\left\|\bH^k_F\right\|=O_P(1) \quad \text{and} \quad  \left\|\hat \Delta^k_F\right\|=O_P\left(\sqrt{\left(1+\frac{m}{n}\right) \alpha_n^{-1}}n^{1/q}\right).
\end{align*}

\end{lem}
\begin{proof}[\textbf{Proof of Lemma \ref{lem: order of H Delta k>r}}]



Let $H_F^k = (n)^{-1}L'\hL^k$. Then, we have
\begin{align*}
\hF^k & = n^{-1} W' \hL^k = n^{-1}(LF' + e)'\hL^K = F (n^{-1}L' \hL^k) + n^{-1} e' \hL^k =  F H^k_F + \hat \Delta^k_F.
\end{align*}

Therefore, we have
\begin{align*}
    &\left\|H^k_F\right\|\leq \left\|\sqrt{n^{-1}}L\right\|\left\|\sqrt{n^{-1}}\hL^k\right\|=O_P(1),
        &\left\|\hat \Delta^k_F\right\|=O_P\left(\sqrt{\left(1+\frac{m}{n}\right) \alpha_n^{-1}}n^{1/q}\right).
\end{align*}

\end{proof}

Recall that $\eIC^*(k)-\eIC^*(r)=\mse(k)-\mse(r)+(r-k)\cdot h^*(n,m)$, we still need to study the difference between $\mse(k)$ and $\mse(r)$ for $k>r$. The following lemma gives a very good bound for any $\mse(k)$.
\begin{lem}\label{lem: order of mse(k)}
Under Assumption \ref{assu: moment cond}, we have the following bounds
\begin{align*}
    u-v_k - \alpha_0 - \alpha_1 v_k^{1/2} < \mse(k) < u-v_k + \alpha_0 + \alpha_1 v_k^{1/2}
\end{align*}
where
\begin{align*}
    &0<u = \frac{1}{nm}\sumi \varepsilon'_i\Xi_{\ii}\varepsilon_{\ii},\\
    &0<v_k=\frac{1}{nm}\sumi \varepsilon'_i  P_{\Xi_{\ii}\tF^k} \varepsilon_{\ii},\\
    &0<\alpha_0=O_P\left( (1/m+1/n) n^{1/q}\right), \\
    &0<\alpha_1=O_P\left(\sqrt{ (1/m+1/n) n^{1/q}}\right) .
\end{align*}
\end{lem}

\begin{proof}[\textbf{Proof of Lemma \ref{lem: order of mse(k)}}]
Note $\bH_F^k \in \Re^{r\times k}$ for $r<k$. This implies there exists a pseudo inverse $\bH_F^{k+} \in \Re^{k \times r}$ such that $\bH_F^k \bH_F^{k+}  = I_r$. In addition, we note that
\begin{align*}
    \| \bH_F^{k+} \| = \| [n^{-1} L'\hL^k]^+\| = \left\| \left[(H')^{-1}n^{-1}(\hL^r - \hDel_L^r)'\hL^k\right]^+ \right\| \leq \| H\|^{-1} \|[(I_r,0) -  n^{-1}\hat \Delta_L^{r'} \hL^k]^+ \| \lessp 1,
\end{align*}
where the last inequality is by the fact that the $r$th singular value of $(I_r,0) -  n^{-1}\hat \Delta_L^{r'} \hL^k$ is bounded away from zero.

By definition, $\tW_{\ii}$ has the following form
\begin{align*}
    \tW_{\ii} &= \Xi_{\ii}\left(FL_i -\left(\hbeta-\beta\right) X_i+ \varepsilon_{\ii}\right) \\
        &= \Xi_{\ii}\left(F\bH^k_F \bH^{k+}_F L_i - \left(\hbeta-\beta\right) X_i+ \varepsilon_{\ii}\right) \\
        &= \Xi_{\ii}\left(\tF^k - \tDel_F^k \right) \bH^{k+}_F L_i - \Xi_{\ii}\left(\hbeta-\beta\right) X_i+ \Xi_{\ii}\varepsilon_{\ii} \\
        &= \Xi_{\ii}\tF^k \bH^{k+}_F L_i - \Xi_{\ii}\tDel_F^k \bH^{k+}_F L_i - \Xi_{\ii}\left(\hbeta-\beta\right) X_i+ \Xi_{\ii}\varepsilon_{\ii}\\
        &= \Xi_{\ii}\tF^k \bH^{k+}_F L_i -a_i - b_i +  \Xi_{\ii}\varepsilon_{\ii}
\end{align*}
and thus we can write the mean-square-error as
\begin{align}
    \mse(k)
    =\,& \frac{1}{nm}\sumi \left(-a_i - b_i +  \Xi_{\ii}\varepsilon_{\ii}\right)' M_{\Xi_{\ii}\tF^k} \left(-a_i - b_i +  \Xi_{\ii}\varepsilon_{\ii}\right)\nonumber\\
    =\,& \frac{1}{nm}\sumi a'_i M_{\Xi_{\ii}\tF^k} a_i + \frac{1}{nm}\sumi b'_i M_{\Xi_{\ii}\tF^k} b_i + \frac{1}{nm}\sumi \varepsilon'_i\Xi_{\ii} M_{\Xi_{\ii}\tF^k} \Xi_{\ii}\varepsilon_{\ii} + \nonumber\\
     & \frac{2}{nm}\sumi  a'_i M_{\Xi_{\ii}\tF^k} b'_i - \frac{2}{nm}\sumi a'_i M_{\Xi_{\ii}\tF^k} \Xi_{\ii}\varepsilon_{\ii} - \frac{2}{nm}\sumi b'_i M_{\Xi_{\ii}\tF^k} \Xi_{\ii}\varepsilon_{\ii} \label{eq: mse(k) in 6 terms}
\end{align}
Now we determine the order of each term above. Since $M_{\Xi_{\ii}\tF^k}$ is a projection matrix, we have
\begin{align*}
    0<\frac{1}{nm}\sumi a'_i M_{\Xi_{\ii}\tF^k} a_i
    &< \frac{1}{nm}\sum_i\|a_i\|_2^2\\
   & = \frac{1}{nm}\sum_i \sum_j \left\{\sum_{l=1}^r\xi_{i,j} [\tDel_F^k \bH^{k+}_F]_{j,l}L_{i,l}\right\}^2 \\
   & \leq  \frac{k}{nm} \sum_i \sum_j \sum_{l=1}^r \xi_{i,j} [\tDel_F^k \bH^{k+}_F]_{j,l}^2 L_{i,l}^2 \\
   & \leq \left( \max_{j,l} \sum_i \xi_{i,j} L_{i,l}^2\right)\frac{r}{nm} \sum_j \sum_{l=1}^r[\tDel_F^k \bH^{k+}_F]_{j,l}^2 \\
   & = \left( \max_{j,l} \sum_i \xi_{i,j} L_{i,l}^2\right)\frac{r}{nm} ||\tDel_F^k \bH^{k+}_F ||_F^2 \\
   & \leq \left( \max_{j,l} \sum_i \xi_{i,j} L_{i,l}^2\right)\frac{r^2}{nm} ||\tDel_F^k \bH^{k+}_F ||^2 \\
   & \lessp n\alpha_n \frac{r^2}{nm} (1+m/n)\alpha_n^{-1}n^{1/q} \\
&\lessp (1/m+1/n) n^{1/q},
\end{align*}
where the second last line follows Lemma \ref{lem: order of H Delta k>r}.

By a similar argument, we have
\begin{align*}
    0<\frac{1}{nm}\sumi b'_i M_{\Xi_{\ii}\tF^k} b_i
    &< \frac{1}{nm}\sumi \|b_i\|_2^2\\
    & \leq \left( \max_{j,l} \sum_i \xi_{i,j} X_{i,l}^2\right)\frac{d^2}{nm} ||\hat \beta - \beta ||^2 \\
& \lessp (1/m+1/n) n^{1/q}.
\end{align*}
We then write the third term as
\begin{align*}
    \frac{1}{nm}\sumi \varepsilon'_i\Xi_{\ii} M_{\Xi_{\ii}\tF^k} \Xi_{\ii}\varepsilon_{\ii}
    &= \frac{1}{nm}\sumi \varepsilon'_i\Xi_{\ii} \varepsilon_{\ii} - \frac{1}{nm}\sumi \varepsilon'_i P_{\Xi_{\ii}\tF_k} \varepsilon_{\ii}\\
    &=u-v_k
\end{align*}
For the cross term with $a_i$, $b_i$, we can use Cauchy-Schwartz inequality and get
\begin{align*}
    \left|\frac{2}{nm}\sumi  a'_i M_{\Xi_{\ii}\tF^k} b_i\right|
    &\leq \frac{2}{nm} \sumi  \|a_i\| \|M_{\Xi_{\ii}\tF^k} b_i\|\\
    &\leq  \left(\frac{2}{nm}\sumi \|a_i\|^2\right)^{1/2} \left(\frac{2}{nm}\sumi \|M_{\Xi_{\ii}\tF^k} b_i\|^2\right)^{1/2}\\
    & \lessp (1/m+1/n) n^{1/q}.
\end{align*}
The last equation is from the previous results. For the other two cross terms, we use the fact $M_{\Xi_{\ii}\tF^k}=I - P_{\Xi_{\ii}\tF^k}$ to split the term, and then either bound it directly or use Cauchy-Schwartz inequality to find the relation. The details are as follows.
\begin{align*}
    \frac{2}{nm}\sumi a'_i M_{\Xi_{\ii}\tF^k}\Xi_{\ii}\varepsilon_{\ii}
    &= \frac{2}{nm}\sumi a'_i \Xi_{\ii}\varepsilon_{\ii} - \frac{2}{nm}\sumi a'_i P_{\Xi_{\ii}\tF^k}\Xi_{\ii}\varepsilon_{\ii}
\end{align*}
For each term above,
\begin{align*}
    \left|\frac{2}{nm}\sumi a'_i \Xi_{\ii}\varepsilon_{\ii} \right|
    &= \frac{2}{nm}\left|\trace\left\{\sumi L'_i \left(H^{k+}_F\right)' \left(\tDel_F^k\right)' \Xi_{\ii}\varepsilon_{\ii}\right\}\right|\\
    &= \frac{2}{nm}\left|\trace\left\{ \left(H^{k+}_F\right)' \left(\tDel_F^k\right)' \sumi \Xi_{\ii}\varepsilon_{\ii} L'_i\right\}\right|\\
    &\leq \frac{2r}{nm} \left\|H^{k+}_F\right\| \left\|\tDel_F^k\right\| \left\|\sumi \Xi_{\ii}\varepsilon_{\ii} L'_i\right\|\\
    &\leq \frac{2r}{nm} \left\|H^{k+}_F\right\| \left\|\tDel_F^k\right\| \left\|\sumi \Xi_{\ii}\varepsilon_{\ii} L'_i\right\|_F\\
    &\leq \frac{2r}{nm} \left\|H^{k+}_F\right\| \left\|\tDel_F^k\right\| \left(\sumj\left\|\sumi \xi_{i,j}\varepsilon_{i,j} L'_i\right\|_2^2\right)^{1/2}\\
    &=O_P\left(\frac{1}{nm}\cdot 1\cdot \sqrt{(1+m/n)\alpha_n^{-1}} n^{1/q} \cdot\sqrt{m\cdot n \alpha_n \log m}\right)\\
    &=O_P\left((1/m+1/n) n^{1/q} \right),
\end{align*}
\begin{align*}
    \left|\frac{2}{nm}\sumi a'_i P_{\Xi_{\ii}\tF^k}\Xi_{\ii}\varepsilon_{\ii}\right|
    &\leq \left(\frac{4}{nm}\sumi \|a'_i\|_2^2\right)^{1/2} \left(\frac{1}{nm}\sumi \varepsilon'_i P_{\Xi_{\ii}\tF^k}\varepsilon_{\ii}\right)^{1/2}\\
    &=O_P\left(\sqrt{ (1/m+1/n) n^{1/q}}\right) v_k^{1/2}.
\end{align*}
Also, for the last cross term in \eqref{eq: mse(k) in 6 terms}, we have
\begin{align*}
    \frac{2}{nm}\sumi b'_i M_{\Xi_{\ii}\tF^k}\Xi_{\ii}\varepsilon_{\ii}
    &= \frac{2}{nm}\sumi b'_i \Xi_{\ii}\varepsilon_{\ii} - \frac{2}{nm\hpi}\sumi b'_i P_{\Xi_{\ii}\tF^k}\Xi_{\ii}\varepsilon_{\ii}
\end{align*}
and each has the bound
\begin{align*}
    \left|\frac{2}{nm}\sumi b'_i \Xi_{\ii}\varepsilon_{\ii} \right|
       &\leq \frac{2p}{nm} \left\|\hat \delta\right\|_F \left(\sumj\left\|\sumi \xi_{i,j}\varepsilon_{i,j} X'_i\right\|_2^2\right)^{1/2}\\
    &=O_P\left(\frac{1}{nm}\cdot \sqrt{\frac{m}{n}}\cdot\sqrt{m\cdot n\log m}\right)\\
    &=O_P\left(\frac{\sqrt{\log m}}{n}\right)
\end{align*}
\begin{align*}
    \left|\frac{2}{nm}\sumi b'_i P_{\Xi_{\ii}\tF^k}\Xi_{\ii}\varepsilon_{\ii}\right|
    &\leq \left(\frac{4}{nm}\sumi \|b'_i\|_2^2\right)^{1/2} \left(\frac{1}{nm}\sumi \varepsilon'_i P_{\Xi_{\ii}\tF^k}\varepsilon_{\ii}\right)^{1/2}\\
    &=O_P\left(\sqrt{ (1/m+1/n) n^{1/q}}\right)  v_k^{1/2}.
\end{align*}
Then we have the result
\begin{align*}
    u-v_k - \alpha_0 - \alpha_1 v_k^{1/2} < \mse(k) < u-v_k + \alpha_0 + \alpha_1 v_k^{1/2}
\end{align*}

for some $0<\alpha _{0}=O_{P}\left( (1/m+1/n)n^{1/q}\right) $, $0<\alpha _{1}=O_{P}\left( \sqrt{(1/m+1/n)n^{1/q}}\right) $.

\end{proof}

Note that the value $u$ does not depend on $k$, so the above lemma implies that the upper and lower bounds for $\mse(k)-\mse(r)$ do not involve the value $u$. As a result, the next step is to find the bound for $v_k$. The difficulty here is from the projection matrix $P_{\Xi_{\ii}\tF^k}$. This matrix varies from different $i$, so it is not possible to get a good bound by the same techniques used in Lemma 4,  \cite{bai2002determining}, or in Lemma C.1, \cite{li2016interact}. Thus, we need a mediator to help us get the desired bounds. Therefore, we would like to introduce the solution proposed by \cite{su2019factor}. In the article, they implement iterative SVD with rank $\br$ by replacing the missing values in $\tW$ with the corresponding estimated values from the previous step. Specifically, suppose
\begin{align}\label{eq:LFdot}
(\dL^{\br}, \dF^{\br}) = \argmin_{L,F} \left\|\Xi \circ (Y - X\hat \beta' - LF') \right\|_F^2,
\end{align}
where $(L,F)$ are chosen so that $L \in \Re^{n \times \br}$, $F \in \Re^{m \times \br}$ and $L'L = I_{\br}$.

\begin{rmk}
The solution ($\dL^{\br}, \dF^{\br}$) following \cite{su2019factor} is only necessary for the proof as a mediator. We do not need to find them when applying the proposed method in this article. The nature of the SVD solution is better for theoretical work. However, as we mentioned in the main article, one needs to implement SVD multiple times until converges to get this solution, and hence it is computationally expensive.
\end{rmk}

Let $\dW$ be the matrix by replacing the missing values in $\tW$ with the corresponding values in $\dL^{\br}\dF^{\br'}$.
\begin{align*}
\dW=\tW + (1_n 1'_m - \Xi)\circ \dL^{\br}\dF^{\br'}
\end{align*}
and denote its SVD (and reduced rank SVD) as
\begin{align*}
\dW = \dU\dD\dV'= \dU_{\br}\dD_{\br}\dV_{\br}' + res
\end{align*}
where $\dU_{\br}\dD_{\br}\dV_{\br}'$ includes the $\br$ largest singular values. Since $\dF^{\br}, \dL^{\br}$ are the converged estimators, so we have $\dU_{\br}\dD_{\br}\dV_{\br}'=\dL^{\br}\dF^{\br'}$.

\begin{lem}\label{lem: order of mse*(k)}
The $\mse$ of the final solution $\dL^{\br}\dF^{\br'}$ is defined and can be written as follows.
\begin{align*}
    \mse^{*}
    &=\frac{1}{nm}\sumi \|\dW_{\ii} - \dF^{\br}\dL^{\br}_i\|_2^2 \left(=\frac{1}{nm}\sumi \|\tW_{\ii} - \Xi_{\ii}\dF^{\br}\dL^{\br}_i\|_2^2\right)\\
    &= \frac{1}{nm}\sumi \dW'_{\ii}(I-P_{\dF^{\br}})\dW_{\ii}= \frac{1}{nm}\sumi \dW'_{\ii} M_{\dF^{\br}} \dW_{\ii}
\end{align*}
and we have
\begin{align*}
    \mse^{*} = u + \alpha_2
\end{align*}
where $u=\frac{1}{nm}\sumi \varepsilon'_i \Xi_{\ii}\varepsilon_{\ii}$ is the same as in $\mse(k)$, and $\alpha_2$ has the bound
\begin{align*}
\alpha_2=O_P\left((1/n+1/m)^{1/2}\alpha_n^{1/2} n^{1/q}\right).
\end{align*}
\end{lem}
\begin{proof}[\textbf{Proof of Lemma \ref{lem: order of mse*(k)}}]
We observe that,
\begin{align*}
    \dW_{\ii} &=\tW_{\ii} + (I - \Xi_{\ii})\dF^{\br}\dL^{\br}_i\\
        &= \Xi_{\ii}\left(FL_i -\left(\hbeta-\beta\right) X_i+ \varepsilon_{\ii}\right) + \left(I - \Xi_{\ii}\right)\dF^{\br}\dL^{\br}_i\\
        &= \Xi_{\ii}\left(FL_i - \dF^{\br}\dL^{\br}_i -\left(\hbeta-\beta\right) X_i+ \varepsilon_{\ii}\right) + \dF^{\br}\dL^{\br}_i\\
        &= \dF^{\br}\dL^{\br}_i + \Xi_{\ii}\varepsilon_{\ii} + \Xi_{\ii}\left(FL_i -     \dF^{\br}\dL^{\br}_i\right) - \Xi_{\ii}\left(\hbeta-\beta\right) X_i \\
        &= \dF^{\br}\dL^{\br}_i + \Xi_{\ii}\varepsilon_{\ii} + a_i - b_i.
\end{align*}
Then we have
\begin{align*}
\mse^{*} = \left(\Xi_{\ii}\varepsilon_{\ii} + a_i - b_i\right)'M_{\dF^{\br}}\left(\Xi_{\ii}\varepsilon_{\ii} + a_i - b_i\right)
\end{align*}
The proof is similar to the one for Lemma \ref{lem: order of mse(k)}. The only difference is the terms involving $a_i$. Note that
\begin{align} \label{eq:a^2}
    \frac{1}{nm}\sumi \left\|a_i\right\|^2_2
    & = \frac{1}{nm}\left\|\Xi\circ\left(FL' - \dF^{\br}\dL^{k'}\right)\right\|_F^2 = O_P\left((1/m+1/n) n^{1/q}\right).
\end{align}

The second equality follows the same argument as in the proof of \citet[Theorems 3.1. and 3.2]{su2019factor}. In particular, their Lemma B.5 is still applicable in our setting because (1) $\{\xi_{i,j}\}_{1\leq i \leq n, 1\leq j \leq m}$ are independent conditional on $X$ and (2) the results in Lemma B.5 still holds if the matrix entries are independent but not identically distributed.

This gives the bounds for $\sumi b'_i M_{\dF^{\br}}a_i$ and $\sumi a'_i M_{\dF^{\br}}a_i$ in $\mse^*$. We also have
\begin{align*}
    \sumi b'_i M_{\dF^{\br}}b_i \lessp (1/m+1/n)n^{1/q}.
\end{align*}

On the other hand, since $P_{\dF^{\br}}$ does not depend on $i$, we can simply write
\begin{align*}
    \left|(nm)^{-1}\sum_i \varepsilon'_{\ii}\Xi_{\ii}P_{\dF^{\br}}a_i\right|
    & = (nm)^{-1}\left|\trace\left\{P_{\dF^{\br}}\sum_i a_i\varepsilon'_{\ii}\Xi_{\ii}\right\}\right|\\
    & \overset{\texti}{\leq} k(nm)^{-1}\left\|P_{\dF^{\br}}\right\| \left\|A(\Xi\circ\varepsilon)\right\|\\
    & \leq k(nm)^{-1}\left\|A(\Xi\circ\varepsilon )\right\|\\
    & \leq k(nm)^{-1} \left\|\Xi\circ\varepsilon\right\|\left\|A\right\|_F\\
    & \overset{\textii}{=} O_P\left( (mn)^{-1/2}  \sqrt{(m+n)\alpha_n n^{1/q}} \sqrt{(1/m+1/n) n^{1/q}}\right) \\ & =O_P\left( (1/m+1/n) \alpha_n^{1/2} n^{1/q}\right),
\end{align*}
where $A \in \Re^{m \times n}$ is the matrix $(a_1, a_2, \cdots, a_n)$. The inequality $\texti$ comes from the fact that $\rank(P_{\dF^{\br}})=k$, and equation $\textii$ are again from \eqref{eq:a^2} and Lemma \ref{lem: part 2}.



Next, we focus on the term $(nm)^{-1}\sum_i \varepsilon'_{\ii}\Xi_{\ii}a_i$. Since $\Xi^2_{\ii}=\Xi_{\ii}$, we have
\begin{align*}
    (nm)^{-1}\left|\sum_i \varepsilon'_{\ii}\Xi_{\ii}a_i \right|
    &= (nm)^{-1}\left|\trace\left\{\sum_i a_i\varepsilon'_{\ii}\Xi_{\ii}\right\}\right|\\
    &= (nm)^{-1}\left|\trace\left\{A \left(\Xi\circ\varepsilon\right)\right\}\right|\\
    &\lessp (nm)^{-1}\left\|A\right\|_F \left\|\Xi\circ\varepsilon\right\|_F\\
    &\leq (nm)^{-1/2} (1/m+1/n)^{1/2} n^{1/q} (m\wedge n)^{1/2}\left\|\Xi\circ\varepsilon\right\| \\
    &\lessp (1/n+1/m)^{1/2}\alpha_n^{1/2} n^{1/q}.
\end{align*}

Last but not least, for the term $(nm)^{-1}\sumi \varepsilon'_{\ii}\Xi_{\ii}P_{\dF^{\br}}\Xi_{\ii}\varepsilon_{\ii}$ we have
\begin{align*}
    (nm)^{-1}\left|\sumi \varepsilon'_{\ii}\Xi_{\ii}P_{\dF^{\br}}\Xi_{\ii}\varepsilon_{\ii}\right|
    &= (nm)^{-1}\left|\trace\left\{\sumi \varepsilon'_{\ii}\Xi_{\ii}P_{\dF^{\br}}\Xi_{\ii}\varepsilon_{\ii}\right\}\right|\\
    &= (nm)^{-1}\left|\trace\left\{ P_{\dF^{\br}}\sumi \Xi_{\ii}\varepsilon_{\ii}\varepsilon'_{\ii}\Xi_{\ii}\right\}\right|\\
    &\leq (nm)^{-1}\left\| P_{\dF^{\br}}\right\|_*\left\|(\Xi\circ\varepsilon)'(\Xi\circ\varepsilon)\right\|\\
    &= O_P\left( (1/m+1/n) \alpha_n n^{1/q}\right)
\end{align*}
because of the fact $\| P_{\dF^{\br}}\|_*=\rank (P_{\dF^{\br}})=\br$ and Lemma \ref{lem: part 2}. Then the result follows.
\end{proof}

Now we can use the bound for $\mse^*$ to find the bound for $v_k$.
\begin{lem}[Bound for $v_k$ and $\mse(k)$ for $k\geq r$]\label{lem: order of v_k mse(k)}
Let $v_k$ be the value defined in Lemma \ref{lem: order of mse(k)}. Under Assumption \ref{assu: moment cond}, for $k\geq r$, we have
\begin{align*}
v_k=O_P\left((1/n+1/m)^{1/2}\alpha_n^{1/2} n^{1/q}\right).
\end{align*}
In particular, the $\mse(k)$ has the form
\begin{align}
    \mse(k) = \frac{1}{nm}\sumi \varepsilon'_{\ii}\Xi_{\ii}\varepsilon_{\ii} + O_P\left((1/n+1/m)^{1/2}\alpha_n^{1/2} n^{1/q}\right). \label{eq: order of mse(k) k>r complete}
\end{align}
\end{lem}
\begin{proof}[\textbf{Proof of Lemma \ref{lem: order of v_k mse(k)}}]
First, by the definitions of $\dL^{\br}$ and $\dF^{\br}$ with $k \leq \br$, we have
\begin{align*}
 \mse^*<\mse(k).
\end{align*}
Then by Lemmas \ref{lem: order of mse(k)} and \ref{lem: order of mse*(k)},
\begin{align*}
    &u+\alpha_2<u-v_k + \alpha_0 + \alpha_1v^{1/2}_k\\
    &\left(v^{1/2}_k\right)^2 -\alpha_1 v^{1/2}_k +\alpha_2 - \alpha_0 <0
\end{align*}
for some $\alpha_0=O_P\left( (1/m+1/n)n^{1/q}\right)$, $0<\alpha_1=O_P\left(\sqrt{ (1/m+1/n) n^{1/q}}\right)$, and $\alpha_2=O_P\left((1/n+1/m)^{1/2}\alpha_n^{1/2} n^{1/q}\right)$. Then we can solve the quadratic form and get
\begin{align*}
    &0<v^{1/2}_k\leq \frac{|\alpha_1| + \sqrt{\alpha^2_1+4|\alpha_2|}}{2} = O_P\left(\sqrt{(1/m+1/n) n^{1/q} + (1/n+1/m)^{1/2}\alpha_n^{1/2} n^{1/q}}\right)\\
    &0<v_k = O_P\left((1/n+1/m)^{1/2}\alpha_n^{1/2} n^{1/q}\right)
\end{align*}
and  \eqref{eq: order of mse(k) k>r complete} follows the bounds given in Lemma \ref{lem: order of mse(k)}.
\end{proof}
\begin{proof}[\textbf{Proof of Theorem \ref{thm: r est eIC} at $g=1$}]

Recall that
\begin{align*}
    \eIC(k) - \eIC(r) = \log\left(\frac{\mse(k)}{\mse(r)}\right) + (k-r)h(n,m).
\end{align*}

{\bf Step 1:} show $P(\eIC(k)- \eIC(r)>0)\overset{P}{\rightarrow}1$ for $k<r$.\\
Note that
\begin{align*}
\mse(k)-\mse(r)= \frac{1}{nm}\sumi \tW^{'}_{\ii}\left(M_{\Xi_{\ii}\tF^k} - M_{\Xi_{\ii}\tF^r}\right)\tW_{\ii} = \frac{1}{nm}\left\{\sumi \tW^{'}_{\ii}\left(P_{\Xi_{\ii}\tF^r}- P_{\Xi_{\ii}\tF^k} \right)\tW_{\ii}\right\}
\end{align*}

By Lemma \ref{lem: order of mse(FH,k)-mse(F,r) k<r}, there exist a positive value $\tau_k$ such that, for $k<r$,
\begin{align*}
P\left(\mse(k) - \mse(r)> \alpha_n \tau_k\right)\rightarrow 1
\end{align*}
Also note that, for any $k$,
\begin{align*}
   0< \mse(k) = \frac{1}{nm}\sumi \tW^{'}_{\ii} M_{\Xi_{\ii}\tF^k}\tW_{\ii}\leq \frac{1}{nm}\| \tW\|_F^2=O_P(\alpha_n).
\end{align*}
Thus, there exist a constant $\iota_k>0$, such that
\begin{align*}
P\left(\frac{\mse(k)}{\mse(r)}>1+\iota_k \right)\rightarrow 1
\end{align*}
and that gives
\begin{align*}
P\left(\log\left(\frac{\mse(k)}{\mse(r)}\right)>\zeta_k \right)\rightarrow 1
\end{align*}
for some $\zeta_k>0$. As a result,
\begin{align*}
    P\left(\eIC(k)-\eIC(r)>0\right)\rightarrow 1
\end{align*}
if $h(n,m)\rightarrow 0$ as $n\wedge m\rightarrow \infty$.\\[6pt]

{\bf Step 2:} show that $P(\eIC(k) - \eIC(r)>0) \rightarrow 1$ for $k>r$.\\
By Lemma \ref{lem: order of v_k mse(k)}, we know that, for $k\geq r$
\begin{align*}
    \mse(k)=\frac{1}{nm}\sumi \sumj \varepsilon^2_{i,j}\xi_{i,j} + O_P\left((1/n+1/m)^{1/2}\alpha_n^{1/2} n^{1/q}\right)
\end{align*}
and note that $\frac{1}{nm}\sumi \sumj \varepsilon^2_{i,j}\xi_{i,j}=O_P(\alpha_n)$. Hence, we have
\begin{align*}
    \frac{\mse(k)}{\mse(r)} - 1 &=O_P\left((1/n+1/m)^{1/2}\alpha_n^{-1/2} n^{1/q}\right),\\
    \log\left(\frac{\mse(k)}{\mse(r)}\right)&=O_P\left((1/n+1/m)^{1/2}\alpha_n^{-1/2} n^{1/q}\right),
\end{align*}
and thus,
\begin{align*}
&    P\left(\eIC(k)-\eIC(r)\leq 0\right) \\
    &= P\left((k-r)h(n,m) +O_P\left((1/n+1/m)^{1/2}\alpha_n^{-1/2} n^{1/q}\right)\leq 0\right).
\end{align*}
Therefore, $P\left(\eIC(k)-\eIC(r)\leq 0\right)\overset{P}{\rightarrow}0\ $ if $\sqrt{\frac{mn \alpha_n}{(m+n) }} h(n,m)\rightarrow \infty$ in a polynomial rate in $n$ as $n\wedge m\rightarrow \infty$.

\end{proof}


Now, if we consider $g>1$ (and thus we bring back the notation $\tL^{k,(g)}$ and $\tF^{k,(g)}$), the key is to prove the analogues of Lemma \ref{lem: order of tDelta_F k<r}. We will show it by induction.

\begin{lem}\label{lem: order of tDelta_F k<r g>1}
Under Assumption \ref{assu: moment cond}, for $k\leq r$, the estimator $\tF^{k,(g)}\in \RR^{m\times k}$ satisfies
\begin{align*}
    \tF^{k,(g)} = F \bH_F^{k,(g)} + \tDel^{k,(g)}_F \qquad \mbox{for all $g\geq 1$ }
\end{align*}
where $\|\bH_F^{k,(g)}\| = O_P(n^{1/q})$, $\sigma_k(\bH_F^{k,(g)'} \bH_F^{k,(g)}) \geq c>0$, and $\tDel^{k,(g)}_F = \tF^{k,(g)} - F \bH^k_F$ satisfies
\begin{align}
    \left\|\tDel^{k,(g)}_F\right\| = O_P\left(\sqrt{(1 + m/n)\alpha_n^{-1}} n^{1/q}\right). \label{eq: order of tDel_F k<r g>1}
\end{align}
\end{lem}
\begin{proof}[\textbf{Proof of Lemma \ref{lem: order of tDelta_F k<r g>1}}]\ \\
Note that the condition is satisfied when $g=1$ by Lemma \ref{lem: order of tDelta_F k<r} and
$$\bH_F^{k,(1)} = \left( \frac{1}{n}\sumi L_i L_i' H_L^k  \pi_i \right)\left(H_L^{k'} \frac{1}{n} \sumi L_iL_i' \pi_i H_L^k\right)^{-1}.$$
Then, we have
\begin{align*}
\sigma_k(\bH_F^{k,(1)'}\bH_F^{k,(1)}) & \gtrsim_P  \alpha_n^{-2}  \sigma_k\left(  \left( \frac{1}{n}\sumi L_i L_i' H_L^k  \pi_i \right)' \left( \frac{1}{n}\sumi L_i L_i' H_L^k  \pi_i \right) \right) \\
& \gtrsim_P  n^{-1/q}  \sigma_k\left(  \left( \frac{1}{n}\sumi L_i L_i' H_L^k  \right)' \left( \frac{1}{n}\sumi L_i L_i' H_L^k  \right) \right) \\
& \gtrsim \alpha_n^{-2}  \sigma_k\left(  \left( \frac{1}{n}\sumi \hat L_i^r \hat L_i^{k'} \right)' H^{-2} \left( \frac{1}{n}\sumi \hat L_i^r \hat L_i^{k'}  \right) - O_P((1/n+1/m)\alpha_n^{-1} n^{1/q}) \right) \\
& \gtrsim n^{-1/q},
\end{align*}
where the first inequality is by the fact that
\begin{align*}
\left    (\frac{1}{n}\sumi L_i L_i' \pi_i\right)^2 \geq \alpha_n^2 n^{-1/q} \left    (\frac{1}{n}\sumi L_i L_i' \right )^2.
\end{align*}
We now assume it is also satisfied by $\tF^{k,(g)}$ and $\bH_F^{k,(g)}$.


Let $\bH_L^{k,(g)} = \left( \frac{F'F \bH^{k,(g)}_F}{m} \right) \left( \frac{\bH^{k,(g)'}_F F'F \bH^{k,(g)}_F}{m} \right)^{-1}$. Then, we have
\begin{align*}
    \tL^{k,(g)}_i - \bH_L^{k,(g)'} L_i  & = \left(\frac{1}{m\pi_i}\tF^{k,(g)'}\Xi_{\ii}\tF^{k,(g)}\right)^{-1} \frac{1}{m\pi_i}\tF^{k,(g)'}\Xi_{\ii}\tW_{\ii} - \bH_L^{k',(g)} L_i \\
    & = \left(\frac{1}{m\pi_i}\tF^{k,(g)'}\Xi_{\ii}\tF^{k,(g)}\right)^{-1}\frac{1}{m\pi_i}\sumj \tF^{k,(g)}_j \xi_{i,j}(\eps_{i,j} - X_i'\hat \delta_j) \\
    &+ \left[\left(\frac{1}{m\pi_i}\tF^{k,(g)'}\Xi_{\ii}\tF^{k,(g)}\right)^{-1}\frac{1}{m\pi_i}\sumj (\tF^{k,(g)}_j \xi_{i,j}F_j')- \bH_L^{k,(g)'} \right] L_i.
\end{align*}

It can be shown that
\begin{align*}
\max_{j'} \sumi \xi_{i,j'} \left\| \left(\frac{1}{m\pi_i}\tF^{k,(g)'}\Xi_{\ii}\tF^{k,(g)}\right)^{-1}\frac{1}{m\pi_i}\sumj \tF^{k,(g)}_j \xi_{i,j}(\eps_{i,j} - X_i'\hat \delta_j)\right\|_2^2 \lessp (1 + n/m) n^{1/q}.
\end{align*}

In addition, similar to Lemma \ref{lem: order of inv tF.Xi.tF - H.F.F.H k<r}, we have
\begin{align*}
    \max_{1\leq i\leq n}\left\| \pi_i\left(\frac{\tF^{k',(g)}\Xi_{\ii}\tF^{k,(g)}}{m}\right)^{-1} - \left( \frac{\bH^{k,(g)'}_F F'F \bH^{k,(g)}_F}{m} \right)^{-1} \right\|=O_P\left(\sqrt{(1/n+1/m)\alpha_n^{-2} } n^{1/q}\right).
\end{align*}
We can also show
\begin{align*}
\left\|\frac{1}{m\pi_i}\sumj (\tF^{k,(g)}_j \xi_{i,j}F_j') - \left( \frac{\bH^{k,(g)'}_F F'F }{m} \right)\right\|=O_P\left(\sqrt{(1/n+1/m)\alpha_n^{-2} } n^{1/q}\right)
\end{align*}
and
\begin{align*}
    \left\| \left( \frac{\bH^{k,(g)'}_F F'F \bH^{k,(g)}_F}{m} \right)^{-1} \right\|=O_P\left(n^{1/q}\right).
\end{align*}

Then, we have
\begin{align*}
\max_{j'} \sumi \xi_{i,j'}\left\| \left[\left(\frac{1}{m\pi_i}\tF^{k,(g)'}\Xi_{\ii}\tF^{k,(g)}\right)^{-1}\frac{1}{m\pi_i}\sumj (\tF^{k,(g)}_j \xi_{i,j}F_j')- \bH_L^{k,(g)'} \right] L_i\right\|_2^2 \lessp (1 + n/m)\alpha_n^{-1} n^{1/q}.
\end{align*}

Therefore, we have
\begin{align*}
    \tL^{k,(g)} = L\bH^{k,(g)}_L + \tDel^{k,(g)}_L
\end{align*}
such that
\begin{align*}
    \left\|\max_{j'} \sumi |\tDel^{k,(g)}_{L,i}|_2^2 \xi_{i,j'}\right\| = O_P\left((1 + n/m)\alpha_n^{-1} n^{1/q}\right).
\end{align*}

We follow the proof of Lemma \ref{lem: order of tDelta_F k<r} and define
\begin{align*}
\bH^{k,(g+1)}_F =  \left( \frac{1}{n}\sumi L_i L_i' \bH_L^{k,(g)}  \pi_i \right)\left(\bH_L^{k,(g)'} \frac{1}{n} \sumi L_iL_i' \pi_i \bH_L^{k,(g)}\right)^{-1}.
\end{align*}

Then, we have
\begin{align*}
   &  \tF^{k,(g+1)}_j - \bH^{k,(g+1)'}_F F_j \\
   & =\left(\sumi \tL_{i}^{k,(g)}\tL_{i}^{k,(g)'} \xi_{i,j}\right)^{-1}\left(\sumi \tL_{i}^{k,(g)}(L_{i}'F_{j}+\hat u_{i,j}) \xi_{i,j}-\tL_{i}^{k,(g)}\tL_{i}^{k,(g)'} \xi_{i,j}\bH^{k,(g+1)'}_FF_{j}\right)\nonumber \\
 & =\left(\sumi \tL_{i}^{k,(g)}\tL_{i}^{k,(g)'} \xi_{i,j}\right)^{-1}\left(\sumi \tL_{i}^{k,(g)}\hat u_{i,j}  \xi_{i,j}\right)\nonumber\\
 & +\left(\sumi \tL_{i}^{k,(g)}\tL_{i}^{k,(g)'} \xi_{i,j}\right)^{-1}\left(\sumi \tL_{i}^{k,(g)} L_{i}'F_{j}  \xi_{i,j}-\tL_{i}^{k,(g)}\tL_{i}^{k,(g)'}\xi_{i,j}\bH^{k,(g+1)'}_FF_{j}\right).
\end{align*}

In addition, we have
\begin{align*}
 & \max_{1\leq j\leq m}\left\|\left(n^{-1}\sumi \tL^{k,(g)}_iL_i'\xi_{i,j}\right) - \left(\bH_L^{k,(g)'} \frac{1}{n} \sumi L_iL_i' \pi_i \right) \right\| \\
 & \leq  \max_{1\leq j\leq m}\left\|\left(n^{-1}\sumi \tDel^{k,(g)}_iL_i'\xi_{i,j}\right) \right\| +   \max_{1\leq j\leq m} \|\left(\bH_L^{k,(g)'} \frac{1}{n} \sumi L_iL_i' (\pi_i-\xi_{i,j}) \right) \\
& =O_P\left(\sqrt{(1/n + 1/m)\alpha_n^{-1}} n^{1/q}\right).
\end{align*}

Therefore, we have
\begin{align*}
 & \max_{1\leq j\leq m}\left\|\left(n^{-1}\sumi \tL^{k,(g)}_i\hL^{k,(g)'}_i\xi_{i,j}\right) - \left(\bH_L^{k,(g)'} \frac{1}{n} \sumi L_iL_i' \pi_i H_L^{k,(g)}\right)
 \right\|  \\
 & \lessp \max_{1\leq j\leq m}\left\|\left(n^{-1}\sumi \tL^{k,(g)}_iL_i'\xi_{i,j}\right) - \left(\bH_L^{k,(g)'} \frac{1}{n} \sumi L_iL_i' \pi_i \right) \right\| +\max_{1\leq j\leq m}\left\|\left(n^{-1}\sumi \tDel^{k,(g)}_{L,i} \tDel^{k,(g)'}_i\xi_{i,j}\right) \right\| \\
& =O_P\left(\sqrt{(1/n + 1/m)\alpha_n^{-1}} n^{1/q}\right).
\end{align*}

Last, we have
\begin{align*}
    \sigma_k(\bH_L^{k,(g)'}\bH_L^{k,(g)}) & = \sigma_k\left(\left( \frac{\bH^{k,(g)'}_F F'F \bH^{k,(g)}_F}{m} \right)^{-1} \left( \frac{F'F \bH^{k,(g)}_F}{m} \right)'\left( \frac{F'F \bH^{k,(g)}_F}{m} \right) \left( \frac{\bH^{k,(g)'}_F F'F \bH^{k,(g)}_F}{m} \right)^{-1}\right) \\
    & \gtrsim_P \left\| \frac{\bH^{k,(g)'}_F F'F \bH^{k,(g)}_F}{m} \right\|^{-2} \sigma_k(\bH^{k,(g)'}_F\bH^{k,(g)}_F) \gtrsim_P n^{-1/q}.
\end{align*}

This implies there exists a constant $c>0$ such that
\begin{align*}
& \bH_L^{k,(g)'}\left( \frac{1}{n} \sumi L_iL_i' \pi_i \right)\bH_L^{k,(g)} \\
& \geq c \bH_L^{k,(g)'}\left( \frac{1}{n} \sumi L_iL_i' \right)\bH_L^{k,(g)} \alpha_n n^{-1/q} \\
& \geq c \alpha_n n^{-1/q} \left( \Sigma_L - o_P(1)\right).
\end{align*}

This implies
 \begin{align*}
& \max_j  \left \|    \left(\sumi \tL_{i}^{k,(g)}\tL_{i}^{k,(g)'} \xi_{i,j}\right)^{-1}\left(\sumi \tL_{i}^{k,(g)} L_{i}'F_{j}  \xi_{i,j}-\tL_{i}^{k,(g)}\tL_{i}^{k,(g)'}\xi_{i,j}\bH^{k,(g+1)'}_FF_{j}\right) \right\| \\
& = \max_j \left\|    \left[\left(\sumi \tL_{i}^{k,(g)}\tL_{i}^{k,(g)'} \xi_{i,j}\right)^{-1}\left(\sumi \tL_{i}^{k,(g)} L_{i}' \xi_{i,j}\right)- \bH^{k,(g+1)'}_F \right] F_{j} \right\| \\
& = O_P\left(\sqrt{(1/n + 1/m)\alpha_n^{-1}} n^{1/q}\right).
 \end{align*}

In addition, by the same arguments in Lemmas \ref{lem: general asymp prop1} and \ref{lem: asymp tF g=1}, we have
\begin{align*}
& \max_j n^{-1} \left\| \sumi \tL_{i}^{k,(g)}\hat u_{i,j}\xi_{i,j}\right\| \\
& \lessp
\max_j n^{-1} \left\| \sumi \tDel_{i}^{k,(g)}\hat u_{i,j}\xi_{i,j}\right\| + \max_j n^{-1} \left\| \sumi L_i \hat u_{i,j}\xi_{i,j}\right\| \\
& \leq n^{-1} \max_j \left( \sumi \|\tDel_{i}^{k,(g)}\|_2^2 \xi_{i,j} \right)^{1/2} \max_j \left( \sumi \hat  u_{i,j}^2 \xi_{i,j} \right)^{1/2} + \max_j n^{-1} \left\| \sumi L_i \hat u_{i,j}\xi_{i,j}\right\| \\
& \lessp \sqrt{ (1/n+1/m) \alpha_n n^{1/q}} .
\end{align*}

Therefore, we have
\begin{align*}
    \left\|\tF^{k,(g+1)} - F \bH^{k,(g+1)}_F \right\| = O_P\left(  ((1+m/n)\alpha_n^{-1} )^{1/2} n^{1/q} \right).
\end{align*}
In addition, we have $\| \bH_F^{k,(g+1)}\| = O_P(n^{1/q})$ and
\begin{align*}
    & \sigma_k(\bH_F^{k,(g+1)'}\bH_F^{k,(g+1)}) \\
    & \gtrsim \left\|\bH_L^{k,(g)'} \frac{1}{n} \sumi L_iL_i' \pi_i \bH_L^{k,(g)}\right\|^{-2} \sigma_k\left(
 \left( \frac{1}{n}\sumi L_i L_i' \bH_L^{k,(g)}  \pi_i \right)'\left( \frac{1}{n}\sumi L_i L_i' \bH_L^{k,(g)}  \pi_i \right) \right) \\
 & \gtrsim_P  n^{-1/q}  \sigma_k\left(\bH_L^{k,(g)'}
 \left( \frac{1}{n}\sumi L_i L_i'   \right)^2 \bH_L^{k,(g)} \right) \\
 & \gtrsim_P  n^{-1/q} \sigma_k\left(\bH_L^{k,(g)'}\bH_L^{k,(g)} \right) \\
 & \gtrsim_P  n^{-1/q}.
\end{align*}

The proof is then completed by induction.
\end{proof}

\begin{lem}\label{lem: order of mse(k,g)-mse(r,g) k>r g>1}
Under Assumption \ref{assu: moment cond}, for $k\geq r$, we have
\begin{align*}
    \mse(k, g) - \mse(r, g) = O_P\left((1/n+1/m)^{1/2}\alpha_n^{1/2} n^{1/q}\right),
\end{align*}
where $\mse(k, g)$ is the analogue of $\mse(k)$ defined by \eqref{eq: def of mse(k)}, in $g^{\rm th}$ iteration. (In fact, $\mse(k)\equiv \mse(k, 1)$.)
\end{lem}
\begin{proof}[\textbf{Proof of Lemma \ref{lem: order of mse(k,g)-mse(r,g) k>r g>1}}]\ \\
Note the definition of $\mse^*$ in Lemma \ref{lem: order of mse*(k)} does not depend on $(k,g)$. Then by definition,
\begin{align*}
    &\mse^*\leq \mse(k, g)\leq \mse(k, 1)=\mse(k)
\end{align*}
for any $k\geq r$ and $g\geq 1$, and thus
\begin{align*}
    &\mse(k,g) - \mse(r, g) \leq \mse(k) - \mse^*\overset{\texti}{=} O_P\left((1/n+1/m)^{1/2}\alpha_n^{1/2} n^{1/q}\right),\\
    &\mse(r, g) - \mse(k, g) \leq \mse(r) - \mse^*\overset{\textii}{=} O_P\left((1/n+1/m)^{1/2}\alpha_n^{1/2} n^{1/q}\right),
\end{align*}
where $\texti$ and $\textii$ are from Lemmas \ref{lem: order of mse*(k)} and Lemma \ref{lem: order of v_k mse(k)}, respectively. Therefore
\begin{align*}
|\mse(k, g) - \mse(r, g)| = O_P\left((1/n+1/m)^{1/2}\alpha_n^{1/2} n^{1/q}\right)
\end{align*}
and the result follows.
\end{proof}

\begin{proof}[\textbf{Proof of Theorem \ref{thm: r est eIC} for $g>1$}]\

Since we already proved that, for any finite $g\geq 2$,
\begin{align*}
    &\mse(k, g) - \mse(r, g)\geq \alpha_n \tau_k>0\quad \mbox{for $k<r$, and}\\
    &\mse(k, g) - \mse(r, g)= O_P\left( (1/n+1/m)^{1/2}\alpha_n^{1/2} n^{1/q}\right) \quad \mbox{for $k>r$}.
\end{align*}
We can follow {\bf Proof of Theorem \ref{thm: r est eIC} at $g=1$}, and have
\begin{align*}
    P(\eIC(k\,|\,g)- \eIC(r\,|\,g)>0)\overset{P}{\rightarrow}1
\end{align*}
if $h(n,m)\rightarrow 0$ and $\sqrt{\frac{mn \alpha_n}{(m+n) }}h(n,m)$ diverges to infinity in a polynomial rate in $n$. Therefore, the statement $P(\hr^{\eIC(g)}=r)\rightarrow 1$ for $g>1$ is also shown.

\end{proof}

\section[Proof of The Inference Results on Coefficient Matrix]{Proof of The Inference Results on $\beta$} \label{app: inference beta}
In this section, we will prove the inference result of $\beta\in \mathbb{R}^{m\times d}$ based on our estimators. For the sake of simplicity of notations, we fix a finite $g>0$ and denote the estimator $\tbeta^{(g)}$, $\tGamma^{(g)}$ as $\tbeta$ and $\tGamma$.

\subsection{A more general result and proof of Theorem \ref{thm: beta hypothesis}}
Let $B\in\RR^{r\times md}$ be a non-random matrix. Recall $\omega_{i,j}=E(X_iX_i'\pi_i)^{-1} X_{i}(\xi_{i,j}\varepsilon_{i,j}+ \pi_i L_{i}'F_{j})$
and $\home_{i,j}=E_n(X_iX_i'\xi_{i,j})^{-1} X_{i}(\xi_{i,j}\hat \varepsilon_{i,j}+ \hpi_i \tilde \Gamma_{i,j})$. Let $\omega_{i}=(\omega_{1}',...,\omega_{m}')'\in\RR^{md}$
and $\home_{i}=(\home_{1}',...,\home_{m}')'\in\RR^{md}$. Consider
\begin{align*}
T_{(B)}^{o}=\left\| n^{-1/2}\sumi \iota_{i}B\omega_{i}\right\|_{\infty}
\end{align*}
and
\begin{align*}
T_{(B)}^{*}=\left\| n^{-1/2}\sumi \iota_{i}B\home_{i}\right\|_{\infty}.
\end{align*}
\begin{prop}
\label{thm: inference beta}Let Assumptions \ref{assu: moment cond}
and \ref{assu: inference} hold. Assume that $r\leq md$. Suppose that the following conditions hold:
\begin{enumerate}
\item there exists a constant $C_{1}>0$ such that the $\ell_{1}$-norm of each row of $B$ is bounded by $C_{1}$.
\item there exist constant $C_{2},C_{3}>0$ such that the diagonal entries of the matrix $B \alpha_n (E\omega_{i}\omega_{i}')B'$ are in $[C_{2},C_{3}]$.
\end{enumerate}
Then
\begin{align*}
\sup_{x\in\RR}\left|P\left(\sqrt{n \alpha_n }\|B(\tbeta-\beta)\|_{\infty}\leq x\right)-P\left(\alpha_n^{1/2} T ^*_{(B)}\leq x\mid\{\home_{i}\}_{i=1}^{n}\right)\right|=o_{P}(1).
\end{align*}
\end{prop}
\noindent With the help of Proposition of \ref{thm: inference beta}, we now prove
Theorem \ref{thm: beta hypothesis}.
\begin{proof}[\textbf{Proof of Theorem \ref{thm: beta hypothesis}}]
Let $W_{\Gcal}$ be the $|\Gcal|\times m$ selection matrix such
that for any $b\in\RR^{m}$, $b_{\Gcal}=Wb$. Clearly, in each row
of $W_{\Gcal}$, all the entries are zero except that one entry is
one. Then let $B=W_{\Gcal}\otimes I_{k}\in\RR^{|\Gcal|k\times mk}$,
where $\otimes$ denotes the Kronecker product and $I_{k}$ is the
$k\times k$ identity matrix. Clearly, each row of $B$ has $\ell_{1}$-norm
equal to one and $|\Gcal|k\leq mk$ (due to $|\Gcal|\leq m$).

We observe that
\begin{align*}
E\omega_{i,j}\omega_{i,j}'= E(X_iX_i'\pi_i)^{-1} \left(EX_{i}X_{i}' \pi_i \varepsilon_{i,j}^{2}\right)  E(X_iX_i'\pi_i)^{-1}+ E(X_iX_i'\pi_i)^{-1} \left(EX_{i}X_{i}'\pi_i^2(L_{i}'F_{j})^{2}\right) E(X_iX_i'\pi_i)^{-1}.
\end{align*}
Let $\sigma_{\min}(\cdot)$ and $\sigma_{\max}(\cdot)$ denote the
minimal and maximal eigenvalue of a symmetric matrix. We now take
$B$ to be the $mk$ by $mk$ identity matrix. Since $E(\varepsilon_{i,j}^{2}\mid X)\geq M_{1}$,
we have that
\begin{align*}
\alpha_n \sigma_{\min}\left(E\omega_{i,j}\omega_{i,j}'\right) & \geq\sigma_{\min}\left[ \alpha_n E(X_iX_i'\pi_i)^{-1} \left(EX_{i}X_{i}' \pi_i \varepsilon_{i,j}^{2}\right)  E(X_iX_i'\pi_i)^{-1}\right] \\
& \geq M_{1}\sigma_{\min}(\alpha_n E(X_iX_i'\pi_i)^{-1}).
\end{align*}
On the other hand, since $X_{i}$, $L_{i}$, $F_{j}$ and $\varepsilon_{i,j}$
are sub-Gaussian, $\sigma_{\max}\left(\alpha_n E\omega_{i,j}\omega_{i,j}'\right)$
is bounded. It follows that diagonal entries of $\alpha_n E\omega_{i,j}\omega_{i,j}'$
are bounded away from zero and infinity. Therefore, we can apply Proposition
\ref{thm: inference beta}, obtaining that under $H_{0}:\ \beta_{j}=\beta_{j}^{o}$
for all $j\in\Gcal$,
\begin{align*}
& \sup_{x\in\RR}\left|P\left(T \leq x\right)-P\left(T ^{*}\leq x\mid\{\home_{i}\}_{i=1}^{n}\right)\right| \\
& = \sup_{x\in\RR}\left|P\left(\sqrt{n\alpha_n}T \leq x\right)-P\left(\sqrt{n\alpha_n}T ^{*}\leq x\mid\{\home_{i}\}_{i=1}^{n}\right)\right| \\
& =\sup_{x\in\RR}\left|P\left(\sqrt{n \alpha_n }\|B(\tbeta-\beta)\|_{\infty}\leq x\right)-P\left(\alpha_n^{1/2} T ^*_{(B)}\leq x\mid\{\home_{i}\}_{i=1}^{n}\right)\right| = o_{P}(1).
\end{align*}
The desired result follows.
\end{proof}

\subsection{Proof of Proposition \ref{thm: inference beta}}
\noindent We first prove three auxiliary results before proving Proposition \ref{thm: inference beta}.
\begin{lem}
\label{lem:  BS 1 aux 1}Let Assumption \ref{assu: moment cond} hold.
Then
\begin{enumerate}
\item $\max_{1\leq j\leq m}n^{-1}\sumi \|X_i\|_2^2\xi_{i,j}(\tGamma_{i,j}-\Gamma_{i,j})^{2}\lessp (n^{-1}+m^{-1})(\log n\vee m)^{2}$.
\item $\max_{1\leq j\leq m}n^{-1}\sumi (\tGamma_{i,j}-\Gamma_{i,j})^{2}\lessp \alpha_n^{-1}(n^{-1}+m^{-1})(\log n\vee m)^{2}$.
\item $\max_{1\leq j\leq m}\|\tbeta_{j}-\beta_{j}\|_{2}\lessp (n\alpha_n)^{-1/2}(\log n\vee m)^{2}$.
\item $\max_{1\leq j\leq m}n^{-1}\sumi \|X_i\|_2^2\left(\xi_{i,j}\heps_{i,j}-\xi_{i,j}\varepsilon_{i,j}\right)^{2}\lessp (m^{-1}+n^{-1})(\log n\vee m)^{2}.$
\end{enumerate}
\end{lem}
\begin{proof}
We proceed in three steps.\\[6pt]
\textbf{Step 1:} show $\max_{j}n^{-1}\sumi\|X_i\|_2^2 \xi_{i,j} (\tGamma_{i,j}-\Gamma_{i,j})^{2}\lessp (n^{-1}+m^{-1})(\log n\vee m)^{2}$.\\

Denote $\kappa_n  =(n^{-1}+m^{-1})\alpha_n^{-3/2} n^{1/q}$. In Theorem \ref{thm: Gamma est}, we have proved
\begin{align*}
    & \left\|\tGamma^{(g)}-\Gamma- \Delta_{i,j}
        \right\|_\infty   =  O_{P}\left(\kappa_n \right)
        \end{align*}
for all $g$, where $\Delta$ is a $n \times m$ matrix with its $(i,j)$th entry
\begin{align*}
    \Delta_{i,j} =  \frac{1}{n} L'_i (EL_iL_i \pi_i)^{-1} \sum_{k=1}^n L_k \xi_{k,j}\varepsilon_{k,j} + \frac{1}{m \pi_i}\sum_{t=1}^m F'_t\xi_{i,t}\varepsilon_{i,t} \Sigma^{-1}_F F_j -  \frac{1}{n}X'_i E(X_iX_i'\pi_i)^{-1} \sum_{k=1}^n \pi_k X_k L_k' F_j.
\end{align*}

Recall that $\max_{i}\|L_{i}\|_{2}\lessp \sqrt{\log n}$ due to the
sub-Gaussian assumption. By Lemma \ref{lem: general asymp prop}, $\max_{j}\|E_{n}L_{i}\xi_{i,j}\varepsilon_{i,j}\|_{2}\lessp \sqrt{ \alpha_n n^{-1}\log m}$,
which means that
\begin{align*}
        &\max_{j}n^{-1}\sumi \xi_{i,j}\|X_i\|_2^2\left|\frac{1}{n}  L'_i (EL_iL_i \pi_i)^{-1} \sum_{k=1}^n L_k \xi_{k,j}\varepsilon_{k,j}\right|^2\\
\leq\,  &\|(EL_iL_i \pi_i)^{-1}\|^2\cdot\left(\max_j n^{-1}\sumi \xi_{i,j} \|L_{i}\|^2_{2}\|X_i\|_2^2\right)\cdot \max_{j}\left\|n^{-1}\sum_{s=1}^{n}L_{s}\xi_{s,j}\varepsilon_{s,j}\right\|^2_{2}\\
=\,& O_P\left(n^{-1}\log m\right).
\end{align*}
In addition, we have $\max_{i}\|\sum_{t=1}^{m}\xi_{i,t}F_{t}\varepsilon_{i,t}\|_{2}\lessp \sqrt{m \alpha_n \log n}$.
By $\max_{j}\|F_{j}\|_{2}\lessp \sqrt{\log m}$, we have
\begin{align*}
&\quad \max_{j}n^{-1}\sumi \xi_{i,j}\|X_i\|_2^2 \left|\frac{1}{m \pi_i}\sum_{t=1}^m F'_t\xi_{i,t}\varepsilon_{i,t} \Sigma^{-1}_F F_j\right|^2\\
&\leq\|\Sigma_{F}^{-1}\|^2\cdot\left(\max_{j}\|F_{j}\|^2_{2}\right)\cdot \left(\max_j n^{-1} \sumi \xi_{i,j}\|X_i\|_2^2\pi_i^{-2}\right)\left(\max_{i}\left\| (m)^{-1}\sum_{t=1}^{m}\xi_{i,t}F_{t}\varepsilon_{i,t}\right\|^2_{2}\right) \\
& =O_P\left( m^{-1}\log m\log n\right).
\end{align*}
The above three displays imply that
\begin{align*}
\max_{j} n^{-1} \sumi \left|\tGamma_{i,j}-\Gamma_{i,j}+\frac{1}{n}X'_i E(X_iX_i'\pi_i)^{-1} \sum_{k=1}^n \pi_k X_k L_k' F_j\right|^2 \lessp (n^{-1}+m^{-1})\log^2(m+n).
\end{align*}
By the elementary inequality of $(a+b)^{2}\leq2a^{2}+2b^{2}$, it follows that
\begin{align*}
    & \max_{j}n^{-1}\sumi \xi_{i,j}\|X_i\|_2^2(\tGamma_{i,j}-\Gamma_{i,j})^{2}\\
\leq\,& 2\max_{j}n^{-1}\sumi \xi_{i,j}\|X_i\|_2^2\left(\frac{1}{n}X'_i E(X_iX_i'\pi_i)^{-1} \sum_{k=1}^n \pi_k X_k L_k' F_j\right)^{2}+O_{P}\left((n^{-1}+m^{-1})\log^2(m+n)\right)\\
=\, & 2\max_{j}n^{-1}\sumi \xi_{i,j}\|X_i\|_2^2 \left(n^{-1}F_{j}'A_{n}E(X_iX_i'\pi_i)^{-1}X_{i}\right)^{2}+O_{P}\left((n^{-1}+m^{-1})\log^2(m+n)\right)\\
=\, &2n^{-2}\max_{j}F_{j}'A_{n}E(X_iX_i'\pi_i)^{-1}(E_n X_iX_i'\|X_i\|_2^2\xi_{i,j})E(X_iX_i'\pi_i)^{-1}A_{n}F_{j}+O_{P}\left((n^{-1}+m^{-1})\log^2(m+n)\right),
\end{align*}
where $A_{n}=\sum_{k=1}^{n} \pi_kL_{k}X_{k}'$. Since $A_{n}=O_{P}((n \alpha_n^2)^{1/2}))$ and $\max_{j}\|F_{j}\|_{2}\lessp \sqrt{\log m}$, the above display implies
\begin{align*}
\max_{j}n^{-1}\sumi \xi_{i,j} \|X_i\|_2^2(\tGamma_{i,j}-\Gamma_{i,j})^{2}
&=  O_{P}\left((n^{-1}+m^{-1})\log^2(m+n)\right).
\end{align*}

The second result can be derived in a similar manner.

\textbf{Step 2:} show $\max_{j}\|\tbeta_{j}-\beta_{j}\|_{2}\lessp (n\alpha_n)^{-1/2}(\log n\vee m)^{2}$.\\
By Theorem \ref{thm: beta est},
\begin{align*}
\max_{j}\left\| \tbeta_{j}-\beta_{j}-n^{-1} \sumi E(X_iX_i'\pi_i)^{-1} X_{i}(\xi_{i,j}\varepsilon_{i,j}+ \pi_i L_{i}'F_{j})\right\|_{2}
\lessp \kappa_n.
\end{align*}
Similar to Step 1, we have $\max_{j}\|E(X_iX_i'\pi_i)^{-1}E_{n}(X_{i}\varepsilon_{i,j}\xi_{i,j})\|_{2}\lessp \sqrt{(n\alpha_n)^{-1}\log m}$.
By $\max_{j}\|F_{j}\|_{2}\lessp \sqrt{\log m}$, we have $\max_{j}\|E(X_iX_i'\pi_i)^{-1}(E_{n} \pi_i X_{i}L_{i}')F_{j}\|_{2}\lessp \sqrt{n^{-1}\log m}$.
Therefore,
\begin{align*}
\max_{j}\|\tbeta_{j}-\beta_{j}\|_{2}\lessp \sqrt{(n\alpha_n)^{-1}\log m}+\kappa_n
\lessp (n\alpha_n)^{-1/2}(\log n\vee m)^{2}.
\end{align*}

\textbf{Step 3:} show $\max_{j}n^{-1}\sumi \|X_i\|_2^2\left(\xi_{i,j}\heps_{i,j}-\xi_{i,j}\varepsilon_{i,j}\right)^{2}\lessp (m^{-1}+n^{-1})(\log n\vee m)^{2}$.\\
By the definition of $\heps_{i,j}$, we observe that
\begin{align*}
\xi_{i,j}\heps_{i,j}-\xi_{i,j}\varepsilon_{i,j}
 & =\xi_{i,j}\left(X_{i}'(\beta_{j}-\tbeta_{j})+(\Gamma_{i,j}-\tGamma_{i,j})\right).
\end{align*}
By the elementary inequality of $(a+b)^{2}\leq 2a^{2}+2b^{2}$,
we have
\begin{align*}
&\max_{j}n^{-1}\sumi \|X_i\|_2^2 \left(\xi_{i,j}\heps_{i,j}-\xi_{i,j}\varepsilon_{i,j}\right)^{2}\\
\leq\, & 2\left(\max_{j}n^{-1}\sumi \|X_i\|_2^2 \xi_{i,j}\left(X_{i}'(\beta_{j}-\tbeta_{j})\right)^{2}\right) + 2\left(\max_{j}n^{-1}\sumi \|X_i\|_2^2 \xi_{i,j}(\Gamma_{i,j}-\tGamma_{i,j})^{2}\right).
\end{align*}
For the first term, we have
\begin{align*}
\max_{j}n^{-1}\sumi \xi_{i,j}\|X_i\|_2^2\left(X_{i}'(\beta_{j}-\tbeta_{j})\right)^{2} \lessp \|E_n (X_iX_i'\|X_i\|_2^2\xi_{i,j}\|) \|\tbeta_j - \beta_j\|_2^2 \lessp (m^{-1}+n^{-1})(\log n\vee m)^{2}.
\end{align*}

The second term is simple because
\begin{align*}
    \max_{j}n^{-1}\sumi \xi_{i,j}\|X_i\|_2^2(\Gamma_{i,j}-\tGamma_{i,j})^{2}     &\lessp (m^{-1}+n^{-1})(\log n\vee m)^{2}. &&\mbox{(Step 1)}
\end{align*}
Therefore, we have
\begin{align*}
& \max_{j}n^{-1}\sumi \|X_i\|_2^2\left(\xi_{i,j}\heps_{i,j}-\xi_{i,j}\varepsilon_{i,j}\right)^{2}
  \lessp  (m^{-1}+n^{-1})(\log n\vee m)^{2}.
\end{align*}
 The proof is complete.
\end{proof}
\begin{lem}
\label{lem:  BS 1}Let Assumption \ref{assu: moment cond} hold. Assume that $r\leq mk$. Suppose that the following conditions hold:
\begin{enumerate}
\item there exists a constant $C_{1}>0$ such that the $\ell_{1}$-norm of each row of $B$ is bounded by $C_{1}$.
\item there exist constant $C_{2},C_{3}>0$ such that the diagonal entries of the matrix  $\alpha_n B(E\omega_{i}\omega_{i}')B'$ are in $[C_{2},C_{3}]$.
\end{enumerate}
Then
\begin{align*}
\left\| B\left(n^{-1}\sumi \omega_{i}\omega_{i}'-\home_{i}\home_{i}'\right)B'\right\|_{\infty}\lessp \alpha_n^{-3/2}(m^{-1/2}+n^{-1/2})(\log n)^{5/2}.
\end{align*}
\end{lem}
\begin{proof}
Let $\zeta=\left\| B\left(n^{-1}\sumi \omega_{i}\omega_{i}'-\home_{i}\home_{i}'\right)B'\right\|_{\infty}$.
We observe that
\begin{align*}
\zeta&\leq C_{1}^{2}\left\| n^{-1}\sumi \omega_{i}\omega_{i}'-\home_{i}\home_{i}'\right\|_{\infty}=C_{1}^{2}\max_{j_{1},j_{2}}\left\| n^{-1}\sumi \omega_{i,j_{1}}\omega_{i,j_{2}}'-\home_{i,j_{1}}\home_{i,j_{2}}'\right\|_{\infty}\\
&\leq C_{1}^{2}\max_{j_{1},j_{2}}\left\| n^{-1}\sumi \omega_{i,j_{1}}\omega_{i,j_{2}}'-\home_{i,j_{1}}\home_{i,j_{2}}'\right\|_{F},
\end{align*}
where $\omega_{i,j}=E(X_iX_i'\pi_i)^{-1} X_{i}(\xi_{i,j}\varepsilon_{i,j}+ \pi_i L_{i}'F_{j})$
and $\home_{i,j}=E_n(X_iX_i'\xi_{i,j})^{-1} X_{i}(\xi_{i,j}\hat \varepsilon_{i,j}+ \hpi_i \tilde \Gamma_{i,j})$.
We proceed in two steps.\\[6pt]
\textbf{Step 1:} show that $n^{-1}\max_{j}\sumi \|\omega_{i,j}-\home_{i,j}\|_{2}^{2}\lessp \alpha_n^{-2}(m^{-1}+n^{-1})(\log(m+n))^{2}$.\\
By the elementary inequality of $\|a+b\|_{2}^{2}\leq2\|a\|_{2}^{2}+2\|b\|_{2}^{2}$,
we notice that
\begin{align*}
 & \left\| \home_{i,j}-\omega_{i,j}\right\|_{2}^{2}\\
 & \lessp \left\| E_n(X_iX_i'\xi_{i,j})^{-1} X_{i}(\xi_{i,j}\heps_{i,j}-\xi_{i,j}\varepsilon_{i,j}+\hpi_i \tGamma_{i,j}-\pi_i \Gamma_{i,j})\right\|_2^2 \\
 & +\left\|(E(X_iX_i'\pi_{i})^{-1}-E_n(X_iX_i'\xi_{i,j})^{-1})X_{i}(\xi_{i,j}\varepsilon_{i,j}+\pi_iL_{i}'F_{j})\right\|_{2}^{2}\\
& \lessp \|E_n(X_iX_i'\xi_{i,j})^{-2}\| \|X_i\|_2^2\xi_{i,j} (\hat \eps_{i,j} - \eps_{i,j})^2 + \|E_n(X_iX_i'\xi_{i,j})^{-2}\| \|X_i\|_2^2\xi_{i,j} |\hpi_i - \pi_i|^2 \Gamma_{i,j}^2 \\
&+ \|E_n(X_iX_i'\xi_{i,j})^{-2}\| \hpi_i^2 (\tGamma_{i,j} - \Gamma_{i,j})^2 + \left\|(E(X_iX_i'\pi_{i})^{-1}-E_n(X_iX_i'\xi_{i,j})^{-1}) \right\|^2 \left\|X_{i}(\xi_{i,j}\varepsilon_{i,j}+\pi_iL_{i}'F_{j})\right\|_{2}^{2} \\
& \equiv A_{1,i,j}^2+\cdots+A_{4,i,j}^2.
\end{align*}
Note that
\begin{align*}
\max_j    \|(EX_iX_i'\pi_i)^{-1}\| \lessp \alpha_n^{-1} \quad \text{and} \quad \max_j \left\|(E(X_iX_i'\pi_{i})^{-1}-E_n(X_iX_i'\xi_{i,j})^{-1}) \right\| \lessp \alpha_n^{-1} (n \alpha_n)^{-1/2} \log ^{1/2} n.
\end{align*}
Therefore, by Lemma \ref{lem:  BS 1 aux 1}, we have
\begin{align*}
 \max_j n^{-1}  \sumi A_{1,i,j}^2   & \lessp n^{-1}\alpha_n^{-2} \max_j \sumi \|X_i\|_2^2 \xi_{i,j} (\hat \eps_{i,j} - \eps_{i,j})^2 \\
& \lessp \alpha_n^{-2}(n^{-1}+m^{-1})(\log n\vee m)^{2}
\end{align*}
and
\begin{align*}
\max_j n^{-1}  \sumi A_{3,i,j}^2 & \lesssim  \alpha_n^{-2} \max_j n^{-1}  \sumi \hpi_i^2 (\tGamma_{i,j} - \Gamma_{i,j})^2 \\
& \lessp n^{1/q} \max_j n^{-1}  \sumi (\tGamma_{i,j} - \Gamma_{i,j})^2 \\
& \lessp \alpha_n^{-1}(n^{-1}+m^{-1})(\log n\vee m)^{2}.
\end{align*}

In addition, by Proposition \ref{prop:pihat}, we have
\begin{align*}
  \max_j  n^{-1}\sumi A_{2,i,j}^2 \lessp n^{-1}(nm\alpha_n)^{-1} n^{1/q} \max_j \sumi
 \|E_n(X_iX_i'\xi_{i,j})^{-2}\| \|X_i\|_2^2\xi_{i,j} \pi_i^2 \Gamma_{i,j}^2
 \lessp (nm)^{-1} n^{1/q}.
\end{align*}

Last, we have
\begin{align*}
    \max_j n^{-1}\sumi A_{4,i,j}^2 \lessp n^{-1} \alpha_n^{-2}(n\alpha_n)^{-1} \log^{1/2} n (n \alpha_n + n \alpha_n^2 \log n) \lessp n^{-1}\alpha_n^{-2}( \log^{1/2} n + \alpha_n \log n).
\end{align*}
Combining the bounds for $A_{1,i,j}^2$ to $A_{4,i,j}^2$, we obtain the desired result.

\textbf{Step 2:} show that the desired result.\\
We observe that
\begin{align*}
\zeta & \lesssim\max_{j_{1},j_{2}}\left\| n^{-1}\sumi \omega_{i,j_{1}}\omega_{i,j_{2}}'-\home_{i,j_{1}}\home_{i,j_{2}}'\right\|_{F}\\
 & \leq\max_{j_{1},j_{2}}n^{-1}\sumi \left\| \omega_{i,j_{1}}\omega_{i,j_{2}}'-\home_{i,j_{1}}\home_{i,j_{2}}'\right\|_{F}\\
 & =\max_{j_{1},j_{2}}n^{-1}\sumi \left\| (\omega_{i,j_{1}}-\home_{i,j_{1}})\omega_{i,j_{2}}'+\home_{i,j_{1}}(\omega_{i,j_{2}}-\home_{i,j_{2}})'\right\|_{F}\\
 & \leq\max_{j_{1},j_{2}}n^{-1}\sumi \left(\left\| (\omega_{i,j_{1}}-\home_{i,j_{1}})\omega_{i,j_{2}}'\right\|_{F}+\left\| \home_{i,j_{1}}(\omega_{i,j_{2}}-\home_{i,j_{2}})'\right\|_{F}\right)\\
 & \leq\max_{j_{1},j_{2}}\sqrt{n^{-1}\sumi \|\omega_{i,j_{1}}-\home_{i,j_{1}}\|_{2}^{2}}\times\sqrt{n^{-1}\sumi \|\omega_{i,j_{2}}\|_{2}^{2}}\\
 & \qquad+\max_{j_{1},j_{2}}\sqrt{n^{-1}\sumi \|\omega_{i,j_{2}}-\home_{i,j_{1}}\|_{2}^{2}}\times\sqrt{n^{-1}\sumi \|\home_{i,j_{1}}\|_{2}^{2}}.
\end{align*}
We observe that $n^{-1}\sumi \|\home_{i,j_{1}}\|_{2}^{2}\leq2n^{-1}\sumi \|\omega_{i,j_{1}}\|_{2}^{2}+2n^{-1}\sumi \|\home_{i,j_{1}}-\omega_{i,j_{1}}\|_{2}^{2}$.
By Step 1, we have that
\begin{align*}
\zeta\lessp \alpha_n^{-1} (m^{-1}+n^{-1})^{1/2}(\log(m+n)) \cdot\sqrt{\max_{j}n^{-1}\sumi \|\omega_{i,j}\|_{2}^{2}}.
\end{align*}
Since $\omega_{i,j}=E(X_iX_i'\pi_i)^{-1} X_{i}(\xi_{i,j}\varepsilon_{i,j}+ \pi_i L_{i}'F_{j})$,
we have
\begin{align*}
\max_{j}\sumi \|\omega_{i,j}\|_{2}^{2} & \leq2\max_{j}\sumi \| E(X_iX_i'\pi_i)^{-1} X_{i}\xi_{i,j}\varepsilon_{i,j} \|_{2}^{2}+2\max_{j}\sumi \| E(X_iX_i'\pi_i)^{-1} X_{i}\pi_i L_{i}'F_{j} \|_{2}^{2}\\
& \lessp n \alpha_n^{-1} + n \log n.
\end{align*}
Thus, we have proved
\begin{align*}
\zeta\lessp \alpha_n^{-1} (m^{-1/2}+n^{-1/2})(\log(m+n))^{2}\times\sqrt{ \alpha_n^{-1} \log n }\lessp \alpha_n^{-3/2}(m^{-1/2}+n^{-1/2})(\log n)^{5/2}.
\end{align*}
\end{proof}
\begin{lem}
\label{lem:  BS 2}Let Assumption \ref{assu: moment cond} hold. Suppose
that the following conditions hold:
\begin{enumerate}
\item there exists a constant $C_{1}>0$ such that the $\ell_{1}$-norm
of each row of $B$ is bounded by $C_{1}$.
\item there exist constant $C_{2},C_{3}>0$ such that the diagonal entries of the matrix $\alpha_n B(E\omega_{i}\omega_{i}')B'$ are in $[C_{2},C_{3}]$.
\end{enumerate}
Assume that $r\leq mk$. Also assume that entries of $X_{i}$ have
bounded sub-Gaussian norm. Then
\begin{align*}
\left\| n^{-1}\sumi B\omega_{i}\omega_{i}'B'-B(E\omega_{i}\omega_{i}')B'\right\|_{\infty}\lessp \sqrt{(n \alpha_n^3)^{-1}\log(n)}.
\end{align*}
\end{lem}
\begin{proof}
Similar to the proof of Lemma \ref{lem:  BS 1}, we observe that
\begin{equation}
\left\| n^{-1}\sumi B\omega_{i}\omega_{i}'B'-B(E\omega_{i}\omega_{i}')B'\right\|_{\infty}\lesssim\max_{j_{1},j_{2}}\left\| n^{-1}\sumi \omega_{i,j_{1}}\omega_{i,j_{2}}'-E\omega_{i,j_{1}}\omega_{i,j_{2}}'\right\|_{F}.\label{eq: lem BS2 eq 1}
\end{equation}
Since the dimension of $\omega_{i,j}$ is bounded, we use Lemma A.1
of \cite{chernozhukov2013gaussian} (applied to each entry of $\omega_{i,j}$)
and obtain
\begin{align}
& E\max_{j_{1},j_{2}}\left\| n^{-1}\sumi \omega_{i,j_{1}}\omega_{i,j_{2}}'-E\omega_{i,j_{1}}\omega_{i,j_{2}}'\right\|_{F} \notag \\
&\lesssim\sqrt{(n\alpha_n^3)^{-1}\log(m^{2} \alpha_n^{-1})}+n^{-1}\log(m^{2}\alpha_n^{-1})\sqrt{E\max_{i,j}\|\omega_{i,j}\|_{2}^{4}},\label{eq: lem BS2 eq 2}
\end{align}
where we use the fact that
\begin{align*}
    \max_{j_{1},j_{2}}E ( \omega_{i,j_{1}}\omega_{i,j_{2}})^2 \lessp \alpha_n^{-3}.
\end{align*}

Recall $\omega_{i,j}=E(X_iX_i'\pi_i)^{-1} X_{i}(\xi_{i,j}\varepsilon_{i,j}+ \pi_i L_{i}'F_{j})$.
Thus,
\begin{align}
E\max_{i,j}\|\omega_{i,j}\|_{2}^{4} & \leq E\max_{i,j}\|E(X_iX_i'\pi_i)^{-1}X_{i}\|_{2}^{4}|\xi_{i,j}\varepsilon_{i,j}+ \pi_iL_{i}'F_{j}|^{4}\nonumber \\
 & \lessp \alpha_n^{-4} E \max_{i,j}\|X_i\|_2^4( \eps_{i,j}^4 \xi_{i,j} + \pi_i^4 \|L_i\|_2^4 \|F_j\|_2^4).\label{eq: lem BS2 eq 3}
\end{align}
We now derive an elementary bound. Let $\{W_{i}\}_{i=1}^{n}$ be random
variables with bounded sub-Gaussian norms, i.e., $P(|W_{i}|>t)\leq\exp(1-t^{2}/K)$
for any $i$ and $t\geq0$. Then by the union bound, for any $z\geq0$,
\begin{align*}
P\left(\max_{i}|W_{i}|>z+\sqrt{K\log n}\right)\leq\sumi P\left(|W_{i}|>z+\sqrt{K\log n}\right)\\
\leq n\exp\left(1-\left[z+\sqrt{K\log n}\right]^{2}/K\right)\leq n\exp\left(1-\left[z^{2}+K\log n\right]/K\right)=\exp(1-z^{2}/K).
\end{align*}
In other words, $\max_{i}|W_{i}|-\sqrt{K\log n}$ is sub-Gaussian.
Thus, $E\left(\max_{i}|W_{i}|-\sqrt{K\log n}\right)^{8}=O(1)$. Since
$(a+b)^{8}\leq2^{7}(a^{8}+b^{8})$ for any $a,b\geq0$, we have that
$E\max_{i}|W_{i}|^{8}=O((\log n)^{4})$. Now we apply this bound to
$W_{i}=\|X_{i}\|_{2}$, obtaining $E\max_{i}\|X_{i}\|_{2}^{8}\lesssim \log^4 n$.
We also apply this bound to $W_{i}=\xi_{i,j}\varepsilon_{i,j}+\pi_i L_{i}'F_{j}$;
since $L_{i}'F_{j}$ is sub-exponential, we only need to slightly
modify the bound and obtain $E\max_{i,j}|\xi_{i,j}\varepsilon_{i,j}+\pi_i L_{i}'F_{j}|^{8}\lesssim(\log(mn))^{8}$.
Therefore, by (\ref{eq: lem BS2 eq 3}), we have
\begin{align*}
E\max_{i,j}\|\omega_{i,j}\|_{2}^{4}\lesssim \alpha_n^{-4}(\log n)^{2}\times(\log(mn))^{4}\overset{\texti}{\lesssim}\alpha_n^{-4}(\log m)^{6},
\end{align*}
where (i) follows by $\log m\asymp\log n$ (due to the assumption
of $n^{1/2}\log(m+n)\lesssim m\lesssim n^{2}$). Thus, by (\ref{eq: lem BS2 eq 1})
and (\ref{eq: lem BS2 eq 2}), we have
\begin{align*}
E\left\| n^{-1}\sumi B\omega_{i}\omega_{i}'B'-B(E\omega_{i}\omega_{i}')B'\right\|_{\infty}
& \lesssim \sqrt{(n \alpha_n^3)^{-1}\log(n)}+(n\alpha_n^2)^{-1}\log^4(n).
\end{align*}
The desired result follows.
\end{proof}
\begin{proof}[\textbf{Proof of Proposition \ref{thm: inference beta}}]\ \\
Define $\alpha_n^{1/2}T_{(B)}=\|Z\|_{\infty}$, where $Z\sim N(0,V_{n})$ and $V_{n}=n^{-1} \alpha_n \sumi EB\omega_{i}\omega_{i}'B'$.\\[6pt]
\textbf{Step 1:} show that $\sup_{x\in\RR}\left|P\left(\sqrt{n}\|B(\tbeta-\beta^{o})\|_{\infty}\leq x\right)-P\left(T_{(B)}\leq x\right)\right|=o_{P}(1)$.\\
Recall that $\omega_{i,j}=E(X_iX_i'\pi_i)^{-1} X_{i}(\xi_{i,j}\varepsilon_{i,j}+ \pi_i L_{i}'F_{j})$
and $L_{i}'F_{j}=\Gamma_{i,j}$. Since the diagonal entries of $\alpha_n B(E\omega_{i}\omega_{i}')B'$
are in $[C_{2},C_{3}]$, it follows that each component of $\alpha_n^{1/2} B\omega_{i}$
has second moment bounded below by $C_{2}$. We have
\begin{align*}
E\|\omega_{i,j}\|_{2}^{3}
& \leq E\left(\|E(X_iX_i'\pi_i)^{-1}X_{i}\|_{2}^{3}\left(|\varepsilon_{i,j}|\xi_{i,j}+ \pi_i|\Gamma_{i,j}|\right)^{3}\right)\lessp \alpha_n^{-2}.
\end{align*}
Since the $\ell_{1}$-norm of each row of $B$ is bounded,
each component of $ \alpha_n^{1/2} B\omega_{i}$ its has third moment bounded by $ C \alpha_n^{-1/2}$ for some constant $C>0$.

Similarly, we have
\begin{align*}
    \max_j E\|\alpha_n^{1/2}\omega_{i,j}\|_{2}^{4}
    &\leq E\left(\|E(X_iX_i'\pi_i)^{-1}X_{i}\|_{2}^{7/2}\left(|\varepsilon_{i,j}| \xi_{i,j}+\pi_i|\Gamma_{i,j}|\right)^{7/2}\right) \lessp \alpha_n^{-1}.
\end{align*}
and
\begin{align*}
    E\max_{j}\|\alpha_n^{1/2} \omega_{i,j}\|_{2}^{9/2}
    &\leq \alpha_n^{9/4} E\left(\|E(X_iX_i'\pi_i)^{-1}X_{i}\|_{2}^{9/2}\max_{j}\left(|\varepsilon_{i,j}| \xi_{i,j}+\pi_i|\Gamma_{i,j}|\right)^{9/2}\right) \lessp \alpha_n^{-9/4} \log^{9/2} n.
\end{align*}
Since the $\ell_{1}$-norm of each row of $B$ is bounded, we have
\begin{align*}
& \max_{r_1 = 1,\cdots,|\mathcal G|k}E    [\alpha_n^{1/2}B_{r_1,\cdot}\omega_{i}]^{3} \lessp \max_{j} E \|\alpha_n^{1/2} \omega_{i,j}\|_{2}^{3} \lessp \alpha_n^{-1/2}\\
& \max_{r_1  = 1,\cdots,|\mathcal G|k}E    [\alpha_n^{1/2}B_{r_1,\cdot}\omega_{i}]^{4} \lessp \max_{j} E \|\alpha_n^{1/2} \omega_{i,j}\|_{2}^{4} \lessp \alpha_n^{-1}\\
&E\max_{r_1 = 1,\cdots,|\mathcal G|k}    [\alpha_n^{1/2}B_{r_1,\cdot}\omega_{i}]^{9/2} \lessp E \max_{j}\|\alpha_n^{1/2} \omega_{i,j}\|_{2}^{9/2} \lessp \alpha_n^{-9/4} \log^{9/2} n.
\end{align*}

Hence, conditions (M.1), (M.2) and (E.2) in Proposition
2.1 of \cite{ChernozhukovChetverikov2017} are satisfied with $q=9/2$
and $B_{n} = C\alpha_n^{-1/2} \log^2 n$. Therefore, by Proposition 2.1 of \cite{ChernozhukovChetverikov2017},
we have that
\begin{align}
& \sup_{x\in\RR}\left|P\left(\left\| n^{-1/2} \alpha_n^{1/2} \sumi B\omega_{i}\right\|_{\infty}\leq x\right)-P\left(\alpha_n^{1/2}T_{(B)}
\leq x\right)\right|\nonumber \\
\lesssim & \,  \left(\frac{(\log(rn))^{9}}{n \alpha_n}\right)^{1/6}+\left(\frac{(\log(rn))^{5}}{n^{1-2/(9/2)} \alpha_n}\right)^{1/3}.\label{eq: thm inference beta eq 3}
\end{align}

For $Z=(Z_{1},...,Z_r)'\in\RR^{r}$, note that $T_{(B)}=\|Z\|_{\infty}=\max_{1\leq j\leq r}|Z_{j}|$
can be written as $\max\{Z_{1},Z_{2},...,Z_r,-Z_{1},-Z_{2},...,-Z_r\}$.
Hence, by Corollary 1 of \cite{chernozhukov2015comparison}, it follows
that for any $\varepsilon>0$,
\begin{equation}
\sup_{x\in\RR}P\left(\alpha_n^{1/2}T_{(B)}\in[x-\varepsilon,x+\varepsilon]\right)\leq\kappa_{1}\varepsilon\sqrt{\log(r/\varepsilon)},\label{eq: thm inference beta eq 3.5}
\end{equation}
where $\kappa_{1}>0$ is a constant depending only on $C_{2},C_{3}$.

By Theorem \ref{thm: beta est},
\begin{align*}
\left\| \tbeta-\beta-n^{-1}\sumi \omega_{i}\right\|_{\infty}\lessp \kappa_n.
\end{align*}
Since the $\ell_{1}$-norm of each row of $B$ is bounded, it follows that
\begin{equation}
\left\| B(\tbeta-\beta)-n^{-1}\sumi B\omega_{i}\right\|_{\infty}\lessp \kappa_n.\label{eq: thm inference beta eq 4}
\end{equation}
Let $M_{n}$ be a sequence such that $(n\alpha_n)^{1/2}\kappa_n\ll M_{n}$
and $M_{n}\sqrt{\log m}=o(1)$. This is possible because $(n\alpha_n)^{1/2}\kappa_n\ll(\log m)^{-1/2}$. The choice of $M_{n}$
and (\ref{eq: thm inference beta eq 4}) imply that
\begin{align*}
\sup_{x\in\RR}\left(P\left(\|\sqrt{n \alpha_n }B(\tbeta-\beta)\|_{\infty}\leq x\right)-P\left(\left\| \alpha_n^{1/2} n^{-1/2}\sumi B\omega_{i}\right\|_{\infty}\leq x+M_{n}\right)\right)\leq o(1).
\end{align*}
By (\ref{eq: thm inference beta eq 3}), we have
\begin{align*}
\sup_{x\in\RR}\left(P\left(\|\sqrt{n \alpha_n}B(\tbeta-\beta)\|_{\infty}\leq x\right)-P\left( \alpha_n^{1/2}T_{(B)}\leq x+M_{n}\right)\right)\leq o(1).
\end{align*}
Therefore,
\begin{equation}
\sup_{x\in\RR}\left(P\left(\|\sqrt{n \alpha_n }B(\tbeta-\beta)\|_{\infty}\leq x\right)-P\left(\alpha_n^{1/2} T_{(B)}\leq x\right)\right)\leq o(1)+\sup_{x\in\RR}P\left(\alpha_n^{1/2} T_{(B)}\in(x,x+M_{n}]\right).\label{eq: thm inference beta eq 6}
\end{equation}
Similarly, (\ref{eq: thm inference beta eq 4}) and the definition
of $M_{n}$ imply
\begin{align*}
\sup_{x\in\RR}\left(P\left(\left\| (n\alpha_n)^{-1/2}\sumi B\omega_{i}\right\|_{\infty}\leq x-M_{n}\right)-P\left(\|\sqrt{n\alpha_n}B(\tbeta-\beta)\|_{\infty}\leq x\right)\right)\leq o(1)
\end{align*}
and thus we can combine it with (\ref{eq: thm inference beta eq 3}),
obtaining
\begin{align*}
\sup_{x\in\RR}\left(P\left(\alpha_n^{1/2} T_{(B)}\leq x-M_{n}\right)-P\left(\|\sqrt{n\alpha_n}B(\tbeta-\beta)\|_{\infty}\leq x\right)\right)\leq o(1),
\end{align*}
which means
\begin{equation}
\sup_{x\in\RR}\left(P\left(\alpha_n^{1/2} T_{(B)}\leq x\right)-P\left(\|\sqrt{n\alpha_n}B(\tbeta-\beta)\|_{\infty}\leq x\right)\right)\leq o(1)+\sup_{x\in\RR}P\left(\alpha_n^{1/2}T_{(B)}\in[x-M_{n},x)\right).\label{eq: thm inference beta eq 7}
\end{equation}
Combining (\ref{eq: thm inference beta eq 6}) and (\ref{eq: thm inference beta eq 7}),
we have
\begin{align*}
&\quad \sup_{x\in\RR}\left|P\left(\|\sqrt{n\alpha_n}B(\tbeta-\beta)\|_{\infty}\leq x\right)-P\left(\alpha_n^{1/2} T_{(B)}\leq x\right)\right|\\
&\leq o(1)+\sup_{x\in\RR}P\left(\alpha_n^{1/2}T_{(B)}\in[x-M_{n},x+M_{n}]\right).
\end{align*}
It follows, by (\ref{eq: thm inference beta eq 3.5}), that
\begin{align*}
\sup_{x\in\RR}\left|P\left(\|\sqrt{n \alpha_n}B(\tbeta-\beta)\|_{\infty}\leq x\right)-P\left(\alpha_n^{1/2}T_{(B)}\leq x\right)\right|\leq o(1)+O(M_{n}\sqrt{\log(r/M_{n})})\overset{\texti}{=}o(1),
\end{align*}
where (i) holds by $r\leq mk$ and the definition of $M_{n}$.\\[6pt]
\textbf{Step 2:} show that $\sup_{x\in\RR}\left|P\left(\alpha_n^{1/2}T_{(B)}\leq x\right)-P\left(\alpha_n^{1/2}T_{(B)}^{*}\leq x\mid\{\home_{i}\}_{i=1}^{n}\right)\right|=o_{P}(1)$.\\
Recall that $\alpha_n^{1/2}T_{(B)}=\|Z\|_{\infty}$, where $Z\sim N(0,V_{n})$ and
$V_{n}=n^{-1}\alpha_n^{1/2}\sumi EB\omega_{i}\omega_{i}'B'$. Also recall
that $\alpha_n^{1/2}T_{(B)}^{*}=\|Z^{*}\|_{\infty}$, where $Z^{*}$ is a mean-zero
Gaussian vector with covariance matrix $\hV_{n}=V_{n}=n^{-1}\alpha_n \sumi B\home_{i}\home_{i}'B'$.
By Lemmas \ref{lem:  BS 1} and \ref{lem:  BS 2}, we have
\begin{align*}
\| V_{n}-\hV_{n}\|_{\infty}\lessp \alpha_n^{-1/2}(m^{-1/2}+n^{-1/2})(\log n)^{5/2}.
\end{align*}
Note that the diagonal entries of $V_{n}$ are in $[C_{2},C_{3}]$.
By Theorem 2 of \cite{chernozhukov2015comparison}, it suffices to
show that
\begin{align*}
\zeta(\log(r/\zeta))^{2}=o_{P}(1),
\end{align*}
where $\zeta=\| V_{n}-V_{n}\|_{\infty}$.
Note that $\zeta(\log(r/\zeta))^{2}=\zeta(\log r-\log\zeta)^{2}\leq\zeta[2(\log r)^{2}+2(\log\zeta)^{2}]$.
Since $r\lesssim m$ and $\log m\asymp\log n$, it follows that
\begin{align*}
\zeta(\log r)^{2}
& \lessp \zeta(\log m)^{2}\lessp \alpha_n^{-1/2}(m^{-1/2}+n^{-1/2})(\log n)^{9/2} = o_{P}(1).
\end{align*}
The desired result follows.
\end{proof}

\section{Pattern of the MSE Values With Different Ranks When The Initial Estimates Are Used}\label{sec: sim misspecify rank}

\subsection{A heuristic argument }\label{sec: argument}
We have an interesting finding that the MSE-based method for rank selection cannot be constructed based on the initial estimates $\hbeta$, $\hL^k$ and $\hF^k$, where $\hL^k$ and $\hF^k$ are rank $k$ SVD estimates of $L$ and $F$, because the MSE value may not be decreasing as $k$ increases when the observation rate of the responses is small. For simplicity of illustration, we let $\pi_i=\pi$, so the observation rate is a constant for all responses. Let $\pi<0.5$. A heuristic argument is given below, and the numerical illustration is provided in Section \ref{sec: num illu}.

The MSE based on the estimates $\hbeta$, $\hL^k$ and $\hF^k$ is given as follows.
\begin{align}
    \mse(k, \hGamma^k)=\frac{1}{nm\hpi}\left\|\Xi\circ\left(Y - X\hbeta^{'} - \hL^k \hF^{k'}\right)\right\|_F^2\label{eq: mse(k,hGamma)}.
\end{align}
Each entry contributing to the Frobenius norm in \eqref{eq: mse(k,hGamma)} is $\xi^2_{i,j}(\hpi W_{i,j}-\hGamma^k_{i,j})^2$, and we know that $\xi_{i,j}W_{i,j}=W_{i,j}$ by definition and $\xi_{i,j}^2=\xi_{i,j}$ since $\xi_{i,j}\in \{0,1\}$. Define $\Delta_k=- 2 \hpi(nm)^{-1}\sumi\sumj \left(W_{i,j}-\hGamma^k_{i,j}\right)\hGamma^k_{i,j}\left(\hpi^{-1}\xi_{i,j}-1\right)$.
We can write \eqref{eq: mse(k,hGamma)} as
\begin{align}
     \mse(k, \hGamma^k)
         =\, & \hpi(nm)^{-1}\sumi\sumj \left[\,\left(W_{i,j}-\hGamma^k_{i,j}\right)^2 + \left(\hGamma^k_{i,j}\right)^2\left(\hpi^{-1}\xi_{i,j}-1\right)^2\,\right] \nonumber\\
         & - 2 \hpi(nm)^{-1}\sumi\sumj \left(W_{i,j}-\hGamma^k_{i,j}\right)\hGamma^k_{i,j}\left(\hpi^{-1}\xi_{i,j}-1\right) \nonumber\\
         =\, &\hpi(nm)^{-1} \left[\,\sum_{l>k}\sigma^2_l(W) + \sumi\sumj \left(\hGamma^k_{i,j}\right)^2\left(\hpi^{-2}\xi_{i,j}-2\hpi^{-1}\xi_{i,j}+1\right)\,\right] + \Delta_k \nonumber\\
         =\, &\hpi(nm)^{-1} \left[\,\|W\|_F^2 + \left\|\hGamma^k\circ \Xi\right\|_F^2\left(\hpi^{-2}-2\hpi^{-1}\right)\,\right] + \Delta_k \nonumber\\
         \overset{\texti}{=\,}&\pi(nm)^{-1} \left[\,
         \|W\|^2_F + \left(\pi^{-1}-2\right)\sum^k_{l=1}\sigma^2_l(W)\,\right]+o_P(1) +\Delta_k \label{eq: mse(k,hGamma) - 2},
\end{align}
where equation $\texti$ can be derived by using the same technique as Lemma \ref{lem: general asymp prop} in Appendix \ref{app: proof of beta est g=1}. In the form of \eqref{eq: mse(k,hGamma) - 2}, if $\pi^{-1}>2$ (or $\pi<0.5$), the first term increases with $k$, so its behavior depends somehow on the value $\Delta_k$. In fact, we have shown in Figure \ref{fig: misspecify rank} in Section \ref{sec: num illu} that in some cases, $\Delta_k$ is small, so the $\mse(k,\hGamma^k)$ value increases as $k$ increases. Therefore, we can not use \eqref{eq: mse(k,hGamma)} in a rank estimation criterion.

\subsection{Numerical illustration}\label{sec: num illu}
In Section \ref{sec: rank est} of the main article, we have discussed the motivation for using $\eIC$ which is based on MSE using the iterative LS estimates. The basic idea is to avoid any missing values instead of padding them as zeros when calculating the MSE. We have also provided an interesting finding in Section \ref{sec: argument} about why the initial estimates $\hGamma^k$ cannot be used to define $\eIC$: $\mse(k, \hGamma^k)$ does not always decrease as $k$ increases when $\pi<0.5$ in Section \ref{sec: argument}.

\begin{figure}[ht]
    \tabfnsymbol \centering
    \includegraphics[width=\linewidth]{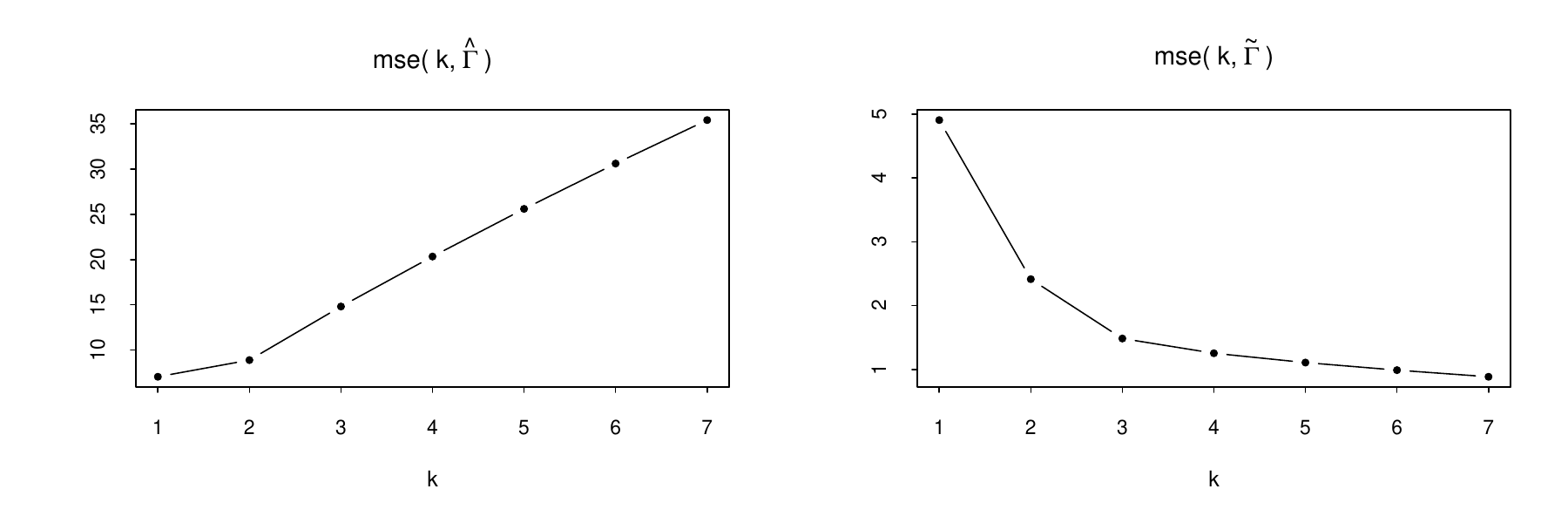}\vspace{-15pt}
    \captionx[Simulation - miss-specified rank estimation criterion]{Values of $\mse(k, \hGamma^k)$ and  $\mse(k, \tGamma^k)$ at step $g=3$, versus different  rank $k$ values. Each point is the average value of 500 simulations. Settings: $n,m=200$, $\pi=0.2$ from DGP 1.}
    \label{fig: misspecify rank}
\end{figure}

Here, we demonstrate the aforementioned phenomenon through simulations to show that the value of $\mse(k,\hGamma^k)$ can increase with $k$ when the initial estimate  $\hGamma^k$  is used. In Figure \ref{fig: misspecify rank}, a plot of $\mse(k,\hGamma^k)$ versus $k$ is shown on the left, and the values are calculated using data generated from DGP 1 with $n,m=200$, $r=3$, $\pi=0.2$. For comparison, the right side shows the value of $\mse(k, \tGamma^k)$, which is used in $\eIC$, where $\tGamma^k$ is the iterative LS estimate obtained at step=3. We can see that $\mse(k,\tGamma^k )$ decreases steadily whereas $\mse(k,\hGamma^k)$ increases with $k$.

\section{Additional Numerical Results}\label{sec: add num}

\subsection{Additional simulation and empirical results}\label{sec: add num sim}
\renewcommand{\thetable}{S\arabic{table}}
\renewcommand{\thefigure}{S\arabic{figure}}
\renewcommand{\arraystretch}{.6}

\begin{figure}[ht]
        \tabfnsymbol \centering
        \includegraphics[width=\linewidth]{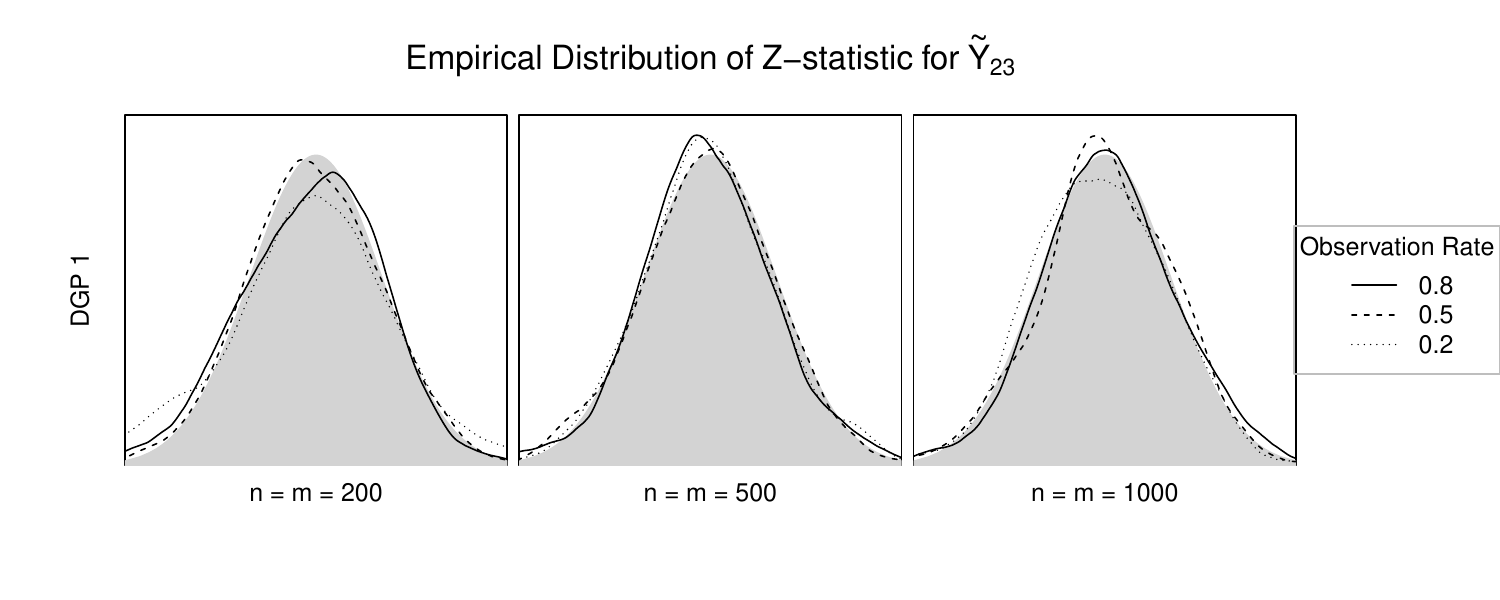}
        \captionx[Sup - empirical distribution of estimators]{The empirical distribution of $\frac{\tY_{ij}-\mu_{ij}}{\hsigma_{n,m}(\tY_{ij})}$ in different simulation settings of DGP1. The shaded area is the density of standard normal distribution.}
        \label{fig: asymp dist beta gamma}
\end{figure}

\begin{figure}[ht]
    \tabfnsymbol \centering
    \includegraphics[width=0.9\linewidth]{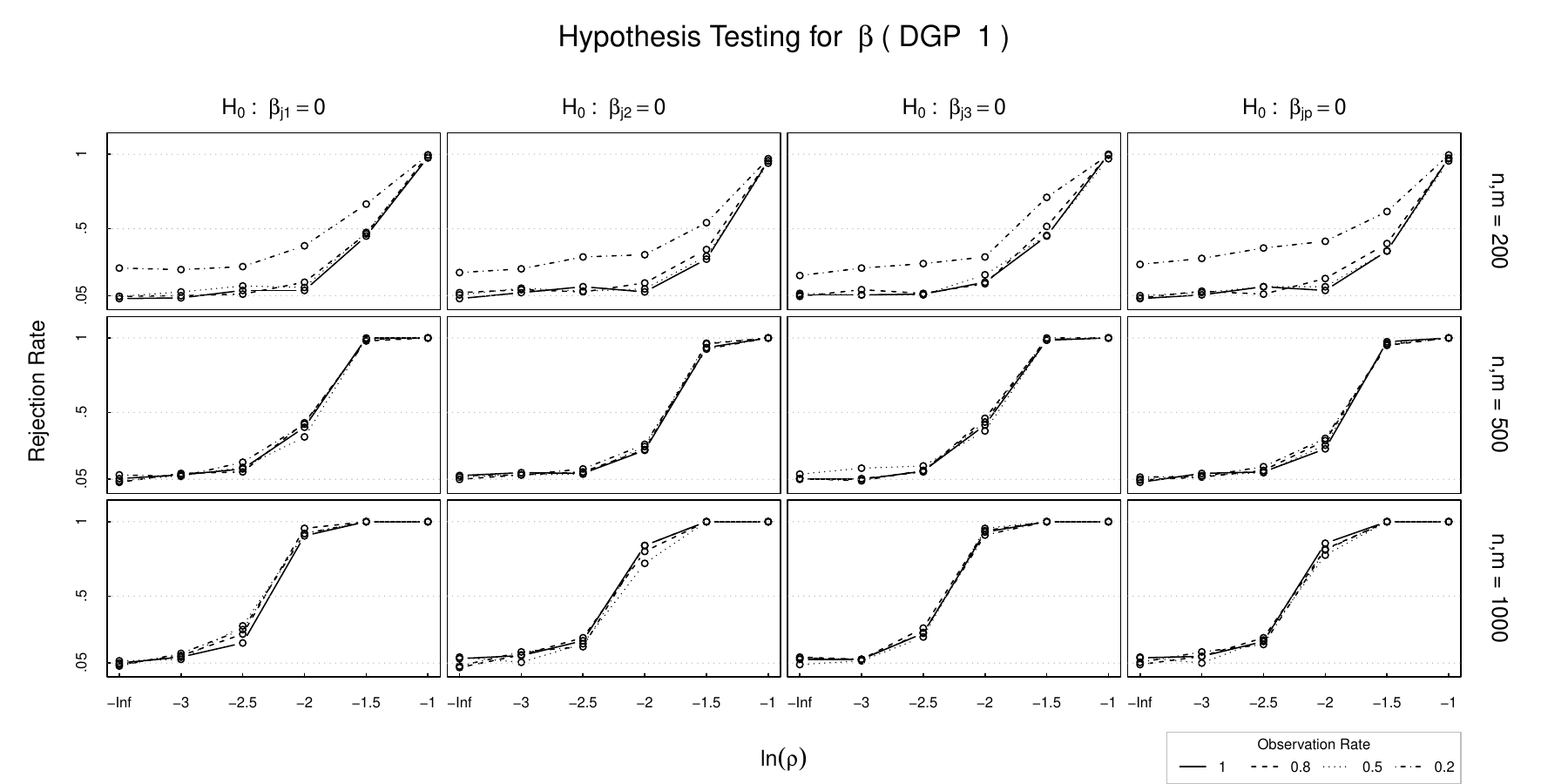}\vspace{-5pt}
    \captionx[Sup - empirical rejection rates of simultaneous testing GDP 1]{Empirical rejection rates at level $\alpha = 0.05$. Each column represents a hypothesis, and each row represents a sample size. When $x = \ln(\rho)=-{\rm Inf}$, the null hypothesis is true. }
    \label{fig: hypothesis test for beta dgp 1}
\end{figure}

\begin{figure}[ht]
    \tabfnsymbol \centering
    \includegraphics[width=\linewidth]{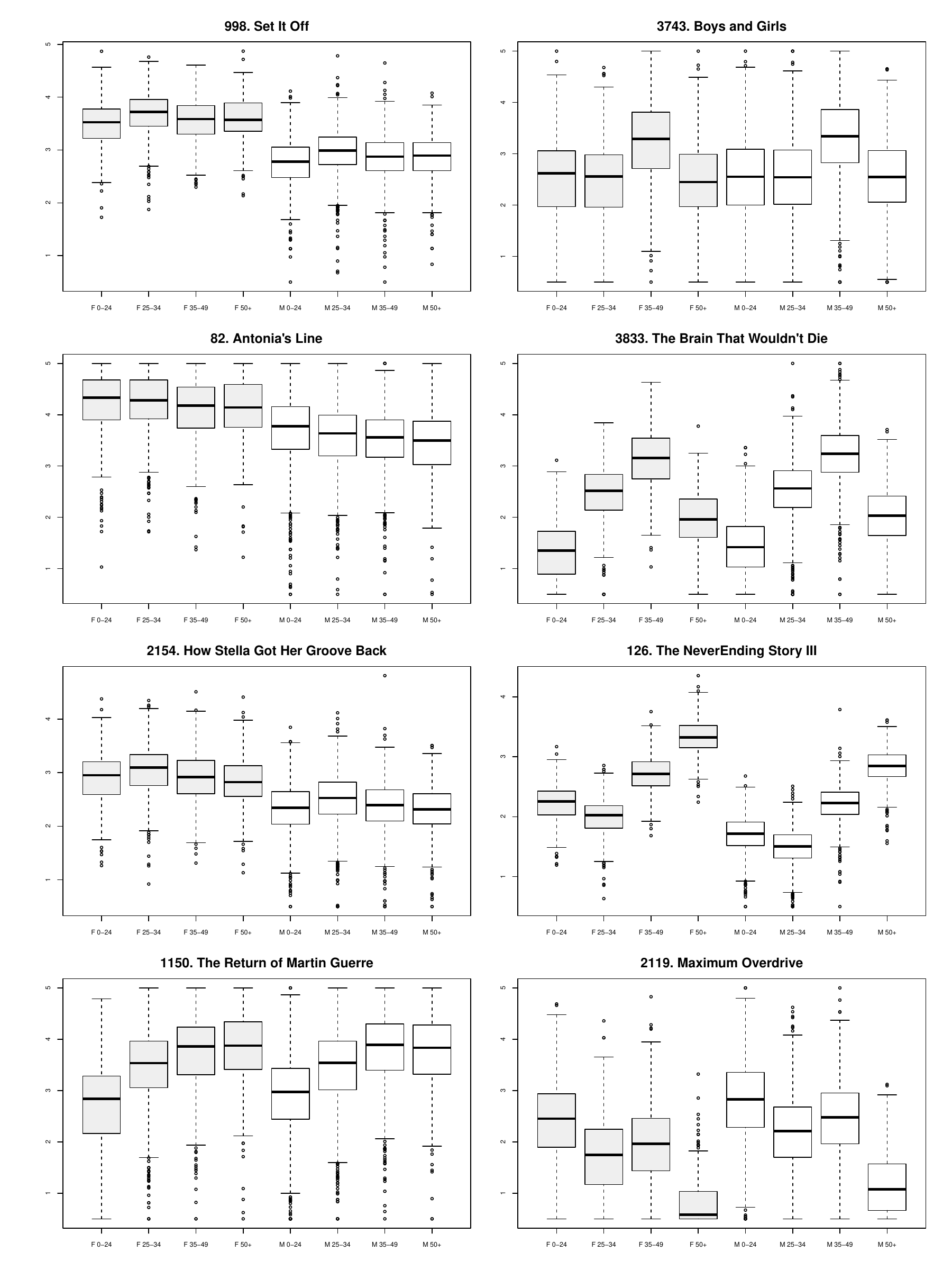}
    \captionx[More boxplot examples of MovieLens data]{Boxplots of the estimated ratings in different gender and age groups for some movies.}\label{fig: ml boxplots plus}
\end{figure}

\begin{figure}[ht]
    \tabfnsymbol \centering
    \includegraphics[width = \linewidth]{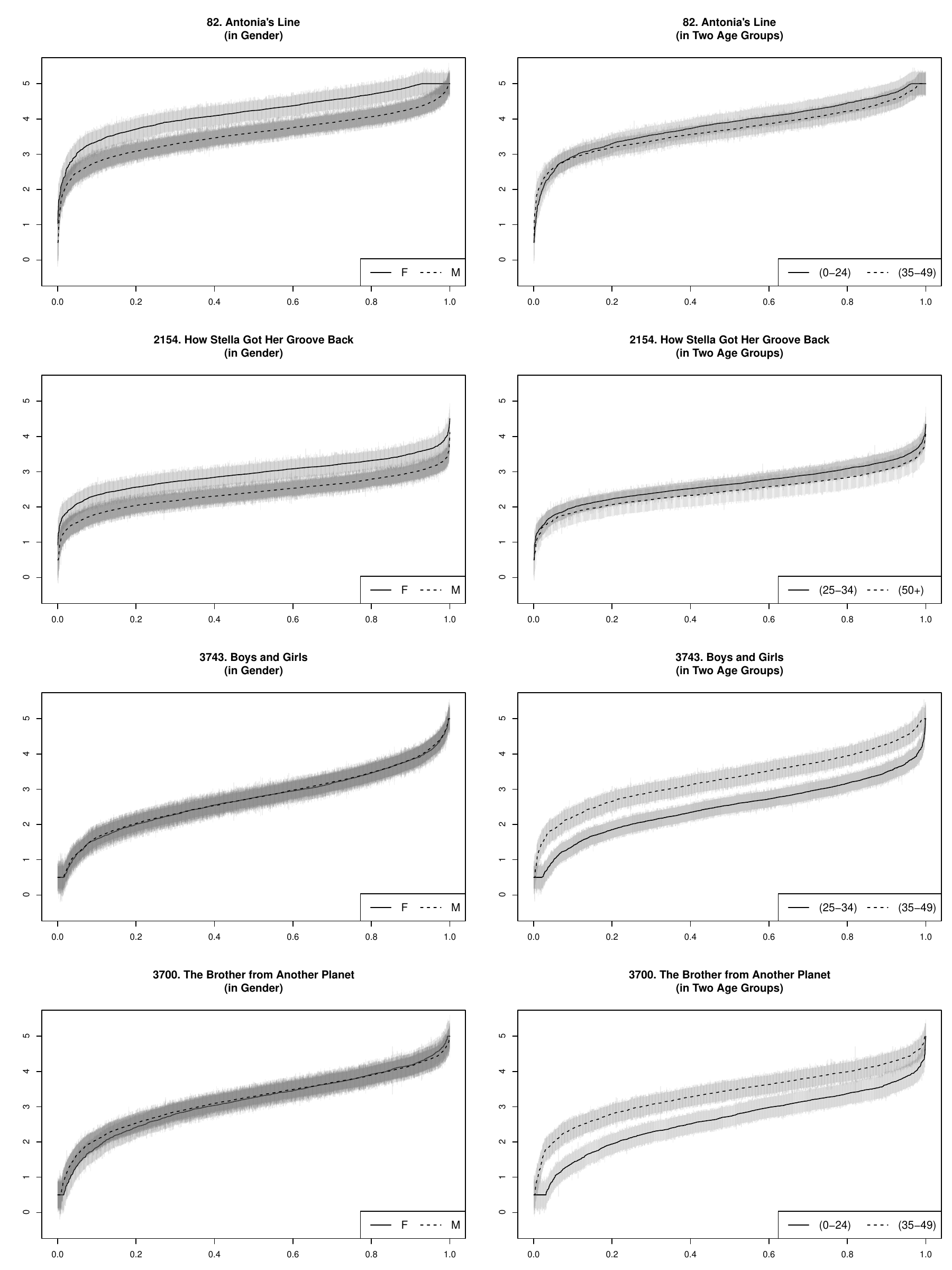}
    \captionx[More examples of estimated ratings and 90\% point-wise CI for MovieLens data]{Estimated ratings and 90\% point-wise confidence intervals in different groups. The $y$-axis is the rating and the $x$-axis is the percentile. Ratings are grouped by gender or age.}\label{fig: ml rating CI plus}
\end{figure}

The tables and figures in this section provide all numerical results for DGP1 in the simulation studies section (Section \ref{sec: simulation}), and additional results of the real data analysis in Section \ref{sec: application}. Because of the space limit, we present all simulation results for DGP2 in the main text, while relegating partial simulation results for DGP1 to Section \ref{sec: add num sim} in the Supplement.
\begin{itemize}
    \item Table \ref{tab: mse beta dgp1}, \ref{tab: mse gamma dgp1}: the full comparison of empirical MSEs between the iterative PCA and iterative LS methods under DGP 1 (Associated with Section \ref{sec: sim performance}, Table \ref{tab: mse dgp1}).
    \item Table \ref{tab: cputime dgp1 full}: full table of computational time comparison between the iterative PCA and iterative LS algorithms. (Associated with Section \ref{sec: sim performance}, Table \ref{tab: cputime dgp1}).
    \item Table \ref{tab: bias and coverage dgp1}: empirical bias and 95\% CI coverage rate of the LS estimator under DGP 1 (Associated with Section \ref{sec: sim performance}, Table \ref{tab: bias and coverage dgp2}).
    \item Table \ref{tab: ml hypo test most sig}: the results of the hypothesis testing for each movie in Section \ref{sec: app insight}. Movies with the top 10 smallest p-values in each test are shown in order.
    \item Figure \ref{fig: asymp dist beta gamma}: empirical distribution of the $Z$-statistic of the selected $\hat{E}(Y_{i,j})$ of DGP 1. (Associated with Section \ref{sec: sim performance}, Figure \ref{fig: asymp dist ey}).
    \item Figure \ref{fig: hypothesis test for beta dgp 1}: rejection rates of hypothesis testing for $\beta$ in different settings under DGP 1. (Associated with Section \ref{sec: sim hypo}, Figure \ref{fig: hypothesis test for beta dgp 2 rank 3}).
    \item Figure \ref{fig: ml boxplots plus}, \ref{fig: ml rating CI plus}: additional boxplots and point-wise CI curves of predicted movie ratings. (Associated with Section \ref{sec: app insight}, Figure \ref{fig: ml boxplots}, \ref{fig: ml rating CI}).
\end{itemize}

\begin{table}[ht]
\tabfnsymbol \centering
    \captionx[Full Table - performance comparison in MSE - part 1/2]{The $\MSE$ of different estimators of $\beta$ in each simulation setting of DGP1.}
    \label{tab: mse beta dgp1}\centering
    \vspace{12pt}{\scriptsize
    \begin{tabularx}{\textwidth}{lllRRRRRRR}\toprule
        \multicolumn{3}{c}{DGP 1} & \multicolumn{1}{c}{Initial} & \multicolumn{3}{c}{iterative PCA} & \multicolumn{3}{c}{iterative LS}\\
        \cmidrule(lr){1-3}\cmidrule(lr){4-4}\cmidrule(lr){5-7}\cmidrule(lr){8-10}
$n,m $  &  $\pi$  &  $\hbeta$  &  $\tbeta_{pca}^{(1)}$  &  $\tbeta_{pca}^{(2)}$  &  $\tbeta_{pca}^{(3)}$  &  $\tbeta_{pca}^{(c)}$  &  $\tbeta_{ls}^{(1)}$  &  $\tbeta_{ls}^{(2)}$  &  $\tbeta_{ls}^{(3)}$\\
\cline{1-10}
$200 $  &  $1  $  &  $ 0.117$  &  $     -$  &  $     -$  &  $     -$  &  $     -$  &  $     -$  &  $     -$  &  $     -$\\
$    $  &  $0.8$  &  $ 0.145$  &  $ 0.119$  &  $ 0.119$  &  $ 0.119$  &  $ 0.119$  &  $ 0.119$  &  $ 0.119$  &  $ 0.119$\\
$    $  &  $0.5$  &  $ 0.230$  &  $ 0.132$  &  $ 0.127$  &  $ 0.126$  &  $ 0.125$  &  $ 0.132$  &  $ 0.126$  &  $ 0.125$\\
$    $  &  $0.2$  &  $ 0.614$  &  $ 0.389$  &  $ 0.281$  &  $ 0.258$  &  $ 0.157$  &  $ 0.389$  &  $ 0.231$  &  $ 0.197$\\
\cline{1-10}
$500 $  &  $1  $  &  $ 0.043$  &  $     -$  &  $     -$  &  $     -$  &  $     -$  &  $     -$  &  $     -$  &  $     -$\\
$    $  &  $0.8$  &  $ 0.054$  &  $ 0.044$  &  $ 0.044$  &  $ 0.044$  &  $ 0.044$  &  $ 0.044$  &  $ 0.044$  &  $ 0.044$\\
$    $  &  $0.5$  &  $ 0.087$  &  $ 0.047$  &  $ 0.047$  &  $ 0.047$  &  $ 0.047$  &  $ 0.047$  &  $ 0.047$  &  $ 0.047$\\
$    $  &  $0.2$  &  $ 0.224$  &  $ 0.084$  &  $ 0.073$  &  $ 0.068$  &  $ 0.057$  &  $ 0.084$  &  $ 0.060$  &  $ 0.058$\\
\cline{1-10}
$1000$  &  $1  $  &  $ 0.023$  &  $     -$  &  $     -$  &  $     -$  &  $     -$  &  $     -$  &  $     -$  &  $     -$\\
$    $  &  $0.8$  &  $ 0.028$  &  $ 0.023$  &  $ 0.023$  &  $ 0.023$  &  $ 0.023$  &  $ 0.023$  &  $ 0.023$  &  $ 0.023$\\
$    $  &  $0.5$  &  $ 0.044$  &  $ 0.024$  &  $ 0.024$  &  $ 0.024$  &  $ 0.024$  &  $ 0.024$  &  $ 0.024$  &  $ 0.024$\\
$    $  &  $0.2$  &  $ 0.110$  &  $ 0.033$  &  $ 0.031$  &  $ 0.030$  &  $ 0.029$  &  $ 0.033$  &  $ 0.029$  &  $ 0.029$
\\
        \bottomrule \\[-5pt]
    \end{tabularx}
    }
\end{table}
\begin{table}[ht]
\tabfnsymbol \centering
    \captionx[Full Table - performance comparison in MSE - part 2/2]{The $\MSE$ of different estimators of $\Gamma$ in each simulation setting of DGP1.}
    \label{tab: mse gamma dgp1}\centering
    \vspace{12pt}{\scriptsize
    \begin{tabularx}{\textwidth}{lllRRRRRRR}\toprule
        \multicolumn{3}{c}{DGP 1} & \multicolumn{1}{c}{Initial} & \multicolumn{3}{c}{iterative PCA} & \multicolumn{3}{c}{iterative LS}\\
        \cmidrule(lr){1-3}\cmidrule(lr){4-4}\cmidrule(lr){5-7}\cmidrule(lr){8-10}
$n,m $  &  $\pi$  &  $\tbeta_{ls}^{(c)}$  &  $\hGamma$  &  $\tGamma_{pca}^{(1)}$  &  $\tGamma_{pca}^{(2)}$  &  $\tGamma_{pca}^{(3)}$  &  $\tGamma_{pca}^{(c)}$  &  $\tGamma_{ls}^{(1)}$  &  $\tGamma_{ls}^{(2)}$\\
\cline{1-10}
$200 $  &  $1  $  &  $     -$  &  $ 0.250$  &  $     -$  &  $     -$  &  $     -$  &  $     -$  &  $     -$  &  $     -$\\
$    $  &  $0.8$  &  $ 0.119$  &  $ 0.430$  &  $ 0.269$  &  $ 0.259$  &  $ 0.258$  &  $ 0.258$  &  $ 0.259$  &  $ 0.258$\\
$    $  &  $0.5$  &  $ 0.125$  &  $ 1.121$  &  $ 0.513$  &  $ 0.372$  &  $ 0.321$  &  $ 0.283$  &  $ 0.305$  &  $ 0.285$\\
$    $  &  $0.2$  &  $ 0.176$  &  $ 7.469$  &  $ 4.186$  &  $ 3.456$  &  $ 3.075$  &  $ 0.419$  &  $ 1.975$  &  $ 1.021$\\
\cline{1-10}
$500 $  &  $1  $  &  $     -$  &  $ 0.096$  &  $     -$  &  $     -$  &  $     -$  &  $     -$  &  $     -$  &  $     -$\\
$    $  &  $0.8$  &  $ 0.044$  &  $ 0.167$  &  $ 0.102$  &  $ 0.099$  &  $ 0.099$  &  $ 0.099$  &  $ 0.099$  &  $ 0.099$\\
$    $  &  $0.5$  &  $ 0.047$  &  $ 0.399$  &  $ 0.182$  &  $ 0.131$  &  $ 0.116$  &  $ 0.108$  &  $ 0.110$  &  $ 0.108$\\
$    $  &  $0.2$  &  $ 0.058$  &  $ 2.354$  &  $ 1.447$  &  $ 1.065$  &  $ 0.835$  &  $ 0.150$  &  $ 0.341$  &  $ 0.162$\\
\cline{1-10}
$1000$  &  $1  $  &  $     -$  &  $ 0.048$  &  $     -$  &  $     -$  &  $     -$  &  $     -$  &  $     -$  &  $     -$\\
$    $  &  $0.8$  &  $ 0.023$  &  $ 0.084$  &  $ 0.052$  &  $ 0.050$  &  $ 0.050$  &  $ 0.050$  &  $ 0.050$  &  $ 0.050$\\
$    $  &  $0.5$  &  $ 0.024$  &  $ 0.195$  &  $ 0.090$  &  $ 0.064$  &  $ 0.057$  &  $ 0.055$  &  $ 0.055$  &  $ 0.055$\\
$    $  &  $0.2$  &  $ 0.029$  &  $ 0.726$  &  $ 0.479$  &  $ 0.334$  &  $ 0.246$  &  $ 0.074$  &  $ 0.085$  &  $ 0.074$
\\
        \bottomrule \\[-5pt]
    \end{tabularx}
    }
\end{table}

\begin{table}[ht]
    \tabfnsymbol \centering
    \captionx[Full Table - computing time]{Computing time in seconds\footnotemark[1] in each setting of DGP1.}\label{tab: cputime dgp1 full}
    \vspace{12pt}{\footnotesize
    \begin{tabularx}{\textwidth}{llRRRRRRR}\toprule
    \multicolumn{2}{c}{} & \multicolumn{3}{c}{Time in sec} & \multicolumn{2}{c}{Number of} & \multicolumn{2}{c}{Ave. time for} \\
    \multicolumn{2}{c}{DGP 1} & \multicolumn{3}{c}{to get estimators} & \multicolumn{2}{c}{iterations} & \multicolumn{2}{c}{1 iteration}\\
    \cmidrule(lr){1-2}\cmidrule(lr){3-5}\cmidrule(lr){6-7}\cmidrule(lr){8-9}
$n,m $  &  $\pi$  &  $\hGamma$  &  $\tGamma^{(c)}_{ls}$  &  $\tGamma^{(c)}_{pca}$  &  $\tGamma^{(c)}_{ls}$  &  $\tGamma^{(c)}_{pca}$  &  $     ls$  &  $    pca$\\
\cline{1-9}  
$200 $  &  $0.8$  &  $   0.04$  &  $   0.16$  &  $   0.38$  &  $   4.02$  &  $   9.29$  &  $   0.04$  &  $  0.041$\\
$    $  &  $0.4$  &  $   0.05$  &  $   0.28$  &  $   1.37$  &  $   7.01$  &  $  33.62$  &  $   0.04$  &  $  0.041$\\
$    $  &  $0.2$  &  $   0.04$  &  $   0.60$  &  $   4.01$  &  $  15.72$  &  $  99.94$  &  $   0.04$  &  $  0.040$\\
\cline{1-9}  
$500 $  &  $0.8$  &  $   0.41$  &  $   0.44$  &  $   3.18$  &  $   3.56$  &  $   7.52$  &  $   0.12$  &  $  0.423$\\
$    $  &  $0.4$  &  $   0.42$  &  $   0.58$  &  $  10.06$  &  $   5.01$  &  $  23.48$  &  $   0.12$  &  $  0.429$\\
$    $  &  $0.2$  &  $   0.41$  &  $   0.80$  &  $  27.54$  &  $   7.40$  &  $  64.53$  &  $   0.11$  &  $  0.427$\\
\cline{1-9}  
$1000$  &  $0.8$  &  $   3.43$  &  $   1.05$  &  $  24.24$  &  $   3.00$  &  $   6.90$  &  $   0.35$  &  $  3.514$\\
$    $  &  $0.4$  &  $   3.03$  &  $   1.16$  &  $  62.22$  &  $   4.03$  &  $  19.88$  &  $   0.29$  &  $  3.130$\\
$    $  &  $0.2$  &  $   2.96$  &  $   1.39$  &  $ 149.12$  &  $   5.52$  &  $  48.80$  &  $   0.25$  &  $  3.056$
\\
    \bottomrule \\[-5pt]
    \tabfootnote{1}{The values are calculated based on 100 simulation replicates}{9}{.95}
    \end{tabularx}%
    }
\end{table}

\begin{table}[ht]
    \tabfnsymbol \centering
    \captionx[Simulation - bias and CI coverage rate]{Average bias and 95\% CI coverage rate for some estimators\footnotemark[1].}\label{tab: bias and coverage dgp1}
    \vspace{12pt}{\footnotesize
    \begin{tabularx}{\textwidth}{llRRRRRRRRRRRR}\toprule
        \multicolumn{2}{c}{DGP 1}  & \multicolumn{6}{c}{Bias ($10^{-2}$)} & \multicolumn{6}{c}{95\% CI coverage rate} \\
        \cmidrule(lr){1-2}\cmidrule(lr){3-8}\cmidrule(lr){9-14}
$n,m $  &  $\pi$  &  $\tGamma_{11}$  &  $\tGamma_{23}$  &  $\tGamma_{35}$  &  $\tY_{11}$  &  $\tY_{23}$  &  $\tY_{35}$  &  $\tGamma_{11}$  &  $\tGamma_{23}$  &  $\tGamma_{35}$  &  $\tY_{11}$  &  $\tY_{23}$  &  $\tY_{35}$\\
\cline{1-14}  
$200 $  &  $0.8$  &  $  -2.2$  &  $  -5.8$  &  $   3.3$  &  $  -0.5$  &  $  -0.0$  &  $  -0.7$  &  $  0.96$  &  $  0.95$  &  $  0.92$  &  $  0.95$  &  $  0.95$  &  $  0.95$\\
$    $  &  $0.4$  &  $   0.4$  &  $   0.1$  &  $   1.0$  &  $  -0.7$  &  $   0.9$  &  $  -2.2$  &  $  0.94$  &  $  0.95$  &  $  0.95$  &  $  0.94$  &  $  0.95$  &  $  0.92$\\
$    $  &  $0.2$  &  $  -3.5$  &  $   4.4$  &  $   2.3$  &  $  -5.7$  &  $   6.0$  &  $   2.6$  &  $  0.88$  &  $  0.89$  &  $  0.91$  &  $  0.88$  &  $  0.87$  &  $  0.90$\\
\cline{1-14}  
$500 $  &  $0.8$  &  $  -0.4$  &  $  -0.2$  &  $  -1.0$  &  $  -0.7$  &  $   0.8$  &  $   1.0$  &  $  0.94$  &  $  0.95$  &  $  0.97$  &  $  0.94$  &  $  0.97$  &  $  0.97$\\
$    $  &  $0.4$  &  $  -2.9$  &  $   1.7$  &  $   0.3$  &  $  -2.2$  &  $   2.2$  &  $  -0.1$  &  $  0.95$  &  $  0.95$  &  $  0.95$  &  $  0.95$  &  $  0.95$  &  $  0.95$\\
$    $  &  $0.2$  &  $   2.4$  &  $  -0.1$  &  $  -2.4$  &  $   0.1$  &  $  -2.1$  &  $  -1.0$  &  $  0.96$  &  $  0.95$  &  $  0.96$  &  $  0.93$  &  $  0.93$  &  $  0.95$\\
\cline{1-14}  
$1000$  &  $0.8$  &  $   1.1$  &  $   0.9$  &  $  -0.9$  &  $   0.4$  &  $   0.5$  &  $  -1.5$  &  $  0.94$  &  $  0.95$  &  $  0.95$  &  $  0.96$  &  $  0.95$  &  $  0.95$\\
$    $  &  $0.4$  &  $  -1.2$  &  $  -0.2$  &  $  -0.8$  &  $  -0.7$  &  $   0.6$  &  $   0.3$  &  $  0.95$  &  $  0.94$  &  $  0.95$  &  $  0.94$  &  $  0.94$  &  $  0.95$\\
$    $  &  $0.2$  &  $   0.9$  &  $   1.7$  &  $   2.7$  &  $   0.2$  &  $   1.9$  &  $  -0.7$  &  $  0.97$  &  $  0.95$  &  $  0.95$  &  $  0.94$  &  $  0.95$  &  $  0.94$
\\
    \bottomrule \\[-10pt]
    \tabfootnote{1}{$\tGamma = \tGamma^{(3)}_{i,j}$ and $\tY_{i,j}=X'_i\tbeta^{(3)}_j + \tGamma^{(3)}_{i,j}$.}{14}{.9}
    \end{tabularx}
    }
\end{table}

    \begin{table}[ht]
    \tabfnsymbol \centering
    \captionx[Most significant movies in different tests of ML data]{Hypothesis testing results for each movie. Movies\footnotemark[1]\footnotemark[2] with the top 10 smallest p-values in each test are shown in order. }\label{tab: ml hypo test most sig}\vspace{0pt}
    {\scriptsize
    \hspace*{-25pt}\begin{tabular}[t]{rll}
    \multicolumn{1}{c}{$ID$} & \multicolumn{1}{c}{movie name} & \multicolumn{1}{c}{$p$-value}\\
    \hline
    &\multicolumn{1}{c}{ $H_0: \beta_{j1}=0$}&\\
    \hline
 1592   &   Air Bud                &  $ <10^{-10}$\\
 203    &   Thanks for Everythin   &  $  <10^{-9}$\\
 1979   &   Friday the 13th Part   &  $  <10^{-8}$\\
 1978   &   Friday the 13th Part   &  $  <10^{-8}$\\
 15     &   Cutthroat Island       &  $  <10^{-8}$\\
 1088   &   Dirty Dancing          &  $  <10^{-8}$\\
 1707   &   Home Alone 3           &  $  <10^{-7}$\\
 2106   &   Swing Kids             &  $  <10^{-6}$\\
 506    &   Orlando                &  $  <10^{-6}$\\
 1201   &   The Good, The Bad an   &  $  <10^{-6}$\\
\hline
    \\
    \tabfootnote{1}{Names are trimmed to 20 characters.}{3}{.35}\\[2em]
    \tabfootnote{2}{Only movies with more than 100 ratings are considered.}{3}{.35}
    \end{tabular}\hspace*{5pt} \begin{tabular}[t]{rll}
    \multicolumn{1}{c}{$ID$} & \multicolumn{1}{c}{movie name} & \multicolumn{1}{c}{$p$-value}\\
    \hline
    &\multicolumn{1}{c}{ $H_0: \beta_{j2}=0$}&\\
    \hline
 3690   &   Porky's Revenge        &  $ <10^{-14}$\\
 3689   &   Porky's II: The Next   &  $ <10^{-12}$\\
 2382   &   Police Academy 5: As   &  $ <10^{-11}$\\
 520    &   Robin Hood: Men in T   &  $ <10^{-11}$\\
 3099   &   Shampoo                &  $ <10^{-10}$\\
 1978   &   Friday the 13th Part   &  $ <10^{-10}$\\
 2379   &   Police Academy 2: Th   &  $  <10^{-9}$\\
 2550   &   The Haunting           &  $  <10^{-9}$\\
 3017   &   Creepshow 2            &  $  <10^{-9}$\\
 1996   &   Poltergeist III        &  $  <10^{-9}$\\
\hline
    &\multicolumn{1}{c}{ $H_0: \beta_{j3}=0$}&\\
    \hline
 2173   &   The Navigator: A Med   &  $ <10^{-20}$\\
 1205   &   The Transformers: Th   &  $ <10^{-19}$\\
 2550   &   The Haunting           &  $ <10^{-17}$\\
 1150   &   The Return of Martin   &  $ <10^{-16}$\\
 1238   &   Local Hero             &  $ <10^{-16}$\\
 3341   &   Born Yesterday         &  $ <10^{-16}$\\
 1592   &   Air Bud                &  $ <10^{-15}$\\
 3690   &   Porky's Revenge        &  $ <10^{-14}$\\
 3099   &   Shampoo                &  $ <10^{-14}$\\
 1985   &   Halloween 4: The Ret   &  $ <10^{-13}$\\
\hline
    &\multicolumn{1}{c}{ $H_0: \beta_{j4}=0$}&\\
    \hline
 65     &   Bio-Dome               &  $ <10^{-31}$\\
 358    &   Higher Learning        &  $ <10^{-28}$\\
 2860   &   Blue Streak            &  $ <10^{-16}$\\
 2119   &   Maximum Overdrive      &  $ <10^{-16}$\\
 2195   &   Dirty Work             &  $ <10^{-13}$\\
 3146   &   Deuce Bigalow: Male    &  $ <10^{-12}$\\
 2296   &   A Night at the Roxbu   &  $ <10^{-10}$\\
 3177   &   Next Friday            &  $ <10^{-10}$\\
 104    &   Happy Gilmore          &  $ <10^{-10}$\\
 2907   &   Superstar              &  $  <10^{-9}$\\
\hline
    \end{tabular}\hspace*{5pt} \begin{tabular}[t]{rll}
    \multicolumn{1}{c}{$ID$} & \multicolumn{1}{c}{movie name} & \multicolumn{1}{c}{$p$-value}\\
    \hline
    &\multicolumn{1}{c}{ $H_0: \beta_{j2}-\beta_{j3}=0$}&\\
    \hline
 2170   &   Wrongfully Accused     &  $ <10^{-14}$\\
 1667   &   Mad City               &  $ <10^{-13}$\\
 1592   &   Air Bud                &  $ <10^{-11}$\\
 502    &   The Next Karate Kid    &  $ <10^{-10}$\\
 1984   &   Halloween III: Seaso   &  $ <10^{-10}$\\
 12     &   Dracula: Dead and Lo   &  $ <10^{-10}$\\
 2162   &   The NeverEnding Stor   &  $  <10^{-9}$\\
 1085   &   The Old Man and the    &  $  <10^{-9}$\\
 1205   &   The Transformers: Th   &  $  <10^{-9}$\\
 3873   &   Cat Ballou             &  $  <10^{-9}$\\
\hline
    &\multicolumn{1}{c}{ $H_0: \beta_{j2}-\beta_{j4}=0$}&\\
    \hline
 358    &   Higher Learning        &  $ <10^{-36}$\\
 2195   &   Dirty Work             &  $ <10^{-35}$\\
 2907   &   Superstar              &  $ <10^{-19}$\\
 65     &   Bio-Dome               &  $ <10^{-19}$\\
 3901   &   Duets                  &  $ <10^{-14}$\\
 3177   &   Next Friday            &  $ <10^{-13}$\\
 2860   &   Blue Streak            &  $ <10^{-12}$\\
 2606   &   Idle Hands             &  $ <10^{-11}$\\
 3225   &   Down to You            &  $ <10^{-11}$\\
 258    &   A Kid in King Arthur   &  $ <10^{-10}$\\
\hline
    &\multicolumn{1}{c}{ $H_0: \beta_{j3}-\beta_{j4}=0$}&\\
    \hline
 2195   &   Dirty Work            &  $ <10^{-27}$\\
 65     &   Bio-Dome              &  $ <10^{-22}$\\
 358    &   Higher Learning       &  $ <10^{-19}$\\
 3177   &   Next Friday           &  $ <10^{-18}$\\
 2907   &   Superstar             &  $ <10^{-16}$\\
 2860   &   Blue Streak           &  $ <10^{-12}$\\
 818    &   A Very Brady Sequel   &  $ <10^{-10}$\\
 3627   &   Carnival of Souls     &  $  <10^{-9}$\\
 122    &   Boomerang             &  $  <10^{-9}$\\
 3693   &   The Toxic Avenger     &  $  <10^{-9}$\\
\hline
    \end{tabular}\vspace{-10pt}
    }
    \end{table}

\clearpage

{\color{black}
\subsection{Comparison of estimation performance with other methods}\label{sec: add num mao comp}
Under a similar model as (\ref{eq: model theta}), \cite{Mao2019} proposed a penalized estimation method to estimate the target matrix $\Theta$ using the Frobenius-norm and nuclear-norm regularization and derived the convergence rate of the resulting penalized estimator. Although they focus on the estimation and our work, on the other hand, provides statistical inference for the matrix completion problem, we also compare the estimation performance of their penalized method and our iterative LS method. We generate the data using same the model given in Table 2 of Section 6 in \cite{Mao2019} with the true rank $r=10$ for the latent row-rank matrix $\Gamma$.

For comparison, we report the empirical root mean squared error (RMSE) for the estimators of $\beta$, $\Gamma$ and $\Theta$ given in model (\ref{eq: model Y}), respectively, as well as the test errors (Test Err.) calculated based on $100$ simulation replicates. The formula for calculating the RMSE and the test error are given in Section 6 of \cite{Mao2019}. In addition, we also report the average value of the absolute bias (Absolute bias) for the estimates of $\beta$, $\Gamma$ and $\Theta$ and the average of the rank estimates based on $100$ simulation replicates. For any estimator $\tilde{\theta}$ of an unknown parameter $\theta$, the absolute bias is calculated by:
Absolute bias($\theta $)$=$$|\text{nsim}^{-1}\sum_{s=1}^{\text{nsim}}\tilde{\theta}^{s}-\theta|$, where $\tilde{\theta}^{s}$ is the estimate of $\theta$ in the $s^{\text{th}}$ simulation replication, and nsim is the number of simulation replicates.
We estimate the rank using the proposed $\eIC$ method with $h(n,m)$ given in (\ref{eq: pen h})  with $C_h=0.15$ and $\delta_h=0.1$. \cite{Mao2019} considered four estimators which have similar performance. For computational convenience, we only report the numerical results of their estimate SVT-1 and denote this method by MCW.

Table \ref{tab: mao comp} summarizes the simulation results for the estimates of unknown parameters obtained by our method (Ours) and the method (MCW) in \cite{Mao2019} under different sample sizes. We see that our method yields slightly smaller RMSEs and test errors than the MCW method at $n,m=800$. However, as the sample size increases to $n,m=1000$ and $1200$, the RMSEs and test errors obtained by our method decrease faster than those values obtained by the MCW method. For example, when the sample size is increased from 800 to 1200, the test error is reduced from 0.4281 to 0.2290 for our method, and it is decreased from 0.4293 to 0.3300 for the MCW method. Our method yields smaller RMSEs than the MCW method when the sample size is large, because our iterative LS estimator is asymptotically unbiased. However, the penalized estimators in general have a nonnegligible bias that can contribute to the RMSE value. From Table \ref{tab: mao comp}, we see that our iterative LS estimator has smaller absolute bias values than the penalized estimator for all cases. Table \ref{tab: mao comp} also shows that our proposed $\eIC$ method can accurately estimate the true rank.
\begin{table}[ht]
\tabfnsymbol \centering
    \captionx[Empirical RMSE and Bias]{
    Empirical root mean squared errors (RMSEs), test errors (Test Err.), average value of absolute biases (Absolute bias), and estimated ranks by the two methods under different sample sizes.}
    \label{tab: mao comp}\centering
    \vspace{12pt}{\scriptsize
    \begin{tabularx}{\textwidth}{llCCCCCCCC}\toprule
        \multicolumn{2}{l}{} & \multicolumn{3}{c}{RMSE} & \multicolumn{3}{c}{Absolute bias} & & \\
        \cmidrule(lr){3-5}\cmidrule(lr){6-8}
 $n, m$   &   Method   &   $\beta$   &   $\Gamma$   &     $\Theta$   &   $\beta$   &   $\Gamma$   &     $\Theta$   &   Test Err.  &      Rank \\
\cline{1-10}
 800    &   Ours   &  $ 0.5149$  &  $ 2.6585$  &  $ 3.4951$  &  $ 0.0413$  &  $ 0.2661$  &  $ 0.3220$  &  $ 0.4304$  &  $  10.00$\\
        &   MCW    &  $ 0.5175$  &  $ 2.6463$  &  $ 3.4997$  &  $ 0.2337$  &  $ 1.9308$  &  $ 2.1944$  &  $ 0.4260$  &  $  51.35$\\
\cline{1-10}
 1000   &   Ours   &  $ 0.4346$  &  $ 2.1471$  &  $ 2.8708$  &  $ 0.0350$  &  $ 0.1885$  &  $ 0.2432$  &  $ 0.2977$  &  $  10.00$\\
        &   MCW    &  $ 0.4669$  &  $ 2.4631$  &  $ 3.2073$  &  $ 0.1955$  &  $ 1.7770$  &  $ 1.9807$  &  $ 0.3716$  &  $  62.06$\\
\cline{1-10}
 1200   &   Ours   &  $ 0.3871$  &  $ 1.8960$  &  $ 2.5598$  &  $ 0.0310$  &  $ 0.1589$  &  $ 0.2102$  &  $ 0.2291$  &  $  10.00$\\
        &   MCW    &  $ 0.4346$  &  $ 2.3627$  &  $ 3.0537$  &  $ 0.1691$  &  $ 1.6987$  &  $ 1.8612$  &  $ 0.3280$  &  $  70.85$
\\
        \bottomrule
    \end{tabularx}
    }
\end{table}

\bibliographystyle{Chicago}
\spacingset{1.1}
{
\small
\bibliography{BibLib}
}

\end{document}